\documentclass[12pt]{article}
\usepackage[mathscr]{eucal}
\usepackage{epsfig,amsfonts}
\usepackage{amsmath}
\usepackage{amsthm,amssymb}
\usepackage{graphicx}
\usepackage{hhline}
\usepackage{mathabx}
\usepackage{relsize}
\usepackage{cite}
\usepackage{psfrag}
\usepackage{comment}
\usepackage{mathrsfs} 
\usepackage{hyperref}
\usepackage{bm}
\usepackage{array}
\usepackage{tikz}

\makeatletter
\@addtoreset{equation}{section}
\makeatother

\def\bea{\begin{eqnarray}}
\def\eea{\end{eqnarray}}
\def\be{\begin{equation}}
\def\ee{\end{equation}}

\def\be{\begin{equation}}
\def\ee{\end{equation}}
\def\bdm{\begin{displaymath}}
\def\edm{\end{displaymath}}
\def\bea{\begin{eqnarray}}
\def\eea{\end{eqnarray}}

\def\s{\sigma}

\def\ri{{\rm i}}
\def\XXint#1#2#3{{\setbox0=\hbox{$#1{#2#3}{\int}$}
    \vcenter{\hbox{$#2#3$}}\kern-.5\wd0}}

\newcommand{\p}{\partial}

\newcommand{\rd}{\mbox{d}}
\newcommand{\re}{\mbox{e}}

\DeclareMathAlphabet{\mathpzc}{OT1}{pzc}{m}{it}

\newcommand{\vp}{\varphi}
\newcommand{\ti}[1]{_{\underline{#1}}}
\newcommand{\g}{\mathfrak{g}}
\newcommand{\dd}{\text{d}}
\def\res{\mathop{\text{res}\,}}
\newcommand{\Ad}{\text{Ad}}
\newcommand{\Lc}{\mathcal{L}}
\newcommand{\Jc}{\mathcal{J}}

\newcommand{\Ac}{\mathcal{A}}
\newcommand{\Pc}{\mathcal{P}}

\newcommand{\Hc}{\mathcal{H}}
\newcommand{\Cc}{\mathcal{C}}
\newcommand{\Id}{\text{Id}}

\newcommand{\C}{\mathbb{C}}
\newcommand{\red}{\text{red}}
\newcommand{\Q}{\mathcal{Q}}
\newcommand{\ze}{\zeta}
\newcommand{\Eh}{\widehat{E}}
\newcommand{\gh}{\widehat{\mathfrak{g}}}
\newcommand{\Zc}{\mathcal{Z}}
\newcommand{\zt}{\widetilde{z}}
\newcommand{\Gt}{\widetilde{\Gamma}}
\newcommand{\vpt}{\widetilde{\varphi}}
\newcommand{\Fc}{\mathcal{F}}

\newcommand{\diag}{\text{diag}}
\newcommand{\Ct}{\mathcal{C}_{G}}
\newcommand{\cL}{{(\rm L)}}
\newcommand{\cR}{{(\rm R)}}
\newcommand{\cLR}{{(\rm L/R)}}
\newcommand{\cM}{{(\rm M)}}
\newcommand{\dI}{{\rm I}}
\newcommand{\dII}{{\rm II}}
\newcommand{\dIII}{{\rm III}}

\newcommand{\gf}{\equiv}
\newcommand{\Wc}{\mathcal{W}}
\newcommand{\Bc}{\mathcal{B}}
\newcommand{\tv}{\begin{array}{c} \otimes \\[-0.35cm] , \end{array}}
\newcommand{\Mt}{M'}
\newcommand{\Nt}{N'}
\newcommand{\Nc}{\widehat{N}}
\newcommand{\Nct}{\widehat{N}'}
\newcommand{\Mc}{\widehat{M}}
\newcommand{\Mct}{\widehat{M}'}
\newcommand{\tp}{\null^t}
\newcommand{\Tp}{\null^{\mathsf{T}}}
\newcommand{\ad}{\text{ad}}
\newcommand{\Dc}{\mathcal{D}}

\newcommand{\Ch}{\mathcal{L}\mathsf{C}_2}
\newcommand{\Oh}{\widehat{O}}
\newcommand{\Cht}[1]{\mathcal{L}\mathsf{C}_{\underline{#1}}}
\newcommand{\Th}{\mathsf{T}_2}
\newcommand{\Kc}{\mathcal{K}}
\newcommand{\Hd}{\mathsf{H}_{2}}
\newcommand{\Kuv}{K_{\scriptscriptstyle{\rm UV}}}
\newcommand{\f}[3]{f_{#1#2}^{{\color{white}#1#2}#3}}
\newcommand{\fu}[3]{f^{#1#2}_{{\color{white}#1#2}#3}}
\newcommand{\Ah}{\widehat{\mathcal{A}}}
\newcommand{\Jq}{\mathsf{J}}
\newcommand{\reg}{\text{reg}}
\newcommand{\Gq}{\Gamma^{({\rm qt})}}
\newcommand{\qt}{({\rm qt})}
\newcommand{\vpq}{\varphi^{({\rm qt})}}
\newcommand{\Zh}{\widehat{\mathcal{Z}}}
\newcommand{\Wh}{\widehat{\mathcal{W}}}
\newcommand{\Wq}{\mathsf{W}}

\newcommand{\Uq}{\mathsf{U}}
\newcommand{\Sc}{\mathcal{S}}
\newcommand{\Qq}{\mathsf{Q}}
\newcommand{\Sq}{\mathsf{S}}
\newcommand{\Aq}{\mathsf{A}}
\newcommand{\hv}{h^{\!\vee}}

\newcommand{\nor}[1]{%
  :\mathrel{\mspace{2mu}#1\mspace{2mu}}:%
}

\makeatletter
\newsavebox{\@brx}
\newcommand{\llangle}[1][]{\savebox{\@brx}{\(\m@th{#1\langle}\)}%
  \mathopen{\copy\@brx\kern-0.5\wd\@brx\usebox{\@brx}}}
\newcommand{\rrangle}[1][]{\savebox{\@brx}{\(\m@th{#1\rangle}\)}%
  \mathclose{\copy\@brx\kern-0.5\wd\@brx\usebox{\@brx}}}
\makeatother

\usetikzlibrary{arrows,matrix,snakes}
\usetikzlibrary{decorations.markings,shapes.geometric,shapes.misc}

\usetikzlibrary{patterns}
\usetikzlibrary{shapes.misc,decorations.pathmorphing,calc}
\tikzset{
  branch cut/.style={
    decorate,decoration=snake,
    to path={
      (\tikztostart) -- (\tikztotarget) \tikztonodes
    }
  }
}
\tikzset{
    partial ellipse/.style args={#1:#2:#3}{
        insert path={+ (#1:#3) arc (#1:#2:#3)}
    }
}

\tikzset{
  pics/torus/.style n args={3}{
    code = {
      \providecolor{pgffillcolor}{rgb}{1,1,1}
      \begin{scope}[
          yscale=cos(#3),
          outer torus/.style = {draw,line width/.expanded={\the\dimexpr2\pgflinewidth+#2*2},line join=round},
          inner torus/.style = {draw=pgffillcolor,line width={#2*2}}
        ]
        \draw[outer torus] circle(#1);\draw[inner torus] circle(#1);
        \draw[outer torus] (180:#1) arc (180:360:#1);\draw[inner torus,line cap=round] (180:#1) arc (180:360:#1);
      \end{scope}
    }
  }
}

\begin{document}

\begin{flushright}
DESY-22-068 \vspace{17pt}
\end{flushright}

\begin{center}
\begin{Large}

\textbf{Integrable sigma models at RG fixed points: \\ quantisation as affine Gaudin models}

\end{Large}

\vspace{1.3cm}
\begin{large}

{Gleb A. Kotousov$^{1}$, Sylvain Lacroix$^{2}$ and J\"org Teschner$^{1,3}$}

\end{large}

\vspace{1.cm}
$^1$ \textit{DESY Theory Group, Notkestrasse 85, 20607 Hamburg, Germany} \vspace{.4cm}\\
$^2$ \textit{Institute for Theoretical Studies, ETH Z\"urich, Clausiusstrasse 47, 8092 Z\"urich, Switzerland} \vspace{.4cm}\\
$^3$ \textit{Department of Mathematics, University of Hamburg,
Bundesstrasse 55,\\ 20146 Hamburg, Germany}\vspace{.55cm}

Emails: {\ttfamily \href{mailto:gleb.kotousov@desy.de}{gleb.kotousov@desy.de}, \href{mailto:sylvain.lacroix@eth-its.ethz.ch}{sylvain.lacroix@eth-its.ethz.ch},\\ \href{mailto:joerg.teschner@desy.de}{joerg.teschner@desy.de} \vspace{0.7cm} }

\end{center}

\begin{center}
\centerline{\bf Abstract} \vspace{.6cm}

\parbox{14cm}{%
The goal of this paper is to make first steps towards the quantisation of integrable non-linear sigma models using the formalism of affine Gaudin models, by approaching these theories through their conformal limits. We focus mostly on the example of the Klim\v{c}\'{i}k model, which is a two-parameter deformation of the Principal Chiral Model on a Lie group $G$. We show that the UV fixed point of this theory is described classically by two decoupled chiral affine Gaudin models, encoding its left- and right-moving degrees of freedom, and give a detailed analysis of the chiral and integrable structures of these models. Their quantisation is then explored within the framework of Feigin and Frenkel. We study the quantum local integrals of motion using the formalism of quantised affine Gaudin models and show agreement of the first two integrals with known results in the literature for $G={\rm SU}(2)$. Evidence is given for the existence of a monodromy matrix satisfying the Yang-Baxter algebra for this model, thus paving the way for the quantisation of the non-local integrals of motion. We conclude with various perspectives, including on generalisations of this program to a larger class of integrable sigma models and applications of the ODE/IQFT correspondence to the description of their quantum spectrum.
}
\end{center}

\vfill

\newpage

\setcounter{tocdepth}{2}
\tableofcontents

\newpage

\section{Introduction}

\paragraph{Integrable NLSMs and their quantisation.} The integrable Non-Linear Sigma Models (NLSMs) in two space-time
dimensions have attracted a lot of attention since several decades. 
One motivation originated from the observation that the ${\rm O}(N)$ symmetric sigma models exhibit
quantum phenomena resembling certain features of Quantum Chromodynamics, \textit{e.g.} asymptotic freedom in the ultraviolet (UV) and a dynamically generated mass gap \cite{Polyakov:1975rr}.
Several subsequent developments have renewed the interest in NLSMs.
A major source of motivation for their study
is the perturbative approach to string theory. 
Moreover, important aspects of the AdS/CFT correspondence can be approached by studying NLSMs with targets  being Anti-de Sitter spaces (see, \textit{e.g.} the review~\cite{Beisert:2010jr}).
NLSMs furthermore offer opportunities to study
profound quantum duality phenomena
such as  the weak/strong coupling dualities  between certain
sigma models and theories with Toda like 
interactions\cite{Fateev:1995ht,Fateev:1996ea}.\footnote{%
The dual theory for the so-called sausage model \cite{FaOZ}, which is a one parameter
integrable deformation of the ${\rm O}(3)$ sigma model, was originally discovered
by Al. Zamolodchikov.}\\

Some NLSMs are known to be integrable on the classical level, 
and a lot of evidence has been accumulated for the conjecture that
the classical integrability of many of these models
survives quantisation. However, a first principle approach to the quantisation 
of the NLSMs demonstrating their integrability does not seem to be available yet. For other classes of integrable QFTs, a lot of progress was achieved by applying the Quantum Inverse Scattering Method \cite{Faddeev:1979gh}. It is based on the construction of quantum monodromy matrices depending on a spectral parameter and satisfying exchange relations of the $R$-matrix form. These relations ensure the commutativity of traces of the monodromy matrices and thus the existence of an infinite number of Integrals of Motion (IMs) in the quantum theory. Moreover, they allow the application of powerful techniques such as the Algebraic Bethe Ansatz and lattice regularisation for the diagonalisation of these IMs.

In the case of NLSMs, the Quantum Inverse Scattering Method faces
serious difficulties 
known as the non-ultralocality problem, which arise from the presence of ambiguities in the Poisson bracket of the monodromy matrices \cite{Maillet:1985fn,Maillet:1985ek}. From this point of view, it 
is thus currently not well-understood 
how to quantise these matrices and how to use them 
for the definition of IMs in the quantised integrable NLSMs. \\

For a few NLSMs there nevertheless exist precise conjectures
about the finite volume spectra which 
have been carefully checked in great detail. 
One of the existing approaches to the study of integrable NLSMs uses the S-matrix bootstrap as a starting point. 
Based on certain plausible assumptions, one can use the factorised S-matrix
as the basis for the Thermodynamic Bethe Ansatz giving 
information on the spectrum of some NLSMs in finite volume, 
see e.g. \cite{GKV} for an application of this method to the ${\rm O}(4)$ sigma model.
By combining this approach with
other methods, significant results have been obtained 
in particular for
a deformation of the ${\rm O}(3)$ sigma model known as the sausage model
\cite{FaOZ,Fateev:2017mug,Bazhanov:2017nzh}. A sophisticated blend of several techniques 
has led to the exact results presented in \cite{Lukyanov:2013wra,Bazhanov:2013cua,Bazhanov:2014joa} on the finite-volume 
spectrum of a NLSM first introduced by Fateev
 in \cite{Fateev:1996ea} 
and which can be regarded as
a two parameter deformation of the ${\rm SU}(2)$ Principal 
Chiral Model.\footnote{This NLSM was introduced to formulate 
a  duality conjecture between a model with three bosonic fields 
and a sigma model. In case of doubt we will always refer to the NLSM when using 
the terminology Fateev model.}
These results strongly support the hope that an integrable 
quantisation  exists for all points in the  parameter spaces
of these NLSMs.

\paragraph{Integrable structure of conformal limits.} In the search for possible starting points for a more direct 
approach to the quantisation of integrable NLSMs
one should distinguish between different limits in their 
parameter spaces. The most natural one may seem to be  the classical limit, which is the starting point of many of the traditional approaches to the quantisation of NLSMs. As mentioned above, one may generically expect to encounter severe difficulties  related to the non-ultralocality problem in these approaches.

A second type of limit sends the couplings of the model to their values 
at an RG fixed point. In fortunate circumstances, the resulting theory
becomes a non-trivial CFT.  One can then exploit the constraints of
conformal symmetry and/or integrability to get detailed information on these
conformal sigma models. This offers an alternative approach to the 
investigation of non-conformal NLSMs based on the expansion 
around the conformal loci. One might hope that the interplay of 
conformal and integrable structures could provide the basis of
powerful techniques for the investigation of quantum integrable 
sigma models.
\bigskip

The investigation of integrable QFTs with the help of the integrable structures
of their conformal limits was initiated in \cite{Zamolodchikov:1989hfa}. One of the hallmarks of two-dimensional conformal invariance is the presence of chiral degrees of freedom in the theory. For instance, a CFT in general possesses local chiral currents $\lbrace \mathsf {W}^{(\rm L)}_s \rbrace$ and $\lbrace \mathsf {W}^{(\rm R)}_s \rbrace$, labelled in particular by an integer $s$ characterising their respective Lorentz spins $+s$ and $-s$. The notations $(\rm L)$ and $(\rm R)$ used here indicate that $\mathsf {W}^{(\rm L)}_s$ and $\mathsf {W}^{(\rm R)}_s$ are respectively left- and right-moving fields, \textit{i.e.} they satisfy
\be
\partial_- \mathsf {W}^{(\rm L)}_s=0\,,\qquad\qquad \partial_+ \mathsf {W}^{(\rm R)}_s=0\qquad\qquad
\big(\partial_\pm=\tfrac{1}{2}(\partial_t\pm\partial_x)\big)\, .
\ee
These sets of chiral fields contain in particular the spin $\pm 2$ components 
$\mathsf {W}^{(\rm L)}_2$ and $\mathsf {W}^{(\rm R)}_2$ of the stress-energy tensor.
Moreover, the modes of the chiral currents $\lbrace \mathsf {W}^{(\rm L)}_s \rbrace$ and $\lbrace \mathsf {W}^{(\rm R)}_s \rbrace$ generate 
two independent operator algebras, called $\mathcal{W}$-algebras, whose commutation relations are encoded in the 
Operator Product Expansions of the currents. They define the algebra of extended conformal symmetry of the theory. In particular, the space of states can be represented as a direct sum of irreducible representations of these $\mathcal{W}$-algebras:
\be\label{ioas981892as}
{\cal H}=\bigoplus_{V^{(\rm L)},V^{(\rm R)}} V^{(\rm L)}\otimes V^{(\rm R)}\, .
\ee

Apart from the conformal structure, the CFT can also have an integrable structure
described by infinite  families of  operators acting 
on each representation $V^{(\rm L)}$ or $V^{(\rm R)}$ that mutually commute. These operators can take the form of both local and non-local IMs, built from the chiral fields of the CFT~\cite{Bazhanov:1994ft,Bazhanov:1996dr,Bazhanov:1998dq}.
For instance, the local IMs are given by
\be\label{oias9812sa}
\mathsf{Q}^{(\rm L)}_s : \, V^{(\rm L)}\mapsto V^{(\rm L)}\,,\qquad\qquad\qquad\qquad
\mathsf{Q}^{(\rm L)}_s=\int_0^{2\pi}\rd x\;\mathsf{F}^{(\rm L)}_{s+1}\,,
\ee
where $\mathsf{F}^{(\rm L)}_{s+1}$ are specific tensor densities built from the left-moving currents $\mathsf {W}^{(\rm L)}_r$ and their derivatives, while the limits in the integration reflect the fact that we are considering the theory on
the space-time cylinder with the space co-ordinate taking values in the standard segment $x\in[0,2\pi)$.  
A similar set $\mathsf{Q}^{(\rm R)}_s$ acts on the second tensor factors
in \eqref{ioas981892as}. 
There is evidence
that these IMs can admit deformations describing 
the integrable structure  of certain massive perturbations 
of these CFTs, suggesting that the integrable 
structure of CFTs can offer a useful starting point for the investigation of
the integrable structures of their massive deformations.

The construction and simultaneous diagonalization of commuting families of local and non-local IMs in a CFT is a well defined  mathematical problem in the representation theory of the chiral algebras, which is however hard to solve in general. In various cases, there nonetheless exists a systematic construction of local IMs, which characterises them as the observables commuting with specific operators called screening charges \cite{Fateev:1987vh,Fateev:1987zh,Feigin:1991qm,Feigin:1993sb}. Moreover, non-local IMs can be built in certain examples with the use of monodromy matrices satisfying quantum exchange relations~\cite{Bazhanov:1994ft,Bazhanov:1996dr,Bazhanov:1998dq}.

\paragraph{ODE/IQFT correspondence.} One of the most powerful instruments for the mathematical description 
of the spectrum of the integrable structures of 
CFTs is known as the ODE/IQFT correspondence \cite{Dorey:1998pt,Bazhanov:1998wj,Bazhanov:2003ni}. This conjectural duality reduces the
problem of calculating the eigenvalues of the local and non-local IMs in an integrable CFT to 
a problem in classical analysis.
The simultaneous eigenstates in a representation $V^{({\rm L,R})}$ are in one-to-one correspondence with a class of Ordinary Differential Equations specified by their monodromy properties.
The corresponding eigenvalues of the IMs are encoded in the connection coefficients of these ODEs. 
A prototype of this scenario is the quantum KdV integrable structure, where
the ODE is the Schr\"{o}dinger equation with ``monster'' potential \cite{Dorey:1998pt,Bazhanov:1998wj,Bazhanov:2003ni} (see also \cite{Conti:2020zft,Conti:2021xzr} for recent results). Other  examples have been found and studied extensively in the literature,
see the review papers \cite{Dorey:2007zx,Dorey:2019ngq} and 
references therein.

There are a few cases where there exists highly non-trivial evidence for the applicability of 
the ODE/IQFT correspondence to the conformal limits of NLSMs, namely the UV fixed-points of the sausage
\cite{Bazhanov:2016glt,Bazhanov:2017xky,Bazhanov:2017nzh} and Fateev 
\cite{Bazhanov:2013cua,Lukyanov:2013wra,Bazhanov:2014joa} models in different regimes of the
parameters.
A particularly striking result of those papers is the existence of a deformation of this description 
replacing the ODEs by closely related PDEs encoding the 
finite volume spectrum of the  sigma models even away from 
the conformal limit. This indicates that such a description is not 
just a peculiarity of the conformal points.
However, for more general NLSMs no such descriptions
seem to be available yet. 

\paragraph{Affine Gaudin models.} The ubiquitous  appearance of Lie theoretic structures 
in the theory of integrable models may offer guidance in the search for a more 
general framework. 
In \cite{Feigin:2007mr}, Feigin and Frenkel proposed an
ambitious conjecture stating that the ODE/IQFT correspondence
 covers a large multi-parametric class of integrable structures defined in terms of an affine Lie algebra $\widehat{\mathfrak{g}}$.
 The theories in this class are called affine Gaudin models (AGMs)\footnote{The classical version of these models also appeared in the independent work \cite{Levin:2001nm}.}. 
In the simplest setting, such a model is built out of $N$ Kac-Moody currents whose modes satisfy the commutation relations of $N$ independent copies of the 
affine algebra $\widehat{\mathfrak{g}}$ at levels $\lbrace k_j \rbrace_{j=1}^N$. These copies are attached to points 
$\lbrace z_j \rbrace_{j=1}^N \subset \mathbb{CP}^1$ on the Riemann sphere, called punctures,
which are considered to be parameters of the theory. It was conjectured in \cite{Feigin:2007mr} that AGMs possess an integrable structure, formed by local and non-local commuting IMs built from the Kac-Moody currents and depending on the punctures $z_j$. Various motivations and checks of this conjecture were presented in that paper. These included the construction of some of these commuting operators, \textit{e.g.} the first non-local IMs extracted from the expansion of the monodromy matrix as well as local charges quadratic in the Kac-Moody currents. The search for a more systematic construction of higher-degree local IMs was further developed in \cite{Lacroix:2018fhf,Lacroix:2018itd}, where more precise conjectures on the form of these operators were formulated and additional examples were explicitly constructed\footnote{Let us also mention that an analog of the Feigin-Frenkel homomorphism for double loop algebras was recently introduced in~\cite{Young:2020ffh}, partly motivated by potential applications to the study of AGMs.}.

It is conjectured in  \cite{Feigin:2007mr} that the eigenvalues of the IMs of AGMs are encoded in
certain differential operators called affine opers, which are associated with the Langlands 
dual affine Lie algebra ${}^{\rm L}\widehat{\mathfrak{g}}$. One of the main proposals of \cite{Feigin:2007mr} is then that these differential operators correspond to the ODEs appearing in the ODE/IQFT correspondence. This 
conjecture, which relies on an analogy with the case of Gaudin models based on finite-dimensional simple Lie algebras \cite{Gaudin76,Gaudin_book83,Feigin:1994in}, has first been checked when the Gaudin model becomes equivalent
to quantum KdV \cite{Feigin:2007mr}. Further checks were performed in \cite{Frenkel:2016gxg} for higher-rank KdV-type integrable structures, building on the extraction of so-called $Q\widetilde{Q}$ systems from affine opers in refs. \cite{Masoero:2015lga,Masoero:2015rcz} as well as the explicit analysis of the 
$\widehat{\mathfrak{sl}}(3)$ case from ref.\cite{Bazhanov:2001xm}.
The results of the previously mentioned works 
\cite{Lacroix:2018fhf,Lacroix:2018itd} where the local IMs are systematically studied and
refs.\cite{Gaiotto:2020fdr,Gaiotto:2020dhf} on the Kondo problems also give evidence
for the conjecture. In the recent paper \cite{Kotousov:2021vih}, restricting to the
$\widehat{\mathfrak{sl}}(2)$ case,
a generalisation of the AGM     is introduced and ODEs describing 
the spectra of these models are proposed, supported by numerical and analytic verifications.\\

The proposal of ref. \cite{Feigin:2007mr} has furthermore inspired the identification of a large class of relativistic two-dimensional integrable field theories containing most of the known classically integrable NLSMs \cite{Vicedo:2017cge,Delduc:2019bcl,Lacroix:2019xeh}, driven in part by the appealing possibility \cite{Vicedo:2017cge} of deriving a (massive) ODE/IQFT correspondence for these models.
The definition 
of these field theories, in the following called relativistic AGMs\footnote{In the main text, we will more precisely refer to these theories as relativistic realisations of AGMs. By a slight abuse of language, and to avoid the use of heavier terminology in the introduction, we will simply use the name relativistic AGM here.}, uses the classical version (or generalisations thereof) of the set-up described in the work of Feigin and Frenkel as a starting point. While the kinematical basis for the definition of relativistic AGMs is essentially identical to the classical limit of the set-up used in \cite{Feigin:2007mr}, one needs to introduce
additional structure 
in order to define models having an interpretation as relativistic field theories on a two-dimensional space-time. Key ingredients  are, in particular,
 the Hamiltonian and momentum observables, 
or equivalently the light-cone Hamiltonians.
The construction of relativistic AGMs 
described in \cite{Vicedo:2017cge,Delduc:2019bcl,Lacroix:2019xeh} furthermore uses
realisations of the
Poisson current algebras appearing in the framework of Feigin and Frenkel
in terms of the Cauchy data on equal time slices of the fundamental fields of NLSMs.

\paragraph{Quantisation of relativistic AGMs in their conformal limits.} Our intention is to make first steps towards a systematic approach to the quantization of integrable NLSMs through their interpretation as relativistic AGMs. Ambitious questions one may wish to address in this context are, in general:
\begin{itemize}
\item[a)] Does there exist a quantisation of the relativistic two-dimensional AGMs preserving their integrability?
\item[b)] Is it possible to apply the ODE/IQFT correspondence from \cite{Feigin:2007mr},
or some massive generalisation thereof, in order to describe the spectra of these quantised models?
\end{itemize}
Loosely speaking, our approach amounts to entering the space of quantised relativistic AGMs through the side entrance provided by their conformal limits, for which we will gather some evidences that the answers to the questions a) and b) above are positive. We will therefore mostly restrict attention to the loci in the 
parameter spaces of these models representing fixed points 
of the RG flow. Our expectation is that, at these fixed points, the classical integrable structure of the relativistic theory
decouples into two separate integrable structures built from
left- and right-moving degrees of freedom respectively. These integrable structures of left- and right-movers are moreover described by two decoupled AGMs, which we will refer to as chiral AGMs. We expect the latter to be quantisable in the sense of Feigin and Frenkel \cite{Feigin:2007mr}, which would thus provide a general framework for the study of conformal limits of integrable NLSMs and their ODE/IQFT correspondence. In contrast, for the relativistic AGMs outside of the conformal loci, new qualitative features seem to be required to develop an extension of this scheme, including the treatment of renormalisation of the parameters and of the dynamical generation of a mass. However, we think that understanding the 
quantisation of the chiral AGMs can offer a useful starting point for the study of these massive cases, for instance through
conformal perturbation techniques.

\paragraph{A case study.}

In this paper we will mostly restrict our attention to the conformal limit of a particular example of relativistic AGM
introduced in  \cite{Klimcik:2008eq,Klimcik:2014bta} that we will refer to as the Klim\v{c}\'{i}k model 
(it is sometimes also called the bi-Yang-Baxter sigma model in the literature). 
It is a two-parameter deformation of the Principal Chiral Model on a Lie group $G$, which coincides with the Fateev model in the case $G={\rm SU}(2)$, as shown in \cite{Hoare:2014pna}. This theory provides a good example where the general program sketched above may be realised, as it exhibits all of its key features while being reasonably simple and well-studied in the literature. In particular, the quantisation of this model has been explored in the works 
\cite{Fateev:1996ea,Bazhanov:2013cua,Bazhanov:2014joa,Bazhanov:2018xzh,Lukyanov:2013wra} 
for the case $G={\rm SU}(2)$, hence providing useful checks of the results obtained from the AGM formalism and a fruitful exchange of ideas and techniques between these frameworks.

\bigskip

The relation between the descriptions of the integrable structure of this model provided
by the formulation as AGM to the more traditional formulations used in the existing literature
turns out to be non-obvious. The basic reason is the use of a gauge symmetry in the 
formulation of the model of interest as AGM, which is a general feature shared by a large
class of AGMs. As observed in many other contexts, the use of a gauge symmetry may
allow one to exhibit some aspects of a quantum theory more transparently, 
while other aspects can be more easily
understood in gauge-fixed representations. A similar situation will be found here. 
While there is a powerful algorithm for the construction of local conserved 
charges in quantised AGMs, it will turn out that a specific gauge fixing is helpful
to employ the results of \cite{Bazhanov:2018xzh} giving evidence for existence
of  Yang-Baxter algebraic structure simplifying the construction 
of non-local conserved quantities considerably. In order to utilise the 
AGMs as a general framework for the quantisation of certain classes of integrable
sigma models one may expect that 
it will be important to understand the relation between
gauged and gauge-fixed formulations in some detail. This is going to 
be an important part of what is done in this work.

\paragraph{Outline.} Let us finally give the outline of the paper, 
summarising its main results along the way.

In Section \ref{Sec:Klim}, we review the definition and properties of the classical Klim\v{c}\'{i}k NLSM on an arbitrary Lie group $G$~\cite{Klimcik:2008eq,Klimcik:2014bta}. This model
can be formulated as a theory on $G\!\times\! G$ with a gauge symmetry by the diagonal subgroup $G_{\text{diag}}$, following \cite{Hoare:2014oua}. In particular, we describe the Lax connection underlying the integrability of this model and discuss both its Lagrangian and Hamiltonian formulations.

The first part of Section \ref{sec2} is devoted to the construction of classical relativistic AGMs and mainly consists of a review of \cite{Vicedo:2017cge,Delduc:2019bcl,Lacroix:2019xeh}. In particular, we recall the definition of the Hamiltonian, the integrable structure and the Lax connection of these theories. The second part concerns the interpretation of the gauged formulation of the Klim\v{c}\'{i}k NLSM on $G \! \times\! G / G_{\text{diag}}$ as a relativistic AGM with 4 punctures, based on \cite{Vicedo:2017cge,Delduc:2015xdm}.

In Section \ref{sec31}, we initiate the analysis of the conformal limit of the Klim\v{c}\'{i}k model on the group ${\rm SU}(2)$, mostly following \cite{Bazhanov:2018xzh}. We start with a review of the RG flow of this theory \cite{Sfetsos:2015nya}.
It turns out that in the IR the model becomes strongly coupled, while in the UV the one-loop RG flow equations
possess a non-trivial fixed point. 
A detailed description of the target space in the UV limit of the sigma model is given. 
Moreover, we discuss the same limit for the Lax connection  and describe how its classical integrable structure splits into two decoupled ones, built from left- and right-moving fields respectively. In particular, we present several convenient choices of gauge for the Lax connection, which are expressed in terms of  non-local chiral fields of the model in the UV limit, such as the (classical) parafermions.

We continue the classical analysis of this UV limit in Section \ref{sec3899891}, now focusing on the local conformal and integrable structures. For instance, we describe the $\mathcal{W}$-algebra underlying the theory, \textit{i.e.} the Poisson algebra formed by its left- or right-moving local currents, as well as a non-trivial relevant subalgebra thereof, called the corner-brane $\mathcal{W}$-algebra. Moreover, we discuss the classical local IMs of the model, which take the form of integrals of particular local chiral densities in this corner-brane $\mathcal{W}$-algebra. Finally, we discuss the characterisation of these $\mathcal{W}$-algebras and local IMs as  (Poisson) commutants of specific sets of classical screening charges. Although the classical local IMs of the model turn out to be very simple, this section will serve as a useful warm-up for a subsequent description of quantum local IMs, which are quite less trivial but admit a similar characterisation in terms of quantum screening charges.

The goal of Section \ref{sec555389i} is to discuss the conformal limit of the Klim\v{c}\'{i}k NLSM from the point of view of AGMs. In particular, we argue in this section that the relativistic AGM with  4 punctures underlying the non-conformal model splits in the UV limit into two decoupled AGMs with 3 punctures, describing respectively the left- and right-moving degrees of freedom of the theory. We formalise this phenomenon by introducing the notion of chiral AGMs, in contrast with the relativistic AGMs describing the non-conformal theories. Moreover, we show that various well-chosen gauge-fixings of these chiral AGMs give rise to the different chiral Lax connections described in Section \ref{sec31} in terms of parafermionic or screening currents. In particular, the use of AGMs in this context allows the construction of generalisations of these current algebras for higher-rank Lie groups $G$. Finally, we discuss the $\mathcal{W}$-algebra and local IMs of the model from the point of view of AGMs. Remarkably, we show that this formalism provides a systematic and efficient construction of chiral local currents of the model. The latter form a $\widehat{\mathfrak{g}} \oplus \widehat{\mathfrak{g}} / \widehat{\mathfrak{g}}$ coset $\mathcal{W}$-algebra, which coincides with the corner-brane algebra of Section \ref{sec3899891} in the case $G={\rm SU}(2)$.\\

Starting from Section \ref{sec666}, we explore the quantisation of the UV fixed-point of the Klim\v{c}\'{i}k model. For instance, Section \ref{sec666} discusses the quantisation of the non-local IMs of this theory. As mentioned earlier, this question is made difficult by the problem of non-ultralocality. However, an observation 
made in \cite{Bazhanov:2018xzh} indicates that the existence of a quantum Yang-Baxter algebra can be consistent with having non-ultralocal Poisson brackets on the classical level. More precisely, evidence was given that there exists a quantum
monodromy matrix satisfying quantum exchange relations, whose classical limit is 
a path ordered exponent over a Lax matrix with a
non-ultralocal Poisson bracket. It was shown that this Lax matrix coincides with the one 
of the Fateev NLSM (${\rm SU}(2)$ Klim\v{c}\'{i}k model) in the conformal limit and in a certain gauge. This means that non-ultralocality may emerge as the result of subtleties in the classical limits of fairly conventional algebraic structures underlying the integrability of many integrable models. We review the results of ref.\cite{Bazhanov:2018xzh} in Section \ref{sec666} and explain
how it would generalise to higher-rank groups.

Section \ref{Sec:QuantLoc} concerns the study of the quantum $\Wc$-algebra and local IMs for the case $G={\rm SU}(2)$, which can be characterised by their commutativity with  quantum screening charges that are described in the previous section. In particular, we use this approach to describe the first few currents of the $\Wc$-algebra, its corner-brane subalgebra and the first two local IMs in the quantum Fateev integrable structure.

Section \ref{Sec:QAGM} is devoted to the quantisation of the chiral AGMs underlying the UV limit of the Klim\v{c}\'{i}k model, in the framework of Feigin and Frenkel \cite{Feigin:2007mr}. After recalling some generalities about quantised AGMs, we discuss the quantum $\widehat{\mathfrak{g}} \oplus \widehat{\mathfrak{g}} / \widehat{\mathfrak{g}}$ coset $\mathcal{W}$-algebra appearing in the theory, focusing mostly on the ${\rm SU}(2)$ case for simplicity. Moreover, the (conjectural) construction of an infinite hierarchy of quantum local IMs in this $\mathcal{W}$-algebra is sketched following the proposal of \cite{Lacroix:2018fhf,Lacroix:2018itd} and the first higher-spin charge in this hierarchy is explicitly constructed in the case $G={\rm SU}(2)$. In particular, it is shown that this local IM coincides with the one found in Section \ref{Sec:QuantLoc} from screening charges, providing strong support to the proposal that the AGM under consideration describes the quantum Fateev integrable structure.

The paper finishes with the discussion of some perspectives in Section \ref{Sec:Discussion}. The main part  
concerns the ODE/IQFT correspondence for the conformal model studied in the paper. In particular, it is observed that the differential equation on the ODE side of the ODE/IQFT
correspondence conjectured in \cite{Lukyanov:2013wra,Bazhanov:2013cua} for the Fateev model in the conformal limit
is correctly predicted by the Feigin-Frenkel scheme. This result suggests that this scheme can be used to derive the ODE/IQFT correspondence for the Klim\v{c}\'{i}k model on an arbitrary Lie group. In another part of the discussion section \ref{Sec:Discussion}, we argue that the quantisation program illustrated here on the example of the Klim\v{c}\'{i}k model should apply to a quite more general class of integrable NLSMs / relativistic AGMs, including for instance models possessing more punctures. In particular, we provide evidence that the conformal limits of these more general models should also be described by decoupled chiral AGMs corresponding to left- and right-moving degrees of freedom. This discussion builds on general results in the literature concerning the light-cone structure of relativistic AGMs \cite{Delduc:2019bcl,Lacroix:2019xeh} and their RG flow~\cite{Delduc:2020vxy,Hassler:2020xyj}. 

\section{The Klim\v{c}\'{i}k model}
\label{Sec:Klim}

\subsection{Lagrangian description}
In the work \cite{Klimcik:2008eq}, Klim\v{c}\'{i}k introduced a model  (sometimes referred to as the bi-Yang-Baxter
sigma model) which is a two parameter deformation of the 
Principal Chiral Model on a semi-simple Lie group $G$. He further proved that this field theory is classically integrable in ref. \cite{Klimcik:2014bta}. As was shown in \cite{Hoare:2014pna}, when $G$ is taken to be
${\rm SU}(2)$, the NLSM coincides with the Fateev sigma model 
introduced in ref. \cite{Fateev:1996ea}.
A lot of the technical aspects of the paper concern the ${\rm SU}(2)$ case.
However, the Lagrangian and Hamiltonian formulations of the theory as well as its Lax connection can be described in a uniform way that does not depend
on the particular choice of the Lie group $G$.

\paragraph{Standard formulation.} The configuration space of the Klim\v{c}\'{i}k model is the set of $G$-valued fields
$g(t,x)$. The deformation is achieved by means of the so-called
Yang-Baxter operator $\hat{R}$, 
a linear operator acting on the Lie algebra $\mathfrak{g}$ that must obey
the modified classical Yang-Baxter equation. The latter has many solutions,
and different choices of $\hat{R}$ generically lead to different classical field theories.
We use the standard Drinfel'd-Jimbo $R$-matrix, which
 is  defined via the Cartan-Weyl decomposition
of the complexified Lie algebra
$\mathfrak{g}^{{\mathbb C}}=\mathfrak{n}_+\oplus\mathfrak{h}\oplus\mathfrak{n}_-$ (here
$\mathfrak{h}$ and $\mathfrak{n}_\pm$ stand for the Cartan and nilpotent subalgebras,
respectively). Namely,
\be\label{Rmat1a}
\hat{R}({\tt h})=0\,,\qquad\qquad \hat{R}({\tt e}_\pm)=\mp\ri\,{\tt e}_\pm
\qquad\qquad \qquad\qquad
\forall\ {\tt h}\in\mathfrak{h}\,, \ {\tt e}_\pm\in\mathfrak{n}_\pm\, .
\ee
The action of the Klim\v{c}\'{i}k model, with deformation parameters $\alpha$ and $\beta$ as in ref. \cite{Klimcik:2014bta} and an overall constant prefactor $K$, reads
\be\label{oaso8912as}
{\cal A}=4K\int \rd t\rd x\
\Big\langle g^{-1}\,\partial_+ g\,,\,
\big(1-\alpha \hat{R}_g-\beta \hat{R}\big)^{-1}\,\big(g^{-1}\partial_- g\big)\Big\rangle\  .
\ee
By the angular brackets $\langle\cdot,\cdot\rangle$ we mean the non-degenerate symmetric form for
the semi-simple Lie algebra $\mathfrak{g}$, while\footnote{%
To fix the normalization of the symmetric form,
if ${\tt t}_a$ is a basis for $\mathfrak{g}$ with
$
[{\tt t}_a,{\tt t}_b]= {f_{ab}}^c\,{\tt t}_c\nonumber
$
then in terms of the structure constants,
\be\label{oias891sa}
\langle {\tt t}_a,{\tt t}_b\rangle=-\tfrac{1}{2h^\vee}\,{f_{ac}}^d\,{f_{bd}}^c\, .\nonumber
\ee
The extra factor, with $h^\vee$ being the dual Coxeter number 
for the Lie algebra $\mathfrak{g}$,
 has been chosen so that the
angular brackets coincide with the negative of the trace over the fundamental representation.\label{ft1}}
\be
{\hat { R}}_{g}=\Ad_g^{-1}\circ\hat{R}\circ \Ad_g \,,\qquad\qquad
\qquad \Ad_g({\tt a})=g\,{\tt a}\,g^{-1}\ \ \ \ 
\forall {\tt a}\in \mathfrak{g}\, .
\ee
Also, the light-cone derivatives $\partial_\pm=\frac{1}{2}\,(\partial_t\pm\partial_x)$ are being used.
\bigskip

In the discussion of reality conditions for the target space background, 
the parameters $\alpha$ and $\beta$ controlling the deformation
are sometimes assumed to be real numbers. 
Contrary to this, we will work in the regime with
$\alpha,\,\beta$ pure imaginary, such that
\be\label{ioas891a}
\alpha=\ri\varepsilon_1\,,\qquad\qquad \beta=\ri\varepsilon_2\,,
\qquad\qquad \qquad \varepsilon_1,\varepsilon_2\in\mathbb{R}  .
\ee
The action \eqref{oaso8912as} describes the propagation of a (1+1)-dimensional field
in a Riemannian manifold with non-trivial torsion potential. 
In the parametric domain \eqref{ioas891a} the metric tensor is
 real, while the components of the B-field are pure imaginary. Note that for
the ${\rm SU}(2)$ case the B-field term in the Lagrangian is a total
derivative and can be ignored.

\paragraph{Gauged formulation.} Important for this work is another formulation of the Klim\v{c}\'{i}k model that was 
 first considered in ref. \cite{Hoare:2014oua}. 
The theory is treated
as a coset NLSM on $G\times G$ gauged by the diagonal subgroup.
Practically speaking, one writes $g=g_1\,g_2^{-1}$ and takes both
$g_1$ and $g_2$ to be independent fields. This effectively doubles
the degrees of freedom. However, there is a redundancy in the
description since the gauge transformation
\be\label{ioas8912sa}
g_1\mapsto g_1\,h\,,\qquad\qquad 
g_2\mapsto g_2\,h\,,
\ee
with $h$ an arbitrary $G$-valued  field,
leaves the action \eqref{oaso8912as} unchanged. 
The gauge dependent fields
$g_1$ and $g_2$ provide one with extra freedom that allows the deformations
governed by $\varepsilon_1$ and $\varepsilon_2$ to be treated in a symmetric way. 
For instance via a simple re-arrangement of terms,
the classical action \eqref{oaso8912as} with
$g=g_1g_2^{-1}$ can be expressed as 
\be\label{oaso8912asaaa}
{\cal A}=4K\int \rd t \, \rd x\
\Big\langle \big(g^{-1}_1\,\partial_+ g_1-g^{-1}_2\,\partial_+ g_2\big)\,,\,
\big(1-\ri\varepsilon_1 \hat{R}_{g_1}-\ri\varepsilon_2 \hat{R}_{g_2}\big)^{-1}\,\big(g^{-1}_1\partial_- g_1-
g^{-1}_2\partial_- g_2\big)\Big\rangle\  .
\ee
This ``gauged'' formulation of the  theory
 turns out to be important for making connection with classical affine Gaudin models.

\paragraph{Lax connection.} The existence of an infinite number of conserved quantities for a classical field theory follows from the 
zero-curvature representation for the Euler-Lagrange equations:
\be\label{ZCR1}
\big[\partial_x + {\cal L}_x(z),\,\partial_t + {\cal L}_t(z) \big]=0\,,
\ee
where $z\in \mathbb{CP}^1$ denotes the spectral parameter. To write the explicit form of the Lax connection of the Klim\v{c}\'{i}k model, it is useful to introduce the currents
\bea\label{currentsK}
{\cal I}_\pm=
2\, \big(1\pm\ri \varepsilon_1\,{\hat { R}}_{ g_1}
\pm\ri \varepsilon_2\,{\hat  { R}_{g_2}}\,\big)^{-1}\, 
\big(\,g^{-1}_1\partial_\pm  g_1-g^{-1}_2\partial_\pm  g_2\,\big)\ .
\eea
One then has
\be\label{ioas8921oi}
{\cal L}_t(z)={\cal L}_+(z)+{\cal L}_-(z)\,,\qquad \qquad {\cal L}_x(z)={\cal L}_+(z)-{\cal L}_-(z)
\ee
with
\be\label{siqjkhdsjh2}
{\cal L}_\pm(z) =
\tfrac{1}{4}\,\big(\,\varepsilon_2^2-\varepsilon_1^2
\pm\ri\varepsilon_2\,\hat{{R}}_{g_2}
\mp\ri\varepsilon_1\,\hat{{R}}_{g_1}\,+\xi\,z^{\pm1}\,\big)\, {\cal I}_\pm
+\tfrac{1}{2}\,\big(\,g_1^{-1}\,\partial_\pm g_1+g_2^{-1}\partial_\pm g_2\,\big)\,,
\ee
 while the constant $\xi$ is defined through the relation
\be\label{oias9812oias}
\xi^2=\big(\,1-(\varepsilon_1+\varepsilon_2)^2\,\big)\,
\big(\,1-(\varepsilon_1-\varepsilon_2)^2\,\big)\, .
\ee

The Lax connection is not invariant with respect to the gauge transformation \eqref{ioas8912sa}. Instead,
one can easily check that
\be\label{oiasi98oias}
\p_\mu + \Lc_\mu(z)\longmapsto h^{-1}\big(\p_\mu+\Lc_\mu(z)\big)h\, .
\ee
The above formula requires some comment. 
In any integrable field theory the Lax connection is not unique, but is defined 
up to a conjugation $
{\bm{\Omega}}^{-1}\,(\p_\mu + \Lc_\mu)\,
{\bm{\Omega}}$.
The specific integrable field theory that we are considering here additionally possesses a gauge symmetry, \textit{i.e.} a local transformation of the fields that preserves the action of the model. Formula \eqref{oiasi98oias}
shows that  the gauge symmetry acts on the Lax connection via conjugation.
The equations of motion of the model are equivalent to the zero-curvature representation \eqref{ZCR1}. The latter
is unchanged under conjugation of $\p_\mu + \Lc_\mu$ so that the
equations of motion are gauge invariant, as expected. 
Moreover, it is a standard result in integrable field theories that the integrals of motion extracted from $\bm{\Omega}^{-1}(\p_\mu+\Lc_\mu)\,\bm{\Omega}$ do not depend on $\bm{\Omega}$, 
thus ensuring in the present case that they 
are also gauge invariant.\bigskip

The zero-curvature representation for the equations of motion of the Klim\v{c}\'{i}k model was 
initially found in ref. \cite{Klimcik:2014bta}.
The corresponding Lax connection is related to the one considered above by
\be\label{ioas912sa}
{{\cal L}}_\mu^{({\rm inv})}(z)=g_2\,{\cal L}_\mu(z)\,g_2^{-1} - \partial_\mu g_2\,g_2^{-1}\,.
\ee
It is straightforward to check that ${{\cal L}}_\mu^{({\rm inv})}$ only depends on the combination
$g=g_1g_2^{-1}$ and hence is gauge invariant. Note that
for $G={\rm SU}(2)$ the Lax connection turns out to be equivalent to the one obtained
in the earlier paper \cite{Lukyanov:2012zt} specialised to the Fateev model
 (see Appendix B  of ref. \cite{Bazhanov:2018xzh} for details).

\subsection{Hamiltonian description}
\label{Sec:KlimHam}

\paragraph{Phase space.} The phase space of a sigma model defined on a group manifold $G$ consists of canonical fields valued in the cotangent bundle $T^\ast G$. The cotangent space $T^\ast_p G$ at any point $p\in G$ can be naturally identified, through the translation by $p^{-1}$, with the cotangent space $T^\ast_{\Id} G$ at the identity. The latter is by construction the dual $\g^\ast$ of the Lie algebra $\g$. This then defines a natural trivialisation $T^\ast G \simeq G \times \g^\ast$ of the cotangent bundle. Moreover, the non-degenerate bilinear form $\langle\cdot,\cdot\rangle$ provides us with a canonical identification of $\g$ and $\g^\ast$, yielding an isomorphism $T^\ast G \simeq G \times \g$. The phase space of canonical fields on $T^\ast G$ can thus be described by a $G$-valued field
$g(x)$ as well as a $\mathfrak{g}$-valued field $X(x)$, which plays the r\^{o}le
of the canonical momentum. Since 
the Klim\v{c}\'{i}k model is being viewed here as a coset NLSM on $G\times G$,
two copies $(g_j(x), X_j(x))$, $j\in\lbrace 1,2 \rbrace$, of these fields are required. 
Using the standard tensorial notations, the canonical Poisson bracket can be expressed as
\begin{subequations}\label{oias9821aaaas}
\bea
\big\{g_j(x)\begin{array}{c} \otimes \\[-0.35cm] , \end{array} 
{ g}_\ell(y)\big\}&=&0\,,\\[0.0cm]
\label{oiasi98xc98cxpoA}
\big\{X_j(x)\begin{array}{c} \otimes \\[-0.35cm] , \end{array} 
g_\ell(y)\big\}&=&\delta_{j\ell}\,\big(1\otimes g_\ell(x)\big)\,{\mathsf C}_{2}\,\delta(x-y)\,,\\[0.1cm]
\label{oiasi98xc98cxpoB}
\big\{{ X}_j(x)\begin{array}{c} \otimes \\[-0.35cm] , \end{array} 
X_\ell(y)\big\}&=& \delta_{j\ell}\, \big[\big(1\otimes X(x)\big),{\mathsf C}_{2}\big]\,\delta(x-y)\,,
\eea
\end{subequations}
with ${\mathsf C}_{2}$ being the quadratic Casimir.
 We take a moment to explain
how the invariant notations look like in a particular  basis
$\{{\tt t}_a\}\in\mathfrak{g}$, with
$[{\tt t}_a,{\tt t}_b]={f_{ab}}^c\,{\tt t}_c$.
We denote by $\{{\tt t}^a\}\in\mathfrak{g}$ the dual basis with respect to the non-degenerate bilinear form $\langle \cdot,\cdot \rangle$, defined by $\langle {\tt t}^a, {\tt t}_b \rangle = \delta^a_{\;\,b}$. The quadratic Casimir then reads
\be
{\mathsf C}_{2}={\tt t}^a\otimes {\tt t}_a\, .
\ee
By expanding the field $X_j$ into components, $X_j=X_{j,a}\,{\tt t}^a$, it is straightforward to check that
 eqs.\;\eqref{oiasi98xc98cxpoA} and \eqref{oiasi98xc98cxpoB} are equivalent to
$$
\big\{X_{j,a}(x),
g_\ell(y)\big\}=\delta_{j\ell}\,g_\ell(x)\ {\tt t}_a\ \delta(x-y)\qquad{\rm and} \qquad
\big\{X_{j,a}(x),
X_{\ell,b}(y)\big\}= \delta_{j\ell}\,{f_{ab}}^c\,X_{j,c}(x)\,\delta(x-y)\,,$$ 
respectively.
\bigskip

\paragraph{Constraint and gauge symmetry.} It follows from the action \eqref{oaso8912asaaa} that the canonical momenta are given by
\be\label{X}
X_1 \approx K\,\big({\cal I}_++{\cal I}_-\big) \qquad \text{ and } \qquad X_2 \approx -K\,\big({\cal I}_++{\cal I}_-\big)\,,
\ee
where the notation $\approx$ will be explained below and ${\cal I}_\pm$ are the currents defined in eq.\;\eqref{currentsK}.
The fact that $X_1$ and $X_2$ turn out to be linearly dependent
arises from the redundancy in the description of
the Klim\v{c}\'{i}k model as a gauged NLSM. In the Hamiltonian
picture, following the framework laid out by Dirac \cite{dirac1964lectures},
 this translates to the system being subject to a constraint
\be\label{ConstKlim}
{\cal C}(x) = X_1(x)+X_2(x)\approx 0\,.
\ee
Although the physically relevant configurations of the canonical fields are only the ones 
lying on the surface ${\cal C}(x) \approx 0$, it will also be useful to consider the full phase space. This is the reason behind the introduction of the symbol ``$\approx$''  which denotes weak equalities, true
only if the constraint is imposed. In contrast, we keep the standard equals sign ``$=$'' for strong equalities, that hold in the full phase space. According to Dirac's formalism, the dynamical quantity ${\cal C}(x)$ plays the r\^{o}le of the generator of the
gauge symmetry \eqref{ioas8912sa}. In particular, it satisfies the first-class condition
\be
\big\{{\cal C}(x)\begin{array}{c} \otimes \\[-0.35cm] , \end{array} 
{\cal C}(y)\big\}\approx 0\, ,
\ee
which ensures that the constraint ${\cal C} (x) \approx 0$ is stable under gauge transformations. Note that under the gauge symmetry, while $g_j$ are transformed by  right multiplication
with $h$, the canonical momenta are conjugated as
\be\label{Eq:GaugeX}
X_1\mapsto h^{-1}\, X_1 \, h\,,\qquad\qquad X_2\mapsto h^{-1}\, X_2 \, h\, .
\ee
The ``physical'' phase space is spanned by the pair of gauge invariant  fields $(g,X)$ with
\be\label{oiasio9812oia}
g=g_1 g_2^{-1}\,,\qquad\qquad\qquad  X=g_2 X_1 g_2^{-1}\, .
\ee
By considering only the gauge invariant dynamical quantities, one recovers
the Hamiltonian description of the Klim\v{c}\'{i}k model as a sigma model on $G$,
with the pair $(g,X)$ encoding the coordinates and canonically conjugate momenta of the
theory. 

\paragraph{Hamiltonian and dynamics.} A direct computation shows that the Hamiltonian associated with the action \eqref{oaso8912asaaa} is given by
\be\label{HKlim}
{\cal H} = \frac{K}{2}\int\rd x\,\Big(
\big\langle {\cal I}_+,{\cal I}_+\big\rangle+\big\langle {\cal I}_-,{\cal I}_-\big\rangle\Big) \,,
\ee
with ${\cal I}_\pm$ being the currents defined in eq.\;\eqref{currentsK}. 
In order to interpret ${\cal H}$ as the generator of the dynamics in the Hamiltonian framework, 
it needs to be expressed in terms of the canonical fields $(g_j,X_j)$. We first do this for the currents ${\cal I}_\pm$. This  step is actually not unique 
since there is the freedom of adding a term which vanishes under the constraint ${\cal C} \approx 0$. 
In order to simplify the comparison with the affine Gaudin model, we will make the following choice:
\be\label{IHam}
{\cal I}_\pm=\frac{1}{4K}\,\Big(
\big(1\,\mp\,2\ri\varepsilon_1\hat{R}_{g_1}\big)\,X_1\,-\,\big(1\,\mp\,2\ri\varepsilon_2\hat{R}_{g_2}\big)\,X_2\,+\,
\big(\varepsilon_1^2-\varepsilon_2^2\big)\big(X_1+X_2\big)
\Big)\vphantom{\bigg)} \pm g_1^{-1}\partial_x g_1 \mp g_2^{-1}\partial_x g_2.
\ee
In particular, one easily checks that this expression is compatible with eq.\;\eqref{X}, \textit{i.e.} is such that $K({\cal I}_+ + {\cal I}_-) \approx X_1 \approx -X_2$ weakly. Reinserting it in \eqref{HKlim} gives ${\cal H}$  in terms of the canonical fields $(g_j,X_j)$. 
As above, one always has the freedom of adding to this expression an arbitrary term proportional to the constraint. 
According to Dirac's formalism, in order to define the dynamics of the model on the full phase space, we introduce the 
``total Hamiltonian''
\be\label{Eq:HamKlim}
{\cal H}_{\rm T}={\cal H} + \int\rd x\,
\big\langle \mu(x)\,,\,{\cal C}(x)\big\rangle,
\ee
where the $\g$-valued field $\mu(x)$ takes the place of a Lagrange multiplier (in particular, a different choice in the canonical expression of the currents ${\cal I}_\pm$ made in
\eqref{IHam} would result in an equivalent Hamiltonian through a redefinition of $\mu$). In the equations of motion, $\partial_t{\cal O}\approx \{ {\cal H}_{\rm T},{\cal O}\}$,
for any gauge invariant observable $\mu$ drops out simply because gauge invariant quantities Poisson commute
with ${\cal C}(x)$.
 However, say, for $g_2(x)$:
\be
g_2^{-1}\partial_t g_2 \approx  \mu
-\tfrac{1}{4}\,\big(1-\varepsilon_1^2+\varepsilon_2^2\,\big)\,\big({\cal I}_+-{\cal I}_-\big)-
\tfrac{\ri}{2}\,\varepsilon_2\,\hat{R}_{g_2}\,\big({\cal I}_++{\cal I}_-\big),
\ee
so that $\mu$ reflects the ambiguity in the choice of  dynamics for the unphysical degrees of freedom.

\paragraph{Lax matrix.} For future reference, we give the formula for the $x$-component of the Lax connection
\eqref{ioas8921oi},\,\eqref{siqjkhdsjh2},
expressed via $g_j$ and $X_j$:
\be\label{ioas8912op}
{\cal L}_x(z)={\cal B}+\frac{\xi}{4}\,\bigg({\cal I}_+\,z - \frac{{\cal I}_-}{z}\bigg)+\lambda(z)\,{\cal C}\, .
\ee
Here, the fields ${\cal I_\pm}$ are given by~\eqref{IHam},
\be
{\cal B}=\frac{1-\varepsilon_1^2+\varepsilon_2^2}{2}\ \bigg(g_1^{-1}\partial_x g_1 - \frac{\ri\varepsilon_1}{2K}\,
\hat{R}_{g_1}X_1\bigg)+\frac{1+\varepsilon_1^2-\varepsilon_2^2}{2}\ 
\bigg(g_2^{-1}\partial_x g_2-\frac{\ri\varepsilon_2}{2K}\,
\hat{R}_{g_2} X_2\bigg)\,,
\ee
the constant $\xi$ is the same as in eq.\;\eqref{oias9812oias}, while $\lambda(z)$ is an arbitrary function of the spectral parameter that takes into account the freedom of adding a term proportional to the constraint in the expression of $\Lc_x(z)$ (in particular, this term has no effect on the zero-curvature representation, which is to be considered weakly).
\bigskip

As was already mentioned, the Lax matrix ${\cal L}_x(z)$ in eq.\;\eqref{ioas8912op} depends on the gauge. In order to obtain a connection
 in terms of the physical fields $(g,X)$ defined in eq.\;\eqref{oiasio9812oia}, one can for instance consider ${\Lc}_x^{({\rm inv})}(z) = g_2 \Lc_x(z)
 g_2^{-1} - \p_x g_2 g_2^{-1}$ (as in eq.\;\eqref{ioas912sa} in the Lagrangian setup). 
This Lax matrix is gauge invariant and can be weakly expressed in terms of $g=g_1 g_2^{-1}$ and 
$X=g_2 X_1 g_2^{-1} \approx - g_2 X_2 g_2^{-1}$ only. It coincides with the one from the work \cite{Klimcik:2014bta}, 
seen in the Hamiltonian framework.
\bigskip

An alternative way of relating $\Lc_x(z)$ with a Lax matrix in terms of the physical fields $(g,X)$ is to impose a gauge fixing condition. The latter should be a second-class constraint, i.e. its Poisson bracket with ${\cal C}$ must define an invertible kernel. A possible choice is
$g_2\equiv \bm{\Omega}(g,X)$ with $\bm{\Omega}$ being some function of $g$ and $X$. Taking into account this gauge fixing and the initial constraint, 
 all the canonical fields $(g_j,X_j)$ may be expressed in terms of $g$ and $X$, namely,
\be
g_1 \equiv g \,\bm{\Omega}(g,X), \qquad g_2 \equiv 
\bm{\Omega}(g,x), \qquad X_1 \equiv - X_2 \equiv \bm{\Omega}(g,X)^{-1} \,X\, \bm{\Omega}(g,X).
\ee
With the  gauge fixing imposed, we then find
\be
{\cal L}_x(z) \equiv \bm{\Omega}^{-1}\,{{\cal L}}_x^{({\rm inv})}(z)\,\bm{\Omega}+\bm{\Omega}^{-1}\partial_x\bm{\Omega}\,,
\ee
where ${{\cal L}}_x^{({\rm inv})}$ is as above. In particular, the condition  $g_2 \equiv \Id$ leads to
 ${\cal L}_x \equiv {{\cal L}}_x^{({\rm inv})}$.

\section{Classical AGMs and their relativistic realisations\label{sec2}}

In this section, we review the construction of classical affine Gaudin models (AGMs) and their relativistic realisations, following the initial construction~\cite{Feigin:2007mr} of Feigin and Frenkel and the subsequent works~\cite{Vicedo:2017cge,Delduc:2019bcl,Lacroix:2019xeh}. Moreover, we will explain how the Klim\v{c}\'{i}k model considered before can be interpreted as such a realisation.

\subsection{Classical affine Gaudin models}
\label{Sec:AGM}

\paragraph{Punctures, levels and twist function.} The main data defining an AGM is the choice of $N$ points $z_1,\ldots,z_N\in\mathbb{CP}^1$ on the Riemann sphere, which we will refer to as \textit{punctures}, and of $N$ non-zero complex numbers $\ell_1,\ldots,\ell_N\in\C^\times$, which we will refer to as \textit{levels}. We encode these parameters in a rational function of the spectral parameter $z$, called the \textit{twist function}:
\be\label{Eq:Twist}
\vp(z) = \sum_{r=1}^N \frac{\ell_r}{z-z_r}.
\ee
For simplicity, we will suppose here that our choice of coordinate $z$ on $\mathbb{CP}^1$ is such that all the punctures $z_r$ are in the finite complex plane $\C$ (see the last paragraph of this subsection for more details on the behaviour of AGMs under a change of coordinate on $\mathbb{CP}^1$).

\paragraph{Kac-Moody currents and Gaudin Lax matrix.} The remaining data entering the definition of an AGM is the choice of a simple Lie algebra $\g$. We attach to each puncture $z_r$ a \textit{$\g^{\C}$-Kac-Moody current} with level $\ell_r$, \textit{i.e.} a $\g^{\C}$-valued field $\Jc_r(x)$ satisfying the Poisson bracket
\be\label{Eq:PbKM}
\big\{ \Jc_r(x)\begin{array}{c} \otimes \\[-0.35cm] , \end{array}
\Jc_s(y)\big\} =  \delta_{rs}\,\Bigl( \big[\big(1\otimes \Jc_r(x)\big),{\mathsf C}_{2}\big]\,\delta(x-y) - \ell_r\, {\mathsf C}_{2}\,\p_x\delta(x-y) \Bigr).
\ee
In this section, we will consider fields on the circle $\mathbb{S}^1$ and thus take the spatial coordinate $x$ to be a periodic variable from $0$ to $2\pi$. In the notations introduced below eq.\;\eqref{oiasi98xc98cxpoB}, expanding $\Jc_r(x) = \Jc_{r,a}(x) \,{\tt t}^a$ in the dual basis $\{{\tt t}^a\}\in\mathfrak{g}$, the above bracket translates to
\be\label{Eq:PbJa}
\big\{ \Jc_{r,a}(x) , \Jc_{s,b}(y)\big\} =  \delta_{rs}\,\Bigl( {f_{ab}}^c\,\Jc_{r,c}(x)\,\delta(x-y) - \ell_r\, \langle {\tt t}_a, {\tt t}_b \rangle\,\p_x\delta(x-y) \Bigr). 
\ee
We will denote by $\Ac$ the Poisson algebra generated by the fields $\Jc_{r,a}(x)$ and equipped with the above Poisson bracket, which defines the unreduced algebra of observables of the AGM (see the next paragraph for the explanation of the adjective unreduced). The Kac-Moody currents $\Jc_r(x)$ are encoded in the \textit{Gaudin Lax matrix}
\be\label{Eq:Gaudin}
\Gamma(z,x) = \sum_{r=1}^N \frac{\Jc_r(x)}{z-z_r}.
\ee

\paragraph{Constraint and gauge symmetry.} In the previous paragraphs, we defined $\vp(z)$ and $\Gamma(z,x)$ in such a way that the 1-forms $\vp(z)\dd z$ and $\Gamma(z,x)\dd z$ have singularities at the punctures $z_r\in\C$. To ensure that the finite points $z_r\in\C$ are the only punctures of the AGM, we need to impose that the 1-forms $\vp(z)\dd z$ and $\Gamma(z,x)\dd z$ are regular at $z=\infty$. For $\vp(z)\dd z$, this is equivalent to supposing that the levels $\ell_r$ satisfy the condition
\be\label{Eq:SumL}
\sum_{r=1}^N \ell_r = 0.
\ee
Similarly, in order to ensure the regularity of $\Gamma(z,x)\dd z$ at infinity, we impose
\begin{equation}
\Ct(x) = \sum_{r=1}^N \Jc_r(x) \approx 0.
\end{equation}
More precisely, we interpret this equation as a \textit{constraint} in the Poisson algebra $\Ac$, following the standard formalism of Dirac~\cite{dirac1964lectures}. In particular, the symbol ``$\approx$'' used here denotes weak equalities, true only when the constraint is imposed. This is similar to the setup considered in the previous section for the Klim\v{c}\'{i}k model: we refer the reader to that section for a more detailed discussion of constrained Hamiltonian field theories.
\bigskip

It is straightforward to check that the Kac-Moody bracket \eqref{Eq:PbKM} implies that the Poisson bracket of the constraint $\Ct$ with itself takes the form\footnote{Note that this bracket would generally include a term proportional to $\bigl(\sum_{r=1}^N \ell_r\bigr)\p_x\delta(x-y)$. This term vanishes due to the condition \eqref{Eq:SumL} imposed on the levels. The condition is thus necessary to ensure that the constraint is first-class.}
\be
\big\{ \Ct(x)\begin{array}{c} \otimes \\[-0.35cm] , \end{array} 
\Ct(y)\big\} = \big[\big(1\otimes \Ct(x)\big),{\mathsf C}_{2}\big]\,\delta(x-y) \approx 0.
\ee
This ensures that the constraint $\Ct \approx 0$ is first-class. In particular, according to the general formalism of Dirac, it can be interpreted as the generator of a \textit{gauge symmetry}.  A straightforward computation shows that this gauge symmetry acts on the Kac-Moody currents as
\be\label{Eq:GaugeGaudin}
\Jc_r \longmapsto h^{-1}\Jc_r\, h + \ell_r\,h^{-1}\p_x h,
\ee
where $h$ is the local $G$-valued gauge parameter (here $G$ is a connected Lie group whose Lie algebra is the simple algebra $\g$ entering the definition of the AGM). Equivalently, one sees that the Gaudin Lax matrix, defined as \eqref{Eq:Gaudin}, transforms as
\be\label{Eq:GaugeGamma}
\Gamma(z) \longmapsto h^{-1}\Gamma(z)\, h + \vp(z)\,h^{-1}\p_x h.
\ee
The algebra of physical observables $\Ac_{\red}$ of the model is defined through the Hamiltonian reduction of the unreduced algebra $\Ac$ with respect to this gauge transformation. This reduced algebra is obtained in two steps: we first impose the constraint $\Ct\approx 0$ and then restrict to gauge invariant quantities, \textit{i.e.} observables that Poisson commute with $\Ct$.

\paragraph{Lax matrix and non-local charges.} We define the \textit{Lax matrix} of the AGM as
\begin{equation}\label{ozkmsamn21}
\Lc(z,x) = \frac{\Gamma(z,x)}{\vp(z)}\,.
\end{equation}
Following the standard formalism of integrable field theories, we then construct the monodromy matrix $M(z)$ of $\Lc(z,x)$ and extract charges from it by evaluating conjugacy invariant functions on $M(z)$. These charges are in general non-local quantities, expressed in terms of the Kac-Moody currents through nested integrals. Moreover, it was argued in~\cite{Feigin:2007mr} that they are pairwise in involution. An alternative proof of this involution property is that the Kac-Moody bracket \eqref{Eq:PbKM} implies
\begin{align}\label{maillet1a}
\big\{ \Lc(z,x)\begin{array}{c} \otimes \\[-0.35cm] , \end{array}
\Lc(w,y)\big\} &=  \big[\big(1\otimes \Lc(w,x)\big),\mathcal{R}(w,z)\big]\,\delta(x-y) - \big[\big(\Lc(z,x)\otimes 1\big),\mathcal{R}(z,w)\big]\,\delta(x-y) 
\nonumber\\
& \hspace{50pt} - \bigl(\mathcal{R}(z,w)+\mathcal{R}(w,z)\bigr)\,\p_x\delta(x-y) \,,
\end{align}
where
\be
\mathcal{R}(z,w) = \frac{\mathsf{C}_2}{w-z}\frac{1}{\vp(w)}\,.
\ee
We recognise here the so-called $r\big/s$ bracket introduced
by  Maillet in refs. \cite{Maillet:1985fn,Maillet:1985ek} with a non-skew-symmetric $\mathcal{R}$-matrix $\mathcal{R}(z,w)$. This bracket ensures the involution of the charges extracted from $M(z)$.

It is clear from eq.\;\eqref{Eq:GaugeGamma} that under a gauge transformation the Lax matrix transforms according 
to $\p_x+\Lc(z) \mapsto h^{-1}\bigl(\p_x+\Lc(z)\bigr)h$. This is similar to the discussion around eq.\;\eqref{oiasi98oias} 
for the Klim\v{c}\'ik model: in particular, this property ensures that the non-local charges extracted 
from conjugacy-invariant functions of $M(z)$ are gauge-invariant. 

\paragraph{Local charges.} Let us introduce the quantity
\begin{equation}\label{Eq:Qz}
\Q(z) = - \frac{1}{2\vp(z)} \int \dd x\;\bigl\langle \Gamma(z,x), \Gamma(z,x) \bigr\rangle.
\end{equation}
We denote by $\ze_1,\ldots,\ze_{N-2}\in\mathbb{CP}^1$ the zeroes of the twist function, or more precisely of the 1-form $\vp(z)\dd z$, which we will suppose are simple. We associate with each of these zeroes a quadratic local charge
\begin{equation}\label{Eq:Qi}
\Q_i = \res_{z=\ze_i} \Q(z) \dd z.
\end{equation}
Starting from the Kac-Moody bracket \eqref{Eq:PbKM}, one then checks that these charges are in involution\footnote{This involution property holds only weakly if one of the zeroes $\ze_i$ or $\ze_j$ is located at infinity in $\mathbb{CP}^1$.}
\begin{equation}\label{Eq:QInv}
\bigl\lbrace \Q_i, \Q_j \bigr\rbrace \approx 0, \qquad \forall\, i,j\in\lbrace 1,\ldots,N-2 \rbrace\, .
\end{equation}
If $\ze_i$ is a finite zero, one easily checks that the local charge $\Q_i$ can be rewritten as
\begin{equation}
\Q_i = -\frac{1}{2\vp'(\ze_i)} \int \dd x\;\bigl\langle \Gamma(\ze_i,x), \Gamma(\ze_i,x) \bigr\rangle.
\end{equation}
Moreover, if $\infty$ is a zero of $\vp(z)\dd z$, one can derive a similar expression for the corresponding charge in terms of the evaluation of $\Gamma(z,x)\dd z$ at $z=\infty$, \textit{i.e.} the evaluation of $-u^{-2}\Gamma(u^{-1},x)\dd u$ at $u=0$ (note that this 1-form is regular at $u=0$ on the constrained surface $\Ct(x) \approx 0$).\bigskip

The quadratic charges $\Q_i$ are in fact the first elements of an infinite hierarchy of local charges in involution, constructed in~\cite{Lacroix:2017isl}\footnote{Technically, the construction of~\cite{Lacroix:2017isl} is restricted to the case of classical Lie algebras $\g$ (of type A, B, C or D). We expect however that a similar construction exists for exceptional algebras.} (building on the work~\cite{Evans:1999mj}). Let us quickly summarise this construction here. The main ingredient is an infinite set of invariant polynomials $\Phi_p$ of degree $p+1$ on $\g$, first constructed in~\cite{Evans:1999mj} and labelled by the positive exponents $p\in\Eh$ of the untwisted affine algebra $\gh$ associated with $\g$ ($\Eh$ is a specific subset of $\mathbb{Z}_{\geq 1}$ which depends on the choice of the Lie algebra $\g$ ; we refer to~\cite{Evans:1999mj,Lacroix:2017isl} for more details). Similarly to the quadratic charges $\Q_i$ defined above, the higher-degree charges are associated with the zeroes $\ze_i$ of $\vp(z)\dd z$. More precisely, if $\ze_i$ is a finite zero, they are defined as\footnote{The prefactor $-\vp'(\ze_i)^{-(p+1)/2}$ differs from the conventions of~\cite{Lacroix:2017isl} and has been introduced for future convenience.}
\begin{equation}\label{Eq:Qip}
\Q_{i,p} = -\frac{1}{\vp'(\ze_i)^{(p+1)/2}} \int \dd x\;\Phi_p\bigl( \Gamma(\ze_i,x) \bigr).
\end{equation}
If $\infty$ is a zero of $\vp(z)\dd z$, one defines the associated charges through a similar expression, in terms of the weak evaluation of $\Gamma(z,x)\dd z$ at $z=\infty$.
\bigskip

For every simple Lie algebra $\g$, the first exponent in $\Eh$ is always $1$. The corresponding invariant quadratic polynomial on $\g$ is simply defined by $\Phi_1({\tt a}) = \frac{1}{2}\,\langle {\tt a},{\tt a}\rangle$. The first charges $\Q_{i,1}$ in the infinite hierarchy then coincide with the quadratic charges $\Q_i$ introduced above. Generalising eq.\;\eqref{Eq:QInv} to higher degree charges, it was proven in~\cite{Lacroix:2017isl} that
\begin{equation}
\bigl\lbrace \Q_{i,p}, \Q_{j,q} \bigr\rbrace \approx 0, \qquad \forall\, i,j\in\lbrace 1,\ldots,N-2 \rbrace, \quad \forall\,p,q\in\Eh\, .
\end{equation}
Moreover, the charges $\Q_{i,p}$ Poisson commute weakly with the constraint $\Ct(x)$ and thus define gauge invariant quantities. Finally, these local charges are also in involution with the non-local charges extracted from the monodromy matrix of $\Lc(z,x)$.

\paragraph{Integrable structure.} Let us summarise the construction so far. In the previous paragraphs, we have constructed non-local charges extracted from the monodromy matrix of $\Lc(z,x)$, as well as local charges naturally associated with the zeroes of $\vp(z)\dd z$. These charges are all in involution and thus generate a Poisson commutative subalgebra $\Zc^{(z_r)}$ of the unreduced algebra of observables $\Ac$: in other words, we have built an integrable structure from the Kac-Moody currents generating $\Ac$. Moreover, the charges in $\Zc^{(z_r)}$ are all gauge-invariant and thus descend to the reduced algebra of observables $\Ac_{\red}$. This defines an integrable structure in the reduced algebra, \textit{i.e.} a Poisson commutative subalgebra $\Zc_{\red}^{(z_r)}$ in $\Ac_{\red}$. This is the main output of the affine Gaudin model construction.

\paragraph{Change of spectral coordinate.} In the previous paragraphs, we have defined the affine Gaudin model through the use of the twist function $\vp(z)$ and the Gaudin Lax matrix $\Gamma(z,x)$, working with an explicit choice of coordinate $z$ on the spectral Riemman sphere $\mathbb{CP}^1$. Let us then end this subsection with a brief discussion on the behaviour of the model under a change of coordinate on $\mathbb{CP}^1$ and in particular on the invariance of the integrable structure. Consider a M\"obius transformation
\begin{equation}
z \longmapsto \zt = \omega(z) = \frac{az+b}{cz+d}
\end{equation}
of the spectral parameter. We denote by $\zt_r=\omega(z_r)$ the position of the punctures of the affine Gaudin model in the new coordinate $\zt$. For simplicity, we will suppose here that none of the $\zt_r$'s are infinite (note however that such a setup will be useful for us later, but for the sake of brevity we postpone its discussion to Subsection \ref{Sec:QuantAGM}). Let us now describe the affine Gaudin model in the new coordinate $\zt$. We impose here that the Kac-Moody currents $\Jc_r(x)$ and their levels $\ell_r$, attached to the punctures of the model, are independent\footnote{Although this seems quite natural, let us note here that the situation is more subtle for affine Gaudin models with higher-order poles at the punctures (see~\cite{Lacroix:2019xeh} for more details).} of the choice of coordinate on $\mathbb{CP}^1$. The twist function and Gaudin Lax matrix of the model in the coordinate $\zt$ are thus
\begin{equation}
\vpt(\zt) = \sum_{r=1}^N \frac{\ell_r}{\zt-\zt_r} \qquad \text{ and } \qquad \Gt(\zt,x) = \sum_{r=1}^N \frac{\Jc_r(x)}{\zt-\zt_r}\,.
\end{equation}
One easily checks that the twist function behaves as a 1-form on $\mathbb{CP}^1$:
\begin{equation}
\vp(z)\dd z = \vpt(\zt)\dd \zt, \qquad i.e. \quad \vp(z) = \vpt\bigl(\omega(z)\bigr) \,\omega'(z)\,.
\end{equation}
Note that the right-hand side of the above equation naively contains a pole at $z=-d/c$, which is not present in the left-hand side. This pole is however proportional to $\sum_{r=1}^N \ell_r$, which we supposed vanishes in eq.\;\eqref{Eq:SumL} to impose the regularity of $\vp(z)\dd z$ at $z=\infty$. This condition is thus necessary to ensure that $\vp(z)\dd z$ behaves as a 1-form.
\bigskip

The situation with the Gaudin Lax matrix is similar but slightly more subtle. Let us start by observing that, since we imposed that the Kac-Moody currents $\Jc_r(x)$ are independent of the choice of coordinate on $\mathbb{CP}^1$, the constraint $\Ct = \sum_{r=1}^N \Jc_r  \approx 0$ is the same for both models in coordinates $z$ and $\zt$. In particular, both these models share the same unreduced and reduced algebras of observables $\Ac$ and $\Ac_{\red}$. Moreover, one easily checks that, up to the constraint $\Ct\approx 0$, the Gaudin Lax matrix also behaves as a 1-form on $\mathbb{CP}^1$ (we note that this property does not hold strongly):
\begin{equation}\label{Eq:Gamma1form}
\Gamma(z,x)\dd z \approx \Gt(\zt,x)\dd \zt, \qquad i.e. \quad \Gamma(z,x) \approx \Gt\bigl( \omega(z), x) \omega'(z).
\end{equation}
In particular, as both $\vp(z)\dd z$ and $\Gamma(z,x)\dd z$ are 1-forms, the Lax matrix $\Lc(z,x)=\Gamma(z,x)/\vp(z)$ behaves as a function on $\mathbb{CP}^1$, \textit{i.e.}
\begin{equation}
\Lc(z,x) \approx \widetilde{\Lc}(\zt,x) = \widetilde{\Lc}\bigl(\omega(z),x).
\end{equation}
Thus, the non-local charges extracted from the monodromy matrices of $\Lc(z,x)$ and $\widetilde{\Lc}(\zt,x)$ coincide weakly. This ensures that the non-local quantities in the affine Gaudin model integrable structure are independent of the choice of coordinate on $\mathbb{CP}^1$.
\bigskip

Let us now turn our attention to the local charges. It is clear that the quantity $\Q(z)\dd z$, defined through eq.\;\eqref{Eq:Qz}, also behaves as a 1-form (weakly). Thus, the quadratic local charges $\Q_i$, introduced as residues of $\Q(z)\dd z$ at the zeroes of $\vp(z)\dd z$, are weakly invariant under a change of spectral parameter (indeed, residues of 1-forms are independent of the choice of coordinate). More generally, the higher-degree local charges $\Q_{i,p}$ are defined in eq.\;\eqref{Eq:Qip} in terms of the evaluations of $\Gamma(z,x)\dd z$ at the zeroes $\ze_i$ of $\vp(z)\dd z$. One easily checks from eq.\;\eqref{Eq:Gamma1form} that under the change of coordinate $z\mapsto \zt=\omega(z)$, these evaluations are multiplied by a factor $\omega'(\ze_i)^{-1}$. In the definition \eqref{Eq:Qip} of the local charges, these terms are compensated by the transformation of the prefactor $\vp'(\ze_i)^{-(p+1)/2}$, ensuring that the local charges are also (weakly) invariant under a change of spectral parameter.
\bigskip

As a conclusion, we thus see that the integrable structures $\Zc^{(z_r)}$ and $\Zc^{(\zt_r)}$ built from the affine Gaudin model in the coordinates $z$ and $\zt$ weakly coincide, \textit{i.e.} $\Zc^{(z_r)}\approx \Zc^{(\zt_r)}$. As a consequence, the integrable structure $\Zc_{\red}^{(z_r)}=\Zc_{\red}^{(\zt_r)}$ in the reduced algebra $\Ac_{\red}$ is independent of the choice of coordinate on $\mathbb{CP}^1$. The affine Gaudin model can thus be seen as associated with the geometric data of a meromorphic 1-form $\vp(z)\dd z$ on $\mathbb{CP}^1$ (here with simple poles). More precisely, the residues $\ell_r$ of this 1-form define uniquely the unreduced algebra $\Ac$ and its Hamiltonian reduction $\Ac_{\red}$, while the positions of its poles (or more precisely the $N-3$ M\"obius invariants built from these positions) determine the integrable structure $\Zc_{\red}$ in $\Ac_{\red}$.

\subsection{Relativistic realisations of classical affine Gaudin models}\label{Sec:Relat}

\paragraph{Realisations.} In the previous subsection, we have defined a classical affine Gaudin model as the data of an integrable structure $\Zc^{(z_r)}$ in the Poisson algebra $\Ac$ built from Kac-Moody currents $\Jc_r(x)$. In order to relate this formal construction to standard 2-dimensional integrable field theories, it is useful to introduce the notion of a realisation of such a model in the algebra $\Fc[T^\ast Q]$ of fields valued in a cotangent bundle $T^\ast Q$. This algebra is generated by canonical fields\footnote{This description relies on the use of a local coordinate chart $(\phi^i)_{i\in\lbrace 1,\ldots,\dim Q\rbrace}$ on the configuration space $Q$: $\phi^i(x)$ are then coordinate fields, while $\pi_i(x)$ are the corresponding conjugate momentum fields describing the fibers of $T^\ast Q$. This is only a local description and a rigorous treatment of $\Fc[T^\ast Q]$ then requires dealing with globality issues. We will not enter into these details here.} $\bigl(\pi_i(x),\phi^j(x)\bigr)_{i,j\in\lbrace 1,\ldots,\dim\,Q\rbrace}$ and is equipped with the canonical bracket
\begin{equation}
\bigl\{ \pi_i(x), \phi^j(y) \bigr\} = \delta_i^{\;j} \,\delta(x-y), \qquad \bigl\{ \pi_i(x), \pi_j(y) \bigr\} = \bigl\{ \phi^i(x), \phi^j(y) \bigr\} = 0.
\end{equation}
A realisation of the Gaudin model in $\Fc[T^\ast Q]$ mainly consists in the choice of a Poisson map $\rho : \Ac \to \Fc[T^\ast Q]$. Concretely, such a realisation is equivalent to constructing $\g^{\C}$-valued currents $\Jc_r^\rho(x)=\rho\bigl(\Jc_r(x)\bigr)$ in $\Fc[T^\ast Q]$ as combinations of the canonical fields $\bigl(\pi_i(x),\phi^j(x)\bigr)$ and their derivatives, in such a way that the canonical bracket implies that these currents satisfy the Kac-Moody bracket \eqref{Eq:PbKM}. Given such a realisation, one can ``transfer'' the integrable structure $\Zc^{(z_r)}$ of the formal affine Gaudin model to $\Fc[T^\ast Q]$ and thus obtain charges in involution $\rho\bigl(\Zc^{(z_r)}\bigr)$ built from the canonical fields $\bigl(\pi_i(x),\phi^j(x)\bigr)$.

Recall that the formal affine Gaudin model in $\Ac$ is subject to the first-class constraint $\Cc_G(x) \approx 0$, generating the gauge symmetry \eqref{Eq:GaugeGaudin}. It is thus natural to also impose a constraint $\Cc(x) = \rho\bigl( \Cc_G(x) \bigr) \approx 0$ in the realisation $\Fc[T^\ast Q]$ and to consider the Hamiltonian reduction $\Fc[T^\ast Q]_{\red}$ with respect to the gauge symmetry it generates. By construction, the integrable structure $\rho\bigl( \Zc^{(z_r)} \bigr)$ is gauge-invariant and descends to $\Fc[T^\ast Q]_{\red}$. Alternatively, one can think of the realisation $\rho : \Ac \to \Fc[T^\ast Q]$ as inducing a realisation $\rho_{\red} : \Ac_{\red} \to \Fc[T^\ast Q]_{\red}$ of the reduced algebra, yielding an integrable structure $\rho_{\red}\bigl( \Zc^{(z_r)}_{\red} \bigr)$ in $\Fc[T^\ast Q]_{\red}$. In principle, one can consider directly realisations of the Poisson algebra $\Ac_{\red}$, without going through the unreduced one $\Ac$: in practice, it is often easier to work with a gauged model in $\Ac$, subject to the constraint $\Cc(x)\approx 0$ and invariant under the corresponding gauge symmetry.

\bigskip

In order to obtain an integrable 2-dimensional field theory in the realisation, one further needs to choose a Hamiltonian $\Hc\in\Fc[T^\ast Q]$ defining the dynamics $\p_t \approx \lbrace \Hc_{\rm T},\cdot\rbrace$ (following Dirac's formalism of constrained systems, the time evolution is generated by the total Hamiltonian $\Hc_{\rm T}$, built from $\Hc$ by adding a Lagrange mulitplier term). This yields a gauged Hamiltonian field theory with fields in $T^\ast Q$ and space-time variables $(t,x)\in\mathbb{R}\times \mathbb{S}^1$. We require that the Hamiltonian (weakly) Poisson commutes with the charges in $\rho\bigl(\Zc^{(z_r)}\bigr)$: by construction, these charges are then conserved and in involution, ensuring that the corresponding 2-dimensional field theory is integrable. In practice, this last condition is often\footnote{As we will see in Subsection \ref{Sec:UVLimAGM}, 
the situation will be slightly more subtle in the case of a chiral realisation.} ensured by choosing the Hamiltonian $\Hc$ to be itself an element of the integrable structure $\rho\bigl( \Zc^{(z_r)} \bigr)$.

\paragraph{Relativistic realisation.}\label{Par:RelReal} We say that a realisation $\rho: \Ac \to \Fc[T^\ast Q]$, with Hamiltonian $\Hc \in \Fc[T^\ast Q]$, is relativistic if
\begin{itemize}
\item[(R1)] the choice of Hamiltonian $\Hc$ is such that the resulting 2-dimensional field theory is relativistic ;
\item[(R2)] the canonical fields $\bigl(\pi_i(x),\phi^j(x)\bigr)$ of $\Fc[T^\ast Q]$ can be reconstructed, up to their initial values at a reference point $x_0\in \mathbb{S}^1$, from the currents $\Jc_r^\rho(x)$ in the realisation $\rho(\Ac)$.
\end{itemize}
The second condition (R2) in this definition ensures that the Kac-Moody currents of the Gaudin model, when seen in the realisation, capture all the degrees of freedom of the field theory (up to potential integration constants\footnote{As we will see on explicit examples, the currents $\Jc_r^\rho(x)$ in the realisation are often expressed in terms of the derivatives $\p_x\phi^i(x)$ of the coordinate fields and thus do not contain the data of their initial conditions $\phi^i(x_0)$.}). As we will see in Subsection \ref{Sec:UVLimAGM}, this is in contrast with what we will define there as chiral realisations, which capture only half of the degrees of freedom of a 2-dimensional theory, corresponding to either left-moving of right-moving fields.\bigskip

It is natural at this point to ask whether there exists a systematic way of constructing relativistic realisations of affine Gaudin models. The results of~\cite{Delduc:2019bcl,Lacroix:2019xeh} imply that, once we have found a realisation $\rho : \Ac \to \Fc[T^\ast Q]$ which satisfies the point (R2) above, there exists a natural choice of Hamiltonian $\Hc$ in $\rho(\Zc^{(z_r)})$ such that the resulting 2-dimensional field theory is relativistic and thus satisfies the point (R1). Let us summarise quickly this construction. Recall the quadratic local charges $\Q_i$ defined in eq.\;\eqref{Eq:Qi}. We introduce
\begin{equation}
\Pc_{\Ac} = \sum_{i=1}^{N-2} \Q_i.
\end{equation}
One checks that this observable (weakly) generates the spatial derivative on Kac-Moody currents, \textit{i.e.} satisfies $\bigl\{ \Pc_{\Ac}, \Jc_r(x) \bigr\} \approx \p_x\Jc_r(x)$. By construction, under the realisation $\rho: \Ac \to \Fc[T^\ast Q]$, its image $\rho(\Pc_{\Ac})$ then generates the spatial derivative on the currents $\Jc_r^\rho(x)$ in $\Fc[T^\ast Q]$ and thus, by the assumption (R2), on all canonical fields $\bigl(\pi_i(x),\phi^j(x)\bigr)$. In other words, $\rho(\Pc_{\Ac})$ is (weakly) the momentum of $\Fc[T^\ast Q]$. This property was the main assumption\footnote{In order to make the comparison with the terminology of~\cite{Delduc:2019bcl,Lacroix:2019xeh} easier, let us note that realisations $\rho$ such that $\rho(\Pc_{\Ac})$ is the momentum of $\Fc[T^\ast Q]$ were called suitable in these works.} made in~\cite{Delduc:2019bcl,Lacroix:2019xeh} to construct a Hamiltonian that makes the realisation relativistic.
\bigskip

Following~\cite{Delduc:2019bcl,Lacroix:2019xeh}, we choose the Hamiltonian $\Hc$ of the realisation as a linear combination of the quadratic charges, in analogy with the above expression for the momentum. More precisely, we define the Hamiltonian as
\begin{equation}\label{Eq:HamReal}
\Hc = \sum_{i=1}^{N-2} s_i \,\rho(\Q_i),
\end{equation}
where the $s_i$'s are constant parameters (in particular, the momentum of $\Fc[T^\ast Q]$ is given by the same expression with all $s_i$'s equal to 1). Having expressed the Hamiltonian and momentum of the theory in terms of the quadaric charges $\rho(\Q_i)$ allows one to study the energy-momentum tensor of the model and thus its space-time symmetries. In particular, it was proven in~\cite{Delduc:2019bcl,Lacroix:2019xeh} that the theory is relativistic if and only if the coefficients $s_i$ all square to 1. We thus take
\begin{equation}\label{Eq:EpsilonRelat}
s_i = +1\, \text{ or } -1, \qquad \forall \, i\in\lbrace 1,\ldots,N-2 \rbrace.
\end{equation}
We will further restrict to cases where there are as many coefficients $s_i$ equal to $+1$ as ones equal to $-1$ (indeed, the study of various examples suggests that choices which do not satisfy this condition lead to rather degenerate models, which in particular do not possess a Lagrangian formulation). Let us note, for completeness, that the total Hamiltonian (which generates the dynamics $\p_t \approx \lbrace \Hc_{\rm T},\cdot\rbrace$ in the unreduced algebra $\Fc[T^\ast Q]$) then takes the form
\begin{equation}\label{Eq:HamRealTot}
\Hc_{\rm T} = \sum_{i=1}^{N-2} s_i \,\rho(\Q_i) + \int \dd x \, \bigl\langle \Cc(x), \mu(x) \big\rangle \,,
\end{equation}
where $\mu$ is a Lagrange multiplier.

\paragraph{Integrable $\bm{\sigma}$-models.} Let us illustrate more concretely the ideas presented in the previous paragraphs by discussing how integrable $\sigma$-models can arise as relativistic realisations of affine Gaudin models. These theories appear when considering realisations of the Kac-Moody currents of the form
\begin{equation}\label{Eq:RealSigma}
\Jc_{r}^\rho(x) = \alpha_{r}^i \bigl( \phi^{\,j}(x) \bigr) \, \pi_i(x) + \beta^{r}_i \bigl( \phi^{j}(x)  \bigr)\, \p_x \phi^{i}(x),
\end{equation}
\textit{i.e.} currents which are linear in the momentum fields $\pi_i(x)$ and the derivatives $\p_x \phi^{i}(x)$ of the coordinate fields, with coefficients which are local functions of the coordinates only.

For these type of realisations, the Hamiltonian \eqref{Eq:HamReal} is quadratic in the momenta and the derivatives of the coordinates and thus yields a Lagrangian density quadratic in the derivatives $\p_t \phi^i$ and $\p_x \phi^i$. Moreover, the relativistic invariance of the theory further ensures that the Lagrangian density takes the form $E_{ij}(\phi^k) \p_+ \phi^i \p_- \phi^j$, in terms of the light-cone derivatives $\partial_\pm=\frac{1}{2}\,(\partial_t\pm\partial_x)$ and a 2-tensor $E_{ij}(\phi^k)$ on $Q$. This is the Lagrangian of a non-linear $\sigma$-model.

More precisely, the theory is a gauged $\sigma$-model on $Q$, invariant under a local action of the group $G$. At the Hamiltonian level, this gauge symmetry is generated by the constraint $\Cc=\sum_{r=1}^N \Jc^\rho_{r}$. For a realisation of the form \eqref{Eq:RealSigma} considered here, it is easy to check that this symmetry shifts the coordinate fields $\phi^{i}(x)$ by functions of the coordinates only (\textit{i.e.} independent of the momenta and the derivatives of the coordinates). At the Lagrangian level, this corresponds to a transformation of the fields $\phi^i$ which does not depend on the derivatives $\p_t\phi^i$ and $\p_x\phi^i$ and thus to a geometric action of the group $G$ on the configuration space $Q$. In particular, the physical target space of the $\sigma$-model is the quotient $Q_{\red}=Q/G$. Equivalently, in the Hamiltonian formulation, the reduction thus amounts to passing from the algebra $\Fc[T^\ast Q]$ to the algebra $\Fc[T^\ast Q_{\red}]$.

\paragraph{Lax connection.} The non-local charges of the formal affine Gaudin model are extracted from the monodromy matrix of the Lax matrix $\Lc(z)$. As explained in the previous subsection, these charges are in involution one with another, as well as with the local ones. In particular, the image of these non-local charges in a relativistic realisation Poisson commute with the Hamiltonian $\Hc$ and are thus conserved. To make the connection with the standard formalism of 2-dimensional integrable field theories, it is useful to come back to the proof that the non-local charges are in involution with the local charges. It was shown in~\cite{Vicedo:2017cge,Lacroix:2017isl} that the Hamiltonian flow generated by these local charges acts on the Lax matrix $\Lc(z)$ by a zero curvature equation, which ensures that the non-local charges extracted from $\Lc(z)$ are invariant under this flow. In particular, in the realisation, the time evolution of the Lax matrix generated by the choice of Hamiltonian takes the form of a zero curvature equation, making the link with the standard Lax formalism of 2-dimensional integrable field theories.
\bigskip

Let us quickly describe the Lax connection of the theory. We will work directly in the realisation, \textit{i.e.} with the Kac-Moody currents $\Jc_r^\rho$ in $\Fc[T^\ast Q]$. Since we are now considering a 2-dimensional field theory, with space-time coordinates $x$ and $t$, we will denote the Lax matrix in the realisation as $\Lc_x(z)$, to stress that it corresponds to the spatial component of the Lax connection. By construction, we have
\begin{equation}\label{Eq:Lax}
\Lc_x(z) = \rho\bigl(\Lc(z)\bigr) = \frac{1}{\vp(z)} \sum_{r=1}^N \frac{\Jc_r^\rho}{z-z_r}.
\end{equation}
This Lax matrix has poles at the zeroes $\ze_1,\ldots,\ze_{N-2}$ of $\vp(z)\dd z$. Taking into account the constraint $\sum_{r=1}^N \Jc_r^\rho \approx 0$, one finds that in the case where all the zeroes $\ze_i$ are finite, the partial fraction decomposition of $\Lc_x(z)$ takes the form
\begin{equation}\label{Eq:LxG}
\Lc_x(z) \approx \mathcal{B}_x + \sum_{i=1}^{N-2} \frac{\mathcal{K}_i}{z-\ze_i},
\end{equation}
where the fields $\mathcal{B}_x$ and $\mathcal{K}_i$ are linear combinations of the Kac-Moody currents $\Jc_r^\rho$. Following~\cite{Vicedo:2017cge,Lacroix:2019xeh}, one finds that the time evolution of $\Lc_x(z)$ generated by the Hamiltonian \eqref{Eq:HamRealTot} takes the form of the zero curvature equation
\begin{equation}\label{ZCEGaudin}
\big[\partial_x + {\cal L}_x(z),\,\partial_t + {\cal L}_t(z) \big] \approx 0\,,
\end{equation}
with the temporal component of the Lax connection given by
\begin{equation}\label{Eq:LtG}
\Lc_t(z) \approx \mathcal{B}_t + \sum_{i=1}^{N-2} \frac{s_i\,\mathcal{K}_i}{z-\ze_i}.
\end{equation}
In this equation, the fields $\mathcal{K}_i$ are the same as in eq.\;\eqref{Eq:LxG}, the coefficients $s_i$ are the ones entering the definition of the Hamiltonian \eqref{Eq:HamRealTot} and $\mathcal{B}_t$ is a field that contains the Lagrange multiplier $\mu$. In the case where one of the zeroes $\ze_i$ is infinite, the corresponding terms in the sums over $i\in\lbrace 1,\ldots,N-2\rbrace$ in the expressions \eqref{Eq:LxG} and \eqref{Eq:LtG} of $\Lc_x(z)$ and $\Lc_t(z)$ take the form $\mathcal{K}_i\,z$ and $s_i\,\mathcal{K}_i\,z$ respectively.\bigskip

Let us end this paragraph with a quick remark. In the definition of a relativistic realisation made at the beginning of this subsection -- see point (R2) in subsection \ref{Par:RelReal} -- we have required that the canonical fields $\bigl( \pi_i, \phi^{j} \bigr)$ of the realisation can be reconstructed from the currents $\Jc_r^\rho$ of the Gaudin model (up to possible integration constants). Since these currents are all contained in the Lax matrix $\Lc_x(z)$, this assumption thus means that the zero curvature equation \eqref{ZCEGaudin} encodes all the equations of motion of the theory. This is a standard requirement in integrable field theories, which ensures that the integrable dynamics concerns all the degrees of freedom of the model and not only a subsector.

\subsection{The Klim\v{c}\'{i}k model as a relativistic realisation of AGM}
\label{Sec:KlimAGM}

In this subsection, we explain how the Klim\v{c}\'{i}k model can be seen as a relativistic realisation of an affine Gaudin model, as first shown in~\cite{Vicedo:2017cge}, based on the results of~\cite{Delduc:2015xdm}. For that, we will use the formulation of the Klim\v{c}\'{i}k model as a gauged $\sigma$-model on $G\times G / G_{\diag}$, as described in Section \ref{Sec:Klim}. In particular, we will use the notations and conventions of that section.

\paragraph{The underlying affine Gaudin model.} As explained in Subsection \ref{Sec:AGM}, a formal affine Gaudin model is defined by the choice of its twist function $\vp(z)$. In this subsection, to obtain the Klim\v{c}\'{i}k model, we will take it to be~\cite{Delduc:2015xdm}
\begin{equation}\label{Eq:TwistKlim}
\vp(z) = \frac{16K}{\xi^2} \frac{z}{z^4+1 - 4 \, \dfrac{\varepsilon_1^2-\varepsilon_2^2}{\xi}\,(z^3+z) + 2\left( 1 + 2\dfrac{(\varepsilon_1^2-\varepsilon_2^2)^2-1}{\xi^2}\right)z^2}\,,
\end{equation}
where $K$, $\varepsilon_1$ and $\varepsilon_2$ are the defining parameters of the Klim\v{c}\'{i}k model introduced in Section \ref{Sec:Klim} and $\xi$ is defined in terms of these parameters by eq.\;\eqref{oias9812oias}. The 1-form $\vp(z)\dd z$ possesses 4 simple poles
\begin{equation}\label{Eq:zrKlim}
z_{1,2} = \frac{1+\varepsilon_1^2-\varepsilon_2^2 \mp 2\varepsilon_1}{\xi}\,, \qquad\;\;  z_{3,4} = -\frac{1+\varepsilon_2^2-\varepsilon_1^2 \mp 2\varepsilon_2}{\xi}\,,
\end{equation}
defining the punctures of the affine Gaudin model, and 2 simple zeroes at $\ze_1=\infty$ and $\ze_2=0$. The levels $\ell_r$ associated with the punctures are determined by computing the residues of $\vp(z)\dd z$ at $z=z_r$ and simply read
\begin{equation}\label{Eq:LevelsKlim}
\ell_{1,2} = \mp \frac{K}{\varepsilon_1}, \qquad \ell_{3,4} = \mp \frac{K}{\varepsilon_2}.
\end{equation}
The formal affine Gaudin model under consideration is described by 4 commuting Kac-Moody currents $\Jc_r(x)$, with levels $\ell_r$.

\paragraph{The realisation.} In the Hamiltonian formulation, the gauged Klim\v{c}\'{i}k model on $G\times G/G_{\diag}$ is defined in the extended algebra of observables $\Fc[T^\ast(G\times G)]$. As explained in Subsection \ref{Sec:KlimHam}, this algebra is described by two $G$-valued fields $g_1(x),g_2(x)$ and two $\g$-valued fields $X_1(x), X_2(x)$, encoding the canonical fields in $T^\ast(G\times G)$ and satisfying the Poisson bracket \eqref{oias9821aaaas}. In order to interpret the Klim\v{c}\'{i}k model as a realisation of the formal affine Gaudin model considered above, we need to realise the Kac-Moody currents $\Jc_r(x)$, $r\in\lbrace 1,\ldots,4\rbrace$, in terms of the fields of $\Fc[T^\ast(G\times G)]$. This realisation is given by~\cite{Delduc:2013fga,Delduc:2015xdm}
\begin{equation}\label{Eq:CurrentsKlim}
\Jc_{1,2}^\rho = \frac{1}{2} X_1 \pm \frac{\ri}{2} \hat{R}_{g_1} X_1 \mp \frac{K}{\varepsilon_1} g_1^{-1}\p_x g_1, \qquad\;\; \Jc_{3,4}^\rho = \frac{1}{2} X_2 \pm \frac{\ri}{2} \hat{R}_{g_2} X_2 \mp \frac{K}{\varepsilon_2} g_2^{-1}\p_x g_2,
\end{equation}
where $\hat{R}:\g\to\g$ is the Yang-Baxter operator \eqref{Rmat1a} entering the definition of the Klim\v{c}\'{i}k model and $\hat{R}_g = \Ad_g^{-1}\circ\hat{R}\circ \Ad_g$, as in Section \ref{Sec:Klim}. Starting from the canonical bracket \eqref{oias9821aaaas} and using the fact that $\hat{R}$ satisfies the modified classical Yang-Baxter equation, one checks that the fields $\Jc_r^\rho$ are commuting Kac-Moody currents with levels $\ell_r$, as announced. We note that, when expressed in terms of the canonical fields $(\pi_i,\phi^{j})$ of $T^\ast(G\times G)$ (working in a choice of coordinate chart $(\phi^{j})$ on the group manifold $G\times G$), the currents $X_{1}$ and $X_2$ are linear in the momentum fields $\pi_i$ while the currents $g_1^{-1}\p_x g_1$ and $g_2^{-1}\p_x g_2$ are linear in the derivatives $\p_x \phi^{\,i}$, so that the Kac-Moody currents are of the form \eqref{Eq:RealSigma}. Moreover, one can prove\footnote{Let us give a sketch of this proof, focusing on the fields $(g_1,X_1)$ and the currents $\Jc_1^\rho$ and $\Jc^\rho_2$. Here, it is useful to perform an analytic continuation and work in the domain where $\ri\varepsilon_1$ is real. In this case, the currents $\Jc_1^\rho$ and $\Jc_2^\rho$ are complex conjugate of one another. Their path-ordered exponentials are then complex conjugate elements of the complexified group $G^{\C}$. Following~\cite{Vicedo:2015pna}, their Iwasawa decompositions, along the factorisation $G^{\C}=G\cdot B_\pm$, then allows one to reconstruct the fields $g_1$ and $X_1$, up to the initial value $g_1(x_0)$ at a reference point $x_0\in \mathbb{S}^1$ (corresponding to integration constants in the path-ordered exponentials).} that the canonical fields $(g_i,X_i)$ can be reconstructed from the data of the Kac-Moody currents $\Jc^\rho_r$, up to integration constants, so that the degrees of freedom of the Gaudin model capture all the fields of $T^\ast(G\times G)$. The realisation under consideration then satisfies the point (R2) in the definition of a relativistic realisation made in 
Subsection \ref{Par:RelReal}.
\bigskip

Formula \eqref{Eq:CurrentsKlim} defines a realisation $\rho: \Ac \to \Fc[T^\ast(G\times G)]$ of the unreduced algebra $\Ac$. In particular, the physical model should be understood as subject to the constraint $\Cc = \sum_{r=1}^4 \Jc_r^\rho \approx 0$. It is straightforward to check from the above expression \eqref{Eq:CurrentsKlim} of $\Jc_r^\rho$ than in the present case, we have $\Cc = X_1 + X_2$. The constraint and gauge symmetry coming from the affine Gaudin model construction thus coincide with the ones \eqref{ConstKlim} and \eqref{ioas8912sa} of the gauged formulation of the Klim\v{c}\'{i}k model. In particular, the physical observables of the realisation are described by the reduced algebra $\Fc[T^\ast(G\times G/G_{\diag})] \simeq \Fc[T^\ast G]$. We will come back to this reduced algebra in the last paragraph of this subsection.

\paragraph{Hamiltonian.} Recall that the Hamiltonian of a (gauged) relativistic realisation can be built from the quadratic charges $\rho(\Q_i)$ associated with the zeroes $\ze_i$ of the twist function. With the choice of twist function  \eqref{Eq:TwistKlim} considered here, the 1-form $\vp(z)\dd z$ has two zeroes $\ze_1=\infty$ and $\ze_2=0$. There are thus two local quadratic charges $\rho(\Q_1)$ and $\rho(\Q_2)$, which can be expressed in terms of the canonical fields $(g_i,X_i)$ by a direct computation (starting from the definition \eqref{Eq:Qi} of these charges in the formal affine Gaudin model and the expression \eqref{Eq:CurrentsKlim} of the realised currents $\Jc_r^\rho=\rho(\Jc_r)$). In particular, one checks that
\begin{equation}
\Pc = \rho(\Q_1) + \rho(\Q_2) = \int \dd x \, \Bigl( \bigl\langle g^{-1}_1 \p_x g_1, X_1 \bigr\rangle + \bigl\langle g^{-1}_2 \p_x g_2, X_2 \bigr\rangle \Bigr)
\end{equation}
is the momentum of the algebra $\Fc[T^\ast(G\times G)]$, in agreement with the general discussion in Subsection \ref{Par:RelReal}.
\bigskip

The Hamiltonian of a relativistic realisation is defined in eq.\;\eqref{Eq:HamRealTot} in terms of the quadratic charges $\rho(\Q_i)$ and coefficients $s_i$, which must be either $+1$ or $-1$ to ensure the relativistic invariance. In the present case, one thus needs to specify a choice of parameters $s_1$ and $s_2$ in $\lbrace +1,-1 \rbrace$. We required in Subsection \ref{Par:RelReal} that there be as many $s_i$'s equal to $+1$ as $-1$, leaving only two possible choices in the present case, which correspond to two opposite Hamiltonians. In what follows, we will choose $s_1 = +1$ and $s_2 = -1$, so that the total Hamiltonian of the gauged model is given by
\begin{equation}
\Hc_{\rm T} = \rho(\Q_1) - \rho(\Q_2) + \int \dd x \, \langle \mu,\Cc \rangle,
\end{equation}
where $\mu$ is a Lagrange multiplier. A direct computation shows that this Hamiltonian coincides with the one \eqref{Eq:HamKlim} of the gauged Klim\v{c}\'{i}k model, up to a redefinition of the Lagrange multiplier $\mu$ (which does not impact the theory).

\paragraph{Lax connection.} Following eq.\;\eqref{Eq:Lax}, the Lax matrix of the theory is given explicitly by
\begin{equation}
\Lc_x(z) = \frac{1}{\vp(z)} \sum_{r=1}^4 \frac{\Jc_r^\rho}{z-z_r}.
\end{equation}
From the explicit expressions \eqref{Eq:TwistKlim}, \eqref{Eq:zrKlim} and \eqref{Eq:CurrentsKlim} of $\vp(z)$, $z_r$ and $\Jc_r^\rho$ considered in this subsection, it is a straightforward computation to check that this Lax matrix coincides with the one \eqref{ioas8912op} of the gauged Klim\v{c}\'{i}k model\footnote{For a specific choice of the function $\lambda(z)$ which was considered as arbitrary in eq.\;\eqref{ioas8912op} since it encoded the freedom of adding any term proportional to the constraint in the Hamiltonian expression of $\Lc_x(z)$.}. A similar computation shows that the temporal component $\Lc_t(z)$ of the Lax connection built from the affine Gaudin model construction (see previous subsection) agrees with the one of the gauged Klim\v{c}\'{i}k model. This achieves the identification of the present relativistic realisation of the affine Gaudin model with the Klim\v{c}\'{i}k model.

\paragraph{Reduced algebra and gauge-invariant currents.} Let us end this subsection with a discussion of the reduced algebra of the model, which describes its physical observables. In terms of the canonical fields $(g_i,X_i)$ of the realisation $\Fc[T^\ast(G\times G)]$, the reduction is obtained by imposing the constraint $\Cc = X_1 + X_2 \approx 0$ and focusing on gauge-invariant observables. As explained in Subsection \ref{Sec:KlimHam}, a generating set of gauge-invariant fields is given by $g=g_1 g_2^{-1}$ and $X = g_2 X_1 g_2^{-1} \approx  -g_2 X_2 g_2^{-1}$. These satisfy the Poisson bracket of canonical fields on one copy of $T^\ast G$ and thus identify the algebra of physical fields of the theory with $\Fc[T^\ast G]$.

Let us now discuss the reduction in terms of the fields of the affine Gaudin model, namely the Kac-Moody currents $\Jc_r^\rho$. Under a gauge transformation with local parameter $h\in G$, the latter transform as $\Jc_r^\rho \mapsto h^{-1}\Jc_r^\rho h + \ell_r\,h^{-1}\p_x h$. To describe the reduction, one needs to build gauge invariant quantities from these currents. Since the field $g_2$ transforms as $g_2 \mapsto g_2h$, one easily checks that the currents
\begin{equation}
{\Jc}_r^{({\rm inv})} = g_2 \Jc_r^\rho g_2^{-1} - \ell_r\,\p_x g_2 g_2^{-1}
\end{equation}
are gauge invariant. Moreover, a straightforward computation starting from the expression \eqref{Eq:CurrentsKlim} of $\Jc_r^\rho$ shows that
\begin{equation}
{\Jc}_{1,2}^{({\rm inv})} = \frac{1}{2} X \pm \frac{\ri}{2} \hat{R}_{g} X \mp \frac{K}{\varepsilon_1} g^{-1}\p_x g,
\end{equation}
in terms of the gauge-invariant canonical fields $(g,X)$. These currents ${\Jc}_1^{({\rm inv})}$ and ${\Jc}_2^{({\rm inv})}$ are also commuting Kac-Moody currents, with levels $\ell_1=-K/\varepsilon_1$ and $\ell_2=+K/\varepsilon_1$. On the other hand, the two other gauge-invariant currents ${\Jc}_3^{({\rm inv})}$ and ${\Jc}_4^{({\rm inv})}$ can be expressed in terms of ${\Jc}_1^{({\rm inv})}$ and ${\Jc}_2^{({\rm inv})}$ by
\begin{equation}
{\Jc}_{3,4}^{({\rm inv})} \approx
 -\frac{\Id \pm \ri \hat{R}}{2} X = -\frac{\Id \pm \ri \hat{R}}{2}\,\bigl( {\Jc}_{1}^{({\rm inv})}+{\Jc}_{2}^{({\rm inv})} \bigr)
\end{equation}
and thus do not carry any additional degrees of freedom. Therefore, the Kac-Moody currents ${\Jc}_1^{({\rm inv})}$ and ${\Jc}_2^{({\rm inv})}$ describe the degrees of freedom of the reduced affine Gaudin model in the algebra $\Fc[T^\ast G]$ and define a ``reduced'' realisation $\rho_{\red} : \Ac_{\red} \to \Fc[T^\ast G]$. In particular, the Klim\v{c}\'{i}k model in its standard (non-gauged) formulation can be described in terms of these two currents: this is the formulation put forward in ref. \cite{Bazhanov:2018xzh} (in particular, the Kac-Moody currents ${\Jc}_{1,2}^{({\rm inv})}$ coincide, up to an overall factor,
with the ones of~\cite[eqs.\;(4.9)-(4.13)]{Bazhanov:2018xzh}, where the labels $1,2$ used here correspond to the ones $\mp$ in that work.).

\section{Klim\v{c}\'{i}k model in conformal limit I: target space and Lax connection for \texorpdfstring{$\bm{G={\rm SU}(2)}$}{G=SU(2)} \label{sec31}}

An important part of the paper concerns the 
investigation of the  integrable structures in the CFT
underlying the UV fixed point of the
Klim\v{c}\'{i}k model. We start by describing the
classical field theory that results when the parameters of the
Klim\v{c}\'{i}k model are taken to be such that 
 the one-loop $\beta$\,-\,function vanishes. Here we focus on
 the target space and the Lax connection in this limit.
The Lie group is set to be ${\rm SU}(2)$ -- the best studied case in
the literature for which  many results about the classical/quantum field theory are known.
 In Section \ref{sec555389i} some aspects of the generalization to arbitrary Lie group $G$ will be 
discussed. The analysis therein will be carried out within the framework
 of affine Gaudin models.
Among the goals of the next several sections  is to  demonstrate
how the results for the ${\rm SU}(2)$ Klim\v{c}\'{i}k model,
based on the approach of the (quantum) inverse scattering method and its variants,
fit with those that we obtain via the Gaudin model formalism.

\subsection{RG flow equations}\label{Sec:RG}
Choosing a local coordinate frame $\lbrace \phi^i \rbrace$ for the target space background,
the classical action of a NLSM takes the form
\be\label{ioas891982sa}
{\cal A}=2\int\rd t\,\rd x \,\big(\mathsf{G}_{ij}(\phi)+\mathsf{B}_{ij}(\phi)\bigr)\,\partial_-\phi^i\,\partial_+\phi^j\,.
\ee
Here, we have decomposed the 2-tensor appearing in the action into a symmetric part $\mathsf{G}_{ij}=\mathsf{G}_{ji}$ and a skew-symmetric part $\mathsf{B}_{ij}=-\mathsf{B}_{ji}$.
The symmetric component $\mathsf{G}_{ij}$ is then interpreted as the metric tensor of the target manifold, while the skew-symmetric one signals the presence of a non-trivial torsion.
More precisely, the corresponding torsion tensor is given by
\be\label{ioasio9812}
\mathsf{H}_{ijk}=\partial_i \mathsf{B}_{jk}+\partial_{k}\mathsf{B}_{ij}+\partial_j \mathsf{B}_{ki}\, .
\ee
A classical field theory of the type \eqref{ioas891982sa} describes  maps
$\{\phi^i(t,x)\}$ from the worldsheet to the target manifold,  which obey a generalized 
version of the Laplace equation.
\bigskip

As long as the curvature of the target space is small,
the quantization of the NLSM \eqref{ioas891982sa} can
be considered within the framework of perturbative 
QFT.  As usual,
the regularization of the UV divergent integrals
in the perturbative expansion introduces
 a scale $\mu$
into the problem.
The $\mu$-dependence of the physical couplings, which
 are encoded  in $\mathsf{G}_{ij}$ and $\mathsf{B}_{ij}$,
 induces a geometric flow of the target space manifold. 
The corresponding RG flow equations were first derived to two loops in ref. \cite{Friedan:1980jf}
for the  case with vanishing torsion tensor. 
The results were extended to a  more general setup in the works \cite{Fradkin:1985ys,Callan:1985ia}.
To the lowest perturbative order, one has 
\bea\label{kasauaua}
\partial_\tau \mathsf{G}_{ij}&=&-\hbar\,
\Big(\mathsf{R}_{ij}-\frac{1}{4}\ {\mathsf{H}_i}^{kl} \,
\mathsf{H}_{klj}+\nabla_i \mathsf{V}_j+\nabla_j \mathsf{V}_i\Big)
+O(\hbar^2)\nonumber\\[-0.2cm]
\\[-0.2cm]
\partial_\tau \mathsf{B}_{ij}&=&-\hbar\,
\Big( \,-\frac{1}{2}\ \nabla_k{\mathsf{H}^k}_{ij}+\mathsf{V}_k \,
{\mathsf{H}^k}_{ij}+\partial_i
 \mathsf{\Lambda}_j-
\partial_j \mathsf{\Lambda}_i\Big)+O(\hbar^2)\ ,\nonumber
\eea
where  
$
\tau=-\frac{1}{2\pi}\,\log(\mu)$ stands for the RG time,
$\mathsf{R}_{ij}$ is the Ricci tensor and $\nabla_i$ the covariant derivative.
The extra vectors $\mathsf{V}_i$ and $\mathsf{\Lambda}_i$ appearing above 
may be arbitrarily chosen. They take into account possible $\mu$-dependent 
reparameterizations of the coordinates as well as transformations of $\mathsf{B}_{ij}$
that leave the torsion tensor \eqref{ioasio9812} invariant.
\bigskip

In the work \cite{Valent:2009nv}
the Ricci tensor and torsion potential 
for a general class of so-called Poisson-Lie sigma models,
which includes the Klim\v{c}\'{i}k model as a particular case, are explicitly computed.
The results were used by the authors of ref. \cite{Sfetsos:2015nya} to show  the 
one loop renormalizability of the
Klim\v{c}\'{i}k model and to derive the
flow of couplings $K$, $\varepsilon_1$ and $\varepsilon_2$ 
entering into the action \eqref{oaso8912as},\,\eqref{ioas891a}.
The one loop RG flow equations read
\be\label{hassasaty}
\partial_\tau \varepsilon_1=-\frac{h^\vee}{4K}\  \hbar \ \varepsilon_1 \,  \xi^2
+O(\hbar^2)\,,\qquad\quad
\partial_\tau
(\varepsilon_2/\varepsilon_1)=O(\hbar^2)\,,\qquad\quad
\partial_\tau(\varepsilon_{{ 1}}/K)=O(\hbar^2)
\ee
with $$\xi^2=\big(\,1-(\varepsilon_1+\varepsilon_2)^2\,\big)\,
\big(\,1-(\varepsilon_1-\varepsilon_2)^2\,\big)\ .$$
Remarkably,  the only dependence on the Lie group $G$
for which the Klim\v{c}\'{i}k model is defined is contained in the dual Coxeter number $h^\vee$\,, 
which enters as an overall factor in the equation for $\partial_\tau \varepsilon_1$.
It appears as a result of our choice of normalization for the symmetric form (see footnote \ref{ft1}).
Upon setting $h^\vee=2$, the RG flow equations 
become equivalent   to the ones derived in the earlier work of Fateev \cite{Fateev:1996ea}
for the ${\rm SU}(2)$ case. 
\bigskip

The model will be considered in the domain with $K>0$, $0<\varepsilon_1+\varepsilon_2<1$
and $|\varepsilon_1-\varepsilon_2|<1$. In the IR limit corresponding to $\tau\to+\infty$ the constant $K$ tends to zero,
the curvature of the target space blows up and the perturbative approach is no longer valid. 
In contrast, it turns out that in  the UV where $\tau\to-\infty$, the curvature of
the manifold remains bounded. In this regime, the classical action can be used as a starting point
for providing a perturbative definition of
the quantum NLSM. The next several sections will be concerned with an analysis of the
Klim\v{c}\'{i}k model in the space of couplings corresponding to the UV fixed point. 
The parameterization 
\be\label{kappdef}
\varepsilon_1=\frac{1}{\sqrt{(1+\kappa^{-1}\,\nu^2)(1+\kappa \nu^2)}}\,, \ \ \ \ \ \ \qquad
\varepsilon_2=\frac{\nu^2}{\sqrt{(1+\kappa^{-1}\,\nu^2)(1+\kappa \nu^2)}}\ \,,
\ee
where $\nu^2>0$ is a RG invariant  and
\be
\kappa=\kappa(\tau)\ : \ \ \ \ 0<\kappa<1
\ee
will be employed.
It is straightforward to check from the differential equations \eqref{hassasaty}
that at high energies the  running coupling $\kappa$ tends to one from below.
We will refer to 
\be\label{Llimitk}
\lim_{\tau\to-\infty}\kappa(\tau)= 1^-\qquad\qquad
{\rm and} \qquad\qquad
 \lim_{\tau\to-\infty} K=\frac{\nu^2\,K_{\scriptscriptstyle{\rm UV}}}{(1+\nu^2)^2}
\qquad\qquad \big(0<K_{\scriptscriptstyle{\rm UV}}<\infty\big)
\ee
as the ``conformal limit'' of the field theory
(the extra  factor multiplying $K_{\scriptscriptstyle{\rm UV}}$ has been chosen for future convenience).

\subsection{Target space of the SU(2) Klim\v{c}\'{i}k model  in the conformal limit\label{sec3kdajkkjsd}}

In studying the  Klim\v{c}\'{i}k model, the ${\rm SU}(2)$
case serves as an important source of intuition.
The corresponding target space manifold 
already exhibits some important  features as $\kappa\to 1^-$, which 
are also expected to be relevant when $G$ is a higher rank group.
Introduce a coordinate frame $(\theta,v,w)$ through 
the Euler decomposition of the group element:
\be\label{uia87887as}
g=\re^{-\frac{\ri v}{2}\,{\tt h}}\,\re^{-\frac{\ri\theta}{2}\,({\tt e}_++{\tt e}_-)}\,
\re^{-\frac{\ri w}{2}\,{\tt h}}\,,
\ee
where
 ${\tt h}$, ${\tt e}_\pm$ obey the standard commutation relations of the $\mathfrak{sl}(2)$
algebra:
\be\label{comm4}
[{\tt h},{\tt e}_\pm]= \pm 2{\tt e}_\pm \ , \qquad [{\tt e}_+,{\tt e}_-]={\tt h}\ .
\ee
The expression for $g$, along with the definition \eqref{oaso8912as},
allows one to  write the ${\rm SU}(2)$ Klim\v{c}\'{i}k model action  
in the form of eq.\;\eqref{ioas891982sa}. As was mentioned before,
in the case under consideration the
second term in the integrand that contains the tensor $\mathsf{B}_{ij}$
is a total derivative and can be ignored. The metric tensor $\mathsf{G}_{ij}$
depends only on $\theta$, since $v$ and $w$  turn out to be Killing coordinates.
We will avoid presenting the formula for $\mathsf{G}_{ij}$ as it exists in
numerous places in the literature see, e.g., appendix A of ref. \cite{Hoare:2014pna}.
\bigskip

There is an important feature of the $\kappa\to 1^-$ limit that needs to be discussed.
It can be motivated by computing the volume of the target space
$V\!ol=\int \sqrt{{\rm det} \,\mathsf{G}_{ij}\vphantom{(}}\,$. A straightforward calculation yields that
\be\label{oias89sd891}
V\!ol\ =\  K^{\frac{3}{2}}\,\nu^{-2}\,(1+\kappa^{-1}\,\nu^2)\,(1+\kappa\,\nu^2)
\int_0^{\pi}\rd \theta \ \frac{8\pi^2\sin(\theta)}{\kappa+\kappa^{-1}-2\cos(\theta)}\ \propto\ 
\log\Big(\frac{1+\kappa}{1-\kappa}\Big)\, .
\ee
Hence in the conformal limit the volume diverges. All this suggests that
the compact target manifold of the ${\rm SU}(2)$ Klim\v{c}\'{i}k model, embedded in Euclidean space, becomes
an infinitely extended object in the parameter domain corresponding to the UV
fixed point. As such, some care is needed in taking $\kappa\to 1^-$.
Different coordinate frames must be introduced to
cover the various regions of the target manifold, which grow to become
an infinite distance apart.  For the ${\rm SU}(2)$ Klim\v{c}\'{i}k model
at least three such regions may be identified. We will define the corresponding
charts as follows. First swap  $\theta\in(0,\pi)$ for $\phi\in(-\infty,+\infty)$ 
using the relation
\be\label{oias9812addddds}
\tan(\tfrac{\theta}{2})=\re^{\phi-\phi_0}\, .
\ee
The constant $\phi_0$ is then set to be
\be\label{oia980cxjhxcj}
\arraycolsep=0.5cm
\begin{array}{ll}
{\rm chart\ I}: & \phi_0=0  \\[0.3cm]
{\rm chart\ II}: & \phi_0=\frac{1}{2} \log\big(\frac{1+\kappa}{1-\kappa}\big)\\[0.3cm]
{\rm chart\ III}: & \phi_0 = \log\big(\frac{1+\kappa}{1-\kappa}\big)\, .
\end{array}
\ee
This way as  $1-\kappa\ll 1$ charts II and III will probe the regions of the target space
where $\theta$ is close to zero. It is easy to see from the integrand in
eq.\;\eqref{oias89sd891} that
this part of the manifold contributes a divergent amount to $V\!ol$.
\bigskip

Consider the action \eqref{oaso8912as},\,\eqref{ioas891a} with $\varepsilon_1$, $\varepsilon_2$
substituted in favour of $\nu^2$ and $\kappa$ as in eq.\;\eqref{kappdef}. 
Plugging in the expression for $g$ \eqref{uia87887as} with
 $\theta$ replaced by $\phi$ according to
 eq.\;\eqref{oias9812addddds}, and taking the $\kappa\to 1^-$ limit yields for the first chart that
\be\label{oias981asas2as}
\lim_{\kappa\to 1^-}{\cal A}= 
{\cal A}_{{\rm cig}}[\phi,\alpha]+ 2K_{\scriptscriptstyle{\rm UV}}\int \rd t\rd x\ 
 \big(\partial_+\chi \,\partial_-\chi\big)\qquad 
\qquad\qquad ({\rm chart\ I}) \, .
\ee
Here
\be\label{oias8912sa}
{\cal A}_{{\rm cig}}[\phi,\alpha]=
2K_{\scriptscriptstyle{\rm UV}}\,\int \rd t\rd x\ \frac{\partial_+\phi\,\partial_-\phi+\partial_+\alpha\,\partial_-\alpha}{1+\re^{2\phi}}
\,,
\ee
while the decoupled boson $\chi$ and the field $\alpha$ are simple linear combinations of  $v$ and $w$:
\be\label{alp1}
\alpha=\frac{\nu^2 v+w}{1+\nu^2}\,,\qquad\qquad\chi=\frac{v-w}{\nu+\nu^{-1}}\  .
\ee
The  matrix elements of 
$g$ \eqref{uia87887as}  in the fundamental representation depend on the Killing coordinates $(v,w)$ through 
the phase factors $\re^{\frac{\ri}{2}(v+w)}=\re^{\ri\alpha+\frac{\ri}{2\nu}\,(1-\nu^2)\,\chi}$ and
 $\re^{\frac{\ri}{2}\,(v-w)}=\re^{\frac{\ri}{2\nu}\,(1+\nu^2)\,\chi}$. As a result, 
the fields $\alpha$ and $\chi$ may always be considered to lie in the interval 
\be\label{globajkkjdsAA}
0\le \alpha<2\pi\,,\qquad\qquad 0\le \chi <\tfrac{4\pi\nu}{1+\nu^2}\, .
\ee
However, while for the field $\alpha$ it is possible to prescribe that $\alpha\sim\alpha+2\pi$, the identifications 
 on $\chi$ are not so straightforward.
For instance, the condition that $v\sim v+4\pi$ translates to
\be\label{globajkkjds}
\big(\chi,\alpha\big)\sim \big(\chi+\tfrac{4\pi\nu}{1+\nu^2},\alpha+\tfrac{4\pi\nu^2}{1+\nu^2}\big)\, .
\ee

\bigskip

The action ${\cal A}_{{\rm cig}}$ defines a field theory
that is often
 referred to as the cigar NLSM \cite{Hamilton,Elitzur:1991cb,Witten:1991yr}. 
This comes from the fact that the  target  manifold can be
embedded into three dimensional space and visualized as the right Hamilton's cigar:
\be
\begin{tikzpicture}
\node at (0,0) {\includegraphics[width=7cm,angle=180,origin=c]{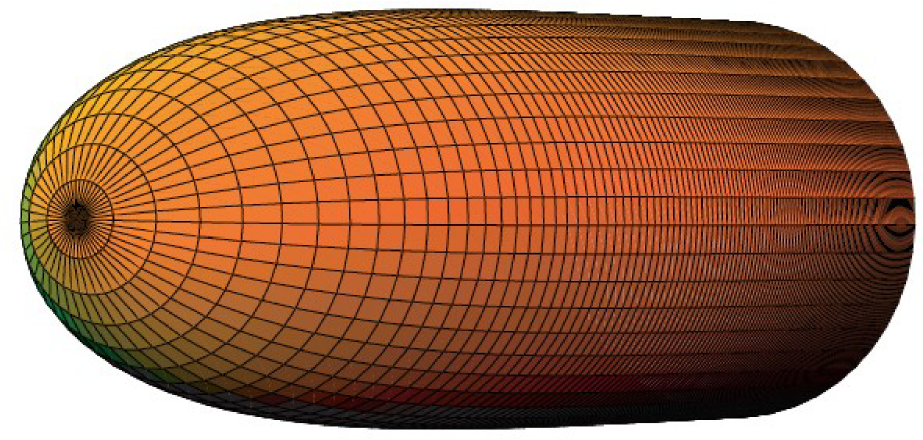}};
\end{tikzpicture}
\nonumber
\ee
The tip corresponds to  $\phi=+\infty$, while for $(-\phi)\gg 1$ the non-trivial term
in the denominator in \eqref{oias8912sa} is  negligible and the target space manifold asymptotically approaches the 
half-infinite cylinder. The coordinate $\alpha$ goes around the circumference of the cigar 
and  belongs to the segment $[0,2\pi)$, in agreement with eq.\;\eqref{globajkkjdsAA}.

\bigskip

The second coordinate chart corresponds to setting $\phi_0=\frac{1}{2}\log\big(\frac{1+\kappa}{1-\kappa}\big)$.
Then the target space asymptotically approaches an infinite cylinder and the action takes the form
$\lim_{\kappa\to 1^-}{\cal A}= {\cal A}_{\rm free}$ with 
\be\label{oias981asas2asBAA}
{\cal A}_{\rm free}=2K_{\scriptscriptstyle\rm UV}\int \rd t\rd x\ 
\Big(\partial_+\phi\,\partial_-\phi+\frac{\partial_+v\,\partial_-v}{1+\nu^{-2}}+
\frac{\partial_+w\,\partial_-w}{1+\nu^2}\, \Big)\, .
\ee
Finally, for chart III one finds that 
\be\label{oias981asas2asB}
\lim_{\kappa\to 1^-}{\cal A}=
{{\cal A}}_{{\rm cig}}[-\phi,\tilde{\alpha}]+
 2K_{\scriptscriptstyle{\rm UV}}\,
\int \rd t\rd x \,\big(\partial_+\tilde{\chi} \,
\partial_-\tilde{\chi}\big)\qquad\qquad
\qquad ({\rm chart\ III}) \,,
\ee
where 
\be\label{alp2}
\tilde{\alpha}=\frac{\nu^2 v-w}{1+\nu^2}\,,\qquad\qquad\qquad\tilde{\chi}=\frac{v+w}{\nu+\nu^{-1}}\  .
\ee
Notice that for the cigar NLSM with action
${{\cal A}}_{{\rm cig}}[-\phi,\tilde{\alpha}]$,
the tip of the target manifold  is located at $\phi=-\infty$, while the infinitely long
cylindrical part extends to $\phi\gg 1$.
\bigskip

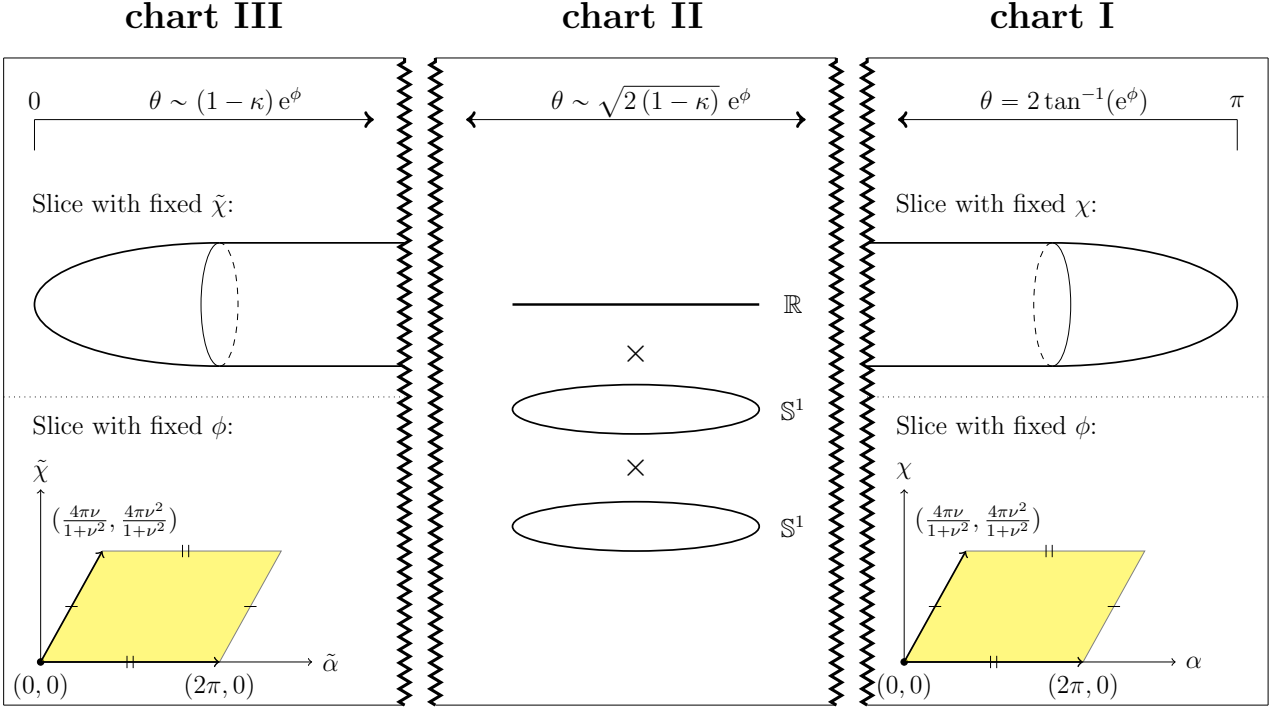
\begin{figure}
\begin{center}
\scalebox{0.82}{
\begin{tikzpicture}
\node at (1.75,4.7) {\Large {\bf chart III}};
\node at (15.5,4.7) {\Large {\bf chart I} };
\node at (8.7,4.7) {\Large {\bf chart II}};
\draw[thick] (2,-1) -- (5,-1);
\draw[thick] (2,1) -- (5,1);
\draw[thick] (2,0) [partial ellipse =90:270:3cm and 1.0cm];
\draw[dashed] (2,0) [partial ellipse =270:450:0.3cm and 1.0cm];
\draw (2,0) [partial ellipse =90:270:0.3cm and 1.0cm];
\node at (0.6,1.6) {Slice with fixed $\tilde{\chi}$:};
\node at (14.6,1.6) {Slice with fixed ${\chi}$:};
\node at (0.6,-2.0) {Slice with fixed $\phi$:};
\node at (14.6,-2.0) {Slice with fixed $\phi$:};
\draw [dotted] (-1.5,-1.5) -- (5,-1.5);
\draw [dotted] (12.5,-1.5) -- (19,-1.5);
\draw[thick] (15.5,-1) -- (12.5,-1);
\draw[thick] (15.5,1) -- (12.5,1);
\draw[thick] (15.5,0) [partial ellipse =270:450:3cm and 1.0cm];
\draw (15.5,0) [partial ellipse =270:450:0.3cm and 1.0cm];
\draw[dashed] (15.5,0) [partial ellipse =90:270:0.3cm and 1.0cm];
\draw (-1.5,4) -- (-1.5,-6.5);
\draw (-1.5,4) -- (5,4);
\draw (-1.5,-6.5) -- (5,-6.5);
\draw[line width = 0.5mm,decorate, decoration={
    zigzag,
    segment length=6,
    amplitude=2.5,post=lineto,
    post length=0pt}] (5,-6.5) -- (5,4);
\draw[line width = 0.5mm,decorate, decoration={
    zigzag,
    segment length=6,
    amplitude=2.5,post=lineto,
    post length=0pt}] (5.5,-6.5) -- (5.5,4);
\draw (5.5,4) -- (12,4);
\draw (5.5,-6.5) -- (12,-6.5);
\draw[line width = 0.5mm,decorate, decoration={
    zigzag,
    segment length=6,
    amplitude=2.5,post=lineto,
    post length=0pt}] (12,-6.5) -- (12,4);
\draw[line width = 0.5mm,decorate, decoration={
    zigzag,
    segment length=6,
    amplitude=2.5,post=lineto,
    post length=0pt}] (12.5,-6.5) -- (12.5,4);
\draw (12.5,4) -- (19,4);
\draw (12.5,-6.5) -- (19,-6.5);
\draw (19,-6.5) -- (19,4);
\node at (8.75,-0.8) {\Large ${\times}$};
\node at (8.75,-2.65) {\Large ${\times}$};
\draw[thick] (8.75,-1.7) [partial ellipse =0:360:2cm and 0.4cm];
\draw[thick] (8.75,-3.6) [partial ellipse =0:360:2cm and 0.4cm];
\draw[line width = 0.4mm] (6.75,0) -- (10.75,0);
\node at (11.3,-1.7) {$\mathbb{S}^1$};
\node at (11.3,-3.6) {$\mathbb{S}^1$};
\node at (11.3,0) {$\mathbb{R}$};
\draw (-1.0,2.5) -- (-1.0,3);
\draw (-1.0,3) -- (4.5,3);
\draw[->, line width = 0.6mm] (4.4,3) -> (4.5,3);
\node at (-1,3.3) {$0$};
\node at (2.1,3.3) {$\theta\sim (1-\kappa)\,\re^{\phi}$};
\draw (6,3) -- (11.5,3);
\draw[->, line width = 0.6mm] (11.4,3) -> (11.5,3);
\draw[->, line width = 0.6mm] (6.1,3) -> (6,3);
\node at (9.0,3.3) {$\theta\sim \sqrt{2\,(1-\kappa)} \ \re^{\phi}$};
\draw (18.5,2.5) -- (18.5,3);
\draw (18.5,3) -- (13,3);
\draw[->, line width = 0.6mm] (13.1,3) -> (13,3);
\node at (18.5,3.3) {$\pi$};
\node at (15.7,3.3) {$\theta=2\tan^{-1}(\re^{\phi})$};
\draw[->] (-0.9,-5.8) -> (3.5,-5.8);
\draw[->] (-0.9,-5.8) -> (-0.9,-3);
\node at (3.8,-5.8) {$\tilde{\alpha}$};
\node at (-0.9,-2.7) {$\tilde{\chi}$};
\filldraw (-0.9,-5.8) circle (1.5pt);
\filldraw[fill = yellow, opacity=0.5] (-0.9,-5.8) -- (2,-5.8) -- (3,-4) -- (0.1,-4) -- cycle;
\node at (-0.9,-6.2) {$(0,0)$};
\node at (2,-6.2) {$(2\pi,0)$};
\node at (0.3,-3.5) {$(\frac{4\pi\nu}{1+\nu^2},\frac{4\pi\nu^2}{1+\nu^2})$};
\draw[->,thick] (-0.9,-5.8) -> (0.1,-4);
\draw[->,thick] (-0.9,-5.8) -> (2,-5.8);
\draw[->] (13.1,-5.8) -> (17.5,-5.8);
\draw[->] (13.1,-5.8) -> (13.1,-3);
\node at (17.8,-5.8) {${\alpha}$};
\node at (13.1,-2.7) {${\chi}$};
\filldraw (13.1,-5.8) circle (1.5pt);
\filldraw[fill = yellow, opacity=0.5] (13.1,-5.8) -- (16,-5.8) -- (17,-4) -- (14.1,-4) -- cycle;
\node at (13.1,-6.2) {$(0,0)$};
\node at (16,-6.2) {$(2\pi,0)$};
\node at (14.3,-3.5) {$(\frac{4\pi\nu}{1+\nu^2},\frac{4\pi\nu^2}{1+\nu^2})$};
\draw[->,thick] (13.1,-5.8) -> (14.1,-4);
\draw[->,thick] (13.1,-5.8) -> (16,-5.8);
\draw (1.4,-4.1) -- (1.4,-3.9);
\draw (1.5,-4.1) -- (1.5,-3.9);
\draw (0.6,-5.7) -- (0.6,-5.9);
\draw (0.5,-5.7) -- (0.5,-5.9);
\draw (2.4,-4.9) -- (2.6,-4.9);
\draw (-0.5,-4.9) -- (-0.3,-4.9);
\draw (15.4,-4.1) -- (15.4,-3.9);
\draw (15.5,-4.1) -- (15.5,-3.9);
\draw (14.6,-5.7) -- (14.6,-5.9);
\draw (14.5,-5.7) -- (14.5,-5.9);
\draw (16.4,-4.9) -- (16.6,-4.9);
\draw (13.5,-4.9) -- (13.7,-4.9);
\end{tikzpicture}
}
\end{center}
\caption{
As  $1-\kappa\ll 1$ the ${\rm SU}(2)$ Klim\v{c}\'{i}k model 
target space becomes a non-compact manifold of length $\propto \log(\frac{1+\kappa}{1-\kappa})$. 
The figure gives a schematic depiction of this manifold, indicating the domains covered by each of  the 
 three coordinate charts introduced in eqs.\;\eqref{oias9812addddds} and \eqref{oia980cxjhxcj}.
In the first and third charts the target space is composed of the cigar and 
a circle of length $4\pi\,\big/(\nu+\nu^{-1})$, which  are glued together in a topologically non-trivial
way. As a result, say, in chart I the coordinates $\chi,\alpha$ obey the compactification conditions
$(\chi,\alpha)\sim(\chi,\alpha+2\pi)$ and $(\chi,\alpha)\sim(\chi+\tfrac{4\pi\nu}{1+\nu^2},\alpha+\tfrac{4\pi\nu^2}{1+\nu^2})$
and the fundamental domain in the $(\chi,\alpha)$ plane at fixed $\phi$ can be chosen to be the parallelogram 
that is shown in the figure.
In chart II
 the target space is asymptotically flat and isomorphic to $\mathbb{R}\times \mathbb{S}^1\times \mathbb{S}^1$.
\label{oias8912}}
\end{figure}

The above analysis of the Klim\v{c}\'{i}k model action for $\kappa\to 1^-$
suggests the following picture for the target space manifold.
Embedded into Euclidean space, it resembles an  infinitely long 3D cylinder $\cong$
$\mathbb{R}\times \mathbb{S}^1\times\mathbb{S}^1$ with 
all the curvature concentrated at the tips.
These are located at $\phi=+\infty$  
and $\phi=-\infty$ in charts I and III, respectively. 
In chart I, the action is that of the cigar NLSM along with an
independent Bose field, see eq.\;\eqref{oias981asas2as}. However,
the corresponding target space is not the Cartesian product 
${\rm cigar}\times \mathbb{S}^1$. The reason is due to the compactification
condition \eqref{globajkkjds}, which involves shifts of both the fields $\alpha$ and $\chi$.
It corresponds to a gluing of the two components in a topologically non-trivial way.
In the coordinate chart III, a
similar phenomenon occurs.
A sketch of the target space background, emphasizing the non-trivial topology, is given in Figure \ref{oias8912}.

\subsection{Conformal limit for the Lax connection\label{sec341}}

The Lax connection  was introduced in Section \ref{Sec:Klim} 
for the  Klim\v{c}\'{i}k model, viewed as a coset sigma model on $G\times G$ gauged by the diagonal 
subgroup. In formulae \eqref{currentsK}-\eqref{oias9812oias}, ${\cal L}_\mu$ is given
as a gauge dependent quantity that is not invariant under the  transformation  $g_j\mapsto g_j\,h$.
In order to simplify the discussion as much as possible,
here we work with the gauge invariant connection 
${\cal L}^{({\rm inv})}_\mu=g_2\,{\cal L}_\mu \,g_2^{-1} - \partial_\mu g_2\,g_2^{-1}$, 
which depends only on the combination $g=g_1\,g_2^{-1}$. 
This is equivalent to fixing the gauge as
 $g_2\equiv \Id $ and $g_1\equiv g$
in formulae \eqref{currentsK}-\eqref{oias9812oias}. This subsection  is meant as a companion to
\ref{Sec:UVLimAGM} and \ref{Sec:GaugeFixChiral}, which discuss the chiral limit in the affine Gaudin model framework.
Indeed, recall that  ${\cal L}_x$  coincides with the Gaudin Lax matrix up to a field independent factor, see eq.\;\eqref{ozkmsamn21}: the chiral limit of ${\cal L}^{({\rm inv})}_\mu$ studied in this subsection thus describes the UV behaviour of the Gaudin Lax matrix in a specific choice of gauge (for $G={\rm SU}(2)$).

\bigskip

To perform the conformal  limit of ${\cal L}^{({\rm inv})}_\mu$ there are a couple of points to keep in mind. 
The first is related to the fact that the target space of the  ${\rm SU}(2)$ Klim\v{c}\'{i}k model becomes non-compact
when $\kappa\to 1^-$. 
As was discussed above, three charts  are needed to describe the target manifold, see Figure \ref{oias8912}. 
In  coordinate charts I and III,  the action of the corresponding NLSM
is the sum of the cigar sigma model action and that of a free boson.
In  chart II  the manifold is asymptotically flat and the dynamics are non-interacting.
It turns out
that in each of the three charts the conformal limit of the Lax connection  is different. 
Focusing on charts I and II
we will consider $\lim_{\kappa\to1^-} {\cal L}^{({\rm inv})}_\mu$ and bring it to a form
that allows for a simple comparison with the results obtained within the Gaudin formalism. In  Subsection \ref{Sec:GaugeFixChiral} a ``chiral'' affine Gaudin model is introduced. It is
explained how the Lax matrix of this model, specialized to $G={\rm SU}(2)$, agrees 
with the conformal limit of 
${\cal L}^{({\rm inv})}_\mu$ provided suitable 
gauge-fixing conditions are imposed on the Kac-Moody currents.

\bigskip

The second point is that the spectral parameter $z$ enters into the Lax connection \eqref{siqjkhdsjh2}
always in combination with $\xi$.
From the definition \eqref{oias9812oias} of this constant, and swapping $\varepsilon_{1,2}$ 
for $\kappa,\nu$ according to eq.\;\eqref{kappdef}, one finds that 
$\xi\sim (1-\kappa)$ as $\kappa$ tends to one. In order to ensure that
a non-trivial dependence on the spectral parameter in the Lax connection remains,
we will assume that $z$ tends to infinity in the conformal limit  such that
 $z^{({\rm L})}=\xi z$ is held fixed. Another possibility, which will be  commented on briefly, is to
keep $z^{({\rm R})}=\xi z^{-1}$ constant. 

\subsubsection{Chart I}
Parameterize the group element $g$ according to  eqs.\;\eqref{uia87887as} and \eqref{oias9812addddds} with $\phi_0=0$.
It is straightforward to see from formula \eqref{siqjkhdsjh2} with $g_2\equiv \Id$ and
$g_1\equiv g$ that the conformal limit of the Lax connection yields:
\be\label{oias98coi}
\lim_{\kappa\to 1^-\atop \xi z\,-\,{\rm fixed}}{\cal L}_+^{({\rm inv})}= {\cal L}_+^{(0)}+z^{({\rm L})}\,{\cal L}_+^{(1)}
\,,\qquad\qquad
\lim_{\kappa\to 1^-\atop \xi z\,-\,{\rm fixed}}{\cal L}_-^{({\rm inv})}={\cal L}_-^{(0)}\qquad\qquad\qquad \big(z^{({\rm L})}=\xi z\,\big)
\ee
with
\be\label{oias128oisd}
{\cal L}_\pm^{(0)}=\frac{\nu^2}{2\,(1+\nu^2)}\ \,(\Id\pm\ri R)\ {\cal I}_\pm^{(\scriptscriptstyle {\rm UV})}\,,\qquad\qquad 
{\cal L}_+^{(1)}=\tfrac{1}{4}\,{\cal I}_+^{(\scriptscriptstyle {\rm UV})}\ .
\ee
Here the currents ${\cal I}_\pm^{(\scriptscriptstyle {\rm UV})}$ are given by the same expressions \eqref{currentsK},
where $g_2\equiv \Id$, $g_1\equiv g$, but $\varepsilon_1$ and $\varepsilon_2$ are substituted for their limiting
values.
To be fully explicit, 
\bea
\frac{{\cal I}_\pm^{(\scriptscriptstyle {\rm UV})}}{1+\nu^{-2}}\,
&=&\Big(\frac{\partial_\pm\phi\mp\ri\partial_\pm\alpha}{1+\re^{2\phi}}\pm\ri\nu\,
\partial_\pm\chi\Big)\,(\pm{\tt h})-
\frac{\ri\partial_\pm\phi\pm\partial_\pm\alpha}{1+\re^{2\phi}}\,
\re^{\phi\pm\ri(\alpha-\nu\chi)}\,{\tt e}_\pm\\[0.2cm]
&&- \Big(\frac{\ri\partial_\pm\phi\pm\partial_\pm\alpha}{1+\re^{2\phi}}+
\ri\nu^2\,\partial_\pm\phi\mp\nu^2\partial_\pm\alpha\mp2\nu\,\partial_\pm\chi\Big)\,
\re^{-\phi\mp\ri (\alpha-\nu\chi)}\,{\tt e}_\mp\,,\nonumber
\eea
where $\alpha$ is the ${\rm U}(1)$ field of the cigar and $\chi$ is the decoupled boson, see \eqref{alp1} for their definitions.
Notice that, since the combination 
$(\Id \pm \ri R)$ appearing in formula \eqref{oias128oisd} sends ${\tt e}_\mp$ to zero,
 ${\cal L}_+^{(0)}$ (${\cal L}_-^{(0)}$) is a  upper (lower) triangular matrix.
\bigskip

An important feature of  the $\kappa\to 1^-$ limit of ${\cal L}_\mu^{({\rm inv})}$
is that the ``$-$'' component of the resulting Lax connection is independent of the 
spectral parameter. 
As a result, 
there exists a similarity transformation with some field dependent matrix 
$\bm{\Omega}$ that does not depend on $z^{({\rm L})}$,
such that
\be\label{oias98cxoi}
\bm{\Omega}\,\big(\partial_-+{\cal L}_-^{(0)}\big)\,(\bm{\Omega})^{-1}=\partial_-\, .
\ee
Then the zero-curvature representation  $[\partial_- + {\cal L}_-(z),\,\partial_+ + {\cal L}_+(z) ]=0$ would 
turn into the condition that the  ``$+$'' component of the flat connection satisfies
$\partial_-{\cal L}_+=0$. In other words, as a consequence of the equations of motion
${\cal L}_+(z)={\cal L}_+(z;t+x)$ is a chiral field that moves to the left. The presence of left and right moving fields
in a 2D (classical) field theory is one of the hallmarks of  conformal invariance.

\bigskip

The matrix $\bm{\Omega}$ is in general  non-local in the sense that its expression  involves
integrals over the fundamental fields. In the case at hand, it requires one to introduce the 
so-called dual field $\alpha^{({\rm D})}$. The cigar NLSM action \eqref{oias8912sa} does not
contain an explicit dependence on $\alpha$. This implies the conservation law
$\partial_+ J_-+\partial_- J_+=0$ for the Noether current $J_\mu=\frac{\partial_\mu\alpha}{1+\re^{2\phi}}$. 
The dual field is defined through  the relation $\partial_\pm\alpha^{({\rm D})}=\pm J_\pm$.
This way, the conservation of $J_\mu$ becomes the condition that the partial derivatives  $\partial_\pm$
acting on $\alpha^{({\rm D})}$ commute, which is necessary  for the definition of the dual field to be  consistent.
Explicitly one has
\be\label{oia98cxz}
\alpha^{({\rm D})}(t,x)=\int_{x_0}^x \frac{\partial_t\alpha}{1+\re^{2\phi}}
\ee
with $x_0$ being some arbitrary reference point.
Then the matrix $\bm{\Omega}$ such that eq.\;\eqref{oias98cxoi}
is satisfied  is given by
\be\label{ioas98cxiui12}
\bm{\Omega}=
\re^{\frac{\pi}{2}\,({\tt e}_+-{\tt e}_-)}\,\big(1+\re^{-2\phi}\big)^{\frac{{\tt h}}{4}}
\, \exp\big(\re^{-\phi-\ri (\alpha+\alpha^{({\rm D})})}\,{\tt e}_-\big)\,
\re^{\frac{\ri}{2}(\alpha^{({\rm D})}+\nu\chi+\frac{\pi}{2})\,{\tt h}}
\ee
(the first field independent factor has been included for future convenience).
\bigskip

Applying the similarity transformation to the other connection component, one finds that:
\be\label{oias98xcoi}
\bm{\Omega}\big(\partial_++{\cal L}_+^{(0)}+z^{({\rm L})}\,{\cal L}_+^{(1)}\big)
\bm{\Omega}^{-1}=\partial_+
+\ 
\frac{1}{4}\left(
\begin{array}{cc}
-\ri (\nu+\nu^{-1})\,z^{({\rm L})}\,\partial_+\chi &
\big(4-(1+\nu^{2})\,z^{({\rm L})}\big)\,\psi_-^{({\rm L})}  \\[0.3cm]
\big(4+(1+\nu^{-2})\,z^{({\rm L})}\big)\,\psi_+^{({\rm L})} & 
\ri(\nu+\nu^{-1})\,z^{({\rm L})}\,\partial_+\chi
\end{array}\right)
\ee
with 
\be\label{oiasoi89123}
\psi^{({\rm L})}_+=\frac{\ri\partial_+\alpha-\partial_+\phi}{\sqrt{1+\re^{2\phi}}}\ 
\re^{\ri(\alpha+\alpha^{({\rm D})})}\ ,\qquad\qquad
\psi^{({\rm L})}_-=-\frac{\ri\partial_+\alpha+\partial_+\phi}{\sqrt{1+\re^{2\phi}}}\ 
\re^{-\ri(\alpha+\alpha^{({\rm D})})}\  ,
\ee
where, for compactness, we have specialized the $\mathfrak{sl}(2)$ generators
to their fundamental representation.
The boson $\chi$ is decoupled in the action \eqref{oias981asas2as} so that the 
Euler Lagrange equations  imply $\partial_-\partial_+\chi=0$ and the field $\partial_+\chi$ is
left moving. 
Also, one can check via a brute force computation that
\be
\partial_-\psi_\pm^{({\rm L})}=0\ .
\ee
Here and below, the superscript ``${\rm\scriptstyle L}$'' will be awarded to fields
that depend only on the combination $t+x$ of the space-time variables 
as a consequence of the equations of motion of the classical field theory.
Similarly, the superscript ``${\rm \scriptstyle R}$'' will be used to denote right moving fields that
are functions of $t-x$.
\bigskip

The non-local quantities $\psi_\pm^{({\rm L})}$ are somewhat special. It turns out that
as a consequence of the canonical structure induced by the cigar NLSM
action  \eqref{oias8912sa}, they satisfy the closed Poisson bracket algebra:
\bea\label{asi7812as}
(K_{\scriptscriptstyle\rm UV})\,\big\{\psi^{({\rm L})}_\pm(x_+),\psi^{({\rm L})}_\pm(y_+)\big\}&=&
-\epsilon(x_+-y_+)\,\psi^{({\rm L})}_\pm(x_+)\, \psi^{({\rm L})}_\pm(y_+)\\[0.2cm]
(K_{\scriptscriptstyle\rm UV})\,\big\{\psi^{({\rm L})}_+(x_+),\psi^{({\rm L})}_-(y_+)\big\}&=&
-\delta'(x_+-y_+)+\epsilon(x_+-y_+)\,\psi^{({\rm L})}_+(x_+)\,\psi^{({\rm L})}_-(y_+)\, .\nonumber
\eea
Since the fields are left moving, the equal time Poisson bracket relations can  be 
re-interpreted as brackets on the light cone with $x_+=t+x$ and $y_+=t+y$.
Also, $\epsilon(x)$ stands for the step function, normalized such that $\partial_x\epsilon(x)=2\delta(x)$.
The quantization of the above Poisson relations leads to an operator algebra \cite{Lepowsky:1984,Fateev:1985mm} that is satisfied by the
fundamental $\mathbb{Z}_n$ Fateev-Zamolodchikov parafermions. In the paper \cite{Fateev:1985mm},
these were introduced for integer ``$n$'' in the construction of the $\mathbb{Z}_n$ CFT models
that describe  the multicritical points of the \
$\mathbb{Z}_n$ statistical
systems (certain generalizations of the $\mathbb{Z}_2$ invariant Ising model). 
For the quantum cigar NLSM, $n$ is identified with the inverse Planck constant
and must be considered to be a generic real number. It is worth mentioning, however, that
the analysis of the critical behaviour of the $\mathbb{Z}_n$ lattice models played
a crucial r\^{o}le in the quantization of the so-called 2D sausage in \cite{Bazhanov:2017nzh}, which is a massive
deformation of the cigar NLSM formally corresponding to the ${\rm SU}(2)$ Klim\v{c}\'{i}k model with $\nu=0$. This is inspite of the fact that the $\mathbb{Z}_n$ CFTs
have no direct relation to the NLSM. In Subsection \ref{Sec:GaugeFixChiral}, 
a generalisation~\cite{Ninomiya:1986dp,Gepner:1987sm,Bardakci:1990lbc} of the Poisson algebra \eqref{asi7812as} for any Lie group $G$ is shown to arise naturally from the 
Gaudin model, see eq.\;\eqref{Eq:BracketBPsi}.  Its quantum version is likewise expected to be important for the study 
of the  quantum Klim\v{c}\'{i}k model for general $G$.

\bigskip
Finally, let us note that keeping the combination $z^{({\rm R})}=\xi z^{-1}$
fixed in taking the conformal limit would yield a Lax connection where 
$\lim_{\kappa\to 1^-}{\cal L}_+^{({\rm inv})}$ is independent of the spectral parameter.
The other connection component, via a transformation similar to the one in the l.h.s. of
eq.\;\eqref{oias98xcoi}, can be brought to a form such that it depends only on the right
moving fields $\partial_-\chi$ and
\be\label{oiasoi89123A}
\psi^{({\rm R})}_+=\frac{\ri\partial_-\alpha-\partial_-\phi}{\sqrt{1+\re^{2\phi}}}\ 
\re^{\ri(\alpha-\alpha^{({\rm D})})}\ ,\qquad\qquad
\psi^{({\rm R})}_-=-\frac{\ri\partial_-\alpha+\partial_-\phi}{\sqrt{1+\re^{2\phi}}}\ 
\re^{-\ri(\alpha-\alpha^{({\rm D})})}\  .
\ee
These right moving classical parafermions Poisson commute with $\psi_\pm^{({\rm L})}$ and
satisfy Poisson bracket relations similar to \eqref{asi7812as}. 
This way, the theory possesses left and right moving fields forming
two independent chiral algebras.

\subsubsection{Chart II}  
For the asymptotically flat domain of the target manifold, \textit{i.e.} chart II,
the field $\theta$ entering into 
 the parameterization of the group element \eqref{uia87887as}
is swapped for $\phi$ such that  $\theta\sim \sqrt{2(1-\kappa)}\ \re^{\phi}$. Then
 the limit $\kappa\to 1^-$
is taken with $\phi$ assumed to be fixed. 
As a result of such a limiting procedure, 
some of the components of ${\cal I}_\pm$ \eqref{currentsK} with $g_2\equiv \Id$ and $g_1\equiv g$
diverge. An explicit computation shows that
\be\label{oias98xciio}
{\cal I}_+^A\sim(1-\kappa)^{\frac{A}{2}}\,,
\qquad\qquad {\cal I}_-^A\sim (1-\kappa)^{-\frac{A}{2}}\,,  \qquad\qquad A=\pm,0
\qquad\qquad (\kappa\to 1^-)\,,
\ee
where the decomposition 
$
{\cal I}_\sigma={\cal I}_\sigma^+\ {\tt e}_++{\cal I}_\sigma^0\ {\tt h}+{\cal I}_\sigma^-\ {\tt e}_-$
w.r.t. the 
$\mathfrak{sl}(2)$ basis is being used ($\sigma=\pm$).
Since the currents ${\cal I}_\pm$ enter into the definition of ${\cal L}_\pm^{({\rm inv})}$, extra care
is needed to ensure that the conformal limit of the Lax connection is well defined.
\bigskip

It is clear that the following limit exists:
\be\label{ozxioasqqq}
{\cal L}_{\scriptscriptstyle {\rm UV}} = 
\lim_{\kappa\to 1^-\atop \xi z\,-\,{\rm fixed}}\, c^{-\frac{\tt h}{4}}\,{\cal L}_+^{({\rm inv})}\,c^{+\frac{\tt h}{4}}
\qquad\qquad {\rm with}\qquad\qquad  c=\frac{1-\kappa}{1+\kappa}
\ee
as the similarity transformation compensates the singular behaviour of 
${\cal I}_+^A$. For the other component, to get a better idea of what is happening, it is useful to 
perform an expansion  
for $1-\kappa\ll 1$:
\be
\, c^{-\frac{\tt h}{4}}\,{\cal L}_-^{({\rm inv})}\,c^{+\frac{\tt h}{4}}=\bigg(\frac{\nu^2}{2\,(1+\nu^2)}\ \,(\Id-\ri R) + O(1-\kappa)^2\bigg)\ 
c^{-\frac{\tt h}{4}}\,{\cal I}_-\,c^{+\frac{\tt h}{4}}\, \qquad\quad (\xi z\,-\,{\rm fixed})\, .
\ee
Although the part of $c^{-\frac{\tt h}{4}}\,{\cal I}_-\,c^{+\frac{\tt h}{4}}$ proportional to ${\tt e}_+$ contains a singularity of
order $(1-\kappa)^{-1}$, it lies in the kernel of $(\Id -\ri R)$ and therefore does not contribute in the limit $\kappa\to 1^-$. 
The ${\tt e}_-$
component
also tends to zero $\sim(1-\kappa)$ as follows directly from formula \eqref{oias98xciio} and the value of 
$c$ given in  \eqref{ozxioasqqq}.  
This way the conformal limit of 
 ${\cal L}_-^{({\rm inv})}$ exists and results in a particularly simple expression:
$$
\lim_{\kappa\to 1^-\atop \xi z\,-\,{\rm fixed}}
c^{-\frac{\tt h}{4}}\ {\cal L}_-^{({\rm inv})}\ c^{+\frac{\tt h}{4}}
= - \tfrac{1}{2}\,(\partial_-\phi+\ri\partial_- w)\, {\tt h}\, .
$$
\medskip

The obtained connection  is somewhat of an intermediate quantity.
The comparison to the Gaudin model will be performed with a Lax connection which
is related to $\partial_++{\cal L}_{\scriptscriptstyle {\rm UV}}$ via a similarity transformation.
An important point is that, working in chart II, the target space is asymptotically flat and the
fields $(v,w,\phi)$ obey the free dynamics.
Consequently, they can be written as a sum of
 a left-moving and right-moving field. In the notation of the work \cite{Bazhanov:2018xzh}, one has
\bea\label{fields11}
&&v(t,x)\asymp \sqrt{1+\nu^{-2}}\ \big(\phi_1^{({\rm L})}(x_+)+{\phi}_1^{({\rm R})}(x_-)\big)\,, \ \ \ \ \ \ \
w(t,x)\asymp \sqrt{1+\nu^2}\ \big(\phi_2^{({\rm L})}(x_+)+{\phi}_2^{({\rm R})}(x_-)\big) \nonumber \\[0.2cm]
&&\phi(t,x)\asymp\phi_3^{({\rm L})}(x_+)+{\phi}_3^{({\rm R})}(x_-)\,,
\eea
where the reason for the extra factors  is to ensure that $\phi_j^{({\rm L})},\phi_j^{({\rm R})}$ 
are all  normalized in the same way, see eq.\;\eqref{oias981asas2asBAA}.
By considering the chiral fields $\phi_j^{({\rm L,R})}$ rather than the original set $(v,w,\phi)$ itself,
one has  the freedom to perform 
more general similarity transformations of $\partial_++{\cal L}_{\scriptscriptstyle {\rm UV}}$
than would otherwise be possible.

\bigskip

Introduce a  family of connections, depending on a parameter $\omega$, as
\be\label{oias98012oisa}
\partial_++{\cal L}_{\rm \scriptscriptstyle UV}^{(\omega)}=
\bm{\Omega}^{(\omega)}
\,\big(\partial_++{\cal L}_{\rm \scriptscriptstyle UV}\big)\,\big(\bm{\Omega}^{(\omega)}\big)^{-1}
\ee 
with 
\be\label{io89fsdss}
\bm{\Omega}^{(\omega)}=\exp\Big(\tfrac{1}{2}\,\omega\,\big(\phi^{({\rm L})}_3-
\ri\sqrt{1+\nu^2}\,\phi_2^{({\rm L})}\big)\,{\tt h}-
\tfrac{1}{2}\,\big(\phi^{({\rm R})}_3+\ri\sqrt{1+\nu^2}\,\phi_2^{({\rm R})}\big)\,{\tt h}\Big)\, .
\ee
Here it is assumed that $(v,w,\phi)$ have been replaced by the chiral fields according to eq.\;\eqref{fields11}.
The second term
in the exponent in $\bm{\Omega}^{({\omega})}$ has been chosen to exactly cancel the dependence
of ${\cal L}_{\rm \scriptscriptstyle UV}^{(\omega)}$ on the right moving bosons $\phi_j^{({\rm R})}$.
 As a result,
\be
\partial_-{\cal L}_{\rm \scriptscriptstyle UV}^{(\omega)}=0\, .
\ee
This has to do with the fact that the corresponding ``$-$'' connection component 
vanishes under the  transformation \eqref{oias98012oisa} and the zero curvature
representation $[\partial_++{\cal L}_+,\partial_-+{\cal L}_-]=0$ with ${\cal L}_-=0$
leads directly to $\partial_-{\cal L}_+=0$.
\bigskip

The resulting family of Lax connections reads
\be\label{Anew}
{\cal L}^{(\omega)}_{\scriptscriptstyle {\rm UV}}= 
-\frac{1}{1-\rho^2}\ \big(\ (1+\nu^2)^{-\frac{1}{2}}\,V_+^{({\omega})}\ {\tt e}_+
+\,(1+\nu^2)^{\frac{1}{2}}\,\rho^2\,V_-^{({\omega})}\ {\tt e}_-\big)-\,
\frac{1}{4}\ 
\bigg(\frac{2\rho^2}{1-\rho^2}+1-\omega\bigg)\,V_0^{({\omega})}\,{\tt h}\, .
\ee 
Here, in line with the notation of  ref. \cite{Bazhanov:2018xzh},
we use the spectral parameter $\rho^2$, which is related to
the original one via the formula
\be\label{oias8912iopds}
z^{({\rm L})}=\frac{4\nu^2}{1+\nu^2}\frac{\rho^2}{1-\rho^2}\ .
\ee
The fields $V^{(\omega)}_\pm$ and $V^{(\omega)}_0$ are expressed in terms of the
left moving bosons as
\bea\label{oi89xciusa}
V_\pm^{({\omega})}&=&\,\big(\ri\,\sqrt{1+\nu^2}\ \partial_+\phi_3^{({\rm L})}
+\partial_+\phi_2^{({\rm L})}\pm\,\nu\  \partial_+\phi_1^{({\rm L})}
\big)\,\re^{\pm \ri\sqrt{1+\nu^2}\,(1-\omega)\,\phi_2^{({\rm L})} \pm(1+\omega)\,\phi_3^{({\rm L})}}\ \nonumber  \\[0.2cm]
V_0^{({\omega})}&=&-2\,
\big(\,\partial_+\phi_3^{({\rm L})}-\ri\,\sqrt{1+\nu^2}\,\partial_+\phi_2^{({\rm L})}\,\big)\,.
\eea
They turn out to satisfy a closed Poisson bracket algebra of $\mathfrak{sl}(2)$ type.
Starting from the Poisson bracket relations of the free chiral Bose fields
\be\label{ioas9821}
(K_{\scriptscriptstyle\rm UV})\,\big\{\phi_j^{({\rm L})}(x_+)\,,\,\phi_\ell^{({\rm L})}(y_+)\big\}=
+\tfrac{1}{4}\, \delta_{j\ell}\, \epsilon(x_+-y_+)
\,,
\ee 
where  $\partial_x\epsilon(x)=2\delta(x)$, one can show that\footnote{%
In Section \ref{sec666} quasi-periodic boundary conditions will be considered, where 
the space coordinate is compactified $x\sim x+2\pi$ and
$\phi_j^{({\rm L})}(x_++2\pi)=\phi_j^{({\rm L})}(x_+)+2\pi P_j$.
In that case $x_+$ and $y_+$ in the Poisson algebra \eqref{PBcurrents1} should
be understood as belonging to the universal cover $x_+,y_+\in\mathbb{R}$ and
the Poisson brackets are modified if either of the light-cone variables lie outside
the segment $[0,2\pi)$, 
see discussion below eq.\,\eqref{oias98jkmaaands}.
The same applies to the Poisson algebra \eqref{asi7812as} if the classical parafermions are
taken to be quasi-periodic fields.}
\bea\label{PBcurrents1}
(K_{\scriptscriptstyle\rm UV})\,\big\{V_+^{({\omega})}(x_+)\,,\,
V_-^{({\omega})}(y_+)\big\}&=&\nu^2\,\delta'(x_+-y_+)-
\big(1+\tfrac{\nu^2}{2}\,(1-\omega)\big)\,V_0^{({\rm \omega})}(x_+)\ 
\delta(x_+-y_+) \nonumber \\[0.3cm]
&-&\big(\,\omega-\tfrac{\nu^2}{4}\,(1-\omega)^2\,\big)\ 
V_+^{(\omega)}(x_+)\ V_-^{(\omega)}(y_+)\ \epsilon(x_+-y_+)
\nonumber \\[0.3cm]
(K_{\scriptscriptstyle\rm UV})\,\big\{V_0^{({\rm \omega})}(x_+)\,,\,
V_\pm^{({\omega})}(y_+)\big\}&=&\mp 
\big(2+\nu^2\,(1-\omega)\big)\,
V_\pm^{({\omega})}(x_+)\ \delta(x_+-y_+)  \\[0.3cm]
(K_{\scriptscriptstyle\rm UV})\,\big\{V_0^{({\rm \omega})}(x_+)\,,\,
V_0^{({\rm \omega})}(y_+)\big\}&=&
2\nu^2\, \delta'(x_+-y_+) \nonumber \\[0.3cm]
(K_{\scriptscriptstyle\rm UV})\,\big\{V_\pm^{({\omega})}(x_+)\,,\,
V_\pm^{({\omega})}(y_+)\big\}&=& 
\big(\,\omega-\tfrac{\nu^2}{4}\,(1-\omega)^2\,\big)\ 
V_\pm^{({\omega})}(x_+)\ 
V_\pm^{({\omega})}(y_+)\ \epsilon(x_+-y_+)
\ .\nonumber
\eea
For general values of $\omega$ the above relations are non-local in the sense that they contain the 
 step function 
$\epsilon(x_+-y_+)$, which is non vanishing when $x_+$ and $y_+$ are distinct.
For the special case when $\omega= \big(\sqrt{1+\nu^{-2}}\pm\nu^{-1})^2$,
the coefficient in front of the step function vanishes and the algebra becomes
 a classical version of the ${\mathfrak{sl}}(2)$ current algebra.

\section{Klim\v{c}\'{i}k model in the conformal  limit II: \texorpdfstring{$\bm{\Wc}$}{W}-algebra and local IMs
for the \texorpdfstring{$\bm{{\rm SU}(2)}$}{SU(2)} case} \label{sec3899891}

Our interest in the conformal limit is partly motivated by the hope that the algebraic structures
underlying the integrability of the Klim\v{c}\'{i}k model can be more easily identified 
there. We are now going to discuss to which extent a paradigm that was 
recognised in integrable field theories of affine Toda type is realised in the 
Klim\v{c}\'{i}k model. In both cases one may exhibit important relations
between  the integrable structure and the extended conformal symmetry
of the field theory.   These  can be described 
as follows.\bigskip

The extension of the conformal symmetry relevant for the models of interest is 
generated by the modes of collections of left-moving and right-moving local fields, each 
generating a Vertex Operator Algebra (VOA). The  VOAs admit a free field 
realisation. It turns out that 
the combinations of the free  fields representing generators of the VOA
can be characterised by a condition on their commutation relations with a distinguished
set of operators called screening charges.\footnote{This terminology 
goes back to the work of Dotsenko and Fateev \cite{Dotsenko:1984nm} on 
free field representations of the minimal models.} 
By considering an extended set of screening charges
one can define a commutative subalgebra of the
VOA. This subalgebra has an infinite number of generators if the model of interest is integrable. The resulting family of commuting operators 
turn out to coincide with the local IMs of the theory.

 \bigskip

The pattern outlined above was exhibited in affine Toda field theories.
In this context,  the 
characterisation of the full set of local conserved quantities using the 
screening charges was achieved by
Feigin and Frenkel in \cite{Feigin:1991qm}. 
For the case of the ${\rm SU}(2)$ Klim\v{c}\'{i}k model there exists evidence 
for a similar but in some respects more subtle picture through the dual representation 
discovered by Fateev \cite{Fateev:1996ea}, see \cite{Bazhanov:2013cua}.

\bigskip

One of our goals will be to exhibit the integrable structures more directly from the 
perspective of the  sigma model formulation. To this aim  we start
by describing the situation on the classical level.
Compared to the affine Toda theories we will see
that the  ${\rm SU}(2)$ Klim\v{c}\'{i}k model  displays at least two new features.

\bigskip

First it  turns out that the local conserved quantities are of somewhat
limited interest at the classical level.
Any classical NLSM of the type \eqref{ioas891982sa}, not necessarily integrable,
possesses two sets of local integrals of motion:
\be\label{iosa89xo}
{\cal Q}_{2m-1}^{({\rm L})}=\int \rd x\ \Big(\mathsf{G}_{ij}(\phi)\,\partial_+\phi^i\,\partial_+\phi^j\Big)^m\,,\qquad\qquad
{{\cal Q}}_{2m-1}^{({\rm R})}=\int \rd x\  \Big(\mathsf{G}_{ij}(\phi)\,\partial_-\phi^i\,\partial_-\phi^j\Big)^m
\ee
with $m=1,2,\ldots\ $.
This is  a simple consequence of classical scale invariance. The integrands in the above formula 
are the $m$-th powers of the Lorentz spin $\pm2$ components of the stress-energy momentum tensor. Since 
these components 
depend only  on $t\pm x$ for $\phi^j$ satisfying the classical equations of motion, it follows immediately that
 $\partial_t\,{\cal Q}_{2m-1}^{({\rm L,R})}=0$
(with suitable boundary conditions imposed). 
For the ${\rm SU}(2)$ Klim\v{c}\'{i}k model at $\kappa\to1^-$ it will turn out that  the characterisation 
based on the classical screening charges
 yields precisely the IMs 
${\cal Q}_{2m-1}^{({\rm L,R})}$ defined in (\ref{iosa89xo}). The fact that these charges
are present in {\it any} sigma model suggests that their existence by itself 
is not directly related to the integrability of the classical field theory.  The situation will change 
drastically in the quantum case. We will later describe the procedure for constructing
the  quantum deformations
of  ${\cal Q}_{2m-1}^{({\rm L,R})}$, which yields
integral expresions over nontrivial
densities  that
are not  functionally dependent.
It is expected that the resulting family of commuting operators appears in the conformal limit of the
local conserved charges for the quantum 
${\rm SU}(2)$ Klim\v{c}\'{i}k 
model.

\bigskip 

 The ${\rm SU}(2)$ Klim\v{c}\'{i}k 
model possesses another interesting feature.
For a generic field theory,  one expects the local 
charges to be given by integrals over densities formed from the
generators of the extended conformal symmetry algebra appearing in
the conformal limit.
In the case at hand with $\kappa\to 1^-$,
it will turn out that the conserved charges can be represented
as integrals over densities belonging to a (Poisson-)VOA which is strictly smaller 
than that describing the extended conformal symmetry.
The relevance of this feature may not be immediately obvious. 
In Subsection \ref{aaaaaaaaaaaa}
it will be related to a remnant of the gauge symmetry in the affine Gaudin model
formulation of the ${\rm SU}(2)$ Klim\v{c}\'{i}k model.

\bigskip

Here we illustrate the basic aspects of this picture 
on the classical level.
The first step is to describe
the space of chiral local  fields
of the ${\rm SU}(2)$ Klim\v{c}\'{i}k model  in the conformal limit.
It follows easily from the relation with the cigar CFT observed above that
these generate a  ${\rm U}(1)\otimes {\cal W}_\infty$ algebra, where 
${\cal W}_\infty$ is the extended algebra of
conformal symmetry of the cigar NLSM. The rest of this section explains
how the extended conformal symmetry and the local IMs  (\ref{iosa89xo})
can be characterised in terms of 
screening charges. The dynamical meaning of the latter 
will be exhibited, governing the leading deviation between solutions
to the equations of motion of the NLSM of our interest 
and the free fields
describing the asymptotic behaviour of the classical solutions at 
infinite times.

\subsection[Classical ${\Wc}_\infty$ algebra]{Classical \texorpdfstring{$\bm{\Wc_\infty}$}{Winfinity} algebra}\label{Sec:WInfinity}

The  quantization of the cigar NLSM results in
a unitary CFT possessing a ${\cal W}_\infty^{({\rm L})}\otimes {{\cal W}}_\infty^{({\rm R})}$ 
 algebra of extended conformal symmetry
with  ${\cal W}_\infty^{({\rm L},{\rm R})}\cong {\cal W}_\infty$ being the ${\cal W}$-algebra studied in ref. \cite{Bakas:1991fs}.
For the classical field theory, the 
${\cal W}_\infty^{({\rm L})}\otimes {{\cal W}}_\infty^{({\rm R})}$ 
 algebra is manifest through the presence
of an infinite number of fields $T_m^{({\rm L})}$ and $T_m^{({\rm R})}$, which
are local tensor densities of
integer Lorentz spin $m$ and $-m$, respectively.
Their special property is that, as a consequence of the equations of motion,
\be\label{oasio9812ddd}
\partial_-T_m^{({\rm L})}=\partial_+{T}_m^{({\rm R})}=0\qquad\qquad\qquad\qquad
\big(2\partial_\pm=\partial_t\pm\partial_x\big)
\ee
so that $T_m^{({\rm L})}=T_m^{({\rm L})}(t+x)$ is a left moving field, while 
$T_m^{({\rm R})}=T_m^{({\rm R})}(t-x)$ is a right moving field.\bigskip

The presence of chiral fields in the cigar NLSM has already been observed in Subsection \ref{sec341}.
However, the classical parafermions  $\psi_\pm^{({\rm L})}$ and $\psi^{({\rm R})}_\pm$
given by eqs.\;\eqref{oiasoi89123} and \eqref{oiasoi89123A}, respectively,
are non-local fields. Their definition involves  the dual field $\alpha^{({\rm D})}$,
which is given by the integral expression \eqref{oia98cxz}. Nevertheless,
one can consider combinations of the classical parafermions  such that the dependence on 
$\alpha^{({\rm D})}$ cancels out. For instance,
\be\label{aosoic98aaaa2}
T_2^{({\rm L})}=\psi_+^{({\rm L})}\,\psi_-^{({\rm L})}=\frac{(\partial_+\phi)^2+(\partial_+\alpha)^2}{1+\re^{2\phi}}
\ ,\qquad\qquad 
T_2^{({\rm R})}=\psi_+^{({\rm R})}\,\psi_-^{({\rm R})}=
\frac{(\partial_-\phi)^2+(\partial_-\alpha)^2}{1+\re^{2\phi}}
\ee
 are  Lorentz spin $\pm2$ local fields, which coincide with the components of the
stress energy momentum tensor of the cigar NLSM.
There exists a left moving spin $+3$ local field, which is independent of $\partial_+ T_2^{({\rm L})}$.
Namely,
\bea\label{aosoic98aaaa3}
T_3^{({\rm L})}&=& \tfrac{1}{2\ri}\,\big(\psi^{({\rm L})}_- \partial_+\psi^{({\rm L})}_+-
\psi^{({\rm L})}_+\partial_+\psi^{({\rm L})}_-\big)  \\[0.2cm]
&=&\frac{1}{1+\re^{2\phi}}\,\Big( 
(\partial_+\alpha)^3+\partial_+\alpha\,(\partial_+\phi)^2-\partial_+\phi\,\partial_+^2\alpha+
\partial_+\alpha\,\partial_+^2\phi+
\frac{(\partial_+\alpha)^3+\partial_+\alpha\,(\partial_+\phi)^2}{1+\re^{2\phi}}\,\Big)\, .\nonumber
\eea
Parity symmetry yields
a right moving Lorentz spin $(-3)$ field, obtained from
$T_3^{({\rm L})}$ via the  replacements $\partial_+\to\partial_-$ and $\psi_\pm^{({\rm L})}\mapsto \psi_\pm^{({\rm R})}$.
Left moving fields of spin $4$ are
$\big(T_2^{({\rm L})}\big)^2$, $\partial^2_+T_2^{({\rm L })}$, $\partial_+ T_3^{({\rm L})}$
and there is an extra one, $T_4^{({\rm L})}$, which is not expressible as a differential polynomial
in terms of the lower spin fields. It takes the form
\be\label{lkkjasknnksmn21}
T_4^{({\rm L})}=\partial_+\psi_+^{({\rm L})}\,\partial_+\psi_-^{({\rm L})}\, .
\ee
Notice that the definition of $T_4^{({\rm L})}$
is ambiguous.
One has the freedom of  adding any linear combination of the spin 4 fields 
$\big(T_2^{({\rm L})}\big)^2$, $\partial^2_+T_2^{({\rm L})}$ and $\partial_+ T_3^{({\rm L})}$
as well as multiplying it by an overall factor. The choice \eqref{lkkjasknnksmn21} for the
spin 4 current as well as \eqref{aosoic98aaaa3} for the spin 3 one is mainly motivated
by the fact that, written in terms of the classical parafermions, their expressions happen to be particularly simple.
In general, for given $m=2,3,\ldots$ the space of left  moving  currents is spanned
by composite fields built from $T^{({\rm L})}_j$ with $j=2,3,\ldots m-1$
and their derivatives, as well as a single  independent field $T_m^{({\rm L})}$.
A similar statement holds true for the right moving fields.
\bigskip

The Lagrangian \eqref{oias8912sa} induces a Poisson structure.
The fields ${T}_m^{({\rm L})}$ and ${{T}}_m^{({\rm R})}$ turn out to
obey a closed Poisson bracket algebra, which can be seen as the classical limit
of the commutation relations for the quantum 
${\cal W}_\infty^{({\rm L})}\otimes {{\cal W}}_\infty^{({\rm R})}$ algebra.
The Poisson Brackets (PB) of any left moving field with a right moving one vanishes.
The first few PB relations involving the set  $\{{T}_m^{({\rm L})}\}_{m=2}^\infty$  follow
from the explicit expressions \eqref{aosoic98aaaa2}-\eqref{lkkjasknnksmn21} as well
as the PBs for the classical parafermions \eqref{asi7812as}. A
straightforward computation results in:
\begin{align}
(K_{\scriptscriptstyle\rm UV})\,\big\{T_2(x_+),T_2(y_+)\big\}&=
-\big(T_2(x_+)+T_2(y_+)\big)\ \delta'(x_+-y_+)\notag\\[0.2cm]
(K_{\scriptscriptstyle\rm UV})\,\big\{T_3(x_+),T_2(y_+)\big\}&= -3\ T_3(x_+)\ \delta'(x_+-y_+)-
\partial T_3(x_+)\ \delta(x_+-y_+)\label{jasususa} \\[0.2cm]
(K_{\scriptscriptstyle\rm UV})\,\big\{T_3(x_+),T_3(y_+)\big\}&=
\tfrac{1}{4}\,\big(T_2(x_+)+T_2(y_+)\big)\ \delta'''(x_+-y_+)\notag\\[1ex]
-2\delta'(x_+&-y_+) \Big(T_4(x_+)
+T_4(y_+) +
T_2(x_+)\, T_2(y_+)-\tfrac{1}{8}\big(\partial^2 T_2(x_+)+\partial^2
T_2(y_+)\big)\Big),\notag
\end{align}
where the shortcut notation $T_m = T_m^{({\rm L})}$ and $\partial = \partial_+$ is being
used in order to  declutter the formula.
The  left moving fields have been treated as a function of one variable $x_+=t+x$ or $y_+=t+y$.
A similar Poisson algebra, obtained from the one above via the replacement $x\to-x$ and $y\to -y$,
is formed by the right moving fields $\{{T}_m^{({\rm R})}\}_{m=2}^\infty$.

\subsection{Free field realization}\label{sec:free}

We are now going to outline how solutions of the equations of motion for the conformal
limits of the sigma model can be parameterised by free field data.  This will  
lead us to a characterisation of the chiral local currents
of the cigar NLSM as differential polynomials in the free fields
 satisfying a simple condition on their Poisson brackets with a 
distinguished observable called the screening charge.

\paragraph{Classical solutions of the cigar NLSM.}
Our starting point is the description of the 
classical solutions to the cigar NLSM equations of motion that were presented in the lecture notes of Lukyanov and Zamolodchikov, published in \cite{Dorey:2019gkd}. It turns out that they can 
be parameterised by a quadruple of four functions --- 
the classical parafermions that were introduced in Subsection \ref{sec341}.
Combined with formulae
 \eqref{oiasoi89123} and \eqref{oiasoi89123A}, which allow one to map 
the values of $\phi(t,x)$ and $\alpha(t,x)$ and their time derivatives
at some initial time slice $t=t_0$ to the  parafermions,
this  solves the Cauchy problem.

\bigskip

As a first step it is useful to observe that the definition of the parafermions
in terms of the cigar sigma model coordinate fields $\phi$ and $\alpha$
given in \eqref{oiasoi89123} and \eqref{oiasoi89123A} can be elegantly reformulated
using the following
 ${\rm SL}(2)$  matrix
\be\label{ioas8912d}
\bm{\omega}=
\left(\begin{array}{cc}
\re^{-\ri{\alpha}^{({\rm D})}}\, \sqrt{1+\re^{-2\phi}} & \re^{-\ri\alpha-\phi} \\[0.2cm]
\re^{\ri\alpha-\phi} & \re^{\ri{\alpha}^{({\rm D})}} \sqrt{1+\re^{-2\phi}}
\end{array}\right)
\ee
with ${\alpha}^{({\rm D})}$ being the dual field defined in \eqref{oia98cxz}.
Namely,
\be\label{ioas8912d22}
\partial_+\bm{\omega}\,\bm{\omega}^{-1}=
\psi_-^{({\rm L})}\,\sigma^++\psi_+^{({\rm L})}\,\sigma^-\,,\qquad\qquad
\bm{\omega}^{-1}\,\partial_-\bm{\omega}=
\psi_-^{({\rm R})}\,\sigma^++\psi_+^{({\rm R})}\,\sigma^-\, ,
\ee
where $\sigma^\pm=\frac{1}{2}(\sigma^x\pm\ri\sigma^y)$
and  $(\sigma^x,\sigma^y,\sigma^z)$ stand for the usual Pauli matrices. 
It is important to keep in mind that the classical parafermions are
left/right moving fields as a consequence of the classical equations of motion,
$\psi_\pm^{({\rm L})}(t,x)=\psi_\pm^{({\rm L})}(t+x)$ and
 $\psi_\pm^{({\rm R})}(t,x)=\psi_\pm^{({\rm L})}(t-x)$. Thus the above formula  
specifies the light-cone derivatives of $\bm{\omega}(t,x)$ in terms of  
four functions of a single variable.
Integrating eq.\;\eqref{ioas8912d22}
allows one to reconstruct the  ${\rm SL}(2)$  matrix directly from $\psi_\pm^{({\rm L})}$, 
 $\psi_\pm^{({\rm R})}$ and an integration constant $\bm{\omega}_0=\bm{\omega}(0,0)$:\footnote{%
Notice that the integrand in the first path ordered exponent in \eqref{omegajdjdjd} 
coincides with the matrix part of the Lax connection \eqref{oias98xcoi}
evaluated at $z^{({\rm L})}=0$.
} 
\be\label{omegajdjdjd}
\begin{aligned}
\bm{\omega}(t,x)&=\overset{\leftarrow}{\cal P}\exp\bigg(\int_{0}^{t+x}\rd x_+\,\big(
\psi_-^{({\rm L})}\,\sigma^++\psi_+^{({\rm L})}\sigma^-\big)\bigg)\ \bm{\omega}_0\\
&\qquad\qquad\times\overset{\rightarrow}{\cal P}\exp\bigg(\int_{0}^{t-x}\rd x_-\,
\big(\psi_-^{({\rm R})}\,\sigma^++\psi_+^{({\rm R})}\sigma^-\big)\bigg)\,.
\end{aligned}\ee
Along with
formula \eqref{ioas8912d}, that enables one to extract the fields $\phi$ and $\alpha$ from $\bm{\omega}(t,x)$, this
provides a way of constructing solutions at least locally to
the cigar NLSM equations of motion  by means of 
the  chiral data $\psi_{\pm}^{({\rm L,R})}$, supplemented by
the integration constant $\bm{\omega}_0$ needed to 
fix the zero modes of $\phi$ and $\alpha$.

\bigskip

For the Cauchy problem, one has at hand the values of 
$\phi$ and $\alpha$ 
and their derivatives $\partial_t\phi$, $\partial_t\alpha$
at some initial time slice.
These are then used to specify   the four functions $\psi_\pm^{({\rm L})}(t+x)$ and
 $\psi_\pm^{({\rm R})}(t-x)$ by means of \eqref{ioas8912d22}, or equivalently 
\eqref{oiasoi89123}\,, \eqref{oiasoi89123A} and \eqref{oia98cxz}, evaluated at $t=t_0$.
One then obtains the solutions $\phi(t,x)$ and $\alpha(t,x)$ at all times
as functionals of the Cauchy data via
eqs.\;\eqref{omegajdjdjd}
and \eqref{ioas8912d}.

\paragraph{Free fields parameterisation and scattering problem interpretation.}
There is another way of parameterising solutions to the cigar NLSM.
It uses the map \eqref{omegajdjdjd}
and \eqref{ioas8912d} from the classical parafermions to the fundamental fields $(\phi,\alpha)$,
but the functions $\psi^{({\rm L})}_\pm$ and $\psi^{({\rm R})}_\pm$ themselves are specified in
a different way. One takes
\be\label{oisa9821sa}
\psi_\pm^{({\rm L})}=- \big(\partial_+\phi^{({\rm L})}\mp\ri\partial_+\alpha^{({\rm L})}\big)\,\re^{\pm
2\ri\alpha^{({\rm L})}}\,,
\qquad
\psi_\pm^{({\rm R})}=
- \big(\partial_-\phi^{({\rm R})}\mp\ri\partial_-\alpha^{({\rm R})}\big)\,\re^{\pm
2\ri\alpha^{({\rm R})}}\ ,
\ee
where $\phi^{({\rm L})},\alpha^{({\rm L})}$ depend only on $t+x$ and
 $\phi^{({\rm R})},\alpha^{({\rm R})}$ on $t-x$, as suggested by the
superscript notation.
The motivation for considering  such a map has to do with the Hamiltonian structure of the cigar NLSM. 
Assuming that the fields $\phi^{({\rm L},{\rm R})}$ and $\alpha^{({\rm L},{\rm R})}$
satisfy the usual PBs for free chiral fields,
\begin{equation}\begin{aligned}\label{ioas9812asaaa}
&(K_{\scriptscriptstyle\rm UV})\,\{\phi^{({\rm L})}(x_+),\phi^{({\rm L})}(y_+)\}=
(K_{\scriptscriptstyle\rm UV})\,\{\alpha^{({\rm L})}(x_+),\alpha^{({\rm L})}(y_+)\}=
+\tfrac{1}{4}\,\epsilon(x_+-y_+)\\[0.2cm]
&(K_{\scriptscriptstyle\rm UV})\,\{\phi^{({\rm R})}(x_-),\phi^{({\rm R})}(y_-)\}=
(K_{\scriptscriptstyle\rm UV})\,\{\alpha^{({\rm R})}(x_-),\alpha^{({\rm R})}(y_-)\}=
-\tfrac{1}{4}\,\epsilon(x_--y_-)\,,
\end{aligned}\end{equation}
where $\partial_x\epsilon(x)=2\delta(x)$, with vanishing Poisson brackets among left
and right moving fields, one may show that 
the parafermionic Poisson algebra \eqref{asi7812as} follows from \eqref{oisa9821sa}.
For this reason, one sometimes refers to  \eqref{oisa9821sa} as providing
 a ``free field'' realization of the classical parafermions. 
A more detailed
discussion of the Hamiltonian formalism for the cigar NLSM can be found in the 
lecture notes of Lukyanov and Zamolodchikov from \cite{Dorey:2019gkd}.

\bigskip

\begin{figure}[h]
\centering
\scalebox{1}{
\begin{tikzpicture}
\node at (0,0) {\includegraphics[width=11cm]{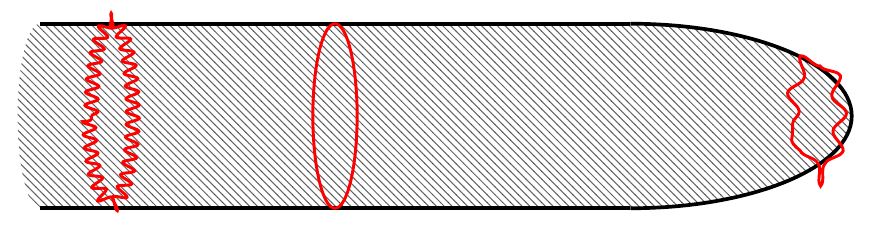}};
\draw[line width = 0.5mm,->] (-1,1.7) -> (0.7,1.7);
\node at (-0.5,2.1) {$P>0$};
\draw[line width = 0.5mm,->] (-4,-1.7) -> (-5.7,-1.7);
\node at (-0.5,2.1) {$P>0$};
\node at (-4.4,-2.1) {$-P$};
\node at (-4.1,1.6) {(C)};
\node at (-1.25,-1.6) {(A)};
\node at (5,-1.3) {(B)};
\end{tikzpicture}
}
\caption{ The scattering of a string in the cigar NLSM.
 Starting in the asymptotically flat domain
the string  (A) approaches the tip with some incoming momentum, 
(B)  scatters and then (C) escapes back to the flat region. 
The initial  (final) states of the scattering process are parameterized by the ``in (out)'' data
 $\phi^{({\rm L},{\rm R})}$ and
$\alpha^{({\rm L},{\rm R})}$.}
\label{fig4}
\end{figure}
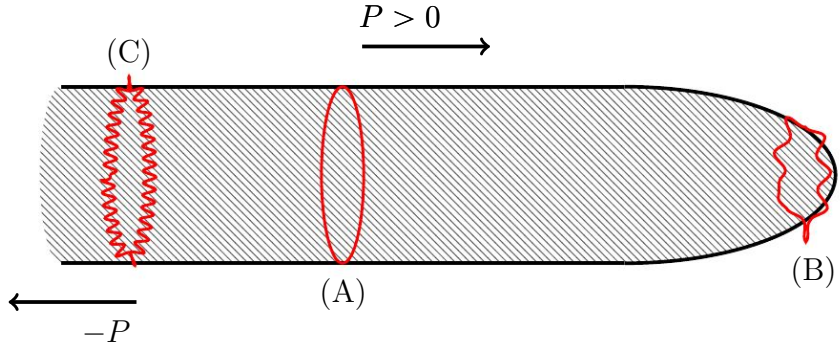

In the present context there exists an intuitive physical interpretation of the 
parameterisation of the classical parafermions using $\alpha^{({\rm L,R})}$ and $\phi^{({\rm L,R})}$. 
It has to do with the fact that the propagation of a classical string on the cigar target space manifold
can be interpreted as a scattering process.
There exists an
 important family of solutions to the cigar NLSM equations of motion such that
for time $t\rightarrow - \infty$ the fields take values
in the semi-infinite cylindrical
end of the cigar, where the
average value
$\phi_0(t)=\int\rd x\, \phi(t,x)$  is large and negative. 
As time increases the classical string approaches
the tip of the cigar,
gets reflected, and  escapes back to the semi-infinite cylindrical
end  for $t\rightarrow + \infty$. An illustration of the scattering process is provided in Figure \ref{fig4}.

\bigskip

Consider the cigar NLSM action \eqref{oias8912sa}. 
In the limits $t\rightarrow \pm\infty$ when
 the zero mode $\phi_0(t)=\int\rd x\, \phi(t,x)$ becomes large and negative
the term $\re^{2\phi}$  in the denominator 
is  negligibly small.
The solutions of the Euler-Lagrange equations  may then be approximated as
\be\label{oias98cxkj21nm2}
\phi(t,x)=\phi^{({\rm L})}(t+x)+\phi^{({\rm R})}(t-x)+\ldots\ ,\qquad
\alpha(t,x)=\alpha^{({\rm L})}(t+x)+\alpha^{({\rm R})}(t-x)+\ldots \,,
\ee
where the ellipses denote terms that are suppressed as $\re^{2\phi_0}$.
The four functions $\phi^{({\rm L},{\rm R})}$ and $\alpha^{({\rm L},{\rm R})}$ can 
therefore be used to
parameterize the degrees of freedom of an asymptotic ``in'' state. 
The relation between the ``in'' data and the values of the fields $\phi(t,x)$ 
and $\alpha(t,x)$ at finite times
is described by a complicated, non-local functional. 
Determining the latter amounts to an explicit integration of the equations of motion.
Formula \eqref{oias98cxkj21nm2} can  be understood as  an approximation
to this functional, valid for $(-\phi_0)\gg 1$ with the next correction 
being  of order $\re^{2\phi_0}$.
\bigskip

It is not hard to see that the asymptotic values of $\phi$ and $\alpha$ for $t\rightarrow \pm \infty$ appearing in \eqref{oias98cxkj21nm2} coincide with the 
fields $\phi^{({\rm L},{\rm R})}$ and $\alpha^{({\rm L},{\rm R})}$ 
previously used in the free field 
realization of the classical parafermions \eqref{oisa9821sa}, as anticipated in the notations.
Indeed,  chiral fields such as $\psi_\pm^{({\rm L})}$ and $\psi_\pm^{({\rm R})}$ can be considered at  $t\rightarrow \pm \infty$
where the suppression of terms of order  $\re^{2\phi_0}$ 
will relate the original
expressions for the classical parafermions \eqref{oiasoi89123},\eqref{oiasoi89123A} 
to \eqref{oisa9821sa}. 
The  procedure for constructing solutions $(\phi,\alpha)$ using formulae \eqref{ioas8912d},
\eqref{omegajdjdjd} with the parafermions as in \eqref{oisa9821sa} 
thereby gets interpreted as an explicit description of the scattering
functional  relating asymptotic data to $\big(\phi(t,x),\alpha(t,x)\big)$ at any time $t$.
\bigskip

The interpretation also  suggests that the map from the free field data 
$(\phi^{({\rm L},{\rm R})},\alpha^{({\rm L},{\rm R})}, \,\bm{\omega}_0)$ to 
solutions $(\phi,\alpha)$ should define a functional that preserves  
the Poisson structure. Indeed, the  Poisson brackets do not depend 
on the choice of time-slice $t$ at which they are evaluated. As a result,
computing them at finite times by means of the canonical bracket
for $(\phi,\alpha)$ should coincide with a calculation performed
at $t\rightarrow \pm \infty$, when  the approximation
$\phi(t,x)\approx\phi^{({\rm L})}(x_+)+\phi^{({\rm R})}(x_-)$ and 
$\alpha(t,x)\approx\alpha^{({\rm L})}(x_+)+\alpha^{({\rm R})}(x_-)$ 
with the chiral fields obeying \eqref{ioas9812asaaa} is applicable.
This could be seen as a ``physical'' explanation of a point
made previously above. Namely, that the PB relations for the 
classical parafermions give the same result \eqref{asi7812as}
whether they are determined via formulae \eqref{oiasoi89123}  
with $\partial_t\phi$ and $\partial_t\alpha$ substituted in terms of the conjugate momenta or via the realization \eqref{oisa9821sa},\eqref{ioas9812asaaa}.

\paragraph{Free field realisation of the $\bm{{\cal W}_\infty}$ algebra and screening charges.}
Consider the local fields $T_m^{({\rm L})}$ that generate the ${\cal W}_\infty$ algebra.
Taking their  expressions \eqref{aosoic98aaaa2}-\eqref{lkkjasknnksmn21}
in terms of the classical parafermions and substituting in the free field realization \eqref{oisa9821sa},
they can all be re-written as differential polynomials in
$\partial_+\phi^{({\rm L})}$ and $\partial_+\alpha^{({\rm L})}$:
\bea\label{oias9812}
T_2^{({\rm L})}&=&\big(\partial_+\phi^{({\rm L})}\big)^2+\big(\partial_+\alpha^{({\rm L})}\big)^2\nonumber \\[0.2cm]
T_3^{({\rm L})}&=& 2\big(\partial_+\alpha^{({\rm L})}\big)^3+
2\partial_+\alpha^{({\rm L})}\,\big(\partial_+\phi^{({\rm L})}\big)^2-\partial_+\phi^{({\rm L})}\,\partial_+^2\alpha^{({\rm L})}+
\partial_+\alpha^{({\rm L})}\,\partial_+^2\phi^{({\rm L})}\nonumber\\[0.2cm]
T_4^{({\rm L})}&=&4\,\big(\partial_+\alpha^{({\rm L})}\big)^4+4\,\big(\partial_+\alpha^{({\rm L})}\big)^2\,
\big(\partial_+\phi^{({\rm L})}\big)^2
+4\,\big(\partial_+\alpha^{({\rm L})}\big)^2\,\partial_+^2\phi^{({\rm L})} \nonumber\\
&&-4\,\partial_+\phi^{({\rm L})}\,\partial_+\alpha^{({\rm L})}\,\partial_+^2\alpha^{({\rm L})}
+\big(\partial_+^2\phi^{({\rm L})}\big)^2+
\big(\partial_+^2\alpha^{({\rm L})}\big)^2  \,.
\eea
An alternative way of obtaining the above formulae is to consider, say, 
$T_2^{({\rm L})}=\big((\partial_+\phi)^2+(\partial_+\alpha)^2\big)/(1+\re^{2\phi})$
at $t\to\pm\infty$. Then the exponential $\re^{2\phi}$ can be neglected and
 $\phi$, $\alpha$  replaced by $\phi^{({\rm L,R})}$ and $\alpha^{({\rm L,R})}$
by means of \eqref{oias98cxkj21nm2} with the ellipses ignored. 
Notice that the Poisson bracket relations \eqref{jasususa} for the ${\cal W}_\infty$
algebra are preserved, i.e., they follow from the free field expressions for $T_m^{({\rm L})}$
combined with the Poisson brackets \eqref{ioas9812asaaa}.

\bigskip

The scattering map relating the free fields to $(\phi,\alpha)$  is highly non-local. 
Arbitrary combinations  of  
$(\phi^{({\rm L},{\rm R})},\alpha^{({\rm L},{\rm R})})$ will not
map to local fields in the NLSM. 
In this regard, the currents $T_m^{({\rm L})}$  \eqref{oias9812}  are distinguished.
They are differential polynomials built from $\partial_+\phi^{({\rm L})}$ and $\partial_+\alpha^{({\rm L})}$ that correspond 
to chiral fields in the cigar sigma model which are local in the fundamental fields $(\phi,\alpha)$. Our next goal will be to find a purely algebraic way of finding the
differential polynomials that possess this special property.
It will be
explained how  they 
all can be characterised 
in terms of their commutation relations with the so-called screening charges.
\bigskip

A key observation is expressed by the following formula displaying
the subleading  corrections in the relation between 
the fields $(\phi^{({\rm L},{\rm R})},\alpha^{({\rm L},{\rm R})})$ and $(\phi,\alpha)$
in the limit $\phi_0\rightarrow -\infty$:
\bea\label{oias9821oix}
\phi(t,x)&\asymp&\phi^{({\rm L})}+\phi^{({\rm R})}+\tfrac{1}{4}\,\re^{2(\phi^{({\rm L})}+\phi^{({\rm R})})}
-\int_0^{t+x}\!\!\!\!\rd x_+ \  V_{\rm cig}^{({\rm L})}(x_+)\;
\int_0^{t-x}\!\!\!\!\rd x_- \ V_{\rm cig}^{({\rm R})}(x_-)+\ldots \\[0.2cm]
\alpha(t,x)&\asymp&\alpha^{({\rm L})}+\alpha^{({\rm R})}+\tfrac{1}{2}\,\re^{2\phi^{({\rm L})}}
\int_0^{t-x}\rd x_- \, V_{\rm cig}^{({\rm R})}(x_-)+\tfrac{1}{2}\,\re^{2\phi^{({\rm R})}}
\int_0^{t+x}\rd x_+\, V_{\rm cig}^{({\rm L})}(x_+)+\ldots\nonumber
\eea
The fields $V_{\rm cig}^{({\rm L/R})}(x_\pm)$ appearing here
are given by
\be\label{mncxjasdjh}
V_{\rm cig}^{({\rm L})}=\partial_+\alpha^{({\rm L})}\,\re^{2\phi^{({\rm L})}}\,,
\qquad\qquad\qquad V_{\rm cig}^{({\rm R})}=\partial_-\alpha^{({\rm R})}\,\re^{2\phi^{({\rm R})}}
\ee
and will turn out to be closely related to the classical analogs of
the screening charges relevant for our problem. Formula \eqref{oias9821oix}
can be verified as follows. One starts with the map between
$(\phi^{({\rm L},{\rm R})},\alpha^{({\rm L},{\rm R})})$ and $(\phi,\alpha)$
given by eqs.\;\eqref{ioas8912d},\;\eqref{omegajdjdjd} and \eqref{oisa9821sa}
and shows that the
 matrix $\bm{\omega}(t,x)$ 
can be brought to the form
\bea\label{sakxmzmnzxas}
\!\!\!\!\!\!\!\!\!\!\bm{\omega}(t,x)&=&\re^{-\ri\alpha^{({\rm L})}(t+x)\,\sigma^z} \,  \re^{(\frac{1}{2}\log(2)-\phi^{({\rm L})}(t+x))\,\sigma^x}\ 
\exp\bigg(
\tfrac{1}{2}\,(\sigma^y+\ri\sigma^z)\,\int_0^{t+x}\rd x_+ \,V_{\rm cig}^{({\rm L})}(x_+)\bigg)\\[0.2cm]
&\times&\exp\bigg(
\tfrac{1}{2}\,(\sigma^y-\ri\sigma^z)\,\int_0^{t-x}\rd x_-\,V_{\rm cig}^{({\rm R})}(x_-)\bigg)\ 
\re^{(\frac{1}{2}\log(2)-\phi^{({\rm R})}(t-x))\,\sigma^x}\re^{\ri\alpha^{({\rm R})}(t-x)\,\sigma^z}\ .\nonumber
\eea
This, itself, is a consequence of the elementary relations
\bea
\partial_+-\psi_-^{({\rm L})}\,\sigma^+-\psi_+^{({\rm L})}\,\sigma^-&=&
\re^{-\ri\alpha^{({\rm L})}\sigma^z}\,\re^{-\phi^{({\rm L})}\sigma^x}\ \big(\,\partial_+-(\sigma^y+\ri\sigma^z)\,
V_{\rm cig}^{({\rm L})}\,\big)\ \re^{+\phi^{({\rm L})}\sigma^x}\,\re^{+\ri\alpha^{({\rm L})}\sigma^z} \nonumber\\[0.2cm]
\partial_-+\psi_-^{({\rm R})}\,\sigma^++\psi_+^{({\rm R})}\,\sigma^-&=&
\re^{-\ri\alpha^{({\rm R})}\sigma^z}\,\re^{+\phi^{({\rm R})}\sigma^x}\ \big(\,\partial_-+(\sigma^y-\ri\sigma^z)\,
V_{\rm cig}^{({\rm R})}\,\big)\ \re^{-\phi^{({\rm R})}\sigma^x}\,\re^{+\ri\alpha^{({\rm R})}\sigma^z} \,. \nonumber
\eea
Since $(\sigma^y\pm\ri\sigma^z)\,V_{\rm cig}^{({\rm L,R})}(x_\pm)$
 commutes with itself
for different values of $x_\pm$, the path ordered exponent becomes an ordinary integral
as in \eqref{sakxmzmnzxas}.
 Note that the constant ${\rm SL}(2)$ matrix $\bm{\omega}_0$ was set to 
$\re^{-\ri\alpha^{({\rm L})}(0)\,\sigma^z}\,\re^{-(\phi^{({\rm L})}(0)+\phi^{({\rm R})}(0)-\log(2))\,\sigma^x}\,
\re^{\ri\alpha^{({\rm R})}(0)\,\sigma^z}$
 in order to simplify the final result, which is equivalent to
fixing the zero modes of
$\phi^{({\rm L})}+\phi^{({\rm R})}$ and $\alpha^{({\rm L},{\rm R})}$.

\bigskip

We now return 
to the question that was previously raised: suppose one has a Lorentz spin $m$
differential polynomial $F_m(x_+)$ in $\partial_+\alpha^{({\rm L})}$ and $\partial_+\phi^{({\rm L})}$.
Which conditions ensure that the free field realization of $F_m(x_+)$ is related to a
 local expression in 
 the fundamental fields $\phi$ and $\alpha$? A necessary requirement is that
the equal-time Poisson brackets  of the polynomial $F_m(t+x)$ with the fundamental fields
$\phi(t,y)$ and $\alpha(t,y)$ should not contain the step function
$\epsilon(x-y)$. The expression for 
$\phi(t,y)$ and $\alpha(t,y)$  in terms of the free fields 
$\phi^{({\rm L},{\rm R})}$ and $\alpha^{({\rm L},{\rm R})}$ can be approximated 
by \eqref{oias9821oix}.
Inspection of that formula   shows that terms proportional to 
the step function
$\epsilon(x-y)$ could originate from the Poisson bracket of $F_m(x_+)$
with $V_{\rm cig}^{({\rm L})}$. It
 takes the form of a sum
 $\sum_{j=0}^m C_j(x_+)\, \delta^{(j)}(x_+-y_+)$ with field dependent  coefficients 
$C_j(x_+)$ so that if  $C_0(x_+)$ does not vanish a subsequent 
integration w.r.t. $y_+$ would yield a step function. Therefore a necessary condition 
for locality is that
\be\label{oias89cjjkaas}
\big\{F_m(x_+)\,,V_{\rm cig}^{({\rm L})}(y_+)\big\}=
\sum_{j=0}^{m}C_j(x_+)\,\delta^{(j)}(x_+-y_+)\qquad {\rm with} \qquad  C_0(x_+)=0\, .
\ee
One can check that the generators $T_m^{({\rm L})}$ of the classical
${\cal W}_\infty^{({\rm L})}$-algebra, explicitly given for $m=2,3,4$ 
in \eqref{oias9812},
satisfy \eqref{oias89cjjkaas}.  It turns out that all solutions 
to the conditions \eqref{oias89cjjkaas} 
represent elements of the ${\cal W}_\infty^{({\rm L})}$-algebra.

\bigskip

We may note that \eqref{oias89cjjkaas}  ensures that the Poisson bracket 
$\{F_m(x_+),V_{\rm cig}^{({\rm L})}(y_+)\}$ is a total derivative w.r.t. 
$y_+$ of an expression that does not contain the $\epsilon$ - function. 
This condition is unchanged under the replacement
\be
V_{\rm cig}^{({\rm L})}(y_+)\mapsto V_{\rm cig}^{({\rm L})}(y_+) + c_{\rm cig}\,
\partial_+\big(\re^{2\phi^{({\rm L})}(y_+)}\big)\,,
\ee
where  $c_{\rm cig}$ is an arbitrary constant, since the Poisson brackets of $F_m(x_+)$
with the added term 
would  automatically be a total derivative in $y_+$. For reasons that will
be explained in the next subsection, it is
convenient to introduce the dynamical quantity
\be\label{aios9821oi111a}
{\cal X}_0 =
\int \rd y_+\;V_+^{(1)}(y_+)\qquad\qquad {\rm with}
\qquad\qquad V_+^{(1)}=\sqrt{1+\nu^2}\,\big(\ri\,\partial_+\phi^{({\rm L})}+\partial_+\alpha^{({\rm L})}\big)\,
\re^{2\phi^{({\rm L})}}\ .
\ee
As discussed above, the requirement on the density \eqref{oias89cjjkaas} is unchanged if $V_{\rm cig}^{({\rm L})}$
\eqref{mncxjasdjh}
is swapped for $V_+^{(1)}$. Moreover, by formally integrating both sides of that equation
w.r.t.  $y_+$,  ignoring any boundary terms, it can be equivalently re-written as
\be\label{oias9821iud}
\{F_m(x_+)\,,\,{\cal X}_0\}\sim 0
\ee
with the notation $\sim$ indicating equality up to  terms
that vanish when $\re^{2\phi^{({\rm L})}}$ is periodic. 
We refer to the obtained formula as the condition of quasi-commutativity with the classical screening 
charge ${\cal X}_0$. Thus one arrives at the conclusion that  
the classical  ${\cal W}_\infty^{({\rm L})}$-algebra 
can  be characterised as the set of differential polynomials  formed out of 
$\partial_+\phi^{({\rm L})}$ and $\partial_+\alpha^{({\rm L})}$
that quasi-commute with
the classical screening charge.

\subsection{Classical screening charges and local IMs\label{oias98oia}}
The integrable structure of the Klim\v{c}\'{i}k model 
induces a set of local and non-local IMs in the field theory underlying its conformal limit.
One may use the previous discussion in order to 
get a Poisson-algebraic characterisation of the local conserved quantities.
The key idea is to use an approximate description of  
the target  of the ${\rm SU}(2)$ Klim\v{c}\'{i}k model
for $\kappa$ close to $1$ as the manifold defined by 
gluing the two cigars associated to chart I
and III to the cylinder representing chart II, see Figure \ref{oias8912}. We will identify 
the local IMs as the dynamical
quantities that are conserved 
in both  of the cigar NLSMs depicted on the left and right sides of that figure.  
The resulting conditions are then
reformulated algebraically  in terms of  a quasi-commutativity condition with the
classical screening charges.  
\bigskip

\bigskip

Consider 
the ${\rm SU}(2)$ Klim\v{c}\'{i}k model as described by  charts I, II and III that were introduced in the previous section. 
For $1-\kappa$ very small  we expect 
that there exist field configurations such that $(\phi,\alpha)$
take values sufficiently far from the tips of both cigars, where
the target space manifold is approximately flat, \textit{i.e.} in chart II.
In that coordinate chart, using the left moving fields
$\phi^{({\rm L})}_1,\phi^{({\rm L})}_2,\phi^{({\rm L})}_3$  defined in eq.\;\eqref{fields11},
the local IMs can be represented in the form 
\be\label{oias98soii}
{\cal Q}_{m-1}^{({\rm L})}=\int \rd x\  F_{m}^{\rm (II)}\big[\phi^{({\rm L})}_1,\phi^{({\rm L})}_2,\phi^{({\rm L})}_3\big]\,,
\ee
where $F_{m}^{\rm (II)}$ is a differential polynomial of Lorentz spin $m$ in the fields
$\partial_+\phi^{({\rm L})}_1,\partial_+\phi^{({\rm L})}_2,\partial_+\phi^{({\rm L})}_3$.
\bigskip

As long as $1-\kappa$ is small but non-vanishing, chart II would
overlap with chart I. Up to corrections of order $(1-\kappa)^2$,
the action in that chart is given by that of the cigar NLSM 
${\cal A}_{{\rm cig}}[\phi,\alpha]$ plus  free field $\chi$, see formula \eqref{oias981asas2as}, where
the fields $\alpha$ and $\chi$ are defined in eq.\;\eqref{alp1}.
We  expect that the local conserved 
quantities of the ${\rm SU}(2)$ Klim\v{c}\'{i}k model in the conformal limit are contained in a subset
of those for the cigar sigma model and free bosonic field:
\be\label{oias98soi}
{\cal Q}_{m-1}^{({\rm L})}=\int \rd x\  F_{m}^{\rm (I)}\big[T_j^{({\rm L})},\partial_+\chi\big]\,.
\ee
Staying away from the tip of the cigar NLSM  allows one to approximate the expressions for 
$T_j^{({\rm L})}$ containing factors $(1+\re^{2\phi})^{-1}$, see for example 
eqs.\;\eqref{aosoic98aaaa2} and \eqref{aosoic98aaaa3}, 
by the 
expressions \eqref{oias9812} involving the free chiral fields 
$\alpha^{({\rm L})}$ and $\phi^{({\rm L})}=\phi_3^{({\rm L})}$. Also, the decoupled boson $\chi$ 
can be written as the sum of a left moving and right moving field,
\be
\chi(t,x)=\chi^{({\rm L})}(t+x)+\chi^{({\rm R})}(t-x)+\ldots
\ee
up to corrections that vanish for $1-\kappa$ small. Comparing eqs.\;\eqref{oias98soii} and \eqref{oias98soi}, with $\alpha^{({\rm L})}$ and $\chi^{({\rm L})}$ expressed in terms of 
$\phi^{({\rm L})}_1$ and $\phi^{({\rm L})}_2$,
we arrive at a first constraint on the form of 
$F_{m}^{\rm (II)}$. 

\bigskip

The key requirement to be imposed is that the local IMs are
simultaneously contained in the set of local conserved quantities of the 
cigar sigma model and free bosonic field associated to chart III. This means that,
 in the overlap of charts II and III, 
${\cal Q}_{m-1}^{({\rm L})}$ admits
a representation of the form
\be\label{oias98soiaaa}
{\cal Q}_{m-1}^{({\rm L})}=\int \rd x\  {F}_{m}^{\rm (III)}\big(\tilde{T}_j^{({\rm L})},\partial_+\tilde{\chi}\big)\,,
\ee
where $\tilde{T}_j^{({\rm L})}$ are the left moving currents for the  cigar with
action 
${{\cal A}}_{{\rm cig}}[-\phi,\tilde{\alpha}]$
\eqref{oias8912sa}  and $\tilde{\chi}$ is   given in eq.\;\eqref{alp2}.
Using the approximate relations  $\tilde{\alpha}\asymp \tilde{\alpha}^{({\rm L})}(x_+)+
\tilde{\alpha}^{({\rm R})}(x_-)$ and  $\tilde{\chi}\asymp \tilde{\chi}^{({\rm L})}(x_+)+
\tilde{\chi}^{({\rm R})}(x_-)$, 
and then swapping  $\big(\tilde{\alpha}^{({\rm L})},\tilde{\chi}^{({\rm L})}\big)$
in favour of
$\big(\phi^{({\rm L})}_1,\phi^{({\rm L})}_2\big)$
a second ansatz for the form of 
$F_{m}^{\rm (II)}$ is obtained. The agreement of the two 
expressions  holds if the densities 
appearing in \eqref{oias98soi} and \eqref{oias98soiaaa}
coincide up to a total derivative, i.e.,
\be\label{oiaio8siujk1}
F_{m+1}^{\rm (I)}\big(T_j^{({\rm L})},\partial_+\chi\big)=
{F}_{m+1}^{\rm (III)}\big(\tilde{T}_j^{({\rm L})},\partial_+\tilde{\chi}\big)+\partial_+(\ldots)\qquad
{\rm in\;\, chart\ II}.
\ee
This requirement turns out to be  sufficiently strong 
to fix the local conserved charges completely.

\bigskip

Let us illustrate  how this works concretely, starting from 
\eqref{oias98soi} on the example of $F_2$. 
First, one needs at hand the relations between the pairs $\big(\alpha^{({\rm L})},\chi^{({\rm L})}\big)$,
$\big(\tilde{\alpha}^{({\rm L})},\tilde{\chi}^{({\rm L})}\big)$ and $\big(\phi_1^{({\rm L})},\phi_2^{({\rm L})}\big)$. 
These can be easily established from the definitions \eqref{alp1}, \eqref{alp2}, \eqref{fields11}
which yields
\begin{equation}\label{I-II-III}
\begin{aligned}
& \alpha^{({\rm L})}=\frac{\nu  \phi_1^{({\rm L})}+\phi_2^{({\rm L})}}{\sqrt{1+\nu^2}}\,,\qquad
\chi^{({\rm L})}=\frac{\phi_1^{({\rm L})}-
\nu \phi_2^{({\rm L})}}{\sqrt{1+\nu^2}}\, ,
\qquad ({\rm chart\ I})\cap ({\rm chart\ II})\\[0.2cm]
&\tilde{\alpha}^{({\rm L})}=
\frac{\nu \phi_1^{({\rm L})}-\phi_2^{({\rm L})}}{\sqrt{1+\nu^2}}\,,
\qquad
\tilde{\chi}^{({\rm L})}= \frac{\phi_1^{({\rm L})}+
\nu\phi_2^{({\rm L})}}{\sqrt{1+\nu^2}}\,,\qquad  ({\rm chart\ II})\cap ({\rm chart\ III})\,.
\end{aligned}
\end{equation}
Then the most general form for $F_2^{\rm (I)}$, ignoring total derivatives
and fixing the overall normalization, is
$F_2^{\rm (I)}=T_2^{({\rm L})}+c_2\,(\partial_+\chi)^2$ with $c_2$ an arbitrary numerical constant
and $T_2^{({\rm L})}=(\partial_+\phi^{({\rm L})})^2+(\partial_+\alpha^{({\rm L})})^2$.
In  a similar way ${F}_2^{\rm (III)}=(\partial_+\phi^{({\rm L})})^2+
(\partial_+\tilde{\alpha}^{({\rm L})})^2+\tilde{c}_2\,(\partial_+\tilde{\chi})^2$.
Re-writing these expressions in terms of $\phi_j^{({\rm L})}$ 
via formula \eqref{I-II-III} one finds that \eqref{oiaio8siujk1} implies
$c_2=\tilde{c}_2=1$. Hence
\be
{\cal Q}_1^{({\rm L})}=\int\rd x_+\,\Big(\big(\partial_+\phi_1^{({\rm L})}\big)^2+
\big(\partial_+\phi_2^{({\rm L})}\big)^2+\big(\partial_+\phi_3^{({\rm L})}\big)^2\Big)\, .
\ee
Notice that the  integrand is the Lorentz spin $+2$ component of the stress-energy momentum tensor
considered in chart II.

\bigskip

We are now going to argue that the requirement \eqref{oiaio8siujk1} is equivalent to the pair 
of conditions 
\be\label{quasi-comm}
\{{\cal Q}_m^{({\rm L})},{\cal X}_0\}\,\sim 0,\qquad\qquad
\{{\cal Q}_m^{({\rm L})},{\cal X}_1\}\,\sim 0\,.
\ee
Here the classical screening charges ${\cal X}_0$ and ${\cal X}_1$ are defined as follows
\be\label{oias9012oia--a}
{\cal X}_0=\int \rd x_+ V_+^{(1)}\,,\qquad\qquad
{\cal X}_1=\int\rd x_+ V_-^{(1)}
\ee
with
\be\label{oias9012oia--b}
V_\pm^{(1)}=\big(\ri\sqrt{1+\nu^2}\ 
\partial_+\phi^{({\rm L})}_3+\partial_+\phi^{({\rm L})}_2
\pm
\nu\partial_+\phi_1^{({\rm L})}\big)\,\re^{\pm2\phi^{({\rm L})}_3}\, .
\ee
Indeed, consider the first of the Poisson quasi-commutativity conditions involving ${\cal X}_0$.
The integrand of the screening charge, with $\phi_1^{({\rm L})}$ and $\phi_2^{({\rm L})}$
swapped for $\chi^{({\rm L})}$ and $\alpha^{({\rm L})}$ according to eq.\;\eqref{I-II-III}
takes the form 
$V_+^{(1)}=\sqrt{1+\nu^2}\,\big(\ri\,\partial_+\phi_3^{({\rm L})}+\partial_+\alpha^{({\rm L})}\big)\,
\re^{2\phi_3^{({\rm L})}}$.
In particular, it contains no dependence on the field $\chi^{({\rm L})}$, while $\alpha^{({\rm L})}$ and
$\phi^{({\rm L})} = \phi^{({\rm L})}_3$ appear in the same combination as for  $V_+^{(1)}$  from \eqref{aios9821oi111a}.
As such, it follows from the discussion in the previous subsection that 
the condition $\{F_{m}^{\rm (I)}(x_+),{\cal X}_0\}\,\sim 0$ 
is solved by densities  of the form 
$F_{m}^{\rm (I)}\big[T_j^{({\rm L})},\partial_+\chi\big]$.
This way, the first formula in \eqref{quasi-comm}  is equivalent to the 
  ansatz \eqref{oias98soi}.
In an analogous way, by noticing that $V_-^{(1)}=
\sqrt{1+\nu^2}\,\big(\ri\,\partial_+\phi_3^{({\rm L})}+\partial_+\tilde{\alpha}^{({\rm L})}\big)\,
\re^{-2\phi_3^{({\rm L})}}$
 one sees that ${\cal X}_1$ is the classical screening charge corresponding to the cigar with action ${{\cal A}}_{{\rm cig}}[-\phi,\tilde{\alpha}]$. Hence, one concludes that $\{{\cal Q}_m^{({\rm L})},{\cal X}_1\}\,\sim 0$ is equivalent to
 the representation \eqref{oias98soiaaa}. In order to satisfy 
both conditions at the same time,  we need to have \eqref{oiaio8siujk1}. 
\bigskip

Formula \eqref{quasi-comm} 
can be analysed directly. 
One starts by considering an arbitrary 
differential polynomial in $\partial_+\phi_j^{({\rm L})}$ 
of the  form
\be
F_{m}=\sum_{i_1,\ldots,i_k\ge 1\atop
j_1,\ldots,j_k=1,2,3}
F_{i_1\ldots i_k}^{j_1\ldots j_k}\ 
\big(\partial_+^{i_1}\phi_{j_1}^{({\rm L})}\big)
\ldots \big(\partial_+^{i_k}\phi_{j_k}^{({\rm L})}\big)\qquad {\rm with}
\qquad i_1+\ldots+i_k=m\,.
\ee
Using the free field Poisson bracket relations for $\phi_j^{({\rm L})}$ \eqref{ioas9821}, 
 one computes the PBs of $F_{m}(x_+)$ with the field $V_+^{(1)}(y_+)$.  
Poisson quasi-commutativity with the screening charge ${\cal X}_0$ requires
that the term in front of the $\delta$-function vanishes,
as in eq.\;\eqref{oias89cjjkaas}. This leads to a set of linear relations among 
$F_{i_1\ldots i_k}^{j_1\ldots j_k}$, which can be used to determine some
of the coefficients. The next step involves the screening charge ${\cal X}_1$.
Imposing a requirement similar to \eqref{oias89cjjkaas} with $V_+^{(1)}$ 
swapped for $V_-^{(1)}$ is too restrictive. The density $F_{m+1}(x_+)$
can only be expressed in terms of $\tilde{T}_j^{({\rm L})}$ and $\partial_+\tilde{\chi}$
up to a total derivative, see \eqref{oiaio8siujk1}, and the Poisson brackets
of this derivative term with $V_-^{(1)}(y_+)$ would lead to a non-zero 
$C_0(x_+)$.
Nevertheless, it must be the case that
\be\label{mncnmjkasmnz}
\big\{F_m(x_+)\,,V_{-}^{(1)}(y_+)\big\}=
\sum_{j=0}^{m}C_j(x_+)\,\delta^{(j)}(x_+-y_+)\qquad {\rm with} \qquad 
C_0(x_+)=\partial_+(\ldots)\, .
\ee
Regarding  the above as a condition on the density
  typically  yields an overconstrained system
of linear equations on the remaining coefficients
$F_{i_1\ldots i_k}^{j_1\ldots j_k}$ which determines them
up to an overall multiplicative factor (that may be zero). 
Applying the above procedure, one finds that 
\be\label{oias891289d}
{\cal Q}_{2m-1}^{({\rm L})}=\int \rd x \ \big({T}_2^{({\rm L})}+(\partial_+\chi)^2\,\big)^m=
\int \rd x \ \big(\tilde{{T}}_2^{({\rm L})}+(\partial_+\tilde{\chi})^2\,\big)^m\qquad\qquad ({\rm chart\ II})
\ee
with $m=1,2,\ldots$, while there are no  local IMs 
labeled by an even subscript that obey \eqref{quasi-comm}.
Note that the integrand in ${\cal Q}_{2m-1}^{({\rm L})}$
is nothing but the $m$-th power of the Lorentz spin $+2$ component of the stress energy momentum tensor for
the   ${\rm SU}(2)$ Klim\v{c}\'{i}k model at $\kappa\to 1^-$, as already mentioned in 
the introduction to this section.

\bigskip

One may find it intriguing to  observe that 
the integrands in \eqref{oias9012oia--a} coincide with the fields 
$V_\pm^{(1)}$ that appear in the connection ${\cal L}_{\scriptscriptstyle {\rm UV}}^{(\omega)}$
at $\omega=1$, see eqs.\;\eqref{Anew} and \eqref{oi89xciusa}. We will  elaborate 
on the implications of this observation in Section \ref{sec666}.

\subsection[Corner-brane ${\cal W}$-algebra]{Corner-brane \texorpdfstring{$\bm{\Wc}$}{W}-algebra}\label{ias9812211aaa}

It follows from our discussion in Subsection \ref{sec:free} that the screening 
charges describe the leading order deviations of the dynamics from the
free field dynamics, as expressed by \eqref{oias9821oix}. We have 
furthermore seen
that there is a Poisson-algebraic characterisation of the 
local charges that remain conserved under these perturbations,
expressed by the conditions \eqref{quasi-comm}. 
These  offer a concise characterisation of the integrable 
structure in terms of the screening charges, representing a classical 
analog of the description developed by Feigin and Frenkel 
for affine Toda theories in \cite{Feigin:1991qm} to our case. 
It seems interesting to note that
the local IMs  ${\cal Q}_m^{({\rm L})}$ possess  another characterisation 
in terms of a second pair of screening charges, 
\begin{equation}\label{oasoio90aaa}\begin{aligned}
&{\cal X}^{\scriptscriptstyle \rm CB}_0=\int \rd x_+\;
V_{+}^{(-1)}(x_+)\,,
\qquad\qquad
{\cal X}^{\scriptscriptstyle \rm CB}_1=\int \rd x_+\;
V_{-}^{(-1)}(x_+)\,
\\[0.2cm]
&V_{\pm}^{(-1)}=
\big(\,\ri\sqrt{1+\nu^2}\ \partial_+\phi_3^{({\rm L})}
+\partial_+\phi_2^{({\rm L})}\pm
\nu\partial_+\phi_1^{({\rm L})}\,\big)\,\re^{\pm 2\ri\sqrt{1+\nu^2}\,\phi_2^{({\rm L})}}.
\end{aligned}\end{equation}
Here the notation  $V_{\pm}^{(-1)}$ reflects the fact that these fields appear
in the matrix elements  of the family of Lax connections ${\cal L}_{\scriptscriptstyle {\rm UV}}^{(\omega)}$ \eqref{Anew}
specialized to $\omega=-1$, compare the above formula  with \eqref{oi89xciusa}.
Poisson quasi-commutativity with ${\cal X}_0^{\scriptscriptstyle \rm CB}$
gives back a ${\cal W}_\infty\otimes {\rm U}(1)$ algebra, where the currents are realized differently 
in terms of the three bosonic fields $\phi^{({\rm L})}_j$. The two conditions 
taken together: 
\be\label{quasi-comm1A}
\{{\cal Q}_m^{({\rm L})},{\cal X}_0^{\scriptscriptstyle \rm CB}\}\,\sim 0,\qquad\qquad
\{{\cal Q}_m^{({\rm L})},{\cal X}_1^{\scriptscriptstyle \rm CB}\}\,\sim 0
\ee
yield the same set of local IMs \eqref{oias891289d}. Although the form of ${\cal Q}_{2m-1}^{({\rm L})}$
is somewhat trivial for the classical field theory, 
let us mention that the analogue of these statements holds in the quantum case \cite{Lukyanov:2012wq}.
There it is far from obvious that the set of local IMs of the quantum Fateev integrable structure,
which are highly non-trivial densities built from three bosonic fields
(see, e.g., Appendix A of \cite{Lukyanov:2012wq}), can be characterized in these two different ways.
\bigskip

Closely related  is the fact that
the integrable structure of the ${\rm SU}(2)$  Klim\v{c}\'{i}k model
has a remarkable feature: the local densities used to construct 
conserved charges can be taken from an honest
subalgebra of ${\cal W}_\infty\otimes {\rm U}(1)$
which is a non-trivial ${\cal W}$-algebra of its own.
At the classical level, its generators $W_m$
are defined through
the pair of conditions 
\be\label{oias982ojkd}
\{W_m(x_+)\,,\,{\cal X}_0\}\sim 0\qquad\quad {\rm and}\qquad\quad
\{W_m(x_+)\,,\,{\cal X}^{\scriptscriptstyle \rm CB}_1\}\sim 0\, .
\ee
Notice that the above  turns out to be a weaker requirement
than simultaneous Poisson quasi-commutativity  with 
${\cal X}_0$ and ${\cal X}_1$   \eqref{quasi-comm} or 
${\cal X}^{\scriptscriptstyle \rm CB}_0$ and  ${\cal X}^{\scriptscriptstyle \rm CB}_1$
\eqref{quasi-comm1A}. The last two are satisfied only by the integrated quantities
${\cal Q}_{2m-1}^{({\rm L})}$, while there exist non-trivial differential polynomials
in $\partial_+\phi_j^{({\rm L})}$ that obey \eqref{oias982ojkd}.
\bigskip

As will be pointed out in Subsection \ref{aaaaaaaaaaaa}, the Poisson-VOA generated by the densities $W_m$
is isomorphic to the  classical chiral algebra of the coset
\begin{equation}\label{coset-W}
\frac{\widehat{\mathfrak{sl}}_{\ell_1^{({\rm L})}}(2)\oplus \widehat{\mathfrak{sl}}_{\ell_2^{({\rm L})}}(2)}
{\widehat{\mathfrak{sl}}_{\ell_1^{({\rm L})}+\ell_2^{({\rm L})}}(2)}
\qquad\qquad
\begin{aligned}
&\ell_1^{({\rm L})}=K_{\scriptscriptstyle\rm UV}/(1+\nu^{-2})\\
&\ell_2^{({\rm L})}=K_{\scriptscriptstyle\rm UV}/(1+\nu^{2})\,.
\end{aligned}
\end{equation}
The quantum version of this statement has been proven  in ref. \cite{Semikhatov:2001zz}.
In order to distinguish the two ${\cal W}$-algebras, whose isomorphism is not obvious, we'll refer to
the one obtained  by means of  the screening
charges as the corner-brane ${\cal W}$-algebra.
This  terminology
comes from the work  \cite{Lukyanov:2012wq}, where 
the algebra appeared in the study of a 2D QFT with boundary interaction
called the ``pillow-brane'' model.
\bigskip

As an illustration, we are going to present   the first two non-trivial 
currents satisfying \eqref{oias982ojkd}. Unlike the case of the classical
 ${\cal W}_\infty\otimes{\rm U}(1)$  algebra, there is no Lorentz spin $+1$ field.
The simplest left moving field turns out to be the Lorentz spin $+2$ component
of the stress-energy momentum tensor:
\be\label{asoi9812}
W_2^{({\rm L})}=K_{\scriptscriptstyle\rm UV}\,\big(\,T_2^{({\rm L})}+(\partial_+\chi)^2\,\big)\, .
\ee
The space of spin $+3$ local fields is spanned by the ``descendent'' $\partial_+ W_2^{({\rm L})}$.
At spin $+4$, one has the non-trivial current
\bea\label{opcx90sa}
W_4^{({\rm L})}&=&K_{\scriptscriptstyle\rm UV}\,\Big(\,T_4^{({\rm L})}-(\nu+\nu^{-1})^2\,\big(T_2^{({\rm L})}\big)^2-
2\,(\nu-\nu^{-1})\,\partial_+\chi\,T_3^{({\rm L})}
-\ri\,(\nu+\nu^{-1})\,\partial_+\chi\,\partial_+
T_2^{({\rm L})} \nonumber \\
&&\qquad\quad- 4\,(\partial_+\chi)^2\,T_2^{({\rm L})}+
2\ri\,(\nu+\nu^{-1})\,\partial^2_+\chi\,T_2^{({\rm L})}+
(\partial_+^2\chi)^2\,\Big)\ .
\eea
Starting directly from the PB relations of the classical
${\cal W}_\infty$ algebra \eqref{jasususa} one can check by direct calculations that the PBs of the fields $W_m^{({\rm L})}$ form a closed algebra,
at least for the first few currents of lowest Lorentz spin (see also 
eqs.\;\eqref{Eq:PbW2} and \eqref{jciuxhjsa1} below).

\bigskip

The appearance of the coset ${\cal W}$-algebra \eqref{coset-W}
in the  ${\rm SU}(2)$  Klim\v{c}\'{i}k model may seem quite
surprising. We will later see that 
the affine Gaudin model  formulation offers a natural 
explanation of this remarkable feature. 

\section{Klim\v{c}\'{i}k model in conformal limit III: as chiral AGM\label{sec555389i}}

The goal of this section is to revisit the conformal limit of the Klim\v{c}\'ik model from the point of view of affine Gaudin models. More precisely, starting from the description made in Section \ref{Sec:KlimAGM} of the Klim\v{c}\'{i}k model as a relativistic realisation of an AGM, we will show that this structure decomposes into two decoupled AGM realisations in the conformal limit, describing the left-moving and right-moving degrees of freedom of the UV fixed-point. In particular, we will explain how the integrable and chiral structures of this conformal limit (described in the previous sections for the ${\rm SU}(2)$ case) naturally arise in the formalism of affine Gaudin models.

\subsection{Conformal limit of the AGM underlying the Klim\v{c}\'{i}k model}
\label{Sec:UVLimAGM}

\paragraph{Twist function and punctures.} Recall that the non-chiral Klim\v{c}\'{i}k model, before taking the conformal limit, is described as a relativistic realisation of the AGM with twist function $\vp(z)$ defined in eq.\;\eqref{Eq:TwistKlim}. In particular, this AGM possesses 4 punctures $z_r$ and 4 levels $\ell_r$, given respectively by eq.\;\eqref{Eq:zrKlim} and eq.\;\eqref{Eq:LevelsKlim}. It is clear from the latter equation and the RG flow \eqref{hassasaty} of the Klim\v{c}\'{i}k model that these levels $\ell_r$ are RG invariants. In contrast, the punctures $z_r$ are not invariant and flow to either $0$ or $\infty$ in the conformal limit. Recall however from Subsection \ref{sec341} that the coordinate $z$ on $\mathbb{CP}^1$ is not the best suited choice of spectral parameter to describe the chiral limit of the Klim\v{c}\'{i}k model: indeed, it was shown in this subsection that one should consider instead the rescaled spectral parameters $z^\cL = \xi\,z$ and $z^\cR = \xi\,z^{-1}$, which allow one to single out respectively the left-moving
 fields and right-moving fields contained in the Lax matrix in the conformal limit. In particular, since the parameter $\xi$ tends to 0 in this limit, these rescalings of the spectral parameter have some important consequences on the structure of the AGM underlying the construction and the conformal limit of its punctures.\\

To describe this more concretely, let us focus for the moment on the spectral parameter $z^\cL = \xi\,z$. In this coordinate, the AGM underlying the Klim\v{c}\'{i}k model is described by the twist function $\xi^{-1}\vp\bigl(\xi^{-1}z^\cL\bigr)$, with poles at the points $(\xi z_1,\xi z_2,\xi z_3,\xi z_4)$. The latter do not all flow to $0$ or $\infty$, contrarily to the unrescaled punctures $z_r$, but rather to $(0,4/(1+\nu^2),0,-4\nu^2/(1+\nu^2))$. It will be useful to consider the UV limit of the twist function for the spectral parameter $z^\cL$, \textit{i.e.} the quantity
\begin{equation}
\vp^\cL(z^\cL) = \lim_{\tau \to -\infty} \xi^{-1} \vp\bigl( \xi^{-1} z^\cL \bigr).
\end{equation}
A direct computation shows that
\begin{equation}\label{Eq:TwistKlimL}
\vp^\cL(z^\cL) = \frac{16\nu^2}{(1+\nu^2)^2} \frac{\Kuv}{z^\cL \left( z^\cL - \frac{4}{1+\nu^2}\right) \left( z^\cL + \frac{4\nu^2}{1+\nu^2}\right)},
\end{equation}
where $\Kuv$ is defined in eq.\;\eqref{Llimitk}. 
The 1-form $\vp^\cL(z^\cL)\dd z^\cL$ has 3 simple poles, located at
\begin{equation}\label{Eq:zL}
z^\cL_1 = \frac{4}{1+\nu^2}, \qquad z^\cL_2 = -\frac{4\nu^2}{1+\nu^2}, \qquad z^\cL_3 = 0.
\end{equation}
This is quite natural: indeed, we have observed above that the punctures $\xi z_r$ of the initial AGM flow to these points in the conformal limit. Note that one could have expected $\vp^\cL(z^\cL)\dd z^\cL$ to have a double pole at $0$, since both $\xi z_1$ and $\xi z_3$ flow to $0$. Recall however that the initial twist function possesses a simple zero at $0$: the two simple poles $\xi z_1$ and $\xi z_3$ therefore collide with this zero, resulting in only a simple pole at $z^\cL_3 = 0$ in the limit. To summarise, the conformal limit makes us pass from a twist function with 4 poles and 2 zeroes to one with 3 poles and 1 zero. This collision procedure and the resulting reduction of the number of punctures will play an important role in the description of the AGM structure underlying the chiral Klim\v{c}\'{i}k model. We will see in Subsection \ref{genAGM} that similar phenomena also appear for more general models. One easily checks that the residues $\displaystyle\ell^\cL_r = \res_{z^\cL={z^\cL_r}} \vp^\cL(z^\cL)\dd z^\cL$ are given by
\begin{equation}\label{Eq:LevelsL}
\ell^\cL_1 = \frac{\Kuv}{1+\nu^{-2}}, \qquad \ell^\cL_2 = \frac{\Kuv}{1+\nu^2}, \qquad \ell^\cL_3 = - \Kuv.
\end{equation}
As one can expect, the first two levels $\ell^\cL_1$ and $\ell^\cL_2$ are respectively equal to the levels $\ell_2$ and $\ell_4$ of the AGM before taking the conformal limit, while the last one $\ell^\cL_3$ is equal to $\ell_{1}+\ell_{3}$.\\

Similar results hold for the AGM considered in the spectral parameter $z^\cR = \xi\,z^{-1}$. In particular, one finds
\begin{equation}
\vp^\cR(z^\cR) = \lim_{\tau \to -\infty} - \frac{\xi}{z^\cR\,^2} \, \vp\left( \frac{\xi}{z^\cR} \right) = -\frac{16\nu^2}{(1+\nu^2)^2} \frac{\Kuv}{z^\cR \left( z^\cR - \frac{4}{1+\nu^2}\right) \left( z^\cR + \frac{4\nu^2}{1+\nu^2}\right)}.
\end{equation}
We note that the function $\vp^\cR$ in fact coincides with the opposite of the function $\vp^\cL$. 
This is a particularity of the Klim\v{c}\'{i}k model, whose action \eqref{oaso8912as},\,\eqref{ioas891a} possesses
the discrete symmetry $\varepsilon_{1,2}\mapsto-\varepsilon_{1,2}$ and $x\mapsto -x$.
As a result,  the 1-form $\vp(z)\dd z$ describing the AGM before the UV limit is odd under the transformation $z\mapsto 1/z$. The 1-form $\vp^\cR(z^\cR)\dd z^\cR$ then has three simple poles $(z^\cR_1,z^\cR_2,z^\cR_3)$ which are equal to their ``left'' counterparts $(z^\cL_1,z^\cL_2,z^\cL_3)$ described above, with residues $(\ell^\cR_1,\ell^\cR_2,\ell^\cR_3)$ equal to $-(\ell^\cL_1,\ell^\cL_2,\ell^\cL_3)$.

Similarly to the above discussion for $\vp^\cL(z^\cL)$, the simple pole $z^\cR_3 = 0$ arises from the collision  in the conformal limit of the two poles $\xi z_2^{-1}$ and $\xi z_4^{-1}$ of $-\xi\,z^\cR\,^{-2} \vp\bigl( \xi\,z^\cR\,^{-1} \bigr)$ with its zero $0$. Let us note that this zero corresponds to the one at $z=\infty$ in the initial twist function $\vp(z)\dd z$: in particular, the zero eliminated by the collision procedure in the spectral parameter $z^\cR$ is different from the one eliminated in the spectral parameter $z^\cL$ (moreover, the poles that collide with this zero are also different in the two different spectral parameters).

\paragraph{Gaudin Lax matrix and realisation.} Our main goal in the rest of this subsection is to interpret $\vp^\cL(z^\cL)$ and $\vp^\cR(z^\cR)$ as the twist functions of two distinct AGMs describing the two chiral halves of the UV fixed-point of the Klim\v{c}\'{i}k model. The formal AGM corresponding to these twist functions are described by Gaudin Lax matrices
\begin{equation}\label{kxznjasjh21}
\Gamma^\cM \bigl( z^\cM \bigr) = \sum_{r=1}^3 \frac{\Jc_r^\cM}{z^\cM - z^\cM_r},
\end{equation}
for $\rm M = \rm L, \rm R$ and where $\Jc_r^\cM$ are Kac-Moody currents with levels $\ell^\cM_r$, which generate the unreduced algebra of observables $\Ac^\cM$ of these AGMs. We now want to find how these formal AGMs can be naturally realised in the UV fixed-point of the Klim\v{c}\'{i}k model. For that, we will consider the conformal limit of the Gaudin Lax matrix
\begin{equation}
\rho\bigl( \Gamma(z) \bigr) = \sum_{r=1}^4 \frac{\Jc_r^\rho}{z-z_r}
\end{equation}
in the relativistic realisation describing the non-chiral model. The Kac-Moody currents $\Jc_r^\rho$ entering this object are realised in terms of the canonical fields $(g_1,X_1,g_2,X_2)$ in $T^\ast(G\!\times\! G)$ as in eq.\;\eqref{Eq:CurrentsKlim}. They depend only on the RG invariants $K/\varepsilon_{1,2}$ and thus are themselves invariant if we assume that the canonical fields are unaffected by the RG flow. We will suppose that this is the case here and will comment further on this assumption in the next paragraph. As for the twist function, we need to perform the conformal limit working with either $z^\cL = \xi\,z$ or $z^\cR = \xi\,z^{-1}$ fixed. Focusing on $z^\cL$ first, we find
\begin{equation}
\lim_{\tau\to -\infty} \rho \bigl( \xi^{-1} \Gamma \bigl( \xi^{-1}z^\cL \bigr) \bigr) = \frac{\Jc_2^\rho}{z^\cL - z^\cL_1} + \frac{\Jc_4^\rho}{z^\cL - z^\cL_2} + \frac{\Jc_1^\rho+\Jc_3^\rho}{z^\cL - z^\cL_3}.
\end{equation}
This suggests to interpret this quantity as the realisation of the formal Gaudin Lax matrix $\Gamma^\cL(z^\cL)$ described above and thus to define a realisation $\rho^\cL_\dI : \Ac^\cL \to \Fc\bigl[T^\ast(G\!\times\! G)\bigr]$ by
\begin{equation}
\rho^\cL_\dI \bigl( \Jc^\cL_1 \bigr) = \Jc_2^\rho, \qquad \rho^\cL_\dI \bigl( \Jc^\cL_2 \bigr) = \Jc_4^\rho, \qquad \rho^\cL_\dI \bigl( \Jc^\cL_3 \bigr) = \Jc_1^\rho+\Jc_3^\rho.
\end{equation}
The index $\dI$ in this equation will be explained in the next paragraph. The fact that $\rho^\cL_\dI$ is a Poisson map is ensured by its definition through a limit of $\Gamma(z)$: indeed, by construction, the fields $\Jc_2^\rho$, $\Jc_4^\rho$ and $\Jc_1^\rho+\Jc_3^\rho$ are Kac-Moody currents in $\Fc\bigl[T^\ast(G\!\times\! G)\bigr]$ with respective levels $\ell^\cL_1 = \ell_2$, $\ell^\cL_2 = \ell_4$ and $\ell^\cL_3 = \ell_1+\ell_3$.

One can perform a similar analysis using the spectral parameter $z^\cR = \xi\,z^{-1}$. This leads us to define the following realisation $\rho^\cR_\dI$ of the Kac-Moody currents $\Jc^\cR_r$:
\begin{equation}
\rho^\cR_\dI \bigl( \Jc^\cR_1 \bigr) = \Jc_1^\rho, \qquad \rho^\cR_\dI \bigl( \Jc^\cR_2 \bigr) = \Jc_3^\rho, \qquad \rho^\cR_\dI \bigl( \Jc^\cR_3 \bigr) = \Jc_2^\rho+\Jc_4^\rho.
\end{equation}
As expected, these fields are Kac-Moody currents in $\Fc\bigl[T^\ast(G\!\times\! G)\bigr]$ with respective levels $\ell^\cR_1 = \ell_1$, $\ell^\cR_2 = \ell_3$ and $\ell^\cR_3 = \ell_2+\ell_4$.

We will refer to $\rho^\cL_\dI$ and $\rho^\cR_\dI$ as chiral realisations, as will be justified later in this subsection. Let us observe that in both these chiral realisations, the sum of all three Kac-Moody currents $\Jc_1^\cLR+\Jc_2^\cLR+\Jc_3^\cLR$ is realised as $\Cc = \Jc_1^\rho + \Jc_2^\rho + \Jc_3^\rho + \Jc_4^\rho = X_1+X_2$, \textit{i.e.} as the constraint of the gauged formulation of the Klim\v{c}\'{i}k model. Thus, the two chiral realisations $\rho^\cL_\dI$ and $\rho^\cR_\dI$ share the same constraint and gauge symmetry in the extended algebra $\Fc\bigl[T^\ast(G\!\times\! G)\bigr]$ as the initial relativistic realisation.

\paragraph{Realisations in different charts.} Before we study the properties of the chiral realisations $\rho_\dI^\cLR$ defined in the previous paragraph, let us explain the origin of the index $\dI$ in this notation. This is related to the assumption made above that the canonical fields $(g_1,X_1,g_2,X_2)$ entering the currents $\Jc_r^\rho$ are not affected by the RG flow. In particular, this means that we chose a way to extract coordinate fields in the gauge invariant quantity $g=g_1 g_2^{-1}$ which does not involve any parameter that flows under the RG. Recall that in the ${\rm SU}(2)$ case, the coordinate fields $(\phi,v,w)$ of the Klim\v{c}\'{i}k model were extracted from $g$ using the Euler angles decomposition \eqref{uia87887as} and the parametrisation \eqref{oias9812addddds}. In particular, this latter equation defined the way we extracted the coordinate $\phi$ from $g$, depending on a parameter $\phi_0$. We then considered various choices of $\phi_0$ in terms of the parameter $\kappa$ in eq.\;\eqref{oia980cxjhxcj}, leading to the three different charts $\dI$, $\dII$ and $\dIII$ for the chiral model in the conformal limit $\kappa\to 1^-$. In particular, the choice $\phi_0=0$ leading to chart $\dI$ corresponds to an extraction of the coordinates $(\phi,v,w)$ in $g$ which does not depend on the parameter $\kappa$ and which is then unaffected by the RG flow: this is the choice that leads to currents $\Jc_r^\rho$ which are RG invariants and thus to the realisations $\rho^\cLR_\dI$ defined in the previous paragraph. These are therefore more precisely realisations of the affine Gaudin models in terms of the fields of the chiral Klim\v{c}\'{i}k model in chart I (and additional non-gauge-invariant degrees of freedom).

This raises the natural question of whether there exist realisations of AGMs corresponding to the charts II and III. We expect the answer to this question to be positive. To construct such realisations, one would first need to express the currents $\Jc_r^\rho$ in terms of the physical fields $(\phi,v,w)$, but with the choices of the parameter $\phi_0$ in terms of $\kappa	$ corresponding to charts II and III in eq.\;\eqref{oia980cxjhxcj}. It seems likely that these currents do not possess a well-defined UV limit when $\kappa	\to 1^-$. Following the results of Subsection \ref{sec341}, we expect that this can be corrected by first performing a well-chosen $\kappa$-dependent conjugation of the currents $\Jc_r^\rho$, ensuring that their limits when $\kappa\to 1^-$ stay finite. Since a conjugation of the currents does not change their Kac-Moody Poisson bracket, we expect this different limiting procedure to produce two other realisations $\rho_\dII^\cLR$ and $\rho_\dIII^\cLR$ of the same underlying formal affine Gaudin models, in terms of the fields in charts II and III respectively. In practice, this procedure can be difficult to implement since the currents do not depend only on the physical field $g=g_1 g_2^{-1}$ but also on extra non-gauge-invariant degrees of freedom. We will avoid this difficulty in the next subsection by considering appropriate gauge-fixed models: in particular, this will allow us to discuss realisations in different charts.

\paragraph{Chirality of the realisations.} Let us now discuss the chirality properties of the realisations $\rho_\dI^\cLR$. For the moment, we will focus on the AGM with spectral parameter $z^\cL$ and the realisation $\rho_\dI^\cL$. The latter was introduced by considering the UV limit of the Gaudin Lax matrix $\xi^{-1} \rho\bigl(\Gamma(\xi^{-1} z^\cL)\bigr)$. Before taking the limit, the dynamic of $\rho\bigl(\Gamma(z)\bigr)$ is equivalent to the one of the Lax matrix $\Lc_x(z)$, which takes the form of a zero curvature equation. We will work with the light-cone representation of this equation, in terms of the Lax connection $\p_\pm + \Lc_\pm(z)$. As already observed in Subsection \ref{sec341}, one checks that in the UV limit, the connection $\p_\pm + \Lc_\pm\bigl( \xi^{-1} z^\cL \bigr)$ in the spectral parameter $z^\cL$ tends to $\p_\pm + \Lc^\cL_\pm\bigl( z^\cL \bigr)$, where the component $\Lc_-^\cL$ is independent of $z^\cL$. As a consequence, this component can be eliminated by a well-chosen gauge transformation with parameter $h$, which acts on the connection as $\p_\pm + \Lc_\pm^\cL \mapsto h^{-1}(\p_\pm +\Lc_\pm^\cL\bigr)h$: indeed, since $\Lc_-^\cL$ is independent of $z^\cL$, one can always choose $h$ such that $h^{-1}(\p_-+\Lc_-^\cL\bigr)h=\p_-$. The remaining component $h^{-1}(\p_++\Lc_+^\cL\bigr)h$ is therefore chiral: more precisely, it is expressed in terms of left-moving fields. In this gauge, the Lax matrix coincides with this light-cone component. We thus deduce that the gauge-transformed Kac-Moody currents $\rho^\cL\bigl(\Jc_r^\cL\bigr)\null^h$ are left-moving.

The above results motivate the following definition of a left-moving chiral realisation of an AGM. Consider a formal AGM with Kac-Moody currents $\bigl(\Jc_1^\cL,\ldots,\Jc_N^\cL\bigr)$ forming an unreduced algebra $\Ac^\cL$ which we realise through a Poisson map $\rho^\cL : \Ac^\cL \to \Fc[T^\ast Q]$. We also suppose that we chose a Hamiltonian $\Hc \in \Fc[T^\ast Q]$ defining the dynamic $\p_t \approx \lbrace \Hc_{\rm T}, \cdot \rbrace$ of a 2d field theory (where $\Hc_{\rm T}$ is built from $\Hc$ by adding a Lagrange multiplier term). We say that this realisation is a left-moving chiral one if:\vspace{9pt} \\
\begin{tabular}{cp{0.89\textwidth}}
(LM1) & the choice of Hamiltonian $\Hc$ is such that the gauge-invariant fields in $\Fc[T^\ast Q]$ decompose into left-moving and right-moving fields ; \vspace{7pt} \\
(LM2) & there exists a gauge transformation with parameter $h$ such that the gauge-transformed Kac-Moody currents $\rho^\cL\bigl(\Jc_r^\cL\bigr)\null^h$ are left-moving. Moreover, all left moving 
gauge-invariant fields in $\Fc[T^\ast Q]$
can be reconstructed from these currents up to integration
constants.
\end{tabular}~\vspace{9pt}\\
It is clear from the previous discussion that the realisation $\rho^\cL_\dI : \Ac^\cL \to \Fc\bigl[T^\ast(G\!\times\! G)\bigr]$ defined above for the UV fixed-point of the Klim\v{c}\'{i}k model is a left-moving chiral realisation with $N=3$ punctures. This notion should be compared with the one of relativistic realisation introduced in Subsection \ref{Par:RelReal}. The condition (R1) in a relativistic realisation that the theory is Lorentz invariant is replaced here by the stronger condition (LM1) that the dynamics decompose the fields into left- and right-movers. Moreover, for a relativistic realisation, the condition (R2) ensured that the fields of the AGM contain all the degrees of freedom of the 2d field theory, while in the case of a chiral realisation the condition (LM2) implies that the fields of the AGM contain only half of the physical degrees of freedom, corresponding to the left-moving fields.

Let us comment on the consequences of this definition. Following the general formalism recalled in Subsection \ref{Sec:AGM}, the main output of the AGM construction is an integrable structure $\Zc^\cL \subset \Ac^\cL$, composed by local and non-local charges in involution built from the Kac-Moody currents $\Jc_r^\cL$. As usual, one can consider the image $\rho^\cL\bigl(\Zc^\cL\bigr)$ of this integrable structure in the realisation $\Fc[T^\ast Q]$. Since the charges in this integrable structure are gauge-invariant (see Subsection \ref{Sec:AGM}), they can equivalently be expressed in terms of the gauge-transformed Kac-Moody currents $\rho^\cL\bigl(\Jc_r^\cL\bigr)\null^h$. By the condition (LM2), this ensures that the integrable structure $\rho^\cL\bigl(\Zc^\cL\bigr)$ inherited from the AGM is built purely from left-moving fields. We will give a more precise description of this left-moving integrable structure for the case of the Klim\v{c}\'{i}k model in the following subsections.\\

It is clear that one can define an analogous notion of a right-moving chiral realisation and that the realisation $\rho^\cR_\dI$ defined in the previous paragraphs for the conformal limit of the Klim\v{c}\'{i}k model belongs to this class. This also motivates the introduction of what we will call a chiral pair of realisations of AGMs. We will define such a pair as the data of a left-moving chiral realisation $\rho^\cL : \Ac^\cL \to \Fc[T^\ast Q]$ and a right-moving chiral realisation $\rho^\cR : \Ac^\cR \to \Fc[T^\ast Q]$ of two AGMs (with observables $\Ac^\cL$ and $\Ac^\cR$) in the same Poisson algebra $\Fc[T^\ast Q]$ such that\footnote{Technically, we also require that the constraints $\sum_{r=1}^N \rho^\cL\bigl( \Jc^\cL_r \bigr)$ and $\sum_{r=1}^N \rho^\cR\bigl( \Jc^\cR_r \bigr)$ coincide in $\Fc[T^\ast Q]$ so that these two realisations share the same gauge symmetry and physical observables.}:\vspace{9pt} \\
\begin{tabular}{cp{0.89\textwidth}}
(CP1) & the Hamiltonian and momentum of the theory take the form $\Hc \approx \Q^\cL + \Q^\cR$ and $\Pc \approx \Q^\cL - \Q^\cR$ with the charges $\Q^\cL$ and $\Q^\cR$ in the left-moving and right-moving integrable structures $\rho^\cL\bigl(\Zc^\cL)$ and $\rho^\cR\bigl(\Zc^\cR)$ respectively ; \vspace{7pt} \\
(CP2) & one can reconstruct, up to initial conditions, all canonical fields in $\Fc[T^\ast Q]$ from the Kac-Moody currents $\rho^\cL\bigl(\Jc^\cL_r\bigr)$ and $\rho^\cR\bigl(\Jc^\cR_r\bigr)$ of the two combined realisations. 
\end{tabular}~\vspace{9pt}\\
The charges $\Q^\cL$ and $\Q^\cR$ introduced in the condition (CP1) are the generators of the light-cone derivatives $\p_+$ and $\p_-$ respectively. The realisations $\rho^\cL_\dI$ and $\rho^\cR_\dI$ introduced above for the Klim\v{c}\'{i}k model define such a pair of chiral realisations.

\paragraph{Gauge symmetries in chiral realisations.} Let us consider an AGM with $N$ Kac-Moody currents forming a Poisson algebra $\Ac^\cL$ and suppose that we are given a left-moving chiral realisation $\rho^\cL : \Ac^\cL \to \Fc[T^\ast Q]$ as defined in the previous paragraph. The model is subject to a first-class constraint and thus admits a gauge symmetry. The latter plays a role in the definition of the left-moving chiral realisation and in particular in the condition (LM2) above, where we require that the realised Kac-Moody currents are left-moving up to a gauge transformation. This might seem surprising at first as one could have expected the definition of a chiral realisation to require that the Kac-Moody currents themselves are left-moving. This is in fact not a well-defined condition since these currents are not gauge-invariant. Indeed, by construction, non-gauge-invariant fields contain unphysical degrees of freedom which, in the dynamics of the model, contribute as arbitrary functions of space and time coordinates $(x,t)$ (these arbitrary functions are encoded in the Lagrange multiplier appearing in the total Hamiltonian of the theory). It thus does not make sense to ask for a non-gauge-invariant field to be left-moving or right-moving, \textit{i.e.} to depend only on the combination $t+x$ or $t-x$. This is why the condition (LM2) in the definition of a left-moving chiral realisation requires the chirality of the Kac-Moody currents up to a well-chosen gauge transformation with parameter $h$: we note that this gauge transformation should then be such that the resulting currents $\rho^\cL\bigl(\Jc_r^\cL\bigr)\null^h$ are in fact gauge-invariant quantities, so that the chirality condition makes sense.\\

Let us also comment on the effects of the constraint and gauge symmetry on the counting of degrees of freedom. To illustrate this, we consider the case of the UV fixed-point of the Klim\v{c}\'{i}k model, focusing on the left-moving chiral realisation $\rho^\cL_\dI : \Ac^\cL \to \Fc\bigl[T^\ast(G\!\times\! G)\bigr]$. The underlying formal AGM possesses three punctures, corresponding to the three Kac-Moody currents generating the algebra $\Ac^\cL$. The number of fundamental fields contained in $\Ac^\cL$ is thus equal to $3\dim \g$. On the other hand, the map $\rho^\cL_\dI$ realises these currents in terms of canonical fields $(g_1,X_1,g_2,X_2)$ in $T^\ast(G\!\times\! G)$, \textit{i.e.} in terms of $4\dim\g$ fields. So far, we have counted all fields in the unreduced algebras $\Ac^\cL$ and $\Fc[T^\ast(G\!\times\! G)]$: to count the physical fields of the theory, we now need to take into account the presence of the constraint and the gauge symmetry. Concretely, the former imposes $\dim\g$ relations between the fields and the latter eliminates an additional $\dim\g$ fields. Thus, after reduction, the number of physical fields in the AGM is equal to $3\dim\g-2\dim\g=\dim\g$ while the number of physical fields in the realisation is equal to $4\dim\g-2\dim\g=2\dim\g$ (corresponding to the gauge-invariant canonical fields $(g,X)$). This is what we expect for a chiral realisation: indeed, the AGM describes only half of the degrees of freedom of the physical 2-dimensional theory, namely the left-moving fields. The remaining $\dim\g$ fields are then described by the right-moving realisation $\rho^\cR_\dI$, so that the chiral pair formed by $\rho^\cL_\dI$ and $\rho^\cR_\dI$ encodes all the fields of the 2-dimensional theory. We note that the reduction of the number of physical fields that occurs when passing from the relativistic realisation of the Klim\v{c}\'{i}k model to the left-moving chiral realisation of its UV fixed-point can be traced back to the passage from 4 Kac-Moody currents to only 3 in the conformal limit, and thus to the collision of poles and zeroes in the twist function of the model, which eliminates one pole.\\

It may seems unnatural to pass through the introduction of non-physical fields $(g_1,X_1,g_2,X_2)$ in the chiral realisation, in particular since the left-moving condition (LM2) concerns only a well-chosen gauge transformation $\rho^\cL\bigl(\Jc_r^\cL\bigr)\null^h$ of the Kac-Moody currents which makes them gauge invariant, and which should thus be expressible in terms of the physical fields in $(g,X)$ only (although this can be difficult to do in practice). One can thus wonder whether there is another way to present chiral realisations that would avoid this issue and would rephrase the conditions (LM1) and (LM2) directly in terms of physical fields. One possible approach is to see the gauge transformation by $h$ appearing in condition (LM2) as bringing the Kac-Moody currents to a particular gauge, satisfying a certain gauge-fixing condition. If one is able to interpret this gauge-fixing as a condition on the Kac-Moody currents, one can then impose it in the formal AGM itself, before taking a realisation, and thus discuss only the realisation of physical fields. This is the subject of the next subsection.

\subsection{Gauge fixing and chiral realisations}
\label{Sec:GaugeFixChiral}

In this subsection, we present a different way of thinking about chiral realisations of AGMs which avoid the introduction of non-physical fields in the realisation. The main idea is to focus on physical degrees of freedom already in the formal AGM, before taking a realisation, by considering gauge-fixing conditions in terms of the Kac-Moody currents defining the AGM. As we will see, this will also allow us to gain some insights on the chiral Poisson algebras underlying the UV fixed-point of the Klim\v{c}\'{i}k model.

\paragraph{Gauge-fixing.} Let us consider an AGM with $N$ Kac-Moody currents $\Jc_r$ generating the Poisson algebra $\Ac$, as in Subsection \ref{Sec:AGM}. The model is subject to the constraint $\Cc = \sum_{r=1}^N \Jc_r \approx 0$ and thus to the gauge symmetry it generates, defining a reduced algebra $\Ac_{\red}$. One efficient way to describe $\Ac_\red$ is to consider a gauge-fixing condition $\Fc \gf 0$ in $\Ac$, where $\Fc$ is an observable built from the Kac-Moody currents $\Jc_r$. Similarly to weak equalities that we denoted with the symbol $\approx$, we will use the symbol $\gf$ to denote equalities that hold under the gauge-fixing condition, \textit{i.e.} when we impose $\Fc \gf 0$ and $\Cc \gf 0$. This condition should be such that it is always attainable by a gauge transformation, \textit{i.e.} starting with any configuration of the Kac-Moody currents $\Jc_r$, there always exists a gauge transformation that brings these currents to a configuration that satisfies the condition $\Fc \gf 0$. We will moreover suppose that this is a complete gauge-fixing condition, \textit{i.e.} that the gauge transformation leading to $\Fc \gf 0$ is unique and thus that there are no residual gauge symmetries. In this case, every gauge orbit contains a unique representative satisfying $\Fc \gf 0$ and one can describe the algebra of reduced observables $\Ac_\red$ as the initial algebra $\Ac$ quotiented by the relations $\Cc \gf 0$ and $\Fc \gf 0$. We will denote this quotient by $\Ac_{\text{GF}}$, which is thus isomorphic to $\Ac_\red$. For simplicity, we will still denote by $\Jc_r$ the image of the Kac-Moody currents in the quotient $\Ac_{\text{GF}}$ (and similarly for other observables): we will keep track of the fact that we are working in the quotient by using the symbol $\gf$ to denote equalities which hold in $\Ac_{\text{GF}}$.

The Poisson structure of $\Ac_\red$ inherited by the Hamiltonian reduction of $\Ac$ can be described in terms of the gauge-fixed algebra $\Ac_{\text{GF}}$ using the Dirac bracket. Indeed, the assumption made above that $\Fc \gf 0$ is a complete gauge-fixing implies that the combined conditions $\Cc \gf 0$ and $\Fc \gf 0$ form a system of second-class constraints in $\Ac$. One can thus define a Dirac bracket $\lbrace \cdot,\cdot \rbrace_D$ on $\Ac$, which is a Poisson bracket ensuring that these constraints $\Cc$ and $\Fc$ Poisson commute with all observables in $\Ac$: this bracket is thus compatible with imposing $\Cc \gf 0$ and $\Fc \gf 0$. It therefore descends to the quotient $\Ac_{\text{GF}}$, making it isomorphic to $\Ac_\red$ as Poisson algebras.

Recall that the definition of AGMs results in the construction of an integrable structure $\Zc^{(z_r)}$ in $\Ac$, composed by charges in involution built from the Kac-Moody currents $\Jc_r$. Since these charges are gauge-invariant, their Dirac bracket coincides with their initial bracket and thus vanishes. The integrable structure $\Zc^{(z_r)}$ thus descends to an integrable structure in the gauge-fixed algebra $\Ac_{\text{GF}}$, which is identified with $\Zc^{(z_r)}_\red$ in the reduced algebra $\Ac_\red$ through the above isomorphism between $\Ac_{\text{GF}}$ and $\Ac_\red$. In conclusion, we can thus describe the Poisson algebra $\Ac_\red$ and the integrable structure $\Zc_\red^{(z_r)}$ of the reduced AGM through the gauge-fixing, as expected.

\paragraph{Chiral realisations through gauge-fixing.} Let us now use the formalism of gauge-fixed AGMs to discuss chiral realisations. We will focus on left-moving realisations, the discussion for right-moving ones being completely analogous. Accordingly, we now consider an AGM with Kac-Moody currents $\Jc_r^\cL$ forming an algebra $\Ac^\cL$. As in the previous paragraph, we can describe the reduced AGM by choosing a gauge fixing $\Fc^\cL \gf 0$ and working with the quotient algebra $\Ac^\cL_{\text{GF}}$, equipped with the Dirac bracket. We can then consider a realisation of this gauge-fixed AGM in terms of canonical fields in a cotangent bundle $T^\ast Q_{\red}$, through a Poisson map $\rho^\cL_{\text{GF}} : \Ac^\cL_{\text{GF}} \to \Fc\bigl[T^\ast Q_{\red}\bigr]$. Since we consider the gauge-fixed model, this realisation is not subject to any constraint or gauge symmetry and the canonical fields in $T^\ast Q_{\red}$ are physical degrees of freedom. We further choose a Hamiltonian $\Hc\in\Fc\bigl[T^\ast Q_{\red}\bigr]$ defining the dynamic $\p_t = \lbrace \Hc, \cdot \rbrace$. We then say that this realisation is a chiral left-moving one if\vspace{9pt} \\
\begin{tabular}{cp{0.87\textwidth}}
(LM1') & the choice of Hamiltonian $\Hc$ is such that the fields in $\Fc[T^\ast Q_{\red}]$ decompose into left-moving and right-moving fields ; \vspace{7pt} \\
(LM2') & the image $\rho^\cL_{\text{GF}}\bigl(\Jc_r^\cL\bigr)$ of the (gauge-fixed) Kac-Moody currents are left-moving and all left-moving fields of the theory can be reconstructed from these currents, up to integration constants. 
\end{tabular}~\vspace{9pt}\\
Compared to the condition (LM2) in the previous subsection, we do not need in (LM2') to require chirality of the Kac-Moody currents up to a gauge transformation, since we have fixed the gauge and are now manipulating purely physical degrees of freedom. As before, such a realisation provides us with an integrable structure $\rho^\cL_{\text{GF}}\bigl( \Zc^\cL \bigr)$ in $\Fc\bigl[T^\ast Q_{\red}\bigr]$. Moreover, by construction, this integrable structure is built from the left-moving fields $\rho^\cL_{\text{GF}}\bigl(\Jc_r^\cL\bigr)$. The notion of left-moving chiral realisation considered here thus corresponds to the gauged-fixed version of the one introduced in the previous subsection.

\paragraph{Applications to the Klim\v{c}\'{i}k model.} In the rest of this subsection, our main goal will be to apply the formalism developed in the previous paragraphs to the UV fixed-point of the Klim\v{c}\'{i}k model. We have identified the AGM underlying the left-moving part of this model in the previous subsection: it is characterised by the twist function $\vp^\cL\bigl(z^\cL)$ in eq.\;\eqref{Eq:TwistKlimL} and is described by three Kac-Moody currents $\Jc_r^\cL$, with levels $\ell_r^\cL$ given by eq.\;\eqref{Eq:LevelsL}. In the following paragraphs, we will discuss different gauge-fixings of this AGM as well as their realisations in terms of the left-moving fields of the UV fixed-point of the Klim\v{c}\'{i}k model, in different charts.

\paragraph{Parafermionic gauge.} In this paragraph, we will consider the following choice of gauge-fixing condition in terms of the Kac-Moody currents:
\begin{equation}\label{Eq:ParaGF}
\Fc^{\cL}_P = \bigl( \hat{R} - \ri) \Jc^\cL_1 - \bigl( \hat{R} + \ri) \Jc^\cL_2 \gf 0,
\end{equation}
where $\hat R : \g^\C \to \g^\C$  is the Yang-Baxter operator \eqref{Rmat1a}. This forces the gauge-fixed current $\Jc^\cL_1$ to belong to the negative Borel subalgebra $\mathfrak{h} \oplus \mathfrak{n}_-$ of $\g^\C$, while the gauge-fixed current $\Jc^\cL_2$ belongs to the positive Borel subalgebra $\mathfrak{h} \oplus \mathfrak{n}_+$ and has a Cartan part which is opposite to the one of $\Jc^\cL_1$. Finally, using the fact that $\Jc_1^\cL+\Jc_2^\cL+\Jc_3^\cL \gf 0$, we find that $\Jc_3^\cL \gf -\Jc_1^\cL -\Jc_2^\cL$ belongs to $\mathfrak{n}_+ \oplus \mathfrak{n}_-$.

To describe more precisely the gauge-fixed AGM, we will use the root decomposition of $\g^\C$. To keep the main text concise, we gather all the relevant definitions, notations and conventions in Appendix \ref{App:Root}. Here, we will mostly need the orthogonal basis $\lbrace {\tt h}_i \rbrace_{i= 1,\ldots,\dim\mathfrak{h}}$ of the Cartan subalgebra $\mathfrak{h}$, as well as the root vectors $\lbrace {\tt e}_\alpha \rbrace_{\alpha\in\Delta}$ forming a basis of $\mathfrak{n}_+ \oplus \mathfrak{n}_-$ ($\Delta$ and $\Delta_\pm$ denote the sets of roots and positive/negative roots). The gauge-fixed currents $\Jc^\cL_r(x)$ introduced above can then be parametrised in terms of $\dim\g$ fields $\lbrace D^\cL_i(x) \rbrace_{i= 1,\ldots,\dim\mathfrak{h}}$ and $\lbrace \Psi^\cL_\alpha(x) \rbrace_{\alpha\in\Delta}$:
\begin{subequations}\label{Eq:BPsi}
\begin{eqnarray}
\Jc^\cL_1(x) &\gf& \frac{1}{2} \sum_{i=1}^{\dim\mathfrak{h}} D^\cL_i(x) \, {\tt h}_i + \sum_{\alpha\in\Delta_+} \Psi^\cL_{\alpha}(x)\,{\tt e}_{-\alpha} \, , \\
\Jc^\cL_2(x) &\gf& -\frac{1}{2} \sum_{i=1}^{\dim\mathfrak{h}} D^\cL_i(x) \, {\tt h}_i + \sum_{\alpha\in\Delta_+} \Psi^\cL_{-\alpha}(x)\,{\tt e}_{\alpha} \, , \\
\Jc^\cL_3(x) &\gf& -\sum_{\alpha\in\Delta} \Psi^\cL_{\alpha}(x)\,{\tt e}_{-\alpha} \, .
\end{eqnarray}
\end{subequations}
One easily checks that this parametrisation indeed satisfies the constraint $\Jc_1^\cL+\Jc_2^\cL+\Jc_3^\cL \gf 0$ and the gauge-fixing condition $\bigl( \hat{R} - \ri \bigr) \Jc^\cL_1 - \bigl( \hat{R} + \ri \bigr) \Jc^\cL_2 \gf 0$.

The fields $\lbrace D^\cL_i(x) \rbrace_{i= 1,\ldots,\dim\mathfrak{h}}$ and $\lbrace \Psi^\cL_\alpha(x) \rbrace_{\alpha\in\Delta}$ generate the gauge-fixed algebra, which we denote in this case by $\Ac^\cL_P$. To describe its Poisson structure, one needs to determine the Dirac bracket of these fields. We perform this computation in details in Appendix \ref{App:Dirac} and simply give the final result here. We find
\begin{subequations}\label{Eq:BracketBPsi}
\begin{align}
\bigl\lbrace D^\cL_i(x), \Psi^\cL_\alpha(y) \bigr\rbrace_P &\gf  0 \,, \\
\bigl\lbrace D^\cL_i(x), D^\cL_j(y) \bigr\rbrace_P &\gf \frac{2\nu^2\Kuv}{(1+\nu^2)^2}\,\delta_{ij}\,\p_x\delta(x-y) \,, \\
\bigl\lbrace \Psi^\cL_\alpha(x), \Psi^\cL_\beta(y) \bigr\rbrace_P &\gf  N^{\alpha,\beta} \, \Psi^\cL_{\alpha+\beta}(x)\, \delta(x-y) - \Kuv\,\delta_{\alpha+\beta,0}\,\p_x\delta(x-y) \label{Eq:BracketPsi} \\
& \hspace{23pt} - \frac{(\alpha,\beta)}{2\Kuv}
\Psi^\cL_\alpha(x) \, \Psi^\cL_\beta(y) \,\epsilon(x-y) \, . \notag
\end{align}
\end{subequations}
In the above bracket, the numbers $N^{\alpha,\beta}$ are defined through the commutation relations of root vectors ${\tt e}_\alpha$ and $(\cdot,\cdot)$ is the standard bilinear form on $\mathfrak{h}^\ast$ (see Appendix \ref{App:Root} for precise definitions).

We will call the Poisson algebra $\Ac^\cL_P$ generated by the fields $D^\cL_i$ and $\Psi^\cL_\alpha$ a parafermionic algebra: the parafermionic currents $\Psi^\cL_\alpha$ were initially introduced in~\cite{Ninomiya:1986dp,Gepner:1987sm} at the quantum level, while their classical bracket \eqref{Eq:BracketPsi} was discussed in~\cite{Bardakci:1990lbc}. A similar analysis can be of course performed for the right-moving AGM: it leads to the description of the gauge-fixed algebra $\Ac^\cR_P$ in terms of $\dim\g$ fields $\lbrace D^\cR_i(x) \rbrace_{i=1,\ldots,\dim\mathfrak{h}}$ and $\lbrace \Psi^\cR_\alpha(x) \rbrace_{\alpha\in\Delta}$ satisfying the same brackets \eqref{Eq:BracketBPsi} as in the left-moving case but with $\Kuv$ replaced by $-\Kuv$.\\

Let us finally discuss the example $G={\rm SU}(2)$. In this case, the Cartan subalgebra is one-dimensional and there is thus a unique Cartan field $D^\cL(x)$ in the decomposition \eqref{Eq:BPsi}. Moreover, there are only two roots $\alpha$ and $-\alpha$, which are opposite one of another, corresponding to two fields $\Psi^\cL_\pm(x)$ in the decomposition \eqref{Eq:BPsi}. The brackets \eqref{Eq:BracketBPsi} now read
\begin{subequations}\label{Eq:BracketBPsiSU2}
\begin{align}
\bigl\lbrace D^\cL(x), \Psi^\cL_\pm(y) \bigr\rbrace_P &\gf  0 \,, \\
\bigl\lbrace D^\cL(x), D^\cL(y) \bigr\rbrace_P &\gf \frac{2\nu^2\Kuv}{(1+\nu^2)^2}\,\p_x\delta(x-y) \,, \\
\bigl\lbrace \Psi^\cL_\pm(x), \Psi^\cL_\pm(y) \bigr\rbrace_P &\gf  - \frac{1}{\Kuv} \Psi^\cL_\pm(x) \, \Psi^\cL_\pm(y) \,\epsilon(x-y) \,, \\
\bigl\lbrace \Psi^\cL_\pm(x), \Psi^\cL_\mp(y) \bigr\rbrace_P &\gf   - \Kuv\,\p_x\delta(x-y)  + \frac{1}{\Kuv}
\Psi^\cL_\pm(x) \, \Psi^\cL_\mp(y) \,\epsilon(x-y) \, .
\end{align}
\end{subequations}

\paragraph{Parafermionic realisation.} Now that we have described a specific gauge-fixing of the AGM under consideration, leading to the parafermionic algebra $\Ac^\cL_P$, let us explain how this algebra can be realised in the UV fixed-point of the Klim\v{c}\'{i}k model. We will focus mostly on the case $G={\rm SU}(2)$ and will speculate on the higher-rank one at the end of the paragraph. We thus consider three fields $D^\cL(x)$ and $\Psi^\cL_\pm(x)$, satisfying the Poisson algebra \eqref{Eq:BracketBPsiSU2}. We have in fact encountered similar fields in the description of the UV fixed-point of the ${\rm SU}(2)$ Klim\v{c}\'{i}k model in Subsection \ref{sec341}. More precisely, using the coordinate fields $(\phi,\alpha,\chi)$ of the chart I of this model, we found that the decoupled free boson $\chi$ induced a left-moving field $\p_+ \chi$ satisfying the same Poisson bracket as $D^\cL$ up to a constant, while the parafermionic fields $\psi^\cL_\pm$ built from the cigar coordinates $(\phi,\alpha)$ in eq.\;\eqref{oiasoi89123} satisfied the Poisson brackets \eqref{asi7812as}, identical to the ones of $\Psi^\cL_\pm$ also up to a constant. We can thus define a realisation of $\Ac^\cL_P$ by\footnote{Here we have expressed the realisation in terms of the light-cone derivatives $\p_+\phi$, $\p_+\alpha$ and $\p_+\chi$ of the fields of the Klim\v{c}\'{i}k model. Replacing time derivatives in favour of the conjugate momenta of these fields yields an expression of the realisation in terms of canonical fields $(\phi,\alpha,\chi,\pi_\phi,\pi_\alpha,\pi_\chi)$.}
\begin{equation}\label{Eq:ParafReal}
\rho^\cL_P\bigl(D^\cL\bigr) = -\frac{2\ri\nu\Kuv}{1+\nu^2} \p_+ \chi \qquad \text{ and } \qquad \rho^\cL_P\bigl(\Psi^\cL_\pm \bigr) = \Kuv\, \psi^\cL_\pm.
\end{equation}
As required in the definition of a left-moving chiral realisation made earlier in this subsection, the image of the gauge-fixed Kac-Moody currents in the realisation, which are expressed in terms of $\rho^\cL_P\bigl(D^\cL\bigr)$ and $\rho^\cL_P\bigl(\Psi^\cL_\pm \bigr)$, are left-moving fields. In particular, this ensures that the Lax matrix
\begin{equation}
\Lc^\cL_P \bigl(z^\cL\bigr) = \frac{\rho^\cL_P\bigl(\Gamma^\cL(z^\cL)\bigr)}{\vp^\cL(z^\cL)}
\end{equation}
induced by this choice of realisation is left-moving. A direct computation shows that
\begin{equation}\label{kxjnasmn121}
\Lc^\cL_P \bigl(z^\cL\bigr) = -\frac{\ri (\nu+\nu^{-1})z^\cL}{4} \p_+ \chi \,{\tt h} + \frac{4-(1+\nu^2)z^\cL}{4} \psi^\cL_-\,{\tt e}_+ + \frac{4+(1+\nu^{-2})z^\cL}{4} \psi^\cL_+\,{\tt e}_-\, .
\end{equation}
This Lax matrix coincides with the one described in eq.\;\eqref{oias98xcoi}. In particular, this ensures that the non-local charges obtained via the Lax matrix from Subsection \ref{sec341} 
coincide with the ones in the AGM integrable structure $\rho^\cL_P\bigl(\Zc^\cL\bigr)$.

Let us note that we are working here in the chart I of the chiral Klim\v{c}\'{i}k model, in terms of the coordinates $(\phi,\alpha)$ of the left cigar and the decoupled free boson $\chi$. There exists a similar realisation of the same underlying AGM in the chart III, in terms of the coordinates $(\phi,\widetilde{\alpha})$ of the right cigar and the decoupled free boson $\widetilde{\chi}$. Finally, one can also find a realisation in terms of the free bosons $(\phi_1,\phi_2,\phi_3)$ of chart II, see eq.\;\eqref{oisa9821sa}.\\

So far, we have focused our attention on the case $G={\rm SU}(2)$. Let us say a few words about the case of an arbitrary simple Lie group $G$. We have described the general gauge-fixed algebra $\Ac^\cL_P$ in the previous paragraph, in terms of the fields $\lbrace D^\cL_i \rbrace_{i= 1,\ldots,\dim\mathfrak{h}}$ and $\lbrace \Psi^\cL_\alpha \rbrace_{\alpha\in\Delta}$ satisfying the bracket \eqref{Eq:BracketBPsi}, which is a natural higher-rank generalisation of the ${\rm SU}(2)$ setup considered above. We expect that this algebra can also be realised in terms of left-moving fields of the UV fixed-point of the Klim\v{c}\'{i}k model on $G$ and that the corresponding Lax matrix coincides with the one of the Klim\v{c}\'{i}k model in a particular choice of gauge. The concrete implementation of these ideas is out of the scope of the present paper and requires further investigation.

Let us however briefly comment on the expected structure of this realisation. It is clear from the Poisson bracket \eqref{Eq:BracketBPsi} that the $\dim\mathfrak{h}$ fields $D_i^\cL$ can be realised in terms of the derivatives $\p_+\chi^i$ of $\dim\mathfrak{h}$ free bosons $\chi^i$, similar to the boson $\chi$ in the ${\rm SU}(2)$ case above. The remaining fields $\Psi_\alpha^\cL$ form a higher-rank parafermionic algebra~\cite{Ninomiya:1986dp,Gepner:1987sm,Bardakci:1990lbc}, which generalises the parafermions of the cigar model. Recall that the latter can be obtained as the UV fixed-point of the sausage model, which itself is identified with the Yang-Baxter deformation of the coset model on ${\rm SU}(2)/{\rm U}(1)$. We expect this higher-rank algebra to describe parafermionic fields in the conformal limit of the Yang-Baxter deformation of the quotient $G/H_0$, where $H_0$ is the maximal torus of $G$ (whose Lie algebra is the intersection of the Cartan subalgebra $\mathfrak{h}$ with the real form $\g$ of $\g^\C$). For $G$ of rank $\ell$, the quotient $G/H_0$ belongs to the class of so-called $\mathbb{Z}_{\ell+1}$-coset spaces, which generalises the class of symmetric spaces, corresponding to $\mathbb{Z}_2$-cosets. The undeformed integrable $\sigma$-models on $\mathbb{Z}_{\ell+1}$-cosets for $\ell>1$ were initially introduced in~\cite{Young:2005jv}, while the construction of their integrable Yang-Baxter deformations was recently described in~\cite{Hoare:2021dix}, generalising the construction of~\cite{Delduc:2013fga} for symmetric spaces. This suggests that the UV fixed-point of the Klim\v{c}\'{i}k model on $G$ can be described as the UV fixed-point of the deformed coset model on $G/H_0$ together with $\dim H_0$ decoupled free bosons. It would be interesting to explore these aspects further.

Let us finally note that another parafermionic algebra defined from a reduction of Kac-Moody currents was studied in~\cite{Bardakci:1990ad}. Although it is different from the one considered in this paragraph, we believe that it corresponds to another gauge-fixing of the same unreduced algebra $\Ac^\cL$ considered here. We will comment more on this aspect and the relation to the so-called $\lambda$-models in Subsection \ref{Sec:Lambda}, as part of the perspectives and possible extensions of the present work.

\paragraph{$\bm\omega$-gauge-fixing.} Let us now consider another gauge-fixing of the algebra $\Ac^\cL$.\footnote{Let us mention that the methods and results described in this paragraph share some similarities with the works~\cite{Kawaguchi:2012ve,Delduc:2013fga,Delduc:2017brb}, which discuss the appearance of semi-classical limits of the finite and affine quantum groups $\mathcal{U}_q(\g)$ and $\mathcal{U}_q(\widehat{\g})$ in certain integrable $\sigma$-models. We note moreover that the results obtained in the present subsection will serve as the starting point in Section \ref{sec666} for the discussion of a quantum affine Borel group structure $\mathcal{U}_q(\widehat{\mathfrak{b}})$ underlying the quantisation of the monodromy of the AGM. It would be interesting to explore in more details the potential relations between these approaches.} For that, we introduce the operator $\hat\Pi = \hat R^2+\Id$, in terms of the Yang-Baxter operator \eqref{Rmat1a}. It is straightforward to check that $\hat\Pi$ is the projector along the Cartan subalgebra $\mathfrak{h}$ in the decomposition $\g^\C = \mathfrak{h} \oplus \mathfrak{n}_+ \oplus \mathfrak{n}_-$. We then consider the gauge-fixing condition
\begin{equation}\label{Eq:OmegaGF}
\Fc^\cL_\omega = \bigl( \hat R - \ri + \ri\, \bigl( 2+\nu^2-(1+\nu^2)\,\omega \bigr)\, \hat\Pi \,\bigr)\, \Jc^\cL_2 - 
\bigl( \hat R + \ri + \ri\, \omega\,\hat\Pi \,\bigr)\, \Jc^\cL_3 \gf 0,
\end{equation} 
where $\omega$ is a constant parameter (the particular dependence on $\omega$ in the above equation has been introduced to make the comparison with the results of Subsection \ref{sec341} easier). In this gauge, the currents $\Jc^\cL_2$ and $\Jc^\cL_3$ are valued in the Borel subalgebras $\mathfrak{h}\oplus\mathfrak{n}_-$ and $\mathfrak{h}\oplus\mathfrak{n}_+$ respectively. Moreover, the current $\Jc^\cL_1$ is valued in the full algebra $\g^\C = \mathfrak{h} \oplus \mathfrak{n}_+ \oplus \mathfrak{n}_-$, except for $\omega=1+2\nu^{-2}$, in which case $\Jc^\cL_1$ has no Cartan component. We will exclude this case here. The physical observables in this gauge can thus be seen as the components of the current $\Jc^\cL_1$. Using the basis $\lbrace {\tt h}_i, {\tt e}_\alpha \rbrace$ of $\g^\C$ introduced in the previous paragraphs (see Appendix \ref{App:Root} for the details), we parametrise these components in terms of $\dim\g$ fields $\lbrace \Xi^\cL_i \rbrace_{i= 1,\ldots,\dim\mathfrak{h}}$ and $\lbrace \Xi^\cL_\alpha \rbrace_{\alpha\in\Delta}$ according to
\begin{equation}\label{Eq:J1Xi}
\Jc^\cL_1(x) \gf -\frac{1}{2} \sum_{i=1}^{\dim\mathfrak{h}} \Xi^\cL_i(x)\,{\tt h_i} - \sum_{\alpha\in\Delta} \Xi^\cL_\alpha(x)\,{\tt e}_{-\alpha}\,.
\end{equation}
The remaining gauge-fixed currents $\Jc^\cL_2$ and $\Jc^\cL_3$ can then be expressed as
\begin{subequations}\label{Eq:J23Xi}
\begin{align}
\Jc^\cL_2(x) &\gf \frac{1+\omega}{2+\nu^2-\nu^2\omega}\;\frac{1}{2}  \sum_{i=1}^{\dim\mathfrak{h}} \Xi^\cL_i(x)\,{\tt h_i} + \sum_{\alpha\in\Delta_+} \Xi^\cL_\alpha(x)\,{\tt e_{-\alpha}} \, , \\
\Jc^\cL_3(x) &\gf \frac{(1+\nu^2)(1-\omega)}{2+\nu^2-\nu^2\omega}\; \frac{1}{2} \sum_{i=1}^{\dim\mathfrak{h}} \Xi^\cL_i(x)\,{\tt h_i} + \sum_{\alpha\in\Delta_+} \Xi^\cL_{-\alpha}(x)\,{\tt e_{\alpha}} \,.
\end{align}
\end{subequations}
One checks that these satisfy the gauge-fixing condition \eqref{Eq:OmegaGF} and $\sum_r \Jc^\cL_r \gf 0$, as expected.\\

One can determine the Poisson structure of the corresponding gauge-fixed algebra $\Ac^\cL_\omega$ by computing the Dirac bracket of the components $\lbrace \Xi^\cL_i \rbrace_{i= 1,\ldots,\dim\mathfrak{h}}$ and $\lbrace \Xi^\cL_\alpha \rbrace_{\alpha\in\Delta}$. The details of this computation are given in Appendix \ref{App:Dirac}. In the end, we find (for $\beta \neq -\alpha$)
\begin{subequations}\label{Eq:PbXi}
\begin{align}
\bigl\lbrace \Xi^\cL_i(x), \Xi^\cL_\alpha(y) \bigr\rbrace_\omega &\gf  \bigl(1-\vartheta\ell^\cL_1\bigr)\alpha({\tt h}_i) \,\Xi^\cL_\alpha(x) \delta(x-y) \,, \\
\bigl\lbrace \Xi^\cL_i(x), \Xi^\cL_j(y) \bigr\rbrace_\omega &\gf  2\ell^\cL_1\bigl(1-\vartheta\ell^\cL_1\bigr)\,\delta_{ij}\,\p_x\delta(x-y) \,, \\
\bigl\lbrace \Xi^\cL_\alpha(x), \Xi^\cL_\beta(y) \bigr\rbrace_\omega &\gf  N^{\alpha,\beta} \, \Xi^\cL_{\alpha+\beta}(x)\, \delta(x-y)   + \frac{1}{2}\, \vartheta \, (\alpha,\beta) \; \Xi^\cL_\alpha(x) \, \Xi^\cL_\beta(y) \,\epsilon(x-y) \, , \\
\bigl\lbrace \Xi^\cL_\alpha(x), \Xi^\cL_{-\alpha}(y) \bigr\rbrace_\omega &\gf  \rho_\alpha^i \,\Xi_i(x)\,\delta(x-y) + \ell^\cL_1\,\p_x\delta(x-y) - \frac{1}{2}\,\vartheta\,(\alpha,\alpha)\; \Xi^\cL_\alpha(x) \, \Xi^\cL_{-\alpha}(y) \,\epsilon(x-y) \, ,
\end{align}
\end{subequations}
where the constants $\rho_\alpha^i$ are defined through $[ {\tt e}_\alpha, {\tt e}_{-\alpha} ] = \sum_i \rho_\alpha^i\,{\tt h}_i$ (see Appendix \ref{App:Root} for details), while the parameters $\ell^\cL_1$ and $\vartheta$ are given by
\begin{equation}\label{Eq:Theta}
\ell^\cL_1 = \frac{\Kuv}{1+\nu^{-2}} \qquad \text{ and } \qquad \vartheta = \frac{1}{\Kuv} \left( \omega - \frac{\nu^2}{4}(1-\omega)^2 \right).
\end{equation}

In the case $G={\rm SU}(2)$, we have one Cartan field $\Xi_0^\cL(x)$ and two root fields $\Xi^\cL_\pm(x)$. The above Poisson bracket then becomes
\begin{subequations}\label{Eq:PbXiSU2}
\begin{align}
\bigl\lbrace \Xi^\cL_0(x), \Xi^\cL_\pm(y) \bigr\rbrace_\omega &\gf  \pm 2\bigl(1-\vartheta\ell^\cL_1\bigr) \,\Xi^\cL_\pm(x) \delta(x-y) \,, \\
\bigl\lbrace \Xi^\cL_0(x), \Xi^\cL_0(y) \bigr\rbrace_\omega &\gf  2\ell^\cL_1\bigl(1-\vartheta\ell^\cL_1\bigr)\,\p_x\delta(x-y) \,, \\
\bigl\lbrace \Xi^\cL_\pm(x), \Xi^\cL_\pm(y) \bigr\rbrace_\omega &\gf   \vartheta\; \Xi^\cL_\pm(x) \, \Xi^\cL_\pm(y) \,\epsilon(x-y) \, , \\
\bigl\lbrace \Xi^\cL_\pm(x), \Xi^\cL_\mp(y) \bigr\rbrace_\omega &\gf  \pm \Xi_0(x)\,\delta(x-y)
 + \ell^\cL_1\,\p_x\delta(x-y) - \vartheta \; \Xi^\cL_\pm(x) \, \Xi^\cL_\mp(y) \,\epsilon(x-y) \, .
\end{align}
\end{subequations}

\paragraph{Free field realisation in the $\bm\omega$-gauge.} Let us now discuss the realisation of the gauge-fixed algebra $\Ac^\cL_\omega$ in the UV fixed-point of the Klim\v{c}\'{i}k model. We will focus first on the ${\rm SU}(2)$ case and will comment on the higher-rank one at the end of the paragraph. We have in fact encountered a Poisson algebra similar to \eqref{Eq:PbXiSU2} in the subsection \ref{sec341}, namely the one \eqref{PBcurrents1} of the left-moving fields $V^{(\omega)}_0$, $V^{(\omega)}_+$ and $V^{(\omega)}_-$ in the asympotic domain of the ${\rm SU}(2)$-Klim\v{c}\'{i}k model (\textit{i.e.} chart II). We then define a realisation $\rho_{\omega}^\cL$ of $\Ac^\cL_\omega$ by
\begin{equation}\label{Eq:RealOmega}
\rho_\omega^\cL\bigl( \Xi^\cL_\pm \bigr) = \frac{\Kuv\, V^{(\omega)}_\mp}{\sqrt{1+\nu^2}}
 \qquad \text{ and } \qquad \rho_\omega^\cL\bigl( \Xi^\cL_0 \bigr) = \Kuv \frac{2+\nu^2(1-\omega)}{2(1+\nu^2)} \, V^{(\omega)}_0.
\end{equation}
One checks from the Poisson brackets \eqref{PBcurrents1} that this indeed defines a realisation of the algebra \eqref{Eq:PbXiSU2}, with the parameters $\ell^\cL_1$ and $\vartheta$ given by \eqref{Eq:Theta}. In the above equation, $V^{(\omega)}_0$ and $V^{(\omega)}_\pm$ should be understood as being defined in terms of the
left-moving fields $(\phi_1^{({\rm L})},\phi_2^{({\rm L})},\phi_3^{({\rm L})})$ by eq.\;\eqref{oi89xciusa}. 
Thus $\rho^\cL_\omega$ defines a left-moving free field realisation of the gauge-fixed algebra $\Ac^\cL_\omega$.

In order to compare the left-moving integrable structure $\rho^\cL_\omega\bigl(\Zc^\cL\bigr)$ induced by this realisation with the one described in Subsection \ref{sec341}, one needs to look at the Lax matrix
\begin{equation}\label{jkajk8912jasA}
\Lc^\cL_\omega \bigl(z^\cL\bigr) = \frac{\rho^\cL_\omega\bigl(\Gamma^\cL(z^\cL)\bigr)}{\vp^\cL(z^\cL)}
\end{equation}
coming from the AGM construction. A direct computation shows that it is given by
\begin{align}\label{jkajk8912jaC}
\Lc^\cL_\omega \bigl(z^\cL\bigr) &= \frac{2\nu^2(\omega-1)-(1+\nu^2)z^\cL}{8\nu^2} \,  V^{(\omega)}_0\,{\tt h} \\
&\hspace{40pt} - \frac{4\nu^2+(1+\nu^2)z^\cL}{4\nu^2\sqrt{1+\nu^2}} \, 
V^{(\omega)}_+\,{\tt e}_+ - \frac{(1+\nu^2)^{\frac{3}{2}}\, z^\cL}{4\nu^2} \, V^{(\omega)}_-\,{\tt e}_- \, . \notag
\end{align}
Through the relation \eqref{oias8912iopds} between the spectral parameters $z^\cL$ and $\rho$, this coincides with the Lax matrix \eqref{Anew} discussed in Subsection \ref{sec341} that
 determines the non-local charges built from the left-moving fields in the asymptotic domain of the ${\rm SU}(2)$ Klim\v{c}\'{i}k model.\\

There are a few interesting  cases of the $\omega$-gauge
that deserve special mention. The first one arises for $\omega=1$. This choice corresponds to $\vartheta=\Kuv^{-1}$ and makes the fields $V^{(\omega)}_\pm$ coincide with the vertex operators entering the definition of the (classical) cigar screening charges \eqref{oias9012oia--a}. This means that there is a particular gauge-fixing of the AGM considered here which is naturally realised in terms of these vertex operators. 
A related point is that for $\omega=-1$, for which $\vartheta=-\frac{1+\nu^2}{\Kuv}$, the components 
of the Kac-Moody currents $\Xi^\cL_\pm$
can be mapped to $V^{(-1)}_\pm$. The r\^{o}le of these fields was discussed in Subsection \ref{ias9812211aaa}.
They appear as the integrands of a second set of classical screening charges
which can be used to
determine the local IMs, see eqs.\;\eqref{oasoio90aaa} and \eqref{quasi-comm1A}.
Also recall that the combination of charges $\int\rd x_+ V_+^{(1)}$ and $\int \rd x_+ V_-^{(-1)}$ defines 
the corner-brane ${\cal W}$-algebra, which is a subalgebra of ${\cal W}_\infty\otimes {\rm U}(1)$. 
An alternative way of  obtaining the local fields belonging to this ${\cal W}$-algebra, 
which is coordinate free and uniform in the choice of 
the Lie group $G$,
will be described in the next subsection.
Note that
setting $\omega=1$ in eq.\;\eqref{Eq:J23Xi}, we see that such a gauge-fixing corresponds to requiring that the 
Kac-Moody currents $\Jc^\cL_2$ and $\Jc^\cL_3$ are valued respectively in the Borel subalgebra $\mathfrak{h}\oplus\mathfrak{n}_-$ and the nilpotent subalgebra $\mathfrak{n}_+$.
On the other hand, for $\omega=-1$ the current $\Jc^\cL_2$ becomes valued in $\mathfrak{n}_-$, while 
 $\Jc^\cL_3$ is an element of the Borel subalgebra $\mathfrak{h}\oplus\mathfrak{n}_+$. 

The last case we would like to discuss 
is $\omega=\bigl(\sqrt{1+\nu^{-2}} \mp \nu^{-1} \bigr)^2$, for which $\vartheta=0$. This corresponds to a particular gauge where there are no $\epsilon$-distributions in the algebra \eqref{Eq:PbXiSU2} and thus to a gauge where the Gaudin fields are local. More precisely, we get that the gauge-fixed field $\Jc^\cL_1$ is a standard Kac-Moody current. We finally note that in this case, the realisation of $\Jc^\cL_1$ in the Klim\v{c}\'{i}k model is particularly simple: indeed, one checks that \eqref{Eq:RealOmega} becomes $\rho^\cL_\omega\bigl(\Jc^\cL_1\bigr)=-\frac{\Kuv}{\sqrt{1+\nu^2}}\,
\bigl(\pm \frac{1}{2} V^{(\omega)}_0\,{\tt h} + V^{(\omega)}_+\,{\tt  e}_+ + V^{(\omega)}_-\,{\tt  e}_- \bigr)$.\\

We expect a realisation similar to \eqref{Eq:RealOmega} to exist for higher-rank groups. Finding its explicit expression would however first require to properly identify the asymptotic domain of the higher-rank Klim\v{c}\'{i}k model, which has not been described yet in the literature. These aspects are out of the scope of the present work and are interesting future perspectives.

\paragraph{Summary and comments.} In this subsection, we have shown that the left-moving fields of the UV fixed-point of the Klim\v{c}\'{i}k model naturally form a chiral realisation of the gauge-fixed AGM with twist function $\vp^\cL(z^\cL)$. In particular, this sheds some light on the chiral algebra underlying this model, \textit{i.e.} the Poisson algebra formed by the left-moving fields of the theory. For instance, the left-moving fields in the charts I, II or III naturally form a parafermionic algebra, which can also alternatively be seen as a well-chosen gauge-fixing of the AGM Poisson algebra. Moreover, the left-moving fields in the asymptotic domain, \textit{i.e.} the chiral free bosons in chart II, can be organised into vertex operators or in a Kac-Moody current, also corresponding to two different gauge-fixings of the AGM Poisson algebra. By construction, all these chiral Poisson algebras are equivalent to one another. We note however that the isomorphisms that relate these algebras are non-local transformations. Moreover, their realisations in general also involve non-local fields.

As a result of having found these realisations, one can construct the corresponding AGM integrable structure in terms of the left-moving fields of the UV fixed-point of the Klim\v{c}\'{i}k model. This integrable structure does not depend on the choice of gauge-fixing (since it is formed by gauge-invariant charges) and we have checked in the previous paragraphs that the corresponding Lax matrices coincide with the ones described in Subsection \ref{sec341}, hence showing that the AGM construction would
 lead to the same non-local integrals of motion as the ones that would be obtained using the Lax matrices from that subsection.

In addition to non-local chiral fields and integrals of motion, we have also discussed in Subsection \ref{sec3899891} the presence of local ones in the UV fixed-point of the Klim\v{c}\'{i}k model. Such objects can also be obtained from the AGM construction: this is the subject of the next subsection.

\subsection[$\Wc$-algebra and local charges]{\texorpdfstring{$\bm{\Wc}$}{W}-algebra and local charges\label{aaaaaaaaaaaa}}

\paragraph{Definition of the $\bm{\Wc}$-algebra.} Let us consider the AGM with twist function $\vp^\cL(z^\cL)$ underlying the left-moving half of the Klim\v{c}\'{i}k model. It is subject to the gauge symmetry $\Jc^\cL_r \mapsto h^{-1} \Jc^\cL_r h + \ell^\cL_r\,h^{-1}\p_x h$ generated by the constraint $\Jc^\cL_1 + \Jc^\cL_2 + \Jc^\cL_3 \approx 0$. In the previous subsections, we have explained how well-chosen gauge-fixed versions of these currents can be realised in terms of the physical left-moving fields of the Klim\v{c}\'{i}k model, allowing us to describe the chiral algebra underlying this theory. We noted however that gauge-fixing is in general a non-local procedure and that these chiral fields are thus non-local observables of the theory. To study local chiral fields, let us introduce the $\Wc$-algebra associated with the AGM. In the formal AGM, with currents $(\Jc^\cL_1,\Jc^\cL_2,\Jc^\cL_3)$, we define it as the set $\Wc^\cL$ of gauge-invariant differential polynomials in the components of these currents (up to the constraint).

To understand the motivation behind this definition, we now consider the realisation of this AGM in the UV fixed-point of the Klim\v{c}\'{i}k model: then the main property of $\Wc^\cL$ is that its image under the realisation is composed of local left-moving fields. To explain why this is the case, let us consider an element $W$ of $\Wc^\cL$. Since $W$ is gauge-invariant, its image in the realisation can be computed in any gauge, and in particular in one where the gauge-fixed Kac-Moody currents are left-moving fields, ensuring that $W$ itself is left-moving. Although these gauge-fixed currents are in general non-local fields, their particular combination that forms $W$ turns out to be local. Indeed, the image of $W$ in the realisation can be computed before gauge-fixing as well, in which case the currents are realised as local fields of the gauged-formulation of the model (see Subsection \ref{Sec:UVLimAGM}). Note that in the previous subsections, we have discussed various left-moving realisations of the AGM in the different charts of the chiral Klim\v{c}\'{i}k model. The $\Wc$-algebra $\Wc^\cL$ can thus be used to generate systematically local left-moving fields in these different charts. Of course, one can also define a $\Wc$-algebra $\Wc^\cR$ in the AGM with twist function $\vp^\cR(z^\cR)$, whose realisation then yields local right-moving fields.

We will give a more explicit description of this $\Wc$-algebra in the next paragraphs. Before that, let us quickly discuss some of its general properties. It is clear that the products and Poisson brackets of gauge-invariant differential polynomials in the currents $\Jc^\cL_r$ are still gauge-invariant differential polynomials: $\Wc^\cL$ is thus a Poisson algebra. Its Poisson structure can be either computed using the initial Kac-Moody bracket for the non-gauge-invariant currents $\Jc^\cL_r$ or using the Dirac bracket of their gauge-fixed versions, since the Dirac bracket of gauge-invariant quantities coincides with the initial one. Let us finally note that in the above definition, we considered the elements of $\Wc^\cL$ as built from the currents $\Jc^\cL_r$ up to the constraint $\Jc^\cL_1 + \Jc^\cL_2 + \Jc^\cL_3 \approx 0$: this avoids considering different combinations of these currents that will coincide for physical configurations that respect this constraint and thus avoids overcounting fields in the $\Wc$-algebra. Since the constraint allows us to express $\Jc^\cL_3$ in terms of $\Jc^\cL_1$ and $\Jc^\cL_2$, we can see the elements of $\Wc^\cL$ as gauge-invariant differential polynomials in the coefficients of $\Jc^\cL_1$ and $\Jc^\cL_2$ only. The gauge symmetry acting on these two currents is equivalently generated by the sum $\Jc^\cL_1+\Jc^\cL_2$, which is itself a Kac-Moody current of level $\ell^\cL_1+\ell^\cL_2$. Elements of $\Wc^\cL$ can thus be regarded as differential polynomials in the components of $\Jc^\cL_1$ and $\Jc^\cL_2$ that Poisson commute with $\Jc^\cL_1+\Jc^\cL_2$ (note however that the latter is not seen here as a constraint). In this sense, and following the general nomenclature on $\Wc$-algebras, $\Wc^\cL$ is then the classical
\begin{equation}
\frac{\widehat\g_{\ell^\cL_1} \oplus \widehat\g_{\ell^\cL_2}}{\widehat\g_{\ell^\cL_1+\ell^\cL_2}}
\end{equation}
coset $\Wc$-algebra.

\paragraph{Explicit construction of the $\bm\Wc$-algebra.} Let us now give a systematic and explicit construction of the fields in $\Wc^\cL$. As explained at the end of the previous paragraph, we can see these fields as gauge-invariant differential polynomials in $\Jc^\cL_1$ and $\Jc^\cL_2$. The key ingredient of our construction is the current
\begin{equation}\label{Eq:B}
\Kc^\cL(x) = \ell^\cL_2 \,\Jc^\cL_1(x) - \ell^\cL_1 \,\Jc^\cL_2(x).
\end{equation}
One easily checks that this current is covariant, in the sense that under a gauge transformation $\Jc^\cL_r \mapsto h^{-1} \Jc^\cL_r h + \ell^\cL_r\,h^{-1}\p_x h$, it simply transforms as
\begin{equation}
\Kc^\cL \longmapsto h^{-1}\Kc^\cL h.
\end{equation}
It is easy to construct elements of $\Wc^\cL$ using this current. Indeed, if $\Phi$ is a conjugacy-invariant polynomial on $\g^\C$, it is clear that the field $\Phi\bigl(\Kc^\cL\bigr)$ is gauge-invariant and thus belongs to the $\Wc$-algebra $\Wc^\cL$. In particular, the quadratic density
\begin{equation}\label{Eq:W2}
W^\cL_2(x) = \frac{1}{2\ell^\cL_1\ell^\cL_2\bigl(\ell^\cL_1+\ell^\cL_2\bigr)} \bigl\langle \Kc^\cL(x), \Kc^\cL(x) \bigr\rangle
\end{equation}
is the simplest example of a field in $\Wc^\cL$. One easily checks from the Kac-Moody bracket of $\Jc^\cL_1$ and $\Jc^\cL_2$ that this field satisfies a closed Virasoro Poisson bracket:
\begin{equation}\label{Eq:PbW2}
\bigl\lbrace W^\cL_2(x), W^\cL_2(y) \bigr\rbrace = - \bigl(  W^\cL_2(x) + W^\cL_2(y) \bigr) \p_x\delta(x-y).
\end{equation}
The field $W_2^\cL$ is the energy-momentum tensor of the theory and thus plays a particular role in the $\Wc$-algebra.\\

The elements of $\Wc^\cL$ that we have constructed so far are polynomials in $\Jc^\cL_1$ and $\Jc^\cL_2$ but do not involve any of their derivatives. To also include such terms in the construction, it will be useful to introduce the covariant derivative
\begin{equation}
\nabla_x = \p_x + \frac{1}{\ell^\cL_1+\ell^\cL_2} \bigl[ \Jc^\cL_1(x) + \Jc^\cL_2(x), \, \cdot \, \bigr].
\end{equation}
The main property of this operator is that the covariant derivative of a covariant current is itself covariant, as one can easily check. Thus, for $p\in\mathbb{Z}_{\geq 0}$, the current $\nabla^p_x\Kc^\cL(x)$ is covariant. This will allow us to construct many elements of $\Wc^\cL$. Let $F : (\g^{\C})^{\oplus m} \to \C$ be a conjugacy-invariant $m$-multilinear form on $\g^\C$ and $p_1,\ldots,p_m \in\mathbb{Z}_{\geq 0}$ be non-negative integers. Then the field
\begin{equation}\label{Eq:WCurrent}
W^\cL_{F \, ; \, p_1,\ldots,p_m}(x) = F\bigl( \nabla^{p_1}_x \Kc^\cL(x), \ldots, \nabla^{p_m}_x \Kc^\cL(x) \bigr)
\end{equation}
belongs to $\Wc^\cL$. It is clear that this family of fields include the examples constructed above from invariant polynomials of $\Kc^\cL$ only. Let us note that the multilinear form $F$ does not have to be symmetric: one can for instance use the completely skew-symmetric form $F : (X,Y,Z) \mapsto \langle X,[Y,Z] \rangle$. We expect fields of the form \eqref{Eq:WCurrent} to span all the $\Wc$-algebra $\Wc^\cL$. In principle, starting from the Kac-Moody bracket of $\Jc^\cL_1$ and $\Jc^\cL_2$, one can compute the Poisson algebra obeyed by these fields: more precisely, the bracket of two such fields should take the form of a linear combination of derivatives $\p_x^k\delta(x-y)$ of the Dirac distribution, with coefficients being themselves built from fields of the form \eqref{Eq:WCurrent}.\\

Under the chiral realisations $\rho^\cL$ considered in Subsection \ref{Sec:GaugeFixChiral}, the Kac-Moody currents $\Jc^\cL_1$ and $\Jc^\cL_2$ in a certain gauge are expressed as left-moving fields in the UV fixed-point of the Klim\v{c}\'{i}k model. These fields typically take the form of linear combinations of the derivatives $\p_+ \phi^a$ of the coordinate fields $\phi^a$ of the $\sigma$-model, with well-chosen coefficients (which can be themselves field dependent). In particular, the currents $\rho^\cL\bigl(\Jc^\cL_1)$ and $\rho^\cL\bigl(\Jc^\cL_2\bigr)$ are spin 1 left-moving fields. Since the spatial derivative $\p_x$ of a left-moving field coincides with its light-cone derivative $\p_+$, we can replace all covariant derivatives acting on these fields by light-cone covariant derivatives
\begin{equation}\label{asjckxnka}
\nabla_+ = \p_+ + \frac{1}{\ell^\cL_1+\ell^\cL_2} \bigl[ \rho^\cL\bigl(\Jc^\cL_1\bigr)(x^+) + \rho^\cL\bigl(\Jc^\cL_1\bigr)(x^+), \, \cdot \, \bigr].
\end{equation}
The image of the field \eqref{Eq:WCurrent} under the realisation then takes the form
\begin{equation}
\rho^\cL\bigl( W^\cL_{F \, ; \, p_1,\ldots,p_m} \bigr) (x^+) = F\bigl( \nabla^{p_1}_+ \Kc^{\rho}(x^+), \ldots, \nabla^{p_m}_+ \Kc^{\rho}(x^+) \bigr),
\end{equation}
where we have abbreviated $\rho^\cL\bigl( \Kc^\cL \bigr)$ as $\Kc^{\rho}$. This is a left-moving field with Lorentz spin $p_1+\ldots+p_m+m$. Although the realised current $\Kc^{\rho}(x^+)$ is in general non-local, the above combination is such that non-local contributions cancel, yielding in the end a local field.

\paragraph{The $\bm\Wc$-algebra for $\bm{\mathfrak{sl}(2)}$.} Let us now turn our attention to the case $\g^\C=\mathfrak{sl}(2)$. The $\Wc$-algebra $\Wc^\cL$ contains only one spin 2 field, namely the energy-momentum tensor $W^\cL_2$ defined in eq.\;\eqref{Eq:W2}. Following the general construction outlined in the previous paragraph, we can also construct a spin 3 field in $\Wc^\cL$ as $\langle \Kc^\cL, \nabla\Kc^\cL\rangle$. Using the ad-invariance of $\langle\cdot,\cdot\rangle$, one checks that the commutator term in the covariant derivative $\nabla\Kc^\cL$ vanishes when contracted with $\Kc^\cL$. This spin 3 field is therefore equal to $\langle \Kc^\cL, \p\Kc^\cL\rangle$ and thus proportional to the derivative $\p W_2^\cL$ of the energy-momentum tensor. There is no new primary spin 3 field.

Let us now turn our attention to spin 4 fields. Up to a constant, the only invariant polynomial of degree 4 on $\g^\C=\mathfrak{sl}(2)$ is the square of the bilinear form $\langle\cdot,\cdot\rangle$. This yields a spin 4 field in $\Wc^\cL$ proportional to $\langle \Kc^\cL, \Kc^\cL \rangle^2$ and thus to $W_2^{\cL\,2}$. Let us now look at the spin 4 fields that involve covariant derivatives of $\Kc^\cL$. We find that there are two, given by 
$\langle \Kc^\cL, \nabla^2 \Kc^\cL \rangle$ and $\langle \nabla \Kc^\cL, \nabla \Kc^\cL \rangle$. A direct computation shows that their sum is proportional to the second derivative $\p^2W^\cL_2$ of the energy-momentum tensor. In the end, there is only one spin 4 field in $\Wc^\cL$ which is not expressed in terms of $W^\cL_2$ and its derivatives, which we can choose as
\begin{equation}\label{mncxnbsad}
W^\cL_4(x) =  \frac{1}{2\ell^\cL_1\ell^\cL_2\bigl(\ell^\cL_1+\ell^\cL_2\bigr)} \ 
\bigl\langle \nabla_x \Kc^\cL(x), \nabla_x \Kc^\cL(x) \bigr\rangle.
\end{equation}
A similar analysis shows that there are three spin 5 fields in $\Wc^\cL$, all expressible in terms of $W^\cL_2$, $W^\cL_4$ and their derivatives, namely $\p^3 W^\cL_2$, $W^\cL_2 \p W^\cL_2$ and $\p W^\cL_4$. For spin 6, we find 6 fields expressed in terms of the lower-spin ones and two new ones.
A summary is provided in Table \ref{tab1}.

\begin{table}[h]
\begin{center}
\scalebox{1.00}{
\begin{tabular}{|c|l|}
\hline
& \\[-0.3cm]
spin &  \\[0.1cm]
\hline
& \\[-0.2cm]
$m=2$ & $W_2=\langle \widehat{\Kc}, \widehat{\Kc} \rangle$ \\[0.4cm]

& \\[-0.2cm]
$m=3$ & $\partial W_2=2\,\langle \widehat{\Kc}, \nabla\widehat{\Kc} \rangle$ \\[0.4cm]

& \\[-0.2cm]
$m=4$ & 
$W_4=\langle \nabla\widehat{\Kc}, \nabla\widehat{\Kc} \rangle$\,, \ 
$W_2^2$\,,\  \ $\partial^2 W_2=2\,\langle \widehat{\Kc}, \nabla^2 \widehat{\Kc} \rangle+
2\,W_4$  \\[0.4cm]
& \\[-0.2cm]
$m=5$ & $W_2\partial W_2$\,,\ \ 
$\partial W_4=2\,\langle\nabla \widehat{\Kc},  \nabla^2\widehat{\Kc}\rangle$\,,\ \
 $\partial^3 W_2=3\partial W_4+2\,\langle \widehat{\Kc},\nabla^3 \widehat{\Kc}\rangle$\\[0.4cm]
& \\[-0.2cm]
$m=6$ & $W_2^3$\,,\ \ 
$W_4 W_2$\,,\ \ 
$(\partial W_2)^2$\,, \ \ 
$\partial^2 W_4=2\,\langle\nabla \widehat{\Kc},
\nabla^3  \widehat{\Kc}\rangle+
2\,\langle\nabla^2  \widehat{\Kc},\nabla^2  \widehat{\Kc}\rangle$\\[0.2cm]
&  \\[-0.2cm]
 &
$W_2\partial^2 W_2$\,,\ \ 
$\partial^4 W_2=2\,\langle\widehat{\Kc},\nabla^4\widehat{\Kc}\rangle+
8\,\langle\nabla\widehat{\Kc},\nabla^3\widehat{\Kc}\rangle+
6\,\langle\nabla^2\widehat{\Kc},\nabla^2\widehat{\Kc}\rangle$
\\[0.2cm]
&  \\[-0.2cm]
 &
$W_{6A}=\langle\nabla^2\widehat{\Kc},\nabla^2\widehat{\Kc}\rangle$\,,\ \ 
$W_{6B}=C\,\langle[\widehat{\Kc},\nabla\widehat{\Kc}],\nabla^2\widehat{\Kc}\rangle$
\\[0.2cm]
\hline
\end{tabular}
}
\caption{\label{tabdsas}
The fields 
from the coset ${\cal W}$-algebra with $\mathfrak{g}^\C=\mathfrak{sl}(2)$ for the first few spins $m=2,3,\ldots,6$.
The simplified notation $W_m = W_m^\cL$ has been introduced to make for easier reading, while
the currents have been expressed in terms of 
$\widehat{\Kc}=C\,\Kc^\cL$ with 
$C=\big(2\ell_1^\cL\,\ell_2^\cL\,(\ell_1^\cL+\ell_2^\cL)\big)^{-\frac{1}{2}}$  
rather than $\Kc^\cL$ 
 in order to reduce the presence of  
$\ell_1^\cL$ and $\ell_2^\cL$ dependent factors. Notice that  the constant
$C$  has been included in the definition of  $W_{6B}$.
\label{tab1}}
\end{center}
\end{table}
\bigskip

One can compute the Poisson bracket obeyed by these various fields starting from the Kac-Moody bracket of $\Jc^\cL_1$ and $\Jc^\cL_2$. We have already pointed out in eq.\;\eqref{Eq:PbW2} that $W^\cL_2$ satisfies a classical Virasoro algebra.
 Similarly, we find
\bea\label{jciuxhjsa1}
\{W_2^\cL(x),W_4^\cL(y)\}&=&\partial_x^2\delta(x-y)\,\partial_x W_2^\cL(x)-
2\partial_x\delta(x-y)\,\big(2W_4^\cL(x)-\partial^2_x W_2^\cL(x)\big)\nonumber \\[0.2cm]
&&\,
-\;\delta(x-y)\,\big(3\partial_x W_4^\cL(x)-\partial^3_x W_2^\cL(x)\big)
 \\[0.4cm]
\{W_4^\cL(x),W_4^\cL(y)\}&=&\partial_x^3\delta(x-y)\,\big(W_4^\cL(x)+W_4^\cL(y)\big) 
-\partial_x\delta(x-y)\,
\big(W_6^\cL(x)+W_6^\cL(y)\big)\, .
\nonumber
\eea
with
\be\label{oias981aaa2osa}
W_6^{({\rm L})}=3\,W_{6A}^{({\rm L})} +
4\,\big(\ell^\cL_2-\ell^\cL_1\big)\,
 W_{6B}^{({\rm L})} +
\frac{1}{\ell_1^\cL+\ell_2^\cL}\,\Big(\big(\partial W_2^{({\rm L})}\big)^2-4\,W_4^{({\rm L})}\,W_2^{({\rm L})}\Big)
\ee
(the notation $W_{6A}^{({\rm L})}$ and $W_{6B}^{({\rm L})}$ is explained in the last line of Table \ref{tab1}).

\paragraph{Realisation in the chiral Klim\v{c}\'{i}k model.} Let us now express how the $\Wc$-algebra $\Wc^\cL$ is realised in the UV fixed-point of the Klim\v{c}\'{i}k model for $G={\rm SU}(2)$. For that, we can use the various types of realisations described in the two previous subsections, which in particular can correspond to different charts of the model. For instance, in chart I (corresponding to the left cigar with coordinates $(\phi,\alpha)$ and the free boson $\chi$), one can use the parafermionic realisation $\rho^\cL_P$ defined in eq \eqref{Eq:ParafReal}. 
The basic building blocks for the fields of the $\Wc$-algebra are the dynamical 
quantity $\Kc^{({\rm L})}$ \eqref{Eq:B}, built out of the Kac-Moody currents, and 
the covariant derivative \eqref{asjckxnka}. Applying the map $\rho^\cL_P$,
one finds that  
\be
\rho^\cL_P({\Kc}^\cL)=4\ell_1^{({\rm L})}\,\ell_2^{({\rm L})}\,{\cal L}_P^{(1)}\,,
\qquad\qquad \qquad\qquad \nabla_+=\partial_++\big[{\cal L}_P^{(0)},\cdot\big]\ .
\ee
Here ${\cal L}_P^{(1)}$ denotes the coefficient of
$z^{({\rm L})}$  in the Lax matrix \eqref{kxjnasmn121},
 while ${\cal L}_P^{(0)}$ is the constant term, i.e.,
$$
{\cal L}_P^{(0)}=\psi_- {\tt e}_++\psi_+ {\tt e}_-\,,\qquad 
{\cal L}_P^{(1)}=-\tfrac{\ri}{4}\,(\nu+\nu^{-1})\,\partial_+\chi \, {\tt h}-
\tfrac{1}{4}\,(1+\nu^2)\,\psi_-{\tt e}_++\tfrac{1}{4}\,(1+\nu^{-2})\,\psi_+\,{\tt e}_-\, .
$$
The  formulae \eqref{Eq:W2} for the spin $2$ current 
and \eqref{mncxnbsad} for the
spin $4$ one yield the expressions \eqref{asoi9812}
and \eqref{opcx90sa}, respectively,
that appeared before in Subsection \ref{ias9812211aaa}.
Notice that, although the matrix elements of
${\cal L}_P^{({\rm L})}$ itself contain the non-local fields $\psi_\pm$,
the currents $W_2$ and $W_4$ turn out to be differential polynomials
in $\partial_+\chi$ and $T_m^{({\rm L})}$. Recall that the latter
are the special combinations of the classical parafermions 
\eqref{aosoic98aaaa2}-\eqref{lkkjasknnksmn21}, which are local in the original
fields $(\phi,\alpha,\chi)$.
 As before, a similar construction also holds for the right-moving W-currents.

So far, we have focused on chart I. Similarly, one can realise the $\Wc$-algebra in the chart III (corresponding to the right cigar with coordinates $(\phi,\widetilde{\alpha})$ and the free boson $\widetilde{\chi}$) and in the chart II (\textit{i.e.} the asymptotic domain where the currents would be expressed in terms of the left moving 
bosons $(\phi_1^{({\rm L})},\phi_2^{({\rm L})},\phi_3^{({\rm L})})$) using the other realisations described in Subsection \ref{Sec:GaugeFixChiral}.
 In conclusion, the AGM construction thus provides an efficient systematic procedure to construct local chiral fields in the different charts of the chiral Klim\v{c}\'{i}k model.

\paragraph{Local charges.} So far, we have used the AGM formalism to describe local chiral fields of the model and the closed Poisson algebra $\Wc^\cL$ that they form. It is also natural to look for local charges built from integrals of chiral fields in $\Wc^\cL$ that are pairwise in involution and thus part of the integrable structure of the model. We have discussed such local charges in the general formalism of AGMs in Subsection \ref{Sec:AGM}: in particular, we have recalled that they are naturally associated with the zeroes of the twist function of the AGM -- see eq.\;\eqref{Eq:Qip}. In the present case, the twist function $\vp^\cL(z^\cL)$ is given by eq.\;\eqref{Eq:TwistKlimL}. The 1-form $\vp^\cL(z^\cL)\dd z^\cL$ then possesses one zero, at $z^\cL=\infty$. Under the constraint $\Jc^\cL_1+\Jc^\cL_2+\Jc^\cL_3 \approx 0$, one checks that the evaluation of the 1-form $\Gamma^\cL(z^\cL)\dd z^\cL$ at $z^\cL=\infty$ is proportional to the current $\Kc^\cL$ introduced in eq.\;\eqref{Eq:B}. For this AGM, the local charges discussed in Subsection \ref{Sec:AGM} then take the form
\begin{equation}
\Q_p^\cL \approx \alpha_p \int \Phi_p \bigl( \Kc^\cL(x) \bigr) \dd x,
\end{equation}
where $p\in \Eh$ runs over the exponents of $\gh$, $\Phi_p$ is a well-chosen invariant polynomial of degree $p+1$ on $\g$ introduced in Subsection \ref{Sec:AGM} and $\alpha_p$ is an overall normalisation factor. In particular, the quadratic local charge $\Q_1^\cL$ is simply given by the integral of the energy-momentum tensor:
\begin{equation}
\Q_1^\cL \approx \int W^\cL_2(x) \, \dd x = \frac{1}{2\ell^\cL_1\ell^\cL_2\bigl(\ell^\cL_1+\ell^\cL_2\bigr)} \int \bigl\langle \Kc^\cL(x), \Kc^\cL(x) \bigr\rangle \, \dd x.
\end{equation}
More generally, the density $\alpha_p \, \Phi_p ( \Kc^\cL )$ of the charge $\Q_p^\cL$ is a local chiral field in the $\Wc$-algebra $\Wc^\cL$ (see previous paragraphs). These particular densities are chosen specifically to ensure the involution of the charges $\Q_p^\cL$.

For the case $\g^\C=\mathfrak{sl}(2)$, the exponents are all odd numbers $2m-1$, $m\in\mathbb{Z}_{\geq 1}$. The corresponding polynomials $\Phi_{2m-1}$ have degree $2m$ and are simply proportional to $\langle \cdot,\cdot \rangle^m$. Thus, the local charges are given by
\begin{equation}
\Q_{2m-1}^\cL \approx \int \bigl.W_2^\cL(x)\bigr.^m\,\dd x.
\end{equation}
They coincide with the ones obtained in Subsection \ref{oias98oia} from screening charges. In particular, they involve only the spin 2 field of the $\Wc$-algebra and none of the other fields. As we will see in Section \ref{Sec:QuantLocalIM}, this will not be the case anymore at the quantum level.

\section{Quantisation of non-local IMs\label{sec666}}

The problem of quantization of the non-local IMs for the affine Gaudin model has not been systematically studied yet
(see, however, ref. \cite{Feigin:2007mr} for some first results).
For the classical field theory a well understood framework exists.
The zero curvature representation for the classical equations of motion
implies  a one parameter family of conserved quantities generated by the
trace of the monodromy $M(z)$. The latter can be thought of as the infinite series:
\be\label{oias9821oias}
\overset{\leftarrow}{{\cal P}}\exp\bigg(-\int_0^{2\pi} \rd x\, {\cal L}(z;x)
\bigg)=1\ -
\int_0^{2\pi}\rd x \,{\cal L}(z;x)+
\!\!\!\!\!\!\!\!\!\!\int\limits_{0<x_2<x_1<2\pi}\!\!\!\!\!\!\!\!\!\!
\rd x_1\rd x_2\  {\cal L}(z;x_1)\,{\cal L}(z;x_2)+\ldots
\ee
with ${\cal L}(z;x)=\frac{1}{\varphi(z)}\,\Gamma(z;x)$ so that the corresponding IMs
 involve
multifold ordered integrations
of  fields built from the Kac-Moody currents ${\cal J}_{i}(x_j)$.
While such classical expressions are well defined,
providing a meaningful definition of the
quantum family of commuting operators runs into problems.
As such, although a formula for the simplest of them corresponding to the first
 term in the expansion of the monodromy
at  $z=z_i$
has been proposed in \cite{Feigin:2007mr},  a systematic and rigorous study
of the full set of non-local charges, including a proof of their mutual commutativity,
is not available at the moment at least for general Lie group $G$ 
(results for $G={\rm SU}(2)$ may be  extracted from the
recent paper \cite{Kotousov:2021vih}). 
\bigskip

One issue can already be traced  at the classical level.
It occurs in the derivation of the Poisson algebra satisfied by the matrix elements of the monodromy.
The $r\big/s$  bracket of the Lax matrix at different spatial points $x$ and $y$
contains a term proportional to the generalized distribution $\delta'(x-y)$, see \eqref{maillet1a}.
In consequence, when calculating the PBs of two path ordered exponents \eqref{oias9821oias},
contact terms arise from the integration of the derivative of the
$\delta$ distribution. These are ambiguous, and their contribution to the final result depends 
on the regularization scheme one employs to treat them. 
This problem was originally studied in detail in  the works of Maillet \cite{Maillet:1985fn,Maillet:1985ek}.
He introduced
a  symmetric point-splitting
prescription  to handle the ambiguities. The computation of the Poisson brackets of the
monodromy matrix then resulted in a ``new integrable canonical structure'', which did not correspond to the classical 
limit of any known quantum algebra at the time. 
In view of this result, it became unclear how to proceed with the quantization of the monodromy matrix even at the formal
algebraic level. The difficulty in quantizing a classically integrable system, where the Poisson brackets
of the Lax connection contain a $\delta'(x-y)$ term,  has since become known as the problem with non-ultralocality.
\bigskip

A full discussion of the Poisson bracket algebra
for $M(z)$  will be reserved for a separate publication.
Here our purpose is to review a proposal presented in ref. \cite{Bazhanov:2018xzh}
that points to  a way of making progress in quantizing the non-local IMs 
for the chiral affine  Gaudin model discussed in the previous section. 
In that work,  using the representation theory of the quantum affine algebra $\mathcal{U}_q\big(\widehat{\mathfrak{sl}}(2)\big)$, 
a  quantum monodromy matrix  satisfying
the Yang-Baxter algebra is constructed. Based on perturbative computations,
it is conjectured that  in the  classical limit the quantum monodromy  becomes
a classical one built from the Lax matrix 
\be\label{Anew-spec}
{\cal L}_{\scriptscriptstyle {\rm UV}}'(\rho)= -
\frac{\rho}{1-\rho^2}\ \big(\ V_+^{}\ {\tt e}_+
+\,V_-^{}\ {\tt e}_-\big)-
\frac{1}{2}\ 
\frac{\rho^2}{1-\rho^2}\,V_0^{}\,{\tt h}\,.
\ee 
Here $V_s$ with $s=+,-,0$ are dynamical variables obeying 
 the Poisson bracket algebra
\begin{align}\label{PBcurrents-remind}
&\big\{V_+^{}(x)\,,\,
V_-^{}(y)\big\}={\nu^2}\ \delta'(x-y)-
V_0^{}(x)\ 
\delta(x-y) -
V_+^{}(x)\ V_-^{}(y)\ \epsilon(x-y)
\nonumber \\[0.2cm]
&\big\{V_0^{}(x)\,,\,
V_\pm^{}(y)\big\}=\mp 
{2} \,V_\pm^{}(x)\ \delta(x-y) \\[0.2cm]
&\big\{V_0^{}(x)\,,\,
V_0^{}(y)\big\}=
{2\nu^2}\ \delta'(x-y) \nonumber \\[0.2cm]
&\big\{V_\pm^{}(x)\,,\,
V_\pm^{}(y)\big\}= 
V_\pm^{}(x)\ 
V_\pm^{}(y)\ \epsilon(x-y)\, .
\nonumber
\end{align}
At this point we note that ${\cal L}_{\scriptscriptstyle {\rm UV}}'$ coincides with the special case $\omega=1$
of the family of flat connections ${\cal L}_{\scriptscriptstyle {\rm UV}}^{(\omega)}$
from eq.\;\eqref{Anew}, up to a similarity transformation
that depends on the spectral parameter $\rho$, but not on the dynamical fields.
In addition, the Poisson brackets for $V_s$ are the same ones that are satisfied by $V_s^{(1)}$ in \eqref{PBcurrents1}
upon setting the inessential constant $K_{\scriptscriptstyle \rm UV}$ to one.
Recall that  ${\cal L}_{\scriptscriptstyle {\rm UV}}^{(\omega)}$ coincides with the Lax matrix obtained from
the AGM
 considered in Section \ref{sec555389i} for the case $G={\rm SU}(2)$ provided that a particular gauge fixing condition 
is imposed on the Kac-Moody currents, see  Subsection \ref{Sec:GaugeFixChiral}. 
The transfer-matrix for the corresponding Gaudin model
is a gauge invariant object and is also unaffected by
the similarity transformation relating ${\cal L}_{\scriptscriptstyle {\rm UV}}'$ to ${\cal L}_{\scriptscriptstyle {\rm UV}}^{(1)}$. 
This way,  the trace of the quantum monodromy matrix from ref. \cite{Bazhanov:2018xzh}
provides a potential definition
of the quantum non-local IMs for this AGM.
\bigskip

A rigorous proof of the conjectures from \cite{Bazhanov:2018xzh}, as well as their generalization to higher rank Lie groups
and a general number of punctures,
would  be valuable for the study of the quantized affine Gaudin model.
The aim of this section is to formulate the main statements concisely, and to discuss some
consequences relevant for our goals.

\subsection[Quantum/classical monodromy for $G={\rm SU}(2)$]{Quantum/classical monodromy for \texorpdfstring{$\bm{G={\rm SU}(2)}$}{G=SU(2)}}
\paragraph{Universal $\bm R$-matrix.}
The construction of the quantum monodromy matrix in \cite{Bazhanov:2018xzh},
which itself is based on the work \cite{Bazhanov:1998dq}, uses the representation theory of the
quantum affine algebra $\mathcal{U}_q\big(\widehat{\mathfrak{sl}}(2)\big)$.
The latter is a Hopf algebra
with generators $\mathsf{X}_0,\mathsf{X}_1$, $\mathsf{Y}_0, \mathsf{Y}_1$, $\mathsf{H}_0$ and
$\mathsf{H}_1$. A 
summary of the set of relations and the definition of the co-product $\Delta$ can be 
found, e.g., in \cite{Bazhanov:1998dq}. For our purposes it is enough to mainly focus on 
the Borel subalgebra ${\cal U}_q(\widehat{\mathfrak{b}}_-)\subset \mathcal{U}_q(\widehat{\mathfrak{sl}}(2))$
generated by the set
 $\{\mathsf{X}_0$, $\mathsf{X}_1$, $\mathsf{H}_0$, $\mathsf{H}_1\}$. 
These operators, amongst themselves, satisfy the commutation relations
\be\label{amnxnbA}
[\mathsf{H}_0,\mathsf{X}_0]=-[\mathsf{H}_1,\mathsf{X}_0]=-2\mathsf{X}_0\,,\qquad
[\mathsf{H}_0,\mathsf{X}_1]=-[\mathsf{H}_1,\mathsf{X}_1]=2\mathsf{X}_1\,,
\qquad [\mathsf{H}_0,\mathsf{H}_1]=0
\ee
as well as the quartic Serre relations
\be\label{Serre1}
\mathsf{X}_i^3\mathsf{X}_j-[3]_q\,\mathsf{X}_i^2
\mathsf{X}_j\mathsf{X}_i+[3]_q\,
\mathsf{X}_i\mathsf{X}_j\mathsf{X}_i^2-
\mathsf{X}_j\mathsf{X}_i^3=0 \qquad\qquad (i,j=0,1) 
\ee
with $[m]_q=(q^m-q^{-m})/(q-q^{-1})$.
Also note that we set the central element  $\mathsf{K}=\mathsf{H}_0+\mathsf{H}_1$ to be zero:
\be
\mathsf{H}_0+\mathsf{H}_1=0\, .
\ee
\smallskip

The algebra  ${\cal U}_q(\widehat{\mathfrak{b}}_-)$  turns out to 
possess an infinite dimensional
commuting subalgebra. It is defined starting from 
 the notion 
of the 
universal $R$-matrix ${\mathcal R}$.
The latter is an element 
of $\mathcal{U}_q(\widehat{\mathfrak{b}}_+)\otimes \mathcal{U}_q(\widehat{\mathfrak{b}}_-)$,
where $\mathcal{U}_q(\widehat{\mathfrak{b}}_+)$ 
is generated by $\mathsf{Y}_0$, $\mathsf{Y}_1$, $\mathsf{H}_0=-\mathsf{H}_1$,
while $\mathcal{U}_q(\widehat{\mathfrak{b}}_-)$ is as above.
It satisfies the Yang-Baxter  equation
\be\label{YBAaaaaa}
 \mathcal{R}_{12}\,\mathcal{R}_{13}\,\mathcal{R}_{23}\,=\,\mathcal{R}_{23}\,
\mathcal{R}_{13}\,\mathcal{R}_{12} 
\ee
with the left and right hand sides   being valued in three tensor copies of  $\mathcal{U}_q\big(\widehat{\mathfrak{sl}}(2)\big)$ and
the lower two indices indicate to which factors of the tensor product the universal ${ R}$-matrix belongs.
It turns out that expressing ${\cal R}$ in the form:
\begin{equation}\label{sann12as}
\mathcal{R}=q^{\frac{1}{2}\mathsf{H}_0\otimes \mathsf{H}_0}\,\overline{\mathcal{R}}
\end{equation}
and requiring  $\overline{\mathcal{R}}$ to be a formal series in the nilpotent generators $\mathsf{X}_0$ and $\mathsf{X}_1$,
is enough to specify  the universal $R$-matrix uniquely. In particular,
using  the  product formula  presented in \cite{Khoroshkin:1994um}, 
one can develop the series expansion
\begin{equation}\begin{aligned}\label{sann12asB}
\overline{\mathcal{R}}\,&=\,\mathrm{id}+(q-q^{-1})\,\big(\mathsf{Y}_0\otimes \mathsf{X}_0+\mathsf{Y}_1\otimes \mathsf{X}_1\big)
+\frac{(q-q^{-1})^2}{1+q^{-2}}\,\big(\mathsf{Y}_0^2\otimes \mathsf{X}_0^2+\mathsf{Y}_1^2\otimes \mathsf{X}_1^2\big)\\
& \qquad+ \frac{q-q^{-1}}{q+q^{-1}}\Big(\mathsf{Y}_0\mathsf{Y}_1\otimes\big(
\mathsf{X}_1\mathsf{X}_0-q^{-2}\,\mathsf{X}_0\mathsf{X}_1\big)+
\mathsf{Y}_1\mathsf{Y}_0\otimes\big(\mathsf{X}_0\mathsf{X}_1-q^{-2}\,\mathsf{X}_1\mathsf{X}_0\big)\Big)+\dots
\end{aligned}\end{equation}
up to any order in the generators.\footnote{%
Here we stick to the conventions of ref. \cite{Bazhanov:1998dq} which 
differ slightly from those of \cite{Bazhanov:2018xzh}, see also footnote 1 of the latter work.}

\paragraph{``Universal'' monodromy matrix.}
Consider the evaluation homomorphism from ${\cal U}_q\big(\widehat{\mathfrak{sl}}(2)\big)$
to the loop algebra ${\cal U}_q\big(\mathfrak{sl}(2)\big)[\lambda,\lambda^{-1}]$ and
 further specify a matrix irrep $\pi$ of 
${\cal U}_q\big(\mathfrak{sl}(2)\big)$. With some abuse of notation, we will use the symbols
${\tt e}_\pm$, ${\tt h}$
for the generators of the quantum algebra specialized to that finite dimensional representation. 
Under this map, the elements of ${\cal U}_q\big(\widehat{\mathfrak{sl}}(2)\big)$
become
\begin{equation}\label{homo1}
\begin{aligned}
& \pi_\lambda\big(\mathsf{X}_0\big)=\lambda^{-1} \,{\tt e}_-\,q^{-\frac{1}{2}{\tt h}},\\
&\pi_\lambda\big(\mathsf{X}_1\big)=\lambda^{-1} \,{\tt e}_+\,q^{+\frac{1}{2}{\tt h}},\end{aligned}
\qquad\qquad\begin{aligned}
& \pi_\lambda\big(\mathsf{Y}_0\big)=\lambda \,q^{+\frac{1}{2}{\tt h}}\,{\tt e}_+,\\
& \pi_\lambda\big(\mathsf{Y}_1\big)=\lambda \,q^{-\frac{1}{2}{\tt h}}\,{\tt e}_-,
\end{aligned}
\qquad\qquad
\pi_\lambda\big(\mathsf{H}_0\big)=-\pi_\lambda\big(\mathsf{H}_1\big)={\tt h}\,,
\end{equation}
where
\be\label{comm5}
[{\tt h},\,{\tt e}_\pm]=\pm\, 2\,{\tt e}_\pm \ , \qquad [{\tt e}_+,\,{\tt e}_-]=\frac{q^{{\tt h}}-q^{-{\tt h}}}{q-q^{-1}} \ .
\ee
By means of such a homomorphism, one can define the ``universal'' monodromy matrix:
\begin{equation}\label{asnbnb21}
\mathsf{M}(\lambda)=(\pi_\lambda\otimes \mathrm{id})(\mathcal{R})\, .
\end{equation}
Up to the factor $q^{\frac{1}{2}\mathsf{H}_0{\tt h}}$, this would be
a formal power series in $\lambda$ with coefficients being
matrices, whose entries are
polynomials in the unspecified generators $\mathsf{X}_0$ and $\mathsf{X}_1$. 
For instance, according to  eqs.\;\eqref{sann12as} and \eqref{sann12asB}, the first few terms
 read
\bea\label{series1AAAAAA}
\mathsf{M}(\lambda)&=&q^{\frac{1}{2}\mathsf{H}_0{\tt h}}\,\bigg[{\rm id}+\lambda\,(q-q^{-1})\,
\big(q^{\frac{{\tt h}}{2}}\,{\tt e}_+\,\mathsf{X}_0+q^{-\frac{{\tt h}}{2}}\,{\tt e}_-\,\mathsf{X}_1\big)
+\frac{\lambda^2\,(q-q^{-1})^2}{1+q^{-2}}\nonumber\\[0.3cm]
&\times & 
\bigg(
 (q^{\frac{{\tt h}}{2}}\,{\tt e}_+)^2\ \mathsf{X}_0^2
+
 (q^{-\frac{{\tt h}}{2}}\,{\tt e}_-)^2\ \mathsf{X}_1^2+
\frac{(q^{\frac{{\tt h}}{2}}\,{\tt e}_+)
(q^{-\frac{{\tt h}}{2}}\,{\tt e}_-)}{q^{2}-1}\ \big(
\mathsf{X}_1\mathsf{X}_0-q^{-2}\,\mathsf{X}_0\mathsf{X}_1\big)
 \nonumber\\[0.3cm]
&+&
\frac{(q^{-\frac{{\tt h}}{2}}\,{\tt e}_-)
(q^{\frac{{\tt h}}{2}}\,{\tt e}_+)}{q^{2}-1}\ 
\big(
\mathsf{X}_0\mathsf{X}_1-q^{-2}\,\mathsf{X}_1\mathsf{X}_0\big)\bigg)+O(\lambda^3)\bigg]\, .
\eea
The 
commutation relations satisfied by the matrix elements of $\mathsf{M}(\lambda)$ follow from eqs.\;\eqref{amnxnbA} and \eqref{Serre1}.
It turns out to be possible to represent them in the form
\begin{equation}\label{RLL}
R(\lambda/\mu) \big[{\mathsf M}(\lambda)\otimes \mathrm{id}\big]\big[\mathrm{id}\otimes{\mathsf M}(\mu)\big]
=
\big[\mathrm{id}\otimes{\mathsf M}(\mu)\big] 
\big[{\mathsf M}(\lambda)\otimes \mathrm{id}\big]
R(\lambda/\mu)
\end{equation}
with  the numerical matrix $R$ being
defined as 
\begin{equation}
R(\lambda/\mu)=(\pi_\lambda\otimes \pi_{\mu})(\mathcal{R})\,.
\end{equation}
These  are 
a direct consequence of applying the
homomorphism  $\pi_\lambda\otimes \pi_\mu\otimes \mathrm{id}$ to both sides
of the Yang-Baxter relation \eqref{YBAaaaaa} 
for the
universal $R$-matrix.  
Note that, in the case when the representations  $\pi_\lambda$ and $\pi_\mu$ are chosen to be   the fundamental ones,
$R(\lambda)$ coincides with the trigonometric solution of the Yang-Baxter equation associated with the six-vertex model.
Namely, up to an overall scalar factor depending on the spectral parameter,  $R(\lambda)$  would be given by
\be\label{jxcbnsdbnjhsa}
R_{\frac{1}{2}\frac{1}{2}}(\lambda)=
\left(\begin{array}{cccc}
q^{-1}\lambda-q\lambda^{-1} & & & \\[0.2cm]
& \lambda - \lambda^{-1} & q^{-1}-q & \\[0.2cm]
& q^{-1}-q &\lambda - \lambda^{-1} & \\[0.2cm]
 & & & q^{-1}\lambda-q\lambda^{-1} 
\end{array}\right)\, .
\ee
The Yang-Baxter algebra \eqref{RLL} yields an infinite family of commuting operators in 
${\cal U}_q(\widehat{\mathfrak{b}}_-)$,
built from the formal generators $\mathsf{X}_0$, $\mathsf{X}_1$ and $\mathsf{H}_0=-\mathsf{H}_1$.
They are
obtained by taking a suitable trace of the monodromy over the matrix representation $\pi$:
\be\label{askjnnm23}
\mathsf{T}(\lambda)={\rm Tr}\big(q^{\frac{1}{2} f(\mathsf{H}_0)\,{\tt h}}\,\mathsf{M}(\lambda)\big)\,.
\ee
Here, as explained in \cite{Bazhanov:1998dq},
the extra term appearing in the trace, with $f(\mathsf{H}_0)$ being any function of the generator $\mathsf{H}_0$,
does not spoil the mutual commutativity of the transfer-matrices,
\be
\big[\mathsf{T}(\lambda),\mathsf{T}(\mu)\big]=0\ .
\ee
\bigskip

The universal monodromy matrix obeys the Yang-Baxter algebra \eqref{RLL}
as a consequence of the commutation relations \eqref{amnxnbA} and the Serre relations \eqref{Serre1}
for the generators of ${\cal U}_q(\widehat{\mathfrak{b}}_-)$. Different representations of these generators give rise
 to  commuting families of operators which are associated to different integrable models.
The work \cite{Bazhanov:1998dq} considers an infinite dimensional representation, where
$\mathsf{X}_{0,1}$ and $\mathsf{H}_{0,1}$ act in the (extended) Fock space of one bosonic field.
In this case the transfer-matrix \eqref{askjnnm23} yields the local and non-local  
IMs of the quantum KdV integrable structure, provided that the extra factor in the trace is set to be
$
q^{\frac{1}{2} f(\mathsf{H}_0)\,{\tt h}}=q^{\frac{1}{2} \mathsf{H}_0\,{\tt h}}
$.
The KdV case serves as a prototype of the construction from
 ref. \cite{Bazhanov:2018xzh}, which is relevant to the ${\rm SU}(2)$ Klim\v{c}\'{i}k model.
The reader may find the review of the basic case contained in Section 2 of that paper
as a helpful companion to what is discussed below.

\paragraph{Quantum monodromy matrix for the Fateev integrable structure.}
The representation of ${\cal U}_q(\widehat{\mathfrak{b}}_-)$ first appearing in
the works \cite{Semikhatov:2001zz,Bazhanov:2013cua} and then considered in ref. \cite{Bazhanov:2018xzh}
takes three independent copies of the Heisenberg algebra as its starting point:
\be\label{phidef2}
[\hat{a}_l^{(j)},\hat{a}_m^{(j)}]=\,\tfrac{l}{2}\,\delta_{l+m,0}\ , \qquad [\hat{\varphi}_0^{(j)},\hat{p}^{(j)}]=\tfrac{\ri}{2}\qquad
\qquad (j=1,2,3)\,.
\ee
One groups the generators into bosonic fields
\be\label{phidef1}
\hat{\varphi}_j(x)=\hat{\varphi}_0^{(j)}+x\,\hat{p}^{(j)}+\ri\,\sum_{m\ne 0} \frac{\hat{a}_m^{(j)}}{m}\,\re^{-\ri m x}
\ee
and then constructs the vertex operators
\be\label{vertex2}
\mathsf{V}_{\pm}=\frac{1}{\sqrt{n\varpi}}\,:\Big( \,\ri \sqrt{\varpi}\, \partial\hat{\varphi}_3+ \sqrt{\tfrac{n+2}{n}}
\ \partial\hat{\varphi}_2\pm\sqrt{\varpi-1}
\,\partial\hat{\varphi}_1\,\Big)\ \re^{\pm\frac{2\hat{\varphi}_3}{\sqrt{n\varpi}}}:\ .
\ee
Here $n>0$ and $\varpi>1$ are considered to be parameters of the theory and we assume the standard 
normal ordering prescriptions (note that the precise way of fixing the ordering of the zero modes $\hat{\varphi}_0^{(3)}$ and $\hat{p}^{(3)}$
in
the definition of the exponential field is not essential).
Then the generators of the Borel subalgebra ${\cal U}_q(\widehat{\mathfrak{b}}_-)$ are
specified to be 
\be\label{asjkmn21}
\mathsf{X}_0=\frac{1}{q-q^{-1}}\int_0^{2\pi} \rd x \,\mathsf{V}_+(x)\,,\qquad\qquad 
\mathsf{X}_1=\frac{1}{q-q^{-1}}\int_0^{2\pi} \rd x \,\mathsf{V}_-(x)\,,
\ee
while
\be\label{h0gen}
 \mathsf{H}_0=-\mathsf{H}_1=-2\ri \sqrt{n\varpi}\, \hat{p}^{(3)} \ .
\ee
In writing these formulae we anticipate the relation between the quantum group parameter $q$ and $(n,\varpi)$, which enter into the 
vertex operators \eqref{vertex2}:
\be\label{oioqiwqw}
q=\exp\Big(-\frac{\ri\pi}{n\varpi}\Big)\, .
\ee

\bigskip

The Heisenberg algebra generated by $(\hat{a}^{(j)}_m,\hat{p}^{(j)})\,_{m\neq 0}^{j=1,2,3}$ possesses a highest weight representation,
which is commonly referred to as the Fock space. 
It is built from the highest state, which is an eigenvector of
the zero mode momenta  $\hat{p}^{(j)}$ and is annihilated by the operators $\hat{a}_m^{(j)}$ with $m\ge 1$. 
The irrep is defined as the linear span
of all possible states obtained by acting on the highest state with
the creation operators $\hat{a}_m^{(j)}$ with $m\le -1$. We will denote
the Fock space as ${\cal F}_{p_1,p_2,p_3}$, where the subscript keeps track of the eigenvalues of
$\hat{p}^{(j)}$. The generators $\mathsf{X}_{0,1}$,
realized as in  \eqref{asjkmn21}, then act as the intertwiners:
\be
\mathsf{X}_0\,:\,{\cal F}_{p_1,p_2,p_3}\mapsto {\cal F}_{p_1,p_2,p_3-\frac{\ri}{\sqrt{n\varpi}}}\,,\qquad\qquad
\mathsf{X}_1\,:\,{\cal F}_{p_1,p_2,p_3}\mapsto {\cal F}_{p_1,p_2,p_3+\frac{\ri}{\sqrt{n\varpi}}}\ .
\ee
In turn, the matrix elements of the quantum monodromy  \eqref{series1AAAAAA}
can be understood as  operators acting in the extended Fock space
$\bigoplus_{m=-\infty}^\infty {\cal F}_{p_1,p_2,p_3+\frac{m\ri}{\sqrt{n\varpi}}}$.
\bigskip

Recall that the generators $\mathsf{X}_0$, $\mathsf{X}_1$ and $\mathsf{H}_0$  
are required to satisfy  \eqref{amnxnbA} and the Serre relations \eqref{Serre1}.
These, in principle, can be investigated from the above definitions and the commutation relations of the Heisenberg algebra \eqref{phidef2}.
While the first requirement \eqref{amnxnbA} is easily established, checking the Serre relations
is subtle.
In fact, in even considering monomials built from $\mathsf{X}_{0,1}$,
such as those entering into eq.\;\eqref{Serre1}, some clarifications are required.
The vertex operators, which are the integrands for $\mathsf{X}_j$, exhibit singular
behaviour whenever two of them come close together:
\be\label{zmnn3hj2a}
\mathsf{V}_{s}(x) \,
\mathsf{V}_{s'}(y)
\sim (x-y)^{-2-2ss'/(n\varpi)}\,\qquad\qquad(s,s'=\pm)\, .
\ee
Upon subsequent integration w.r.t. $x$ and $y$ 
a divergent result would  generically be obtained. Nevertheless,
it is possible to define renormalised monomials in $\mathsf{X}_j$  in the following way.
%in the following way, which could be thought
%of as an analytic regularization of the UV divergences.
%One staggers the integration variables
%One staggers the integration variables by a small imaginary number $\propto \,\ri\eta$ such that 
 %if $\mathsf{V}_s(x)$ occurs to the left of $\mathsf{V}_{s'}(y)$ in the integral,  then
%$\Im m(x)<\Im m(y)$. 
One may, for example,  shift one of  the integration contours by a small imaginary amount $\propto \,\ri\eta$ such that 
 if $\mathsf{V}_s(x)$ occurs to the left of $\mathsf{V}_{s'}(y)$ in the integral,  then
$\Im m(x)<\Im m(y)$. 
Taking the limit $\eta\to 0^+$ one expects to get a well defined result,
which could be thought of as a renormalized version of the original monomial.  To illustrate, consider one of 
the simplest cases:
\be\label{jhas781hj}
(q-q^{-1})^2\,{\mathsf X}_0\,{\mathsf X}_1=
\lim_{\eta\to 0^+}\int_{0}^{2\pi} \rd y\;
\int_{0-\ri\eta}^{2\pi-\ri\eta}\rd x\ 
{\mathsf V}_+(x)\,{\mathsf V}_-(y)\ .%\\[0.3cm]
%&=&
\ee
For fixed $y$ belonging to the open interval $y\in(0,2\pi)$, the integration contour
for the $x$ variable can be made to coincide with subsets of $[0,2\pi]$ outside of the
segment $[y-\eta,y+\eta]$. The latter may be replaced by $C_\eta(y)$,
the part of the circle  of radius $\eta$ with center $y$ that 
is contained in the lower half of the complex $x$ plane. This way,
\be\label{kjaskj9023}
(q-q^{-1})^2\,{\mathsf X}_0\,{\mathsf X}_1=\lim_{\eta\to 0^+}\int_{\eta}^{2\pi-\eta}\rd y\,\bigg(\int_{0}^{y-\eta}\rd x +
\int_{y+\eta}^{2\pi}\rd x+\int_{C_\eta(y)}\rd x\bigg)\ \mathsf{V}_+(x)\,\mathsf{V}_-(y)\ .
\ee
In order to compute
the integration over the semi-circle, one uses the OPE
$\mathsf{V}_+(x)\,\mathsf{V}_-(y)$, where only the singular part needs to be taken into account.
In view of eq.\eqref{zmnn3hj2a}, it gives a  contribution $\propto\,\eta^{-1+\frac{2}{n\varpi}}$.
The singularities developed by the integral over  $C_\eta(y)$ 
in 
the limit  $\eta\to 0^+$
exactly cancel the singularities   
occurring in the first two integrals in the round brackets in 
\eqref{kjaskj9023}, when
upper and lower integration endpoints, respectively, approach the point $y$.
\bigskip

To verify the Serre relations \eqref{Serre1} one could represent
monomials formed out of the operators $\mathsf{X}_j$
in a similar way as illustrated by the example in eq.\,\eqref{kjaskj9023}. The result can
be represented by a multifold integral over 
$\mathsf{V}_{s_1}(x_1)\ldots\mathsf{V}_{s_4}(x_4)$, where  the integration
variables run 
over the  combination of segments
$x_i\in[0,2\pi]\cap \{x_i\,:\ |x_i-x_j|>\eta\}$
along with subtractions that 
cancel the UV divergences.
Then, as explained in ref.\cite{Bazhanov:1998dq}, by means of the 
braiding relations 
\be\label{ksakjnm23}
\mathsf{V}_{s}(x)\,\mathsf{V}_{s'}(y)=
q^{2ss'}\,\mathsf{V}_{s'}(y)\,\mathsf{V}_{s}(x)\,,
\qquad\qquad x>y\qquad (s,s'=\pm)
\ee
the order of the vertex operators is rearranged in each integral and the result is expressed in terms of the 
``renormalized'' ordered integrals
\be\label{Iordered}
\mathsf{J}_{\rm ren}(s_1,\ldots,s_m) = \lim_{\eta\to 0^+}
\Big(\mathsf{J}_{\eta}(s_1,\ldots, s_m)+\ldots\Big)
\ee
with
\be
 \mathsf{J}_{\eta}(s_1,\ldots, s_m)=
\!\!\!\!\!\!\!\!\int\limits_{\eta<x_m<\ldots<x_2<x_1<2\pi-\eta\atop |x_i-x_j|>\eta}
\hspace{-1cm}{\rm d}x_1\ldots{\rm d}x_m\, \mathsf{V}_{s_1}(x_1)\ldots \mathsf{V}_{s_m}(x_m)
\ee
and the   ``$\ldots$'' represent counterterms ensuring the existence of the limit $\eta\to 0^+$.
It has been observed in  ref.\cite{Bazhanov:1998dq} that once the monomials in 
$\mathsf{X}_j$ are rewritten  using the ordered integrals, the terms $\mathsf{J}_\eta(s_1,\ldots, s_m)$  cancel each other
in the Serre relations.
Since the counterterms are  determined by the singular part
of  $\mathsf{J}_\eta$, one expects that they will mutually
cancel each other in the Serre relations as well.

\bigskip

As was first pointed out in the work \cite{Bazhanov:1998dq}, 
 the full series expansion  \eqref{series1AAAAAA}
 may be brought to the form 
\be\label{Mseries}
{\mathsf M}(\lambda)=\,
\re^{-\frac{\pi{\tt h}}{\sqrt{n\varpi}}\,\hat{p}^{(3)}}\ \sum\limits_{m=0}^\infty\ \lambda^m\!\!\!\sum\limits_{s_1\ldots s_m=\pm}
\big(q^{\frac{{\tt h}}{2}s_1}{\tt e}_{\,s_1}\big)\ldots 
\big(q^{\frac{{\tt h}}{2}s_m}{\tt e}_{\,s_m}\big)\ \mathsf{J}_{\rm ren}(s_1,\ldots,s_m)\ .
\ee
If $\mathsf{J}_{\rm ren}(s_1,\ldots,s_m)$ had coincided 
with the usual ordered integral, this would be
 recognized as a path ordered exponent:
\be\label{Mordered1}
\mathsf{M}(\lambda)=\ 
\re^{-\frac{\pi{\tt h}}{\sqrt{n\varpi}}\,\hat{p}^{(3)}}\ \ \overset{\leftarrow}{{\cal P}}\exp\bigg(\lambda
\int_0^{2\pi}{\rm d}x\ \Big(\mathsf{V}_+\,q^{\frac{{\tt h}}{2}}\,{\tt e}_++
\mathsf{V}_-\,
q^{-\frac{{\tt h}}{2}}\,{\tt e}_-\Big)\bigg)\ .
\ee
In the present case the quantum monodromy can still be understood as a path ordered exponent
over $\lambda\big(\mathsf{V}_+\,q^{\frac{{\tt h}}{2}}\,{\tt e}_++
\mathsf{V}_-\,
q^{-\frac{{\tt h}}{2}}\,{\tt e}_-\big)$, though one must keep in mind that such a representation
contains UV divergences that need to be regularized in
order to make $\mathsf{M}(\lambda)$ well defined.

\paragraph{Main conjecture.}
The main conjecture of ref. \cite{Bazhanov:2018xzh} concerns the classical 
limit of the
quantum monodromy $\mathsf{M}(\lambda)$ described above.
This limit is interpreted as sending $n\,\propto\, \hbar^{-1} \to\infty$ so that the 
deformation parameter $q$ \eqref{oioqiwqw} tends to one.
The fields $\phi_i^{({\rm L})}=\hat{\varphi}_i/(n\varpi)^\frac{1}{2}$ become 
canonically normalized classical fields satisfying the Poisson bracket relations \eqref{ioas9821}
with $K_{\scriptscriptstyle \rm UV}=1$. We keep the superscript ``${{\rm L}}$''
as a reminder that $\phi_i^{({\rm L})}$ will be identified with the left  moving chiral bosons
appearing in the asymptotically flat domain of the chiral ${\rm SU}(2)$ Klim\v{c}\'{i}k model.
 Similar to their  quantum counterparts
$\hat{\varphi}_i$, they should be taken to be  quasi-periodic:
\be\label{classfield2}
\phi_j^{({\rm L})}(2\pi)-\phi_j^{({\rm L})}(0)=2\pi P_j\ \ \ \ \ \ (i=1,2,3)\, ,
\ee
where the dynamical quantity $P_j$ appears in the $n\to\infty$ limit of the
operators $\hat{p}_j$ from eq.\;\eqref{phidef1}.
As it follows from \eqref{vertex2},\;\eqref{asjkmn21}
the classical limit of $\mathsf{X}_0$ and $\mathsf{X}_1$
yields
\be\label{chiclass}
{\cal X}_0=\lim\limits_{n\to \infty}\,(q-q^{-1})\,\mathsf{X}_0=\int_0^{2\pi}{\rm d}x\,V_{+}(x) \ , \ \ \ \ \ \
 {\cal X}_1=\lim\limits_{n\to \infty}\,(q-q^{-1})\,\mathsf{X}_1=\int_0^{2\pi}{\rm d}x\,V_{-}(x)\ ,
\ee
where
\be\label{cvertex1}
V_\pm=\big(\ri\,\sqrt{1+\nu^2}\,\partial_+\phi_3^{({\rm L})}
+\partial_+\phi_2^{({\rm L})}
\pm\nu\ \partial_+\phi_1^{({\rm L})}
\big)\,\re^{\pm 2\phi_3^{({\rm L})}}
\ee
and
 \be
 \nu=\lim\limits_{n\to\infty}\sqrt{\varpi-1}\qquad\qquad\qquad\qquad  (\varpi-1>0) \nonumber
 \ee
(we use the derivative $\partial_+$ rather than $\partial$ since the classical fields
$\phi_i^{({\rm L})}$ are being understood as functions of $t+x$).
Notice that the formula for $V_\pm$ in terms of the free chiral fields $\phi^{({\rm L})}_j$ coincides with 
the first line of \eqref{oi89xciusa} for $\omega=1$. Together with the extra field
$$
V_0=-2\,
\big(\,\partial_+\phi_3^{({\rm L})}-\ri\,\sqrt{1+\nu^2}\,\partial_+\phi_2^{({\rm L})}\,\big)
$$
these dynamical variables obey the classical Poisson algebra \eqref{PBcurrents-remind} 
(in view of the quasi-periodic boundary conditions \eqref{classfield2} 
the fields $V_\pm(x)$ are single valued functions on the 
universal cover $x\in\mathbb{R}$. In obtaining formula \eqref{PBcurrents-remind}  it
is being assumed that both $x,y\in[0,2\pi)$, see also discussion below eq.\,\eqref{oias98jkmaaands}).
At this point, we are ready to formulate the main statement of ref. \cite{Bazhanov:2018xzh}.
\bigskip

\noindent
{\bf Conjecture}:
The classical limit of the quantum monodromy matrix $\mathsf{M}(\lambda)$ described by eqs.\;\eqref{series1AAAAAA},\,
\eqref{phidef2}-\eqref{oioqiwqw} is given by
\be\label{kasnbnb3asd2}
\lim_{n\to\infty} \mathsf{M}(\lambda)=
 \re^{-\pi P_3\, {\tt h}}\ \overset{\leftarrow}{{\cal P}}\exp\bigg(
\frac{1}{1-\rho^2}\int_0^{2\pi}\rd x\  \Big(\rho\,\big(\,V_+(x)\,{\tt e}_++ V_-(x)\,{\tt e}_-\,\big)+
\frac{\rho^2}{2}\ V_0(x)\,{\tt h}\Big)\bigg)\, .
\ee
Here $V_\pm$ and $V_0$ are as above with 
$\phi_j^{({\rm L})}=\lim_{n\to\infty} \hat{\varphi}_i/(n\varpi)^\frac{1}{2}$
and ${\tt h}$, ${\tt e}_\pm$ stand for the generators of the $\mathfrak{sl}(2)$ algebra \eqref{comm4}.
The spectral parameters $\rho$ and $\lambda$ are related via a power series
of the form
\be\label{kasnbnb3asd3}
\rho=\lambda\,\big(1+O(\lambda^2)\big)\, ,
\ee
where the higher order coefficients are non-universal in the sense that they depend 
on the details of the regularization needed to assign a precise meaning to $\mathsf{M}(\lambda)$
and its classical limit. Notice that the  monodromy 
\eqref{kasnbnb3asd2} is nothing but the path ordered integral of 
$-{\cal L}_{\scriptscriptstyle {\rm UV}}'(\rho)$ from eq.\,\eqref{Anew-spec}
up to the extra factor $\re^{-\pi P_3{\tt h}}$.
\bigskip

The following comment is in order here.
 Consider  the  exponent in the r.h.s. of eq.\;\eqref{kasnbnb3asd2}.
Suppose one were to drop the term $\propto \rho^2\, V_0(x)$ and remove the factor
 $\frac{1}{1-\rho^2}$ appearing in front of the integral. Then the resulting classical monodromy matrix 
would coincide literally with the classical limit of the 
path-ordered exponent representation for the quantum monodromy \eqref{Mordered1} with $\rho=\lambda$.
As it happens, the expression \eqref{Mordered1} requires  UV
regularization, \textit{e.g.} by introducing a UV cut-off and performing explicit subtractions of the singular part
as described above. 
The modified form of the  r.h.s. of \eqref{kasnbnb3asd2}, including the appearance
of the extra field $V_0(x)$, could be thought of as the result of anomalous contributions 
coming from the counterterms that remain finite in the  double limit where $\hbar\to0$ and the UV regulator goes away.

\bigskip

\paragraph{The Sklyanin exchange relations.}
The main conjecture has an  important corollary.
Since it implies that the classical limit of $\mathsf{M}(\lambda)$ is well defined,
one can take the same limit for both sides of the quantum exchange relations \eqref{RLL}.
As  was first observed  by Sklyanin \cite{Sklyanin:1979gh},   assuming that the
numerical matrix goes as  
$R(\lambda)= {\rm const}\times\big({\rm id}+\ri\hbar\, r(\lambda)+O(\hbar^2)\big)$ and using the correspondence
principle $[\cdot,\cdot]\mapsto \ri\hbar\{\cdot,\cdot\}$,
one obtains the Poisson algebra
\be\label{manzhnjsmnx}
\{M(\lambda)\,\overset{\otimes}{,}\,M(\mu)\}=[M(\lambda)\otimes M(\mu),r(\lambda/\mu)]
\ee
for the classical monodromy $M(\lambda)=\lim_{n\to\infty}\mathsf{M}(\lambda)$
(recall that the  Planck constant   $\hbar\propto\, n^{-1}$).
Note that for the
$\mathfrak{sl}(2)$ case, where the $R$ matrix
specialized to the fundamental representation is proportional to $R_{\frac{1}{2}\frac{1}{2}}(\lambda)$
from eq.\,\eqref{jxcbnsdbnjhsa}, the matrix $r(\lambda)$ is given by\footnote{%
There is an ambiguity in the definition of the classical $R$-matrix.
An additive shift by the identity matrix $r(\lambda)\mapsto r(\lambda)+f(\lambda)\,{\rm id}$,
where $f(\lambda)$ is any function of $\lambda$,
has no effect on the Sklyanin exchange relations \eqref{manzhnjsmnx}.
This ambiguity is fixed in formula \eqref{zmnkjas} by requiring that the trace of $r(\lambda)$ be zero.}
\be\label{zmnkjas}
r(\lambda)=\frac{1}{\lambda-\lambda^{-1}}\,\Big({\tt e}_+\otimes {\tt e}_-+{\tt e}_-\otimes {\tt e}_++\tfrac{1}{4}\,(\lambda+\lambda^{-1})\,
{\tt h}\otimes {\tt h}\Big)\, .
\ee
Since the  quantum monodromy is built from
an associative algebra,
the Jacobi identity for the commutator $[\cdot,\cdot]$
holds. It automatically follows that
the Jacobi identity must  be valid for the Poisson algebra satisfied by 
the classical limit of $\mathsf{M}(\lambda)$. 
This is equivalent to the classical Yang-Baxter equation for the matrix $r(\lambda)$
entering into eq.\,\eqref{manzhnjsmnx}. The expression \eqref{zmnkjas} is easily
recognized to be the trigonometric solution to that equation associated to  $\mathfrak{sl}(2)$.
\bigskip

That the Poisson algebra of the classical monodromy matrix obeys
the Sklyanin exchange relations as well as the Jacobi
identity is a non-trivial statement in
the case at hand. This is because $M(\lambda)$ takes the form
of a path-ordered exponent of the connection ${\cal L}_{\scriptscriptstyle {\rm UV}}'(\rho)$ \eqref{Anew-spec},
where the fields $V_\pm(x)$, $V_0(x)$ entering therein  obey the non-ultralocal algebra \eqref{PBcurrents-remind}. 
As was mentioned in the preamble, the presence of the $\delta'(x-y)$ term introduces ambiguities
in the computation of the Poisson brackets for the matrix elements of the monodromy. These generically
lead to a violation of the Jacobi identity and a modification to the algebra \eqref{manzhnjsmnx}.
The main conjecture essentially implies that there exists a regularization procedure such that
the Jacobi identity and Sklyanin exchange relations are preserved. 
Note that a way to obtain the Sklyanin exchange relation in a non-ultralocal system was proposed in the work \cite{Duncan:1989vg}.  
For the example of the Principal Chiral Model, 
the authors obtained the Poisson algebra \eqref{manzhnjsmnx} where to handle the contact terms  
 a certain ``retarded'' monodromy matrix was introduced.

\paragraph{Perturbative checks.} 
For the reader's convenience, we recall and elaborate on the main argument
presented in ref.\cite{Bazhanov:2018xzh} in support of the  conjecture \eqref{kasnbnb3asd2}.
In that work, the series expression  \eqref{series1AAAAAA} is used
as a starting point for taking the classical limit,  
where the monomials in $\mathsf{X}_j$ are
 regularized in some way, \textit{e.g.} by means of deforming contours as explained above. In general,
each  term in the expansion for $\mathsf{M}(\lambda)$ takes the form of
a polynomial in the non-commutative variables
 $\mathsf{X}_0$ and $\mathsf{X}_1$ with coefficients depending on the deformation parameter $q$.
To take $n\,\propto\,\hbar^{-1}\to \infty$
one should expand $q=\re^{-\frac{\ri\pi}{n\varpi}}$ 
for large $n$, express the result in terms of commutators and then replace
the commutators with Poisson brackets using 
 the correspondence principle $ [\,\cdot\,,\,\cdot\,]\mapsto -\frac{2\pi\ri}{n\varpi}\ \{\,\cdot\,,\,\cdot\,\}$.
It is easy to see that with this procedure the first few terms shown in \eqref{series1AAAAAA} become
 \bea\label{Cseries1}
&&\hspace{-1cm}\lim\limits_{n\to \infty}{\mathsf M}(\lambda)=\re^{-\pi P_3\, {\tt h}}\ \bigg[1+
\lambda\,
({\cal X}_0\,{\tt e}_++{\cal X}_1\,{\tt e}_-\,)\,+\\[0.3cm]
&&\hspace{-1cm}\,\hspace{0.0cm}\tfrac{1}{2}\,\lambda^2\,\Big(
{\cal X}_0^2\ {\tt e}_+^2+{\cal X}_1^2\  {\tt e}_-^2 
+\big({\cal X}_0{\cal X}_1+\{{\cal X}_1,{\cal X}_0\}\big)\ {\tt e}_+{\tt e}_-+
\big({\cal X}_0{\cal X}_1+\{{\cal X}_0,{\cal X}_1\}\big)\ {\tt e}_-{\tt e}_+\Big)+\ldots\ \bigg]\,,
\nonumber
\eea
where ${\tt h},{\tt e}_\pm$ satisfy
 the commutation relations of the $\mathfrak{sl}(2)$ algebra.
Developing the series expansion for the classical monodromy matrix quickly
becomes cumbersome at higher orders.
In ref. \cite{Bazhanov:2018xzh}, a typical computation is illustrated for a term
that contributes to \eqref{Cseries1}
at fourth order in $\lambda$. 
Supported by perturbative computations up to $\lambda^5$, it
is expected that
the series \eqref{Cseries1} exists and can  in  principle be  extended to all orders in
the spectral parameter.
\bigskip

In computing the Poisson brackets
that appear in the series expansion  \eqref{Cseries1}
one meets a familiar issue.
The integrands of ${\cal X}_0$ and ${\cal X}_1$
satisfy the $\mathfrak{sl}(2)$ like current algebra given in eq.\;\eqref{PBcurrents-remind} and those formulae
contain a term proportional to the derivative of the delta function.
As such when 
evaluating  the PBs  involving ${\cal X}_0$, ${\cal X}_1$
 one  encounters the problem of contact terms that was mentioned before. 
The appearance of the distribution $\delta'(x-y)$ and the associated 
ambiguity in the evaluation of the series \eqref{Cseries1} have an interpretation
in the context of the current discussion. They are the classical analogues
of the singular behaviour of the vertex operators \eqref{zmnn3hj2a}  as well as
the freedom  in the choice of the UV regularization scheme needed to 
give meaning to products of $\mathsf{X}_j$. In principle, once the
 quantum operator $\mathsf{M}(\lambda)$ is precisely defined, 
the corresponding classical expression is fixed completely. 
\bigskip

There exists a way
of dealing with the contact terms directly  at the level of the classical field theory,
which goes along the following line.
When evaluating nested PBs involving ${\cal X}_0$ and ${\cal X}_1$,
one assumes that the Jacobi identity holds true 
and uses it to reduce the types and number of ambiguous integrals that arise.
By means of the Jacobi identity and skew-symmetry it is expected that
 all of the Poisson brackets  appearing in the series expansion may be brought to the form
\be\label{basicPB}
\{{\cal X}_{s_1},\{{\cal X}_{s_2},\{{\cal X}_{s_3},\{\ldots,\{{\cal X}_{s_{m-1}},{\cal X}_{s_m}\}\ldots\}\  \ \  \ \ \ \ \ \qquad
(s_1,\ldots,s_m=0,1)\ .
\ee 
To give an example, substituting 
\be
A=\{{\cal X}_0,\{{\cal X}_0,{\cal X}_1\}\}\,,\qquad\qquad
B={\cal X}_0\,,\qquad\qquad C={\cal X}_1
\ee
into
\be
\{A,\{B,C\}\}+\{C,\{A,B\}\}+\{B,\{C,A\}\}=0
\ee
gives rise to the identity
$$
\{\{{\cal X}_0,\{{\cal X}_0,{\cal X}_1\}\},\{{\cal X}_0,{\cal X}_1\}\}=
\{{\cal X}_1,\{{\cal X}_0,\{{\cal X}_0,\{{\cal X}_0,{\cal X}_1\}\}\}\}+
\{{\cal X}_0,\{{\cal X}_1,\{{\cal X}_0,\{{\cal X}_1,{\cal X}_0\}\}\}\}\ ,
$$
which swaps a nested PB to a sum of two that are of the above type.
\bigskip

In evaluating the reduced class of
Poisson brackets \eqref{basicPB} only three types of ambiguous integrals arise:
\bea\label{asmnxzaks}
I_1&=&\int_0^{2\pi}\rd x_1\rd x_2\  \delta'(x_1-x_2) \nonumber\\[0.2cm]
I_2&=&\int_0^{2\pi}\rd x_1 \rd x_2 \rd x_3\  \sigma(x_1-x_2)\,\delta'(x_2-x_3)\,F(x_1)\\[0.2cm]
I_3&=&\int_0^{2\pi}\rd x_1\rd x_2\rd x_3\rd x_4\  \delta'(x_2-x_4)\,\sigma(x_1-x_2)\,\sigma(x_2-x_3)\,
F(x_1)\,G(x_3)\ .\nonumber
\eea
Here $F(x)$ and $G(x)$ are arbitrary functions and $\sigma(x)$ stands for  the Heaviside step function
so that, in particular, the $\epsilon(x)$ term in the Poisson algebra \eqref{PBcurrents-remind} can be expressed
as $\epsilon(x)=\sigma(x)-\sigma(-x)$. 
For the first integral it is natural to require that
 $\delta'(x_1-x_2)$  be anti-symmetric. This way,
\be\label{amb1}
I_1=0\, .
\ee
For $I_2$ some additional care is needed. 
To illustrate the issue, let's take the domain of $x_2$ and $x_3$ to
be $(\eta_{2A},2\pi-\eta_{2B})$ and $(\eta_{3A},2\pi-\eta_{3B})$, respectively, where
the infinitesimally small regulators $\eta_{mA}$ and $\eta_{mB}$ with $m=2,3$ are taken to be positive. The last condition
ensures that $|x_2-x_3|<2\pi$ over the full integration domain. 
This is important, since the Poisson algebra \eqref{PBcurrents-remind} that was used to
evaluate the nested PBs \eqref{basicPB} and led to the ambiguous integrals \eqref{asmnxzaks}
is valid when $|x-y|<2\pi$ and otherwise requires modification to take into account that the fields
$V_\pm(x)$ are quasi-periodic.
Then, integrating  
the expression for $I_2$ \eqref{asmnxzaks} w.r.t. $x_3$ one finds that
\be\label{nasjh832jhds}
I_2=-\int_0^{2\pi}\rd x_1\int_{\eta_{2A}}^{2\pi-\eta_{2B}}\rd x_2\, \sigma(x_1-x_2)\,F(x_1)\,
\big(\delta(x_2-2\pi+\eta_{3B})-\delta(x_2-\eta_{3A})\big)\ .
\ee
A case by case study  yields that $I_2=0$
 if both $\eta_{3A}<\eta_{2A}$ and $\eta_{3B}<\eta_{2B}$ or 
 $\eta_{3A}>\eta_{2A}$ and $\eta_{3B}>\eta_{2B}$; 
$I_2=-\int_0^{2\pi}\rd x_1 \, F(x_1)$ for
$\eta_{3B}>\eta_{2B}$ and $\eta_{3A}<\eta_{2A}$; while
$I_2=\int_0^{2\pi}\rd x_1 \, F(x_1)$ when 
$\eta_{3B}<\eta_{2B}$ and
$\eta_{3A}>\eta_{2A}$.
 This way, the value of $I_2$ is ambiguous and depends on the precise treatment of 
the endpoints of the integration domain. In what follows, we  set
\be\label{amb2}
 I_2=\gamma \int_0^{2\pi}\,\rd x\ F(x) \,,
\ee
where  $\gamma$ is some undetermined constant.
For $I_3$, there is no issue of contact terms and an
integration by parts w.r.t. $x_2$ or $x_4$ results in zero so that
\be\label{amb3}
I_3=0\ .
\ee
Note that the above formulae also specify the value of any ambiguous integrals
that can be related to $I_2$ and $I_3$ via anti-symmetry of $\delta'(x_1-x_2)$.\footnote{%
Defining the Poisson algebra for  ${\cal X}_{0,1}$, along the lines of
the above discussion, would require the specification of 
all possible integrals involving $\delta'(x-y)$ that could appear in nested PBs of ${\cal X}_{0,1}$.
This would lead to the introduction of further undertermined constants parameterizing the ambiguities, 
similar to  $\gamma$ from eq.\,\eqref{amb2}. However, 
skew-symmetry and the Jacobi identity imposes constraints on
the constants that may even fix 
them completely. A systematic study  is beyond the scope of the present work.}
\bigskip

In order to appreciate how the series for the classical monodromy matrix
\eqref{Cseries1} could organize into a path ordered exponent, it is
worthwhile performing an explicit computation up to order $\lambda^2$. 
To this end, introduce the ordered integrals over the classical fields
\be
J_{\rm cl}(s_1,s_2,\ldots,s_m)=\!\!\!\!\!\!\!\!\int\limits_{0<x_m<\ldots<x_2<x_1<2\pi}
\hspace{-1cm}{\rm d}x_1\ldots{\rm d}x_m\, {V}_{s_1}(x_1)\ldots {V}_{s_m}(x_m)\
\ee
with $s_j\in\{-1,0,1\}$. The coefficients of ${\tt e}_\pm$ and ${\tt e}_\pm^2$  in \eqref{Cseries1}
are expressed in terms of  these by means of the elementary identities:
\be
{\cal X}_0^m=m!\,{J}_{\rm cl}(+,+,\ldots,+)\, ,\qquad\qquad
{\cal X}_1^m=m!\,{J}_{\rm cl}(-,-,\ldots,-)\ ,
\ee
while 
\be
{\cal X}_0\,{\cal X}_1=J_{\rm cl}(+,-)+J_{\rm cl}(-,+)\, .
\ee
The non-trivial part concerns the evaluation of $\{{\cal X}_0,{\cal X}_1\}$.
The Poisson brackets of ${\cal X}_0$ and ${\cal X}_1$ are obtained from those of  $V_+(x)$ and $V_-(y)$,
presented in the first line of eq.\;\eqref{PBcurrents-remind}, by means of integration w.r.t. $x$ and $y$.
A brief inspection of that formula shows that there are three terms that need to be taken into account.
One of them is proportional to $\delta'(x-y)$. It contributes nothing to the end result
since the  integrated quantity $\int\rd x\rd y \,\delta'(x-y)$ was prescribed to be zero, see eqs.\;\eqref{asmnxzaks}
and \eqref{amb1}. The part multiplying the delta function yields $-\int\rd x V_0(x)=-J_{\rm cl}(0)$. For the
last term, one uses the fact that by definition $\epsilon(x)=\sigma(x)-\sigma(-x)$ and hence
$$\int_0^{2\pi}\rd x\int_0^{2\pi} \rd y\  V_+(x)\,V_-(y)\,\epsilon(x-y)=J_{\rm cl}(+,-)-J_{\rm cl}(-,+)\, .$$
As a result, one finds 
 \bea
\lim\limits_{n\to \infty}{\mathsf M}(\lambda)&=&\re^{-\pi P_3\, {\tt h}}\ \Big[1+
\lambda\,\big(J_{\rm cl}(+)+J_{\rm cl}(-)\big)+
\lambda^2\,\big(J_{\rm cl}(+,+)\,{\tt e}_+^2 + J_{\rm cl}(-,-)\,{\tt e}_-^2 \nonumber \\[0.2cm]
&+&
J_{\rm cl}(+,-)\,{\tt e}_+\,{\tt e}_-+J_{\rm cl}(-,+)\,{\tt e}_-{\tt e}_+
+\tfrac{1}{2}\,J_{\rm cl}(0)\,{\tt h}\big)+\ldots\Big]\, .
\nonumber
\eea
Notice that the expression in the square brackets coincides with the expansion of a path-ordered exponent for the 
connection ${\cal L}=-\lambda\,(V_+{\tt e}_++V_-{\tt e}_-)-\tfrac{1}{2}\lambda^2\,V_0(x)+O(\lambda^3)$.
This  provides a check of the conjecture \ref{kasnbnb3asd2} with $\rho=\rho(\lambda)$ as in \eqref{kasnbnb3asd3}
to second order in $\lambda$.

\bigskip
The main piece of evidence for the conjecture
is an explicit evaluation of the series \eqref{Cseries1}
up to and including $\lambda^5$  that was carried out in
the work  \cite{Bazhanov:2018xzh}.
In the computation of the higher order terms, in addition to $I_1$,
ambiguous integrals
of the form $I_2$ and $I_3$ \eqref{asmnxzaks} arise, which leads to the
undetermined constant $\gamma$ \eqref{amb2} entering into the result.
The latter turns out to be consistent with the conjecture \eqref{kasnbnb3asd2},
where the spectral parameter $\rho$ is related to $\lambda$ as 
\be\label{kasjhnm21}
\rho=\lambda\,\big(1-\nu^2\gamma\lambda^2-\gamma\nu^2\,(1-2\gamma\nu^2)\,\lambda^4+O(\lambda^6)\big)\, .
\ee

\subsection{Some comments for the case of higher rank Lie groups\label{cxnbbsdbv3}} 
While the work \cite{Bazhanov:2018xzh} proposes a direction for
quantizing the non-local IMs of the three punctured affine Gaudin model for $G={\rm SU}(2)$,
we would not have reviewed it here if the arguments were not readily adaptable  to
the case of arbitrary Lie group $G$. Indeed, the main  ingredients  required
to carry out the first
few steps of the generalization are already contained in Subsection \ref{Sec:GaugeFixChiral}.
\bigskip

Consider the Poisson bracket relations \eqref{PBcurrents-remind} satisfied by the components of the Lax matrix
\eqref{Anew-spec}.
They are essentially the ${\rm SU}(2)$ case of the general algebra \eqref{Eq:PbXi} specialized to $\omega=1$.
In  simplified notation
$\Xi_\alpha = \Xi_\alpha^{({\rm L})}$,
$\Xi_i = \Xi_i^{({\rm L})}$ and setting the inessential constant
$K_{\rm\scriptscriptstyle UV}$ to be one, the PB relations for general Lie group read:
\bea\label{mnnzxnbasnbabni}
\bigl\lbrace \Xi_i(x), \Xi_\alpha(y) \bigr\rbrace &=&  
\frac{\alpha({\tt h}_i)}{1+\nu^2}\, \,\Xi_\alpha(x)\, \delta(x-y)  \nonumber\\
\bigl\lbrace \Xi_i(x), \Xi_j(y) \bigr\rbrace&=&  \frac{2\nu^2\,\delta_{ij}}{(1+\nu^2)^2}\,\,\delta'(x-y)  \\
\bigl\lbrace \Xi_\alpha(x), \Xi_\beta(y) \bigr\rbrace &=&  N^{\alpha,\beta} \, \Xi_{\alpha+\beta}(x)\, \delta(x-y)   + \frac{1}{2} \, (\alpha,\beta) \; \Xi_\alpha(x) \, \Xi_\beta(y) \,\epsilon(x-y) \nonumber\\
\bigl\lbrace \Xi_\alpha(x), \Xi_{-\alpha}(y) \bigr\rbrace &=&
\frac{\nu^2}{1+\nu^2}\,
\delta'(x-y) +  \rho_\alpha^i \,\Xi_i(x)\,\delta(x-y)
- \frac{1}{2}\,(\alpha,\alpha)\; \Xi_\alpha(x) \, \Xi_{-\alpha}(y) \,\epsilon(x-y) \nonumber
\eea
The conventions for the Lie algebra generators ${\tt h}_i$ and ${\tt e}_\alpha$ used in this paper
 as well as the
meaning of the symbols $\alpha({\tt h}_i)$, $(\alpha,\beta)$ and $N^{\alpha,\beta}$
 are explained in Appendix \ref{App:Root}.
Also note that, in our normalization, $\rho_\alpha^i=\frac{1}{2}\,\alpha({\tt h}_i)$.

\bigskip

Recall that the dynamical variables $\Xi_\alpha$ and $\Xi_j$ possess the interpretation of
the components of the Kac-Moody currents in the so-called $\omega$ gauge for $\omega=1$,
 see eqs.\;\eqref{Eq:J1Xi} and \eqref{Eq:J23Xi}. The  Lax matrix
$\frac{1}{\varphi^{({\rm L})}(z^{({\rm L})})}\,
\Gamma^{({\rm L})}(z^{({\rm L})})$, with the gauge fixing imposed, is expressed linearly in terms of them.
Similar to the ${\rm SU}(2)$ case,
 it is convenient to perform 
an extra similarity transformation to the Lax matrix by a non-dynamical spectral parameter dependent matrix.
The latter involves the Weyl coweight ${\tt h}_\delta$, whose defining property
is that 
\be
\big[{\tt h}_\delta,{\tt e}_\alpha\big]=\ell(\alpha)\,{\tt e}_\alpha\,\qquad\qquad\forall {\tt e}_\alpha\in\mathfrak{n}_\pm\,,
\ee
with $\ell(\alpha)$ denoting the level of the root $\alpha\in\Delta$ (see Appendix \ref{App:Root}
for additional details). 
Then introduce
\be\label{kdmnncxhjd1}
{\cal L}_{\scriptscriptstyle {\rm UV}}'(\rho)=
 \frac{1}{\varphi^{({\rm L})}(z^{({\rm L})})}\ \big((1+\nu^2)^{\frac{1}{h}}\rho\,\big)^{{\tt h}_\delta}\ 
\Gamma^{({\rm L})}(z^{({\rm L})})\ \big((1+\nu^2)^{\frac{1}{h}}\rho\,\big)^{-{\tt h}_\delta}\, .
\ee
Here the spectral parameter $z^{({\rm L})}$ is swapped for $\rho$ according to 
\be
z^{({\rm L})}=\frac{4\nu^2}{1+\nu^2}\,\frac{\rho^h}{1-\rho^h}\,,
\ee
which is a simple generalization of eq.\;\eqref{oias8912iopds}, while
$h$ is the Coxeter number
(e.g. $h=N$ for $G={\rm SU}(N)$). 
By substituting the expressions for the gauge-fixed Kac-Moody currents 
\eqref{Eq:J1Xi},\,\eqref{Eq:J23Xi} into the 
 general formula for the Gaudin Lax matrix  \eqref{kxznjasjh21} and \eqref{Eq:zL},
and making use of the definition \eqref{Eq:TwistKlimL}   for the twist function,
${\cal L}_{\scriptscriptstyle {\rm UV}}'$ can be written directly in terms of the component
 fields $\Xi_\alpha$ and $\Xi_i$. This yields 
\be\label{akjsjk2aassa}
{\cal L}_{\scriptscriptstyle {\rm UV}}'(\rho)\equiv\frac{1}{\rho^h-1}\,\sum_{\alpha\in\Delta_+}\!\!\Big(
\big((1+\nu^2)^{\frac{1}{h}}\rho\,\big)^{\ell(\alpha)}\,\Xi_{-\alpha}\,{\tt e}_\alpha+
\big((1+\nu^2)^{\frac{1}{h}}\rho\,\big)^{h-\ell(\alpha)}\,\Xi_{\alpha}\,{\tt e}_{-\alpha}\Big)
+\frac{(1+\nu^2)\,\rho^h}{2\,(\rho^h-1)}\,\sum_{i=1}^{\dim\mathfrak{h}}\Xi_i\,{\tt h}_i
\ee
with the first summation being over all the positive
roots $\Delta_+$. 
\bigskip

In the usual treatment of the affine Gaudin model, the fields are 
assumed to be periodic functions of $x\sim x+2\pi$. The $G={\rm SU}(2)$ case
motivates one to consider
a generalization  of these boundary conditions, since the classical fields $\phi_i^{({\rm L})}$ entering
into the connection had to be quasi-periodic \eqref{classfield2}. We expect it to be possible to broaden the notion of the
affine Gaudin model and arrange so that the components 
$\Xi_\alpha$ and $\Xi_j$ of the gauge fixed currents obey
\be\label{askjjh12oiaaasa}
\Xi_\alpha(x+2\pi)=\exp\bigg(-4\pi\sum_{j=1}^{\dim \mathfrak{h}}
 \rho_\alpha^j\,P_j\bigg)\,\Xi_\alpha(x)\qquad\qquad {\rm and}
\qquad\qquad \Xi_j(x+2\pi)=\Xi_j(x)\, .
\ee
In turn, the Lax matrix ${\cal L}_{\scriptscriptstyle {\rm UV}}'$ also becomes quasi-periodic:
\be\label{oias98jkmaaands}
{\cal L}_{\scriptscriptstyle {\rm UV}}'(\rho;x+2\pi)=
\re^{2\pi P_j\,{\tt h}_j}\,{\cal L}_{\scriptscriptstyle {\rm UV}}'(\rho,x)\, \re^{-2\pi P_j\,{\tt h}_j}
\ee
(an implicit summation over ``$j$'' is being implied in each exponent).
Note that the derivation of the Poisson bracket algebra \eqref{mnnzxnbasnbabni} 
assumed periodic boundary conditions. For the case of 
quasi-periodic fields $\Xi_\alpha(x)$, they
remain valid for $x,y\in[0,2\pi)$ and can be extended from that domain to the universal cover $x,y\in\mathbb{R}$
in a way that is consistent with eq.\,\eqref{askjjh12oiaaasa}. Among other things, 
$\delta(x-y)$ and $\delta'(x-y)$  should be considered as being periodic distributions,
while their coefficients in \eqref{mnnzxnbasnbabni} would be modified by factors, e.g.,
$\exp\big(2\,(x-y)\,\sum_{j} \rho_\alpha^j\,P_j\big)$ in 
 the first line of that equation.
As for the step function, its extension is given by
 $\epsilon(x)=2m+1$ for $2\pi m<x<2\pi(m+1)\ \ (m\in \mathbb{Z})$.
A quick analysis of the Poisson bracket algebra \eqref{mnnzxnbasnbabni}
under the shift $x\to x+2\pi$,
focusing only on the $\epsilon(x-y)$ term, shows that
the quasi-periodicity factors $P_j$ must be  treated as dynamical variables such that
\be\label{mncxbnasbnq}
\{2\pi P_j,\Xi_\alpha(x)\}=\,-\tfrac{1}{2}\,\alpha({\tt h}_j)\,\Xi_\alpha(x)\,,\qquad\qquad
\{2\pi P_i\,,\,\Xi_j(x)\}=0\,,
\qquad\qquad \{P_i,P_j\}=0\ .
\ee
\bigskip

In a  2D integrable classical field theory defined on the space-time cylinder with $x\sim x+2\pi$,
the transfer-matrix $T(z)$ is usually introduced as the Wilson loop integrated along the constant time-slice
 $0\le x-x_0< 2\pi$. 
An important property is that $T(z)$ should be independent of the arbitrary
constant $x_0$, i.e., the integration contour needs to close.
In the case when the connection is quasi-periodic as in \eqref{oias98jkmaaands}
this requires the insertion of an extra factor inside the trace.
Choosing some  representation for $G$,
we take the monodromy and transfer-matrix to be
\be\label{kjasi821kj}
T(\rho)={\rm Tr}\big(\re^{-\pi P_j\,{\tt h}_j}\,M(\rho)\big)\,,\qquad\qquad
M(\rho)=\re^{-\pi P_j\,{\tt h}_j}\
\overset{\leftarrow}{{\cal P}}\exp\bigg(-\int_0^{2\pi}\rd x \, {\cal L}_{\scriptscriptstyle {\rm UV}}'(\rho)\bigg)
\ee
(note that the index $j$ in $ P_j\,{\tt h}_j$ is being summed over).
\bigskip

The monodromy matrix $M(\rho)$ is our candidate, which is expected to appear in the
classical limit of a  suitable representation of the Yang-Baxter algebra \eqref{RLL}.
Exploring this further in any detail is beyond the scope of the present paper. 
Nevertheless, we mention some first pieces of evidence, specializing to the case $G={\rm SU}(N)$
in order to avoid unnecessary technical details. 
Let $\alpha_j$ be the $N-1$ simple roots of the Lie algebra and denote by $\alpha_0$ 
the negative root with minimum level, i.e.,
\be
\alpha_0=-\alpha_1-\alpha_2-\ldots-\alpha_{N-1}\, .
\ee
The inner products $A_{ij}=(\alpha_i,\alpha_j)$ with $i,j=0,1,\ldots,N-1$ are encoded in the extended Cartan matrix:
\be
\begin{tikzpicture}
\node at (0,0) {$A_{ij}=\left(
\begin{array}{ccccccc}
2 & -1 & 0 & 0 & \cdots & 0 & -1 \\[0.2cm]
-1 & 2 & -1 & 0 & \cdots & 0 & 0 \\[0.2cm]
0 & -1& 2 & -1 & \cdots & 0  & 0\\[0.2cm]
\vdots & \vdots & \vdots & \vdots & \ddots & \vdots & \vdots\\[0.2cm] 
0 & 0 & 0 & 0 & \cdots & 2 & -1 \\[0.2cm]
-1 & 0 & 0 & 0 & \cdots & -1 & 2
\end{array}
\right)\ .$};
\end{tikzpicture}
\nonumber
\ee
Also, introduce the notation 
\be
{\cal X}_j=\, (1+\nu^2)^{\frac{1}{N}}\,\int_0^{2\pi}\rd x \, \Xi_{-\alpha_j}(x)\,,\qquad\qquad j=0,1,2,\ldots, N-1\, .
\ee
Then one can check that the first two terms in the expansion of the classical 
monodromy from eq.\;\eqref{kjasi821kj} coincide with 
\bea\label{series1}
M(\rho)&=&\re^{-\pi P_j\,{\tt h}_j}\,\bigg[1+\lambda\sum_{j=0}^{N-1}
{\tt e}_j \mathcal{X}_j
+ \frac{\lambda^2}{2}\bigg(\, \sum_{j=0}^{N-1}({\tt e}_j)^2\,\mathcal{X}_j^2
\ +\ \!\!\!\!\!\sum_{i\ne j\atop(\alpha_i,\alpha_j)=0}\!\!\!\!
{\tt e}_i{\tt e}_j\,\mathcal{X}_i\mathcal{X}_j\nonumber
\\[0.3cm]
&+&\!\!\!\!\sum_{i\ne j\atop(\alpha_i,\alpha_j)=-1}\!\!\!\!
{\tt e}_i{\tt e}_j\,\big(\mathcal{X}_i\mathcal{X}_j+2\{\mathcal{X}_j,\mathcal{X}_i\}\big)\bigg)
+O(\lambda^3)\bigg]
\,,
\eea
where
\be
\rho=\lambda\,\big(1+O(\lambda^{N})\big)\, .
\ee
It turns out that the series \eqref{series1} appears in the classical limit of the universal monodromy matrix
built from the quantum affine algebra ${\cal U}_q\big(\widehat{\mathfrak{sl}}(N)\big)$, similar to what
was discussed for the $\mathfrak{sl}(2)$ case. One should check and eventually prove
that it coincides term by term with $M(\rho)$ from eq.\;\eqref{kjasi821kj}. This would be the first step
for applying   the quantum inverse scattering method to the problem of 
quantization of the non-local IMs for the affine Gaudin model with Lax matrix $\Gamma^{({\rm L})}/\vp^{({\rm L})}$.

\section{Quantum local IMs and \texorpdfstring{$\bm\Wc$}{W}-algebra for \texorpdfstring{$\bm{G={\rm SU}(2)}$}{G=SU(2)}}\label{Sec:QuantLoc}
The classical  algebra of extended conformal symmetry and the classical
local IMs were discussed in Section \ref{sec3899891}.  It was explained how
they admit a representation in terms of free chiral fields
$\partial_+\phi^{({\rm L})}_j$, which is determined through the quasi-commutativity condition  \eqref{quasi-comm}.
The classical screening charges entering into that formula 
are given by eqs.\;\eqref{oias9012oia--a},\;\eqref{oias9012oia--b}.
One may notice that they
coincide with ${\cal X}_0$ and ${\cal X}_1$
from \eqref{chiclass} and \eqref{cvertex1}, which appear in the classical limit
of the generators of the Borel subalgebra $\mathsf{X}_0,\mathsf{X}_1\in
{\cal U}_q(\widehat{\mathfrak{b}}_-)\subset{\cal U}_q\big(\widehat{\mathfrak{sl}}(2)\big)$.
This way, accepting the results of the previous section,
one is led to conclude that the full classical integrable structure is determined by
${\cal X}_0$ and ${\cal X}_1$. The non-local IMs are given by the trace of the
power series \eqref{Cseries1}, which involves the classical screening charges and their Poisson brackets, while
the local IMs are obtained via the quasi-commutativity condition that was mentioned before.
\bigskip

For the quantum case the situation is analogous. The r\^{o}le of
the screening charges is played by $\mathsf{X}_0$ and $\mathsf{X}_1$,
which are the operators built from the free chiral fields $\hat{\varphi}_j$  that are
given by eqs.\;\eqref{vertex2} and \eqref{asjkmn21}.  These 
go into the
power series expansion for the monodromy $\mathsf{M}(\lambda)$, see 
eq.\;\eqref{series1AAAAAA}, through which the non-local
IMs are obtained by taking an appropriate trace \eqref{askjnnm23}.
Note that, at this stage, the generating function $\mathsf{T}(\lambda)$  
is meant as a formal power series expansion in the spectral parameter $\lambda$,
without reference to convergence,
whose operator valued coefficients mutually commute.
The quantum local IMs are determined from $\mathsf{X}_0$ and $\mathsf{X}_1$ in a way that
is a straightforward generalization of what was discussed for the classical case in Subsection \ref{oias98oia}.
It is useful to re-phrase this in the language of Operator Product Expansions (OPEs).
One starts with a general ansatz for a Lorentz spin $m$ differential  polynomial:
\be\label{asnbczbnnb21}
\mathsf{F}_{m}=\sum_{i_1,\ldots,i_k\ge 1\atop
j_1,\ldots,j_k=1,2,3}
F_{i_1\ldots i_k}^{j_1\ldots j_k}\ 
\big(\partial^{i_1}\hat{\varphi}_{j_1}\big)
\ldots \big(\partial^{i_k}\hat{\varphi}_{j_k}\big)\qquad {\rm with}
\qquad i_1+\ldots+i_k=m\,.
\ee
Using the free field OPEs
\be\label{mncnaskjx}
\hat{\varphi}_i(x)\,\hat{\varphi}_j(y)=-\tfrac{1}{2}\,\delta_{ij}\,\log(x-y)+\reg\,,
\ee
where ``$\reg$'' stands for terms that are regular in $x-y$,
one requires that the OPE of $\mathsf{F}_{m}(x)$ with the integrand
of the screening charge $\mathsf{X}_0$ obeys
\be\label{asnb12nbsa}
\mathsf{F}_m(x)\,\mathsf{V}_+(y)=\sum_{j=0}^{m-1}\frac{\mathsf{C}_j(y)}{(x-y)^{j+1}}+\reg\qquad\qquad
{\rm with} \qquad\qquad \mathsf{C}_0(y)=0\, .
\ee
This is easily recognized as the quantum version of \eqref{oias89cjjkaas}.
The result is 
a set of fields  generating a quantum algebra of extended conformal symmetry.
A particular subset of them  enter as densities for the local IMs.
To determine the latter one needs to make use of the second
screening charge $\mathsf{X}_1$. One imposes the further condition  that
\be\label{asnb12nbsaaaaa}
\mathsf{F}_m(x)\,\mathsf{V}_-(y)=\sum_{j=0}^{m}\frac{\mathsf{C}_j(y)}{(x-y)^{j+1}}+\reg\,,\qquad\qquad
{\rm where} \qquad\qquad \mathsf{C}_0(y)=\partial_y(\ldots)
\ee
in direct analogy with eq.\;\eqref{mncnmjkasmnz}. This then fixes the coefficients $F_{i_1\ldots i_k}^{j_1\ldots j_k}$
entering into the density $\mathsf{F}_m$ up to an overall multiplicative factor
 (with the possibility that $\mathsf{F}_m=0$). The resulting local IM,
\be
\mathsf{Q}_{m-1}=\int_0^{2\pi}\rd x\  \mathsf{F}_{m}(x)\,,
\ee
can be said to ``quasi-commute'' with the screening charges:
\be\label{asnnbnb32aaaa}
[\mathsf{Q}_{m-1},\mathsf{X}_0]\sim 0\,,\qquad\qquad\qquad [\mathsf{Q}_{m-1},\mathsf{X}_1]\sim 0
\ee
in the sense that the commutator of $\mathsf{Q}_{m-1}$ with their integrands  $\mathsf{V}_+(y)$
and  $\mathsf{V}_-(y)$ is a total derivative in $y$.
\bigskip

In what follows we perform an explicit analysis of eqs.\;\eqref{asnb12nbsa},\;\eqref{asnb12nbsaaaaa}
and derive the formula for the first non-trivial local IM of the Fateev integrable structure. 
This will be important for the next section, where it is checked that the local IMs constructed from the affine Gaudin model
following refs. \cite{Lacroix:2018fhf,Lacroix:2018itd} coincides with those
obtained through the quasi-commutativity condition with $\mathsf{X}_0$ and $\mathsf{X}_1$.
Note that the extended algebra of conformal symmetry and the local IMs 
 have  been studied in refs. \cite{Semikhatov:2001zz,Lukyanov:2012wq} and our discussion
uses the results of those papers.

\subsection[Quantum ${\cal W}$-algebra and local IMs]{Quantum \texorpdfstring{$\bm{\Wc}$}{W}-algebra and local IMs}

Consider the first condition \eqref{asnb12nbsa}. In finding the densities
$\mathsf{F}_m(x)$, it is useful to introduce the fields $\hat{\chi}$ and $\hat{\alpha}$, which are 
related to $\hat{\varphi}_1$ and $\hat{\varphi}_2$ via the orthogonal transformation:
\be
\left(\begin{array}{c}
\hat{\chi} \\[0.2cm]
\hat{\alpha}
\end{array}
\right)=\frac{1}{\sqrt{n\varpi+2}}\ 
\left(\begin{array}{cc}
\sqrt{n+2} & -\sqrt{n(\varpi-1)} \\[0.2cm]
\sqrt{n(\varpi-1)} & \sqrt{n+2}
\end{array}\right)\
\left(\begin{array}{c}
\hat{\varphi}_1 \\[0.2cm]
\hat{\varphi}_2
\end{array}\right)\ .
\ee
It is easy to see from eq.\;\eqref{vertex2} that
$\mathsf{V}_+\,\propto\, \partial\hat{\alpha}\,\exp(\frac{2\hat{\varphi}_3}{\sqrt{n\varpi}})
\ +\ {\rm total\ derivative}$, so that any differential polynomial in $\partial\hat{\chi}$ 
trivially satisfies \eqref{asnb12nbsa}.
By means of the free field OPE \eqref{mncnaskjx} and repeated use of Wick's theorem,
one  analyses that  condition  for the local densities
that are built from the two remaining fields $\partial\hat{\alpha}$ and
$\partial\hat{\varphi}_3$. Limiting oneself to fields with Lorentz spin $m=4$
or lower, one obtains the currents:
\bea\label{Weq1}
\mathsf{T}_2&=&\big(\partial\hat{\varphi}_3\big)^2+
\big(\partial\hat{\alpha}\big)^2-\tfrac{1}{\sqrt{n\varpi}}\ \partial^2\hat{\varphi}_3\nonumber\\[0.4cm]
\mathsf{T}_3&=& \tfrac{2}{3}\,\big(3+2/(n\omega)\big)\,\big(\partial\hat{\alpha}\big)^3 +
2\,\partial\hat{\alpha}\,\big(\partial\hat{\varphi}_3\big)^2
-(n\varpi+2)\,(n\varpi)^{-\frac{1}{2}} \ \partial\hat{\varphi}_3\,\partial^2\hat{\alpha}
 \nonumber \\[0.2cm]
&+&(n\varpi)^{\frac{1}{2}}\ \partial^2\hat{\varphi}_3\,\partial\hat{\alpha}+
\tfrac{1}{6}\,\big(1+2/(n\varpi)\big)\,\partial^3\hat{\alpha}  \\[0.4cm]
\mathsf{T}_4&=&4\,\big(1+1/(n\varpi)\big)\,\big(\partial\hat{\alpha}\big)^4+4\,\big(\partial\hat{\alpha}\big)^2\,
\big(\partial\hat{\varphi}_3\big)^2+4(n\varpi)^{-\frac{1}{2}}
\,(n\varpi+1)\,\big(\partial\hat{\alpha}\big)^2\,
\partial^2\hat{\varphi}_3
\nonumber \\[0.2cm]
&+&\big(1+2/(n\varpi)\big)\,\Big((n\varpi+1)\,\big(\partial^2\hat{\alpha}\big)^2+
n\varpi\,\big(\partial^2\hat{\varphi}_3\big)^2-4\sqrt{n\varpi}\,
\partial\hat{\varphi}_3\,\partial\hat{\alpha}\,
\partial^2\hat{\alpha}-\tfrac{1}{6}\,\sqrt{n\varpi}\,\partial^4\hat{\varphi}_3\Big)\nonumber
\eea
(every composite field appearing above is assumed to be normal ordered).
The formulae above should be compared with eq.\;\eqref{oias9812}, which lists the 
first few classical local left moving fields
of the cigar NLSM in the free field realization. Taking into account that
$\alpha^{({\rm L})}=\hat{\alpha}/(n\varpi)^{\frac{1}{2}}$ and $\phi^{({\rm L})}=\hat{\varphi}_3/(n\varpi)^{\frac{1}{2}}$ 
become canonically normalized fields
in the classical limit, one finds that
\be\label{askj21nas}
T_m^{({\rm L})}=\lim_{n\to\infty}(n\omega)^{-\frac{m}{2}}\,\mathsf{T}_m\,.
\ee
This way, the currents $\mathsf{T}_m$ are the first few members of the quantum ${\cal W}_\infty$-algebra.
\bigskip

Let's turn to the second condition \eqref{asnb12nbsaaaaa}.
There is an alternative way of stating it. 
Ignoring total derivatives, the screening charges $\mathsf{V}_+$
and $\mathsf{V}_-$ \eqref{vertex2} are related to each other by the transformation 
$\hat{\varphi}_1\mapsto-\hat{\varphi}_1$ and $\hat{\varphi}_3\mapsto -\hat{\varphi}_3$. Therefore,
eq.\;\eqref{asnb12nbsaaaaa} is equivalent to  $\mathsf{F}_m(x)$ being invariant under 
a simultaneous flipping of the signs of $\hat{\varphi}_1$ and $\hat{\varphi}_3$ up to a total derivative.
An  analysis yields that $\mathsf{Q}_{m-1}$ with odd $m$ vanish, while
the first few members of the ``even'' local IMs are given by
\be\label{asnbnb12asj}
\mathsf{Q}_1=-\int_0^{2\pi}\rd x\, \Big(\mathsf{T}_2+\big(\partial\hat{\chi}\big)^2\Big)\,,\qquad
\qquad\qquad
\mathsf{Q}_3=\int_0^{2\pi}\rd x \  \mathsf{F}_4(x)
\ee
with
\bea\label{asnbnbnb21}
\mathsf{F}_4&=&C_1\,:\mathsf{T}_2^2:+C_2\,\mathsf{T}_4+C_{3}\,(\partial^2\hat{\chi})^2-
C_4\,C_5\,\mathsf{T}_3\,\partial\hat{\chi} \\[0.2cm]
&-&
2 \,C_5^2\,(2+3n\varpi)\,
\Big(
3 n\varpi\,
\mathsf{T}_2\,(\partial\hat{\chi})^2
+\tfrac{1}{2}\,(4+3n\varpi)\,(\partial\hat{\chi})^4\Big)\, . \nonumber 
\eea
Here the constants $C_j$ read
\be\label{asnnmnbzxjkas}\arraycolsep=0.5cm
\begin{array}{ll}
C_1=n\,\varpi\,(3n+4)\,(2+3n-3n\varpi)\,,
& 
C_4=\frac{12 n\varpi}{\sqrt{2+n\varpi}}\,\big(2+2n-n\varpi\big)
\\[0.4cm]
C_2=\frac{2n\varpi}{2+n\varpi}\,\big(2+3n\,(n+2)\,-n\,(3n+5)\,\varpi\big)\,, &
C_5=\big(\frac{n\,(n+2)\,(\varpi-1)}{2+n\varpi }\big)^{\frac{1}{2}}
\\[0.4cm]
C_{3}=(2+3n\varpi)\,\big(4+10n+5n^2-n(7+5n)\varpi\big)\,,
\end{array}
\ee
while $\nor{\mathsf{T}_2^2}$ stands for the 
normal ordered field which can  be defined as the
first regular term in the OPE of $\mathsf{T}_2(x)\mathsf{T}_2(0)$.
Note that in the classical limit the second line of \eqref{asnbnbnb21}, as well as the
term $C_1\nor{\mathsf{T}_2^2}$ in the first line, dominate. As a result,
the local IM $\mathsf{Q}_3$ becomes proportional to the integral of the square of 
$T_2^{({\rm L})}+(\partial_+\chi)^2$  and, in particular, contains no dependence on the other classical 
fields $T_3^{({\rm L})}$ or $T_4^{({\rm L})}$.
This was remarked upon at the end of Subsection\,\ref{oias98oia}.

\subsection[Quantum corner-brane ${\cal W}$-algebra]{Quantum corner-brane \texorpdfstring{$\bm{\Wc}$}{W}-algebra}\label{Sec:CB}

In Subsection \ref{ias9812211aaa} 
the classical corner-brane ${\cal W}$-algebra was described.
Its fields are certain combinations of $T_m^{({\rm L})}$ and 
$\partial_+\chi^{({\rm L})}$ that are determined via a commutativity condition
with the classical screening charge ${\cal X}^{\scriptscriptstyle \rm CB}_1$ \eqref{oasoio90aaa}.
There is a direct analogue of this in the quantum case. The quantum corner-brane ${\cal W}$-algebra
is generated by the differential polynomials in $\partial\hat{\varphi}_j$ which, together with $\mathsf{X}_0$,
also quasi-commute with the screening charge:
\be\label{Xcb}
\mathsf{X}_1^{\scriptscriptstyle \rm CB}=\int_0^{2\pi}
\big( \,\ri \sqrt{n\varpi}\ \partial\hat{\varphi}_3+ \sqrt{n+2}
\,\partial\hat{\varphi}_2-\sqrt{n(\varpi-1)}
\,\partial\hat{\varphi}_1\,\big)\,\re^{-\frac{2\ri\hat{\varphi}_2}{\sqrt{n+2}}}\ .
\ee
By analysing the condition similar to \eqref{asnb12nbsa} but with $\mathsf{V}_+(y)$ replaced by the
integrand of $\mathsf{X}_1^{\scriptscriptstyle \rm CB}$, one obtains the spin 2 field:
\bea
\mathsf{W}_2&=&-\mathsf{T}_2-(\partial\hat{\chi})^2+\ri\,\Big(\frac{n\varpi +2}{n\,(n+2)\,(\varpi-1)}\Big)^{\frac{1}{2}}
\,\partial^2\hat{\chi} \\[0.3cm]
&=&-(\partial\hat{\varphi}_1)^2-(\partial\hat{\varphi}_2)^2-
(\partial\hat{\varphi}_3)^2+\frac{\ri}{\sqrt{n(\varpi-1)}}\,\partial^2\hat{\varphi}_1
-\frac{\ri}{\sqrt{n+2}}\,\partial^2\hat{\varphi}_2+\frac{1}{\sqrt{n\varpi}}\,\partial^2\hat{\varphi}_3\nonumber
\eea
The extra minus sign here as compared with $\mathsf{T}_2$ is a matter of convention. It
was introduced so that  $\mathsf{W}_2$ satisfies the usual 
 stress-energy momentum tensor  OPE:
\be\label{w2w2}
\mathsf{W}_2(x)\,\mathsf{W}_2(y)=
\frac{c}{2\,(x-y)^4}+\frac{2\mathsf{W}_2(y)}{(x-y)^2}+\frac{\partial \mathsf{W}_2(y)}{x-y}\,
+\reg
\ee
with the central charge being
\be\label{akskj2jk1kjsa}
c=3+6\,\bigg(\,\frac{1}{n\,\varpi}-\frac{1}{n\,(\varpi-1)}-\frac{1}{n+2}\,\bigg)\ .
\ee
\bigskip

As in the classical case, there is no independent spin 3 field.
A straightforward but somewhat lengthy computation yields that that the space of Lorentz spin 4 fields
is spanned by $\nor{\mathsf{W}_2^2}$, which is defined as the first regular term in the OPE
$\mathsf{W}_2(x)\mathsf{W}_2(0)$,  $\partial^2\mathsf{W}_2$ and
\bea
\mathsf{W}_4^{({\rm PB})}&=&\mathsf{F}_4+\partial\mathsf{O}_3
\eea
with 
\bea
\mathsf{O}_3&=&2\ri C_5\,(2+3n\varpi)\,\big(n\varpi\,\mathsf{T}_2\,\partial\hat{\chi}+
(\tfrac{4}{3}+n\varpi)\,(\partial\hat{\chi})^3\big)+n\,(n+2)\,(\varpi-1)\,(4+5n\varpi)\,\partial^2\hat{\chi}\,
\partial\hat{\chi}\nonumber\\[0.2cm]
&-&\tfrac{\ri}{3}\,C_5\,(2+n\varpi)\,\partial^3\hat{\chi}+
\frac{2\ri n\varpi\,(2+2n-n\varpi)}{\sqrt{2+n\varpi}}\, \mathsf{T}_3
+2n\varpi\,(n+1)\,(n+1-n\varpi)\,\partial\mathsf{T}_2\ .
\eea
Here the field $\mathsf{F}_4$ is the same one that appears in \eqref{asnbnbnb21}, while the value of the constant
$C_5$ is given in eq.\;\eqref{asnnmnbzxjkas}.
The normalization of $\mathsf{W}_4^{({\rm PB})}$ has been chosen so that it exactly coincides with the
spin $4$ current presented in Appendix A in the work \cite{Lukyanov:2012wq} on the ``Pillow Brane'' model.
The parameters $(\alpha_1,\alpha_2,\alpha_3)$ from that paper should be set to
$\alpha_1=-\sqrt{n\varpi}$, $\alpha_2=-\ri\sqrt{n(\varpi-1)}$ and $\alpha_3=-\ri\sqrt{n+2}$, while the fields
$\hat{\varphi}_j$ are related to the triple $(X,Y,Z)$ used in that work as
$\hat{\varphi}_1= Y$, $\hat{\varphi}_2=- Z$ and $\hat{\varphi}_3=- X$.
Note that in the classical limit
\be
\lim_{n\to\infty} n^{-4}\,\big(\mathsf{{ W}}_4^{({\rm PB})}
+9\,n^3\,\varpi\,(\varpi-1)\,\nor{\mathsf{W}_2^2}\big)=
-2\nu^2\,(1+\nu^2)^2\big(3W_4^{({\rm L})}+9\,(1+\nu^2)\,\big(W_2^{({\rm L})}\big)^2+
\partial^2_+W_2^{({\rm L})}\big)
\ee
where $W_4^{({\rm L})}$ and $W_2^{({\rm L})}$ are the classical corner-brane ${\cal W}$-algebra currents given in eqs.\;\eqref{asoi9812},\;\eqref{opcx90sa} and $\nu^2=\lim_{n\to\infty} \varpi-1$.
\bigskip

The fields $\mathsf{W}_m(x)$ form a closed  algebra. 
The OPE encoding the commutation relations between the Fourier
modes of the currents follow from the explicit formulae 
presented above and the 
free field OPEs  \eqref{mncnaskjx}. Rather
than working with $\mathsf{W}_4$, it is useful to consider a slightly different
spin 4 current:
\be\label{xcmnhjasnbc}
\mathsf{W}_{4,P}=\mathsf{W}_4^{({\rm PB})}-\delta_1\,\nor{\mathsf{W}_2^2}-\delta_2\,\partial^2\mathsf{W}_2
\ee
with
\bea\label{Eq:Delta1CB}
\delta_1&=&\frac{15n^2}{\beta}\,(n+2)\,(3n+4)\,(1-\varpi)\,\varpi\,\big(4+6n+
9n^2\,(1-\varpi)\,\varpi\big) \\[0.2cm]
\delta_2&=&-\frac{4n^2}{\beta}\,(n+2)\,(1-\varpi)\,\varpi\,\big(18+24n\,(n+2)
+n^2\,(35n+46)\,(1-\varpi)\,\varpi\big)\nonumber
\eea
and
\be\label{asknm21mnmn21}
\beta=n\,\big(30\,(n+2)+37n^2\,(1-\varpi)\,\varpi+
44n\,(1-\varpi)\,\varpi\big)\, .
\ee
Although the expression for $\mathsf{W}_4$ in terms of the free fields $\hat{\varphi}_j$
is significantly more complex than that for $\mathsf{W}_4^{({\rm PB})}$, it has the advantage
of being a primary field. Namely,
\be\label{w2w4}
\mathsf{W}_2(x)\mathsf{W}_{4,P}(y) =  \frac{4\mathsf{W}_{4,P}(y)}{(x-y)^2} + \frac{\p \mathsf{W}_{4,P}(x)}{x-y} + 
\reg\,.
\ee
The OPE of $\mathsf{W}_{4,P}$ with itself gives a cumbersome result that was first presented in ref. \cite{Semikhatov:2001zz}, see Appendix
A of that work.  We reproduce it here in our notations in order to make a later comparison with the Gaudin model. It reads
\bea\label{aksjj2nnaaaaaddda}
&&\Wq_{4,P}(x)\Wq_{4,P}(y) = \frac{\gamma\,c}{4(x-y)^8} + \frac{\gamma\,\bigl( \Wq_2(x)+\Wq_2(y) \bigr)}{(x-y)^6} + \frac{\gamma}{(x-y)^4} \Bigl( \beta_1 \bigl(\nor{\Wq_2(x)^2}+\nor{\Wq_2(y)^2} \bigr) \hspace{15pt} \notag \\
&& \hspace{10pt} + \beta_2 \bigl(\p^2\Wq_2(x)+\p^2\Wq_2(y) \bigr) + \beta_3 \bigl(\Wq_{4,P}(x)+\p^2\Wq_{4,P}(y) \bigr) \Bigr) + \frac{\Wq_6(x)+\Wq_6(y)}{(x-y)^2} + \reg \, ,
\eea
where
\be\arraycolsep=0.5cm
\begin{array}{ll}\label{betas}
\beta_1 =- \frac{21n^2}{\beta} (n+2)(\varpi-1)\varpi\,, & 
\beta_2=-
\frac{1+3\beta_1}{10} \\[0.5cm]
\beta_3 = \frac{4}{\gamma} \,\Big( \frac{\beta}{210}\, \big(147 - 97 \beta_1 - 144 \beta_1^2 \big) - 48 \beta_1 + 12 \Big) & 
\end{array}
\ee
and
\be\label{gamma}
\gamma = \frac{160n^2}{\beta} \,
(1-n)(n+2)(3n+4)(\varpi-1) \varpi \bigl(9+3n-n^2(\varpi-1)\varpi\bigr)\bigl(4+6n-9n^2(\varpi-1)\varpi\bigr)\,. 
\ee
Recall that the central charge and the constant $\beta$ were defined in eqs.\;\eqref{akskj2jk1kjsa} and 
\eqref{asknm21mnmn21}, respectively. Also $\mathsf{W}_6$ appearing in the
last line of the OPE \eqref{aksjj2nnaaaaaddda} is a  Lorentz spin 6 field belonging to the quantum corner-brane 
${\cal W}$-algebra that is not expressible as a differential polynomial in $\mathsf{W}_2$ and $\mathsf{W}_4$.
\bigskip

Finally, let us note that all the local IMs from the Fateev integrable structure can be written
in terms of the fields $\mathsf{W}_m$ and their ``descendents''. In particular, for the first few IMs
$\mathsf{Q}_1$ and $\mathsf{Q}_3$ presented in \eqref{asnbnb12asj}, one finds that
\be\label{Eq:W4CB}
\mathsf{Q}_1=\int_0^{2\pi}\rd x\ \mathsf{W}_2\,,\qquad\qquad
\mathsf{Q}_3=\int_0^{2\pi}\rd x\ \mathsf{W}_4^{({\rm PB})}\ .
\ee

\section{Quantised AGM: \texorpdfstring{$\bm\Wc$}{W}-algebra and local IMs}\label{Sec:QAGM}

The goal of this section is to discuss the quantisation of affine Gaudin models and their applications to the quantum UV-fixed point of the Klim\v{c}\'{i}k model: it is thus meant as a quantisation of the setup described in the classical case in Section \ref{sec555389i}. Quantised AGMs\footnote{Here and below, we use the terminology of ``quantised'' affine Gaudin models rather than ``quantum'' affine Gaudin models to avoid any confusion: indeed, the algebraic structure underlying this quantum model is still a standard affine algebra $\gh$, and not a quantum (deformed) one ${\cal U}_q(\gh)$.} were initially introduced in the work~\cite{Feigin:2007mr} of Feigin and Frenkel, motivated by an analogy with Gaudin models based on finite algebras instead of affine ones. In particular, it was argued in~\cite{Feigin:2007mr} that quantised AGMs can provide a systematic formalism to study integrable structures in conformal field theories. Building on these results and ideas, the study of quantised AGMs and their integrable structure was further pursued in the works~\cite{Lacroix:2018fhf,Lacroix:2018itd}, which focused on the description of local quantum charges in these models. In this section, we will review the results of~\cite{Feigin:2007mr,Lacroix:2018fhf,Lacroix:2018itd} and will discuss their applications to the quantisation of the Klim\v{c}\'{i}k model at the UV-fixed point. In particular, we will use the formalism of quantised AGMs to describe the quantum $\Wc$-algebra underlying this model and the quantum local IMs therein, focusing mainly on the ${\rm SU}(2)$ case. As we will see, these results agree with the ones obtained from free field realisations and screening charges in Section \ref{Sec:QuantLoc} and offer interesting perspectives for future applications in higher rank groups or more general models.

\subsection{Quantising affine Gaudin models}
\label{Sec:QuantAGM}

\paragraph{Reminder: change of spectral parameter in a classical AGM.} Our goal in this subsection is to describe the quantisation of an AGM with $N$ punctures $z_r$, Kac-Moody currents $\Jc_r$ and levels $\ell_r$. For that, it will be useful to first recall a few facts about the ``geometrical'' behaviour of this classical model with respect to the spectral parameter $z\in\mathbb{CP}^1$, as discussed in Subsection \ref{Sec:AGM}. In particular, recall that the twist function behaves as a 1-form $\vp(z)\dd z$ on $\mathbb{CP}^1$. In the previous sections, we have always supposed that our choice of spectral parameter $z$ is such that none of the punctures $z_r$ is situated at infinity. This requires the 1-form $\vp(z)\dd z$ to be regular at $z=\infty$ and thus the residues $\ell_r$ to satisfy the condition \eqref{Eq:SumL}, which we recall here for convenience:
\begin{equation}\label{Eq:SumL2}
\sum_{r=1}^N \ell_r = 0.
\end{equation} 

The second main ingredient defining an AGM is the Gaudin Lax matrix, which also defines a 1-form $\Gamma(z,x)\dd z$ on $\mathbb{CP}^1$. To ensure the regularity of this 1-form at infinity, we imposed that $\Cc_G(x) = \sum_{r=1}^N\Jc_r(x)$ vanishes, which we interpreted as a constraint $\Cc_G(x) \approx 0$ in the Poisson algebra $\Ac$. Moreover, the condition on the levels \eqref{Eq:SumL2} recalled above ensures that $\Cc_G$ is a first-class constraint and thus that the symmetry it generates is a gauge symmetry. This played an important role in the construction of the AGM. In particular, the integrable structure $\Zc^{(z_r)}\subset\Ac$ built in this model is formed by gauge invariant charges and is (weakly) independent of the choice of spectral parameter.

\paragraph{Sending a puncture at infinity.} In the previous sections, we have always supposed for simplicity that our choice of coordinate $z$ on $\mathbb{CP}^1$ is such that the punctures $z_r$ are finite. It will in fact be useful for quantisation to discuss quickly what happens if we perform a change of spectral parameter which sends a puncture to infinity. Up to relabelling, we can always suppose that this is the $N$-th puncture. We will  now work with a coordinate $z$ such that $z_N = \infty$. The twist function and Gaudin Lax matrix in this coordinate then take the form
\begin{equation}
\vp(z) = \sum_{r=1}^{N-1} \frac{\ell_r}{z-z_r} \qquad \text{ and } \qquad \Gamma(z,x) = \sum_{r=1}^{N-1} \frac{\Jc_r(x)}{z-z_r}\,.
\end{equation}
In particular, the level $\ell_N$ and current $\Jc_N$ associated with the $N$-th puncture do not appear explicitly in these expressions, since $z_N=\infty$. However, one has
\begin{equation}
\res_{z=\infty} \vp(z)\dd z = - \sum_{r=1}^{N-1} \ell_r\,.
\end{equation} 
Since the $N$-th puncture $z_N$ is at infinity, this residue should coincide with $\ell_N$, hence $\ell_N = -\sum_{r=1}^{N-1} \ell_r$. We thus recover in this way the condition \eqref{Eq:SumL2}. A similar statement holds for the Gaudin Lax matrix: more precisely, we want to ensure that the residue of $\Gamma(z,x)\dd z$ at infinity coincides with the Kac-Moody current $\Jc_N(x)$ attached to the $N$-th puncture $z_N=\infty$. We treat this as a constraint in the algebra $\Ac$ generated by $\Jc_1,\ldots,\Jc_N$, which then reads
\begin{equation}\label{Eq:ConstPunctureInfinity}
\Jc_N(x) \approx -\sum_{r=1}^{N-1} \Jc_r(x).
\end{equation}
We thus recover the constraint $\Cc_G(x)\approx 0$.\\

Working with the coordinate $z$ such that the puncture $z_N$ is at infinity presents a slight advantage. Indeed, the Gaudin Lax matrix is formally independent of the Kac-Moody current $\Jc_N$, which only enters in the model through the constraint \eqref{Eq:ConstPunctureInfinity}. Thus, the Lax matrix $\Lc(z,x)$ of the model, which allows one to build non-local conserved charges, as well as the local charges built from $\Gamma(z,x)$ at the zeroes of $\vp(z)$ (see Subsection \ref{Sec:AGM}), are directly expressed in terms of $N-1$ Kac-Moody currents $\Jc_1,\ldots,\Jc_{N-1}$ only. In contrast, in another coordinate where all the punctures are finite, all of the $N$ currents $\Jc_r$ enter the explicit definition of these charges. Of course, the dependence of these charges on $\Jc_N$ can be removed weakly, by imposing the constraint to express $\Jc_N$ in terms of the $N-1$ remaining currents: the charges obtained in this way then coincide with the ones directly built from the coordinate in which $z_N=\infty$. Working in this coordinate thus has the advantage of describing the integrable structure directly in terms of $N-1$ Kac-Moody currents, without having to impose the constraint explicitly.

Let us also discuss how the gauge symmetry is implemented in the model with $z_N=\infty$. The gauge invariant observables are the ones which weakly Poisson commute with the constraint $\Cc_G = \sum_{r=1}^N \Jc_r$. Since the observables in this model are naturally built from the currents $\Jc_1,\ldots,\Jc_{N-1}$ only, their Poisson bracket with $\Cc_G(x)$ coincides with their bracket with the current
\begin{equation}\label{Eq:JDiag}
\Jc_{\diag}(x) = \sum_{r=1}^{N-1} \Jc_r(x).
\end{equation}
We can then characterise the physical observables of the model with $z_N=\infty$ as the quantities built from $\Jc_1,\ldots,\Jc_{N-1}$ which Poisson commute with $\Jc_{\diag}$. Note however that $\Jc_{\diag}$ is not treated here as a constraint: indeed, it is not set to zero. We note also that the setup considered here formally resembles an AGM with $N-1$ finite punctures, in which we do not impose that the residues $-\sum_{r=1}^{N-1}\ell_r$ of $\vp(z)\dd z$ and $-\Jc_{\diag}$ of $\Gamma(z,x)\dd z$ vanish, but where we focus on observables invariant under the diagonal symmetry generated by $\Jc_\diag$.\footnote{This is closer to the historical vision of Gaudin models associated with finite algebras, as in the foundational references~\cite{Gaudin76,Gaudin_book83}, where these models were seen as spin chains with long-range interactions and a global diagonal symmetry.} This is reminiscent of the discussion of the classical $\Wc$-algebra in Subsection \ref{aaaaaaaaaaaa}: as we shall see, this will also be useful later in this section for the construction of the quantum $\Wc$-algebra and the local integrals of motion therein.

\paragraph{Quantum Kac-Moody currents.} Consider a basis $\lbrace {\tt t}_a \rbrace$ of $\g$, the dual basis $\lbrace {\tt t}^a \rbrace$ with respect to the bilinear form $\langle\cdot,\cdot\rangle$ and the decomposition $\Jc_r = \Jc_{r,a} {\tt t}^a$ of the classical Kac-Moody current. The Poisson bracket of these components is then given by eq.\;\eqref{Eq:PbJa}. This Poisson algebra has a natural quantisation $\Ah$. Namely, $\Ah$ is a non-commutative algebra generated by fields $\Jq_{r,a}(x)$ satisfying the commutation relations
\begin{equation}\label{Eq:ComKM}
\bigl[ \Jq_{r,a}(x), \Jq_{s,b}(y) \bigr] = -2\pi\, \delta_{rs} \left(  \f abc\,\Jq_{r,c}(x)\,\delta(x-y) + \ri\,k_r\,\eta_{ab}\,\p_x\delta(x-y)\right) \; ,
\end{equation}
where $\eta_{ab}=\langle{\tt t}_a,{\tt t}_b\rangle$, $k_r$ are constant numbers that we will call quantum levels and the factors $2\pi$ and $\ri$ have been introduced for future convenience. To take the classical limit of this algebra, we consider $\hbar \to 0$ with
\begin{equation}\label{Eq:ClassicalLim}
k_r = -\frac{2\pi\ell_r + O(\hbar)}{\hbar} \qquad \text{ and } \qquad \Jq_{r,a} = -\frac{2\pi\Jc_{r,a} + O(\hbar)}{\ri\hbar}.
\end{equation}
It is then straightforward to check that
\begin{equation}
\bigl[ \Jc_{r,a}(x), \Jc_{s,b}(y) \bigr] = \ri \hbar \, \bigl\lbrace \Jc_{r,a}(x), \Jc_{s,b}(y) \bigr\rbrace + O(\hbar^2)  \, ,
\end{equation}
with the Poisson bracket $\lbrace \Jc_{r,a}, \Jc_{s,b} \rbrace$ given by eq.\;\eqref{Eq:PbJa}. This ensures that the semi-classical limit of the quantum algebra $\Ah$ is indeed $\Ac$.\\

Using the language of Vertex Operator Algebras, the commutation relation \eqref{Eq:ComKM} obeyed by the quantum Kac-Moody currents $\Jq_{r,a}(x)$ can be rephrased in terms of the following OPE (Operator Product Expansion):
\begin{equation}\label{Eq:OpeJ}
\Jq_{r,a}(x)\Jq_{s,b}(y) = \delta_{rs} \left( \frac{ \ri\,\f abc\,\Jq_{r,c}(y)}{x-y} + \frac{k_r\,\eta_{ab}}{(x-y)^2} \right) + \reg \;,
\end{equation}
where $\reg$ designates terms which are regular when $x\to y$.

\paragraph{Quantum twist function and Gaudin Lax matrix.} Let us introduce
\begin{equation}\label{Eq:QuantumGaudin}
\vpq(z) = \sum_{r=1}^N \frac{k_r}{z-z_r} \qquad \text{ and } \qquad \Gq_a(z,x) = \sum_{r=1}^N \frac{\Jq_{r,a}(x)}{z-z_r}\, .
\end{equation}
We call $\vpq(z)$ the quantum twist function and $\Gq(z,x)$ the quantum Gaudin Lax matrix. It is clear from eq.\;\eqref{Eq:ClassicalLim} that the behaviour of these objects in the classical limit $\hbar \to 0$ is given by
\begin{equation}\label{Eq:ClassicalLimGaudin}
\vpq(z) = -\frac{2\pi\vp(z)+O(\hbar)}{\hbar} \qquad \text{ and } \qquad \Gq(z,x) = -\frac{2\pi\Gamma(z,x)+O(\hbar)}{\ri\hbar},
\end{equation}
where $\vp(z)$ and $\Gamma(z,x)$ are the twist function and Gaudin Lax matrix of the classical AGM, as defined in eq.\;\eqref{Eq:Twist} and \eqref{Eq:Gaudin}.

\paragraph{Change of spectral parameter in quantised AGMs.} Let us now discuss the behaviour of the quantised AGM under a change of spectral parameter $z \mapsto \zt$ with $z=\omega(\zt)$. This question was touched upon indirectly in the work~\cite{Lacroix:2018fhf}, which explored quantised AGMs starting from the conjecture of Feigin and Frenkel~\cite{Feigin:2007mr} stating that the spectrum of these models is encoded in objects called affine opers (we will discuss some of the results of~\cite{Lacroix:2018fhf} and affine opers later in this section and the next). In particular, the behaviour of affine opers under a change of spectral parameter was studied in~\cite{Lacroix:2018fhf} and suggests that the twist function is no longer a 1-form at the quantum level. More precisely, it should transform according to the following rule:
\begin{equation}\label{Eq:ChangeZQuant}
\widetilde{\vp}^{\qt}(\zt\,) = \vpq\bigl( \omega(\zt) \bigr)\, \omega\,'(\zt) + \hv \; \frac{\omega\,''(\zt)}{\omega\,'(\zt)},
\end{equation}
where $\hv$ is the dual Coxeter number of the Lie algebra $\g$. The first term in this equation corresponds to the standard transformation rule of a 1-form under the change of coordinate $z\mapsto \zt=\omega^{-1}(z)$. The second term breaks this property and in fact makes $-\vpq(z)\dd z$ behave as the component of a meromorphic connection on the line bundle $\Omega^{\,\hv}$ over $\mathbb{CP}^1$, where $\Omega^{k}$ denotes the $k$-th tensor power of the canonical line bundle. This term can be interpreted as a quantum correction: indeed, considering the classical limit $\hbar\to 0$ with the scaling property \eqref{Eq:ClassicalLimGaudin}, we easily see that this term is subdominant in the limit of eq.\;\eqref{Eq:ChangeZQuant}, which reads
\begin{equation}
\vpt(\zt) = \vp\bigl( \omega(\zt) \bigr)\, \omega\,'(\zt) + O(\hbar),
\end{equation}
recovering that the classical twist function $\vp(z)$ behaves as a 1-form when $\hbar = 0$.\\

The transformation rule \eqref{Eq:ChangeZQuant} has an important consequence. Let us consider the behaviour of the quantum twist function around infinity. For that, we consider the change of variable $z \mapsto \zt = z^{-1}$. Applying the rule \eqref{Eq:ChangeZQuant}, we find that the twist function in the coordinate $\zt$ is given by
\begin{equation}
\widetilde{\varphi}^{\qt}(\zt) = -\frac{1}{\zt^{\,2}}\vpq\left( \frac{1}{\zt} \right) - \frac{2\hv}{\zt}.
\end{equation}
Using the expression \eqref{Eq:QuantumGaudin} of $\vpq(z)$ in terms of the quantum levels $k_r$, we find that, around $\zt=0$,
\begin{equation}
\widetilde{\vp}^{\qt}(\zt) = - \frac{1}{\zt} \left( \sum_{r=1}^N k_r + 2\hv + O(\zt) \right).
\end{equation}
The fact that the AGM does not possess a puncture at $z=\infty$ and thus that the twist function $\widetilde{\vp}^{\qt}(\zt)$ is regular at $\zt=0$ then implies
\begin{equation}\label{Eq:SumK}
\sum_{r=1}^N k_r = -2\hv.
\end{equation}
This is the quantum equivalent of the condition \eqref{Eq:SumL2} on the classical levels $\ell_r$. In fact, it is clear from the scaling behaviour \eqref{Eq:ClassicalLim} of $k_r$ in the classical limit $\hbar \to 0$ that the quantum condition \eqref{Eq:SumK} reduces to eq.\;\eqref{Eq:SumL2} in this limit, as expected. We will discuss  the algebraic interpretation of this quantum condition in the following paragraph.

\paragraph{Quantum gauge symmetry.} Recall that in the classical case, we consider the AGM up to the gauge symmetry generated by the constraint $\Cc_G \approx 0$: in particular, the condition \eqref{Eq:SumL2} on the classical levels $\ell_r$ ensures that this constraint is first-class. The quantum equivalent of $\Cc_G$ is the field
\begin{equation}\label{Eq:QuantumGaugeGen}
\sum_{r=1}^N \Jq_r (x),
\end{equation}
which is a Kac-Moody current with level $\sum_{r=1}^N k_r$. It is thus natural to consider the quantised AGM as gauged with respect to the symmetry generated by this field, which is then seen as a quantum constraint. At the quantum level, this constraint and gauge symmetry should be treated using the BRST formulation. In particular, the consistency of this formulation requires the BRST charge built from the generator of the symmetry to square to zero. For the case of a symmetry generated by a Kac-Moody current, as here, this condition has been studied in~\cite{Hlousek:1986ux}, where it was proven that it requires the level of the current to be equal to $-2\hv$. In the case at hand, since the generator \eqref{Eq:QuantumGaugeGen} has level $\sum_{r=1}^N k_r$, we find that the condition for the consistency of the BRST formulation coincides with the condition \eqref{Eq:SumK} found above by considering the geometrical behaviour of the quantum twist function. This is the quantum equivalent of the classical condition \eqref{Eq:SumL2}, which ensures on the one hand that the classical constraint is first-class and on the other hand that the classical twist function behaves as a 1-form.

\paragraph{Setting a puncture to infinity.} As mentioned above, imposing the constraint and the gauge symmetry at the quantum level requires the use of the BRST formalism. There is however a way to avoid these technical complications. Indeed, recall from earlier discussions that, in the classical case, considering a setup in which one of the puncture $z_N$ is at infinity allows us to build the AGM from the currents $\Jc_1,\ldots,\Jc_{N-1}$ only and thus avoids having the constraint to be explicitly imposed. We will follow a similar approach in the quantum case and thus suppose from now on that we chose a coordinate $z$ on $\mathbb{CP}^1$ such that $z_N=\infty$.\footnote{For completeness, one should then prove that the quantised AGM constructed in this setup is equivalent to the one that would be constructed in a coordinate where none of the punctures are at infinity. This would in particular require a treatment in the BRST formalism and is beyond the scope of this paper.} By construction, the quantum Gaudin Lax matrix $\Gq(z,x)$ in this coordinate then involves only the Kac-Moody currents $\Jq_1,\ldots,\Jq_{N-1}$. Observables built from these currents are then gauge invariant if they commute (or equivalently have regular OPE) with the diagonal generator
\begin{equation}
\Jq_{\diag}(x) = \sum_{r=1}^{N-1} \Jq_r(x).
\end{equation}
Since the modes of the Kac-Moody current $\Jq_r$ form the affine algebra $\gh_{k_r}$ with level $k_r$, these gauge-invariant observables then naturally form a
\begin{equation}
\frac{\gh_{k_1}\oplus\ldots\oplus\gh_{k_{N-1}}}{\gh_{k_1+\ldots+k_{N-1}}}
\end{equation}
coset algebra.

\paragraph{Quantum integrable structure.} Recall from Subsection \ref{Sec:AGM} that the main output of the classical AGM formalism is the construction of an integrable structure $\Zc^{(z_r)}$, formed by gauge-invariant charges in involution built from the classical Kac-Moody currents. It is natural to expect that a quantum equivalent $\Zh^{(z_r)}$ of this integrable structure exists in the quantised AGM described above, corresponding to commuting operators built from the quantum Kac-Moody currents $\Jq_r$. This was conjectured and motivated in~\cite{Feigin:2007mr}, using an analogy with Gaudin models based on finite algebras, for which the quantum integrable structure is well understood. In particular, some of the commuting operators in $\Zh^{(z_r)}$ were discussed in~\cite{Feigin:2007mr}, including certain non-local charges and quadratic local charges (as well as higher degree local charges in specific examples). Building on the results and ideas of~\cite{Feigin:2007mr}, the systematic construction of higher degree local charges was also considered in~\cite{Lacroix:2018fhf,Lacroix:2018itd}, resulting in some conjectures (and first checks thereof) on the form of these charges. We will discuss some of these results as well as a few new ones in Subsection \ref{Sec:QuantLocalIM} and Appendix \ref{App:LocalIM} below.

\paragraph{Quantum realisations.} So far, we have described a formal quantised AGM, whose algebra of physical observables is a Vertex Operator Algebra built from (the quotient of) abstract quantum Kac-Moody currents $\Jq_r$. In order to apply this formalism to the description of integrable 2-dimensional quantum field theories, and in particular integrable $\s$-models, one needs to consider realisations of this algebra in terms of the fields of these theories, as we did in the classical setup. A natural question at this point is whether any classical realisation of an AGM will be upgraded to a quantum realisation in the above sense: we expect that this is in fact not the case in general. Indeed, there is no guarantee that classical Kac-Moody currents of a 2-dimensional field theory will survive quantisation and obey quantum OPEs as in eq.\;\eqref{Eq:OpeJ}.

There is however one important setup where we expect such a quantisation to hold, namely chiral realisations in conformal models. In this case, the currents of the AGM are realised as left- or right-moving fields of the theory (up to gauge symmetry, see Section \ref{sec555389i}). The quantisation of this theory defines a 2-dimensional CFT whose chiral algebras of left- and right- moving fields are described by Vertex Operator Algebras: we then expect these chiral algebras to contain a realisation of the corresponding quantised AGM. In this case, we then interpret the quantum fields describing the AGM as functions of the light-cone variable $x+t$ or $x-t$. We will see examples of this construction in the next subsection, for the case of the UV fixed-point of the ${\rm SU}(2)$ Klim\v{c}\'{i}k model. In Section \ref{sec555389i}, we have denoted the quantities in classical chiral AGMs with a label $\cL$ or $\cR$, to emphasise that they are realised in terms of left-moving or right-moving fields: to lighten the notations, we will drop this label in the present section. It is however important to keep in mind that we expect these quantised AGMs to only be realised in terms of chiral fields of 2-dimensional CFTs. 

\subsection[Quantum $\Wc$-algebra]{Quantum \texorpdfstring{$\bm\Wc$}{W}-algebra}
\label{Sec:QuantW}

\paragraph{Generalities.} As explained in the previous subsection, the physical observables of a quantised AGM with $N-1$ finite punctures and an infinite one $z_N=\infty$ can be seen as the quantities built from the Kac-Moody currents $\Jq_1,\ldots,\Jq_{N-1}$ which commute with the current $\Jq_{\diag}$. Such observables are in general non-local in the currents. We can however restrict to local fields satisfying this property, yielding the quantum $\Wc$-algebra $\Wh$ underlying the model. More precisely, $\Wh$ is defined as the set of normal ordered differential polynomials in the components of $\Jq_1,\ldots,\Jq_{N-1}$ which have a regular OPE with $\Jq_{\diag}$. Such a field $\Wq$ is thus built as a linear combination of monomials of the form
\begin{equation}\label{Eq:MonomW}
\nor{\p^{p_1}\Jq_{r_1,a_1}\cdots\p^{p_m}\Jq_{r_m,a_m}}
\end{equation}
(with $p_k\in\mathbb{Z}_{\geq 0}$, $r_k \in\lbrace 1,\ldots,N-1\rbrace$ and $a_k\in\lbrace 1,\ldots,\dim\g \rbrace$) and satisfies
\begin{equation}
\Jq_{\diag,a}(x) \Wq(y) = \reg, \qquad \forall \, a\in\lbrace 1,\ldots,\dim\g \rbrace.
\end{equation}
We will sometime call these fields W-currents. We note that $\Wh$ is a Vertex Operator Subalgebra, in the following sense: for any two fields $\Wq$ and $\Wq'$ in $\Wh$, $\p\Wq$ and the normal ordered product $\nor{\Wq\,\Wq'}$ belong to $\Wh$ and the coefficients of the OPE $\Wq(x)\Wq'(y)$ are elements of $\Wh$.

Moreover, $\Wh$ is a graded Vertex Operator Algebra. The corresponding grading is defined by assigning the weight one to Kac-Moody components $\Jq_{r,a}$ and to the derivative $\p$, such that the monomial \eqref{Eq:MonomW} has grade $\sum_{k=1}^m (p_k+1) = p_1+\ldots+p_m+m$. The algebra $\Wh$ then decomposes into a direct sum
\begin{equation}
\Wh = \bigoplus_{p\in\mathbb{Z}_{\geq 2}} \Wh_p,	
\end{equation}
where $\Wh_p$ is the subspace of W-currents of grade $p$, which is of finite dimension. Moreover, if $\Wq_p$ and $\Wq_q$ have respective gradings $p$ and $q$, then $\p\Wq_p$ has grading $p+1$ and $\nor{\Wq_p\Wq_q}$ has grading $p+q$. Finally, let us observe that under a quantum chiral realisation of the AGM in a 2-dimensional CFT, this grading coincides with the Lorentz spin of the W-currents if the latter are left-moving fields and to minus their spin if they are right-moving. By a slight abuse of language, we will sometimes refer to a W-current with grade $p$ as a spin $p$ field, having in mind applications to left-moving realisations.

\paragraph{The $\bm{(\gh\oplus\gh)/\gh}$ coset $\bm\Wc$-algebra.} Let us now specialise our setup by considering an AGM with $N=3$ punctures, which is the AGM that underlies one of the chiral halves of the UV fixed-point of the Klim\v{c}\'{i}k model. In this case, the $\Wc$-algebra $\Wh$ is then composed by the differential polynomials in two currents $\Jq_1$ and $\Jq_2$ which commute with $\Jq_{\diag}=\Jq_1+\Jq_2$. It thus forms a $(\gh_{k_1}\oplus\gh_{k_2})/\gh_{k_1+k_2}$ coset $\Wc$-algebra. In Subsection \ref{aaaaaaaaaaaa}, we have given a systematic description of the classical version of this $\Wc$-algebra, by building classical W-currents from covariant fields and their covariant derivatives. This construction does not seem to straightforwardly generalise to the quantum case considered here: indeed, these classical expressions generally receive quantum corrections which are not directly expressed in terms of the covariant fields.

Let us illustrate this with a simple example, namely the quantum spin 2 W-current $\Wq_2$. To find it, one can start with an ansatz as a linear combination of the fields $\eta^{ab}\nor{\Jq_{1,a}\Jq_{1,b}}$, $\eta^{ab}\nor{\Jq_{2,a}\Jq_{2,b}}$ and $\eta^{ab}\nor{\Jq_{\diag,a}\Jq_{\diag,b}}$, where $\eta^{ab}$ denotes the inverse of $\eta_{ab}=\langle{\tt t}_a, {\tt t}_b\rangle$. A direct computation shows that the regularity of the OPE $\Jq_{\diag,a}(x)\Wq_2(y)$ fixes the corresponding coefficients (up to a global factor), yielding:
\begin{equation}\label{Eq:QuantumW2}
\Wq_2 = \frac{\eta^{ab}}{2}\left(\frac{1}{k_1+\hv}\, \nor{\Jq_{1,a}\Jq_{1,b}} \,+\, \frac{1}{k_2+\hv} \,\nor{\Jq_{2,a}\Jq_{2,b}} \, -\; \frac{1}{k_1+k_2+\hv} \,\nor{\Jq_{\diag,a}\Jq_{\diag,b}} \right) \, .
\end{equation}
We recognise here the energy-momentum tensor of the Goddard-Kent-Olive (GKO) construction~\cite{Goddard:1984vk,Goddard:1986ee}. The overall normalisation has been fixed so that $\Wq_2$ satisfies the standard Virasoro OPE:
\begin{equation}\label{Eq:Vir}
\Wq_2(x) \Wq_2(y) = \frac{c}{2(x-y)^4} + \frac{2\Wq_2(y)}{(x-y)^2} + \frac{\p \Wq_2(y)}{x-y} + \reg \,,
\end{equation}
where the central charge $c$ reads
\begin{equation}\label{Eq:cGKO}
c = \dim\g\left( \frac{k_1}{k_1+\hv} + \frac{k_2}{k_2+\hv} - \frac{k_1+k_2}{k_1+k_2+\hv} \right).
\end{equation}
To study the classical limit of this W-current, let us recall that the levels $k_r$ and currents $\Jq_r$ satisfy the asymptotic property \eqref{Eq:ClassicalLim} when $\hbar \to 0$. We then find
\begin{equation}\label{Eq:ClassLimW2}
\Wq_2 = \frac{2\pi\,W_2+O(\hbar)}{\hbar}, \qquad \text{ where } \qquad W_2 = \frac{\bigl\langle \Jc_1,\Jc_1 \bigr\rangle}{2\ell_1} + \frac{\bigl\langle \Jc_2,\Jc_2 \bigr\rangle}{2\ell_2} - \frac{\bigl\langle \Jc_{\diag},\Jc_{\diag} \bigr\rangle}{2(\ell_1+\ell_2)}
\end{equation}
is the classical spin 2 W-current. In particular, we see that the shifts of the levels by $\hv$ in eq.\;\eqref{Eq:QuantumW2} can be interpreted as quantum corrections. Moreover, one checks that the OPE \eqref{Eq:Vir} gives back, in the classical limit, the Poisson bracket \eqref{Eq:PbW2}. Finally, using $\Jc_{\diag}=\Jc_1+\Jc_2$, we find that $W_2$ is proportional to $\langle\Kc,\Kc\rangle$, where $\Kc=\ell_2\,\Jc_1-\ell_1\,\Jc_2$ is the covariant classical current introduced in eq.\;\eqref{Eq:B}: we thus recover the classical construction \eqref{Eq:W2} of $W_2$ in terms of covariant fields. However, at the quantum level, the expression \eqref{Eq:QuantumW2} for $\Wq_2$ does not seem to be directly expressible in terms of a quantum version of the covariant current $\Kc$. It would be interesting to explore whether a less naive quantisation of the classical procedure outlined in Subsection \ref{aaaaaaaaaaaa} can be found, that would allow a more systematic construction of the quantum $(\gh_{k_1}\oplus\gh_{k_2})/\gh_{k_1+k_2}$ coset $\Wc$-algebra.

\paragraph{Quantum corner-brane $\bm\Wc$-algebra.} Let us finally specialise the setup even more by considering $\g^\C=\mathfrak{sl}(2)$. In this case, $\Wh$ becomes the corner-brane $\Wc$-algebra~\cite{Semikhatov:2001zz,Lukyanov:2012wq}. Its classical version has been discussed in terms of screening charges in Subsection \ref{ias9812211aaa} and from the point of view of Gaudin models and coset algebras in Subsection \ref{aaaaaaaaaaaa}. Its quantisation from screening charges was described in Subsection \ref{Sec:CB}. We finally discuss its quantisation as a $\widehat{\mathfrak{sl}}(2)\oplus\widehat{\mathfrak{sl}}(2)/\widehat{\mathfrak{sl}}(2)$ coset $\Wc$-algebra in this paragraph and the appendix \ref{App:QuantumW}\footnote{Most of the OPE computations presented in this paragraph and the appendix were performed using the Mathematica package ``OPEdefs'' by K. Thielemans~\cite{Thielemans:1991uw}.}. It was proven in~\cite{Semikhatov:2001zz} that these two quantisations are equivalent: we refer to this work for a more detailed treatment. For the purpose of the present article, it will be enough to describe the first elements of this algebra and compare them with the free field realisation studied in Subsection \ref{Sec:CB}.\\

This $\Wc$-algebra contains a unique spin 2 field, the energy-momentum tensor $\Wq_2$ described in the previous paragraph for an arbitrary Lie algebra $\g$, as well as a unique spin 3 field $\p\Wq_2$. The obvious spin 4 fields are $\nor{\Wq_2^2}$ and $\p^2\Wq_2$. In the classical case, we found in Subsection \ref{aaaaaaaaaaaa} that there exists an additional independent spin 4 field, built in eq.\;\eqref{mncxnbsad} from the covariant current $\Kc$. As for the energy-momentum tensor $\Wq_2$ discussed in the previous paragraph, this construction in terms of $\Kc$ does not directly generalise to the quantum case. We can however determine the quantum spin 4 field $\Wq_4$ by a brute force computation, solving the linear equation $\Jq_{\diag}(x)\Wq_4(y)=\reg$ starting with a general ansatz for $\Wq_4$ in terms of the Kac-Moody currents $\Jq_1$ and $\Jq_2$. This does not define uniquely $\Wq_4$, since we can freely add to it any linear combination of $\nor{\Wq_2^2}$ and $\p^2\Wq_2$ as well as multiply it by an overall constant. Here, we will consider a choice that makes the expression of $\Wq_4$ relatively simple and which ensures that we recover the classical spin 4 current \eqref{mncxnbsad} in the classical limit. Since the end result of this computation is rather convoluted, we give the full expression of $\Wq_4$ in Appendix \ref{App:QuantumW}, eq.\;\eqref{Eq:W4Quant}-\eqref{Eq:Coeff} (see also eq.\;\eqref{Eq:ClassLimW4} for a discussion of its classical limit).\\

From this expression, one can moreover compute the OPE of $\Wq_2$ with $\Wq_4$. We find
\begin{equation}
\Wq_2(x)\Wq_4(y) = \frac{\alpha_1\,c}{(x-y)^6} + \frac{\alpha_2\,\Wq_2(y)}{(x-y)^4} + \frac{\alpha_1\,\p\Wq_2(y)}{(x-y)^3} +  \frac{4\Wq_4(y)}{(x-y)^2} + \frac{\p \Wq_4(y)}{x-y} + \reg \, ,
\end{equation}
where $c$ is the central charge \eqref{Eq:cGKO} and $\alpha_1$ and $\alpha_2$ are constant parameters defined in terms of the levels $k_1$ and $k_2$, whose explicit expressions we give in Appendix \ref{App:QuantumW}, eq.\;\eqref{Eq:Alpha12}. The presence of poles of order 6, 4 and 3 in this OPE means that $\Wq_4$ is not a primary field. One can however construct a primary spin 4 field $\Wq_{4,P}$ from $\Wq_4$, hence satisfying
\begin{equation}\label{Eq:OpeW2W4p}
\Wq_2(x)\Wq_{4,P}(y) =  \frac{4\Wq_{4,P}(y)}{(x-y)^2} + \frac{\p \Wq_{4,P}(y)}{x-y} + \reg \, ,
\end{equation}
by adding to it a well-chosen linear combination of the descendant fields $\nor{\Wq_2^2}$ and $\p^2\Wq_2$. More precisely, we find
\begin{equation}
\Wq_{4,P}(x) = \Wq_4(x) + \alpha_3\,\nor{\Wq_2(x)^2} + \alpha_4\,\p^2\Wq_2(x),
\end{equation}
with the parameters $\alpha_3$ and $\alpha_4$ defined by eq.\;\eqref{Eq:Alpha34}.\\

Let us finally describe the OPE of the primary field $\Wq_{4,P}$ with itself. A direct computation shows that
\bea
&&\Wq_{4,P}(x)\Wq_{4,P}(y) = \frac{\gamma\,c}{4(x-y)^8} + \frac{\gamma\,\bigl( \Wq_2(x)+\Wq_2(y) \bigr)}{(x-y)^6} + \frac{\gamma}{(x-y)^4} \Bigl( \beta_1 \bigl(\nor{\Wq_2(x)^2}+\nor{\Wq_2(y)^2} \bigr) \hspace{15pt} \notag \\
&& \hspace{10pt} + \beta_2 \bigl(\p^2\Wq_2(x)+\p^2\Wq_2(y) \bigr) + \beta_3 \bigl(\Wq_{4,P}(x)+\p^2\Wq_{4,P}(y) \bigr) \Bigr) + \frac{\Wq_6(x)+\Wq_6(y)}{(x-y)^2} + \reg \, ,\label{Eq:OpeW4pW4p}
\eea
where $\Wq_6$ is a spin 6 W-current whose explicit expression we will not need and where the parameters $\beta_i$ and $\gamma$ are given by
{\small
\begin{align}\label{Eq:Beta}
\beta &= 176 - 44 (k_1+2)^2 - 44 (k_2+2)^2 - k_1 k_2 (37 k_1 + 37 k_2 + 192) \, \notag ,\\
\beta_1 &= -\frac{21}{\beta} (k_1+2) (k_2+2) (k_1 + k_2 + 2) \, , \notag \\
\beta_2 &= - \frac{1+3\beta_1}{10}\,, \qquad\quad \beta_3 = \frac{4}{\gamma} \left( \frac{\beta}{210} (147 - 97 \beta_1 - 144 \beta_1^2 ) - 48 \beta_1 + 12 \right) \,,\\
\gamma &= -\frac{160}{\beta}(k_1\!-\!1)(k_1\!+\!2)(3k_1\!+\!4)(k_2\!-\!1)(k_2\!+\!2)(3k_2\!+\!4)(k_1\!+\!k_2\!+\!2)(k_1\!+\!k_2\!+\!5)(3k_1\!+\!3k_2\!+\!8) \, \notag.
\end{align}}

\paragraph{Free field realisation in the $\bm{{\rm SU}(2)}$ chiral Klim\v{c}\'{i}k model.} The OPEs \eqref{Eq:Vir}, \eqref{Eq:OpeW2W4p} and \eqref{Eq:OpeW4pW4p} of $\Wq_2$ and $\Wq_{4,P}$ then take the exact same form as the ones \eqref{w2w2}, \eqref{w2w4} and \eqref{aksjj2nnaaaaaddda} found in Subsection \ref{Sec:CB} using free field realisations and screening charges. Recall that in this subsection, the central charge $c$ and the coefficients $\gamma$ and $\beta_i$ were expressed in terms of the two parameters $n$ and $\varpi$ entering the definition of the screening charges -- see eqs.\;\eqref{akskj2jk1kjsa},\;\eqref{betas} and \eqref{gamma}, while in the present case they are expressed in terms of the Kac-Moody levels $k_1$ and $k_2$. These expressions exactly coincide under the identification
\begin{equation}\label{Eq:ParamW}
k_1 = n(\varpi-1)-2, \quad k_2 = n \qquad \Leftrightarrow \qquad n=k_2, \quad \varpi = \frac{k_1+k_2+2}{k_2} \, .
\end{equation}
(Note that the dual Coxeter number of $\mathfrak{sl}(2)$ is given by $\hv=2$.) This provides an explicit isomorphism, at least for the first W-currents, between the corner-brane $\Wc$-algebra defined in terms of screening charges and the $\widehat{\mathfrak{sl}}(2)_{k_1}\oplus\widehat{\mathfrak{sl}}(2)_{k_2}/\widehat{\mathfrak{sl}}(2)_{k_1+k_2}$ coset $\Wc$-algebra, as expected in general from the reference~\cite{Semikhatov:2001zz}.\footnote{Let us mention for completeness that this $\Wc$-algebra is also isomorphic to a reduction of the affine superalgebra $\widehat{D}(2|1;\alpha)$ by a maximal nilpotent subalgebra, as shown in~\cite{Semikhatov:2001zz}.}

Recall that the description of the corner-brane $\Wc$-algebra in terms of screening charges in Subsection \ref{Sec:CB} uses chiral free fields $(\hat\varphi_1,\hat\varphi_2,\hat\varphi_3)$. The latter are interpreted as the quantisation of the classical left-moving free fields $(\phi^\cL_1,\phi^\cL_2,\phi^\cL_3)$ of the asymptotic domain of the ${\rm SU}(2)$ Klim\v{c}\'{i}k model (in particular, the screening charges considered in Sections \ref{sec666} and \ref{Sec:QuantLoc} are the quantum versions of the classical screening charges discussed in Section \ref{sec3899891} for the ${\rm SU}(2)$ Klim\v{c}\'{i}k model). The results of the present subsection and of~\cite{Semikhatov:2001zz} thus provide us with a free field realisation of the $\widehat{\mathfrak{sl}}(2)_{k_1}\oplus\widehat{\mathfrak{sl}}(2)_{k_2}/\widehat{\mathfrak{sl}}(2)_{k_1+k_2}$ coset $\Wc$-algebra in terms of the left-moving quantum fields describing the asymptotic domain of the quantum ${\rm SU}(2)$ chiral Klim\v{c}\'{i}k model. As usual, a similar realisation exists in terms of right-moving free fields.

\subsection{Quantum local IMs}
\label{Sec:QuantLocalIM}

Our goal in this subsection is to discuss local Integrals of Motion (IMs) in quantised AGMs and the application of these results to the AGM underlying the quantum UV fixed-point of the ${\rm SU}(2)$ Klim\v{c}\'{i}k model. For that, it will be useful to first revisit briefly the construction of such local IMs in the classical case.

\subsubsection{Revisiting local charges in classical AGMs}
\label{Sec:RevisitClassIM}

Let us consider a classical AGM with $N$ punctures, twist function $\vp(z)$ and Gaudin Lax matrix $\Gamma(z,x)$. To ease the comparison with the quantum case, we will use here the setup considered in Subsection \ref{Sec:QuantAGM}, where we have chosen a spectral parameter such that $z_N=\infty$, \textit{i.e.} one of the punctures is located a infinity. In this case, the classical observables are built from only $N-1$ Kac-Moody currents $\Jc_1,\ldots,\Jc_{N-1}$ and their gauge invariance is equivalent to their Poisson commutativity with the diagonal current $\Jc_{\diag}(x)$ defined in eq.\;\eqref{Eq:JDiag}.

The construction of local charges in involution in classical AGMs was described in~\cite{Lacroix:2017isl} and recalled around eq.\;\eqref{Eq:Qip}. In the present setup, it can be rephrased as follows. For every exponent $p\in \Eh$ of the affine Lie algebra $\gh$, we introduce a density
\begin{equation}\label{Eq:ClassicalS}
\Sc_{p+1}(z,x) = \tau_p^{a_1\ldots a_{p+1}}\, \Gamma_{a_1}(z,x)\cdots\Gamma_{a_{p+1}}(z,x),
\end{equation}
built by contracting a symmetric invariant $(p+1)$-tensor\footnote{In the notations of eq.\;\eqref{Eq:Qip}, this tensor is in one-to-one correspondence with the invariant polynomial $\Phi_p$ through $\Phi_p(X) = \tau_p^{a_1 \ldots a_{p+1}}X_{a_1} \ldots X_{a_{p+1}}$ for any $X=X_a\,{\tt t}^a \in \g^\C$.} $\tau_p$ on $\g^\C$ with the components of the Gaudin Lax matrix $\Gamma(z,x)$. The local charges are then defined as the quantities
\begin{equation}\label{Eq:Qip2}
\Q_{i,p} = - \frac{1}{\vp'(\ze_i)^{(p+1)/2}} \int \dd x\;\Sc_{p+1}(\ze_i,x)\, ,
\end{equation}
where $\ze_1,\ldots,\ze_{N-2} \in \C$ are the zeroes of the twist function $\vp(z)$. The main properties obeyed by the densities $\Sc_{p+1}(z,x)$ are the following:
\begin{itemize}
\item[(i)] the Poisson bracket $\bigl\lbrace \Jc_{\diag}(x), \Sc_{p+1}(z,y) \bigr\rbrace$ is proportional to $\vp(z)$ ;
\item[(ii)] the Poisson bracket of two densities take the form 
\begin{equation}\label{Eq:PbSS}
\bigl\lbrace \Sc_{p+1}(z,x), \Sc_{q+1}(w,y) \bigr\rbrace = \Ac^{(0)}_{p,q}(z,w\,;y)\,\delta(x-y) + \Ac^{(1)}_{p,q}(z,w\,;y)\,\delta'(x-y),
\end{equation}
where $\Ac^{(0)}_{p,q}(z,w\,;y)$ is the sum of a derivative with respect to $y$ and terms proportional to $\vp(z)$ and $\vp(w)$, with coefficients regular at $(z,w)=(\ze_i,\ze_j)$ for $i,j\in\lbrace 1,\ldots,N-2 \rbrace$.
\end{itemize}
These properties follow from the Poisson brackets
\begin{align}
&\bigl\lbrace \Jc_{\diag,a}(x), \Gamma_b(z,y) \bigr\rbrace = \f abc\; \Gamma_c(z,y) \,\delta(x-y) - \eta_{ab}\, \vp(z)\,\delta'(x-y) \, , \label{Eq:PbGamma}\\
&\bigl\lbrace \Gamma_a(z,x), \Gamma_b(w,y) \bigr\rbrace = -\f abc\, \frac{\Gamma_c(z,y)-\Gamma_c(w,y)}{z-w} \,\delta(x-y) + \eta_{ab}\, \frac{\vp(z)-\vp(w)}{z-w}\,\delta'(x-y) \notag
\end{align}
and require a specific choice of invariant tensors $\tau_p$ in the definition \eqref{Eq:ClassicalS} of $\Sc_{p+1}(z,x)$, see~\cite{Lacroix:2017isl}.

The property (i) ensures that the evaluation $\Sc_{p+1}(\ze_i,x)$ at a zero $\ze_i$ of $\vp(z)$ Poisson commutes with $\Jc_{\diag}$ and thus is gauge invariant. In other words, the density $\Sc_{p+1}(\ze_i,x)$ belongs to the classical $\Wc$-algebra underlying the AGM (formed by gauge invariant differential polynomials in the Kac-Moody currents $\Jc_1,\ldots,\Jc_{N-1}$). Similarly, the property (ii) ensures that the local charges $\Q_{i,p}$ defined in eq.\;\eqref{Eq:Qip2} are in involution. Indeed, integrating the bracket \eqref{Eq:PbSS} over $x$ and $y$ and evaluating it at $(z,w)=(\ze_i,\ze_j)$, we find that the Poisson bracket of $\Q_{i,p}$ and $\Q_{j,q}$ is proportional to the integral of $\Ac^{(0)}_{p,q}(\ze_i,\ze_j \,;y)$, which vanishes by the conditions obeyed by $\Ac^{(0)}_{p,q}(z,w \,;y)$ in point (ii).

\subsubsection{Local charges in quantised AGMs}
\label{Sec:ConjLocal}

\paragraph{Conjectured form of the quantum charges.} Let us now consider the quantised AGM, described by quantum Kac-Moody currents $\Jq_1,\ldots,\Jq_{N-1}$. We would like to construct the quantisation of the local charges $\Q_{i,p}$ discussed above, which would then form commuting operators built from local differential polynomials in these currents. This question was studied in~\cite{Feigin:2007mr,Lacroix:2018fhf,Lacroix:2018itd}. The approach followed in these references was inspired by the conjecture, first formulated in~\cite{Feigin:2007mr}, that the spectrum of these charges is encoded in certain objects called affine opers. Studying these affine opers and their properties, it was conjectured in~\cite{Lacroix:2018fhf} that the quantum local charges take the following form\footnote{This conjecture can also be motivated by the so-called ODE/IQFT correspondence. We will come back briefly to this point in Subsection \ref{Sec:ODE/IQFT}.}:
\begin{equation}\label{Eq:ConjQ}
\Qq_{\gamma,p} = \int  \Wq_{\gamma,p+1}(x) \, \dd x \,, \qquad \Wq_{\gamma,p+1}(x) = \oint_\gamma \Pc(z)^{-p/\hv} \,\Sq_{p+1}(z,x) \, \dd z \, .
\end{equation}
Let us explain the different ingredients entering this formula. The quantity $\Pc(z)$ is defined (up to a global factor) from the quantum twist function $\vpq(z)$ by the condition\footnote{Let us briefly comment on this definition of $\Pc(z)$. As we have mentioned in Subsection \ref{Sec:QuantAGM}, the transformation rule \eqref{Eq:ChangeZQuant} of the quantum twist function $\vpq(z)$ under a change of spectral parameter means that $\p_z - \vpq(z)$ is a connection on the line bundle $\Omega^{\,\hv}$. The condition \eqref{Eq:dlogP} can then be interpreted as defining $\Pc(z)$ as a (multivalued) section of the bundle $\Omega^{\,\hv}$ which is flat with respect to this connection. In particular, this means that the object $\Pc(z)\,(\dd z)^{\hv}$ behaves as an order $\hv$ differential with respect to the spectral parameter, \textit{i.e.} that $\Pc(z)^{1/\hv}\dd z$ is a 1-form.\label{FootnoteP}}
\begin{equation}\label{Eq:dlogP}
\p_z \log \Pc(z) = \vpq(z).
\end{equation}
Explicitly, in terms of the punctures $z_r$ and levels $k_r$, this function is then given by
\begin{equation}\label{Eq:P}
\Pc(z) = \prod_{r=1}^{N-1} (z-z_r)^{k_r},
\end{equation}
for a specific choice of the overall constant undetermined by eq.\;\eqref{Eq:dlogP}. In particular, since the levels $k_r$ are not necessarily integers, the function $\Pc(z)$ is in general multi-valued.

The integration contour $\gamma$ in eq.\;\eqref{Eq:ConjQ} is a so-called Pochhammer contour. It is a closed path in $\mathbb{CP}^1$ whose main characteristic is that the function $\Pc(z)$ possesses a single-valued branch along $\gamma$, ensuring that the object $\Pc(z)^{-p/\hv}$ appearing in the integral \eqref{Eq:ConjQ} is unambiguously defined along the integration contour. For generic values of the levels $k_r$, the set $P$ of equivalence classes of independent Pochhammer contours $\gamma$ (up to smooth deformations which do not change contour integrations over $\gamma$) is of size $N-2$.

Finally, the quantity $\Sq_{p+1}(z,x)$ appearing in eq.\;\eqref{Eq:ConjQ} is a local spin $p+1$ density built as a differential polynomial in the Gaudin Lax matrix components $\Gq_a(z,x)$, their derivatives with respect to $z$ and the quantum twist function. More precisely, this density is expected to take the form
\begin{equation}\label{Eq:QuantumS}
\Sq_{p+1}(z,x) = \tau_p^{a_1\ldots a_{p+1}}\, \nor{\Gq_{a_1}(z,x)\cdots\Gq_{a_{p+1}}(z,x)} + \, \dots \; ,
\end{equation}
where $\tau_p$ is the invariant tensor discussed in Subsection \ref{Sec:RevisitClassIM} and the dots represent additional ``corrections'' of the form
\begin{equation}\label{Eq:Corr}
\p_z^{\alpha_1} \vpq(z) \cdots \p_z^{\alpha_n} \vpq(z) \; \nor{ \p_z^{\beta_1} \p_x^{\gamma_1} \Gq_{a_1}(z,x) \cdots \p_z^{\beta_m} \p_x^{\gamma_m} \Gq_{a_m}(z,x) }\,,
\end{equation}
where $m,n,\alpha_i,\beta_i,\gamma_i\in\mathbb{Z}_{\geq 0}$ are non-negative integers such that $m+n\leq p$ and $m+\sum_{k=1}^m \gamma_k=p+1$ (so that these corrections are of spin $p+1$).

To justify and consolidate this conjecture, we need to discuss two main points, following~\cite{Lacroix:2018fhf,Lacroix:2018itd}. The first one is the classical limit of the charges $\Qq_{\gamma,p}$ and its relation to the classical ones $\Q_{i,p}$ described in Subsection \ref{Sec:RevisitClassIM}. The second one is the gauge invariance and commutativity of these quantum charges and the corresponding requirements on the densities $\Sq_{p+1}(z,x)$.

\paragraph{Classical limit.} We start with the discussion of the classical limit. Recall from eq.\;\eqref{Eq:ClassicalLimGaudin} that the quantum Gaudin Lax matrix $\Gq(z,x)$ and twist function $\vpq(z)$ are of order $O(\hbar^{-1})$ when $\hbar \to 0$. The first term in the density \eqref{Eq:QuantumS} is thus of order $O(\hbar^{-p-1})$. Similarly, the corrective term \eqref{Eq:Corr} is of order $O(\hbar^{-m-n})$. Since we supposed $m+n \leq p$ earlier, we thus see that this term is subdominant in the classical limit $\hbar \to 0$, allowing us to interpret the dots in eq.\;\eqref{Eq:QuantumS} as representing quantum corrections. Using the explicit asymptotic \eqref{Eq:ClassicalLimGaudin} of $\Gq(z,x)$ in the limit $\hbar \to 0$, we then find
\begin{equation}
\Sq_{p+1}(z,x) = \left(\frac{2\pi\ri}{\hbar}\right)^{p+1}\,\Bigl( \Sc_{p+1}(z,x) + O(\hbar) \Bigr),
\end{equation}
where $\Sc_{p+1}(z,x)$ is the classical density defined in eq.\;\eqref{Eq:ClassicalS}.

This is suggestive of a relation between the classical limit of the quantum charges $\Qq_{\gamma,p}$ and the classical charges $\Q_{i,p}$. To establish this more precisely, we need to understand the classical limit of the Pochhammer integral in eq.\;\eqref{Eq:ConjQ}. Using the expression \eqref{Eq:P} of $\Pc(z)$ and the classical limit \eqref{Eq:ClassicalLim} of the levels $k_r$, we find that the term $\Pc(z)^{-p/\hv}$ in this integral obeys the following asymptotic behaviour:
\begin{equation}
\Pc(z)^{-p/\hv} = \exp\left( \frac{2\pi p}{\hv\hbar} \Bigl( \rho(z) + O(\hbar) \Bigr) \right), \qquad \text{ where } \qquad \rho(z) = \sum_{r=1}^{N-1} \ell_r\,\log(z-z_r)
\end{equation}
satisfies $\p_z \rho(z) = \vp(z)$. The limit $\hbar\to 0$ of the integral \eqref{Eq:ConjQ} can then be treated using the well-known saddle-points method (see \textit{e.g.}~\cite{Reshetikhin:1994qw} for a reference on saddle-points in the context of Pochhammer integrals). In particular, this method implies that the integral localises at the extrema of the function $\rho(z)$, which are nothing but the zeroes $\ze_i$ of the classical twist function $\vp(z)=\p_z \rho(z)$. Thus, with appropriate rescalings by powers of $\hbar$ and prefactors, the classical limit of the quantum charge $\Qq_{\gamma,p}$ is expressed as a linear combination of the classical local charges $\Q_{i,p}$ defined in eq.\;\eqref{Eq:Qip2}. We note moreover that the number of Pochhammer contours $\gamma\in P$ coincides with the number $N-2$ of zeroes of $\vp(z)$, so that the overall counting of local charges agrees at the classical and quantum levels.

\paragraph{Intermezzo: twisted derivatives.} In order to discuss the gauge invariance and commutativity of the quantum charges, we will need to introduce an additional key ingredient, the so-called twisted derivative $D_{z,p}$. If $f(z)$ is a meromorphic function of $z$, we define its twisted derivative of degree $p$ as
\begin{equation}\label{Eq:TwistedDer}
D_{z,p} f(z) = \p_z f(z) - \frac{p}{\hv} \vpq(z)  f(z).
\end{equation}
The fundamental property of this derivative is the following. For any Pochhammer countour $\gamma\in P$ and any meromorphic function $f(z)$ regular along $\gamma$, we have
\begin{equation}\label{Eq:IntTwisted}
\oint_\gamma \Pc(z)^{-p/\hv} D_{z,p} f(z) \, \dd z = 0.
\end{equation}
The key observation to prove this fact is that $\Pc(z)^{-p/\hv} D_{z,p} f(z) = \p_z \bigl( \Pc(z)^{-p/\hv} f(z) \bigr)$, as one easily checks using $\p_z\Pc(z)=\vpq(z)\Pc(z)$. The integrand in the above equation is thus a total derivative. The facts that $\Pc(z)$ is single-valued on the Pochhammer countour $\gamma$ and that the latter is a closed path then ensure that the integral vanishes, as claimed.

\paragraph{Properties of the densities.} Recall that the densities $\Sc_{p+1}(z,x)$ of the classical local charges satisfy the points (i) and (ii) discussed in Subsection \ref{Sec:RevisitClassIM}. These conditions ensured the gauge invariance and involution of the charges and could be derived from the fundamental Poisson brackets \eqref{Eq:PbGamma} obeyed by the Gaudin Lax matrix. In the quantum case, the Poisson bracket of the theory is replaced by the OPE \eqref{Eq:OpeJ} of the quantum Kac-Moody currents $\Jq_r$. One checks that this OPE implies the following quantum equivalents of the brackets \eqref{Eq:PbGamma}:
\begin{align}
& \Jq_{\diag,a}(x)  \Gq_b(z,y) \ = \frac{\ri \f abc}{x-y}\; \Gq_c(z,y) + \frac{\eta_{ab}}{(x-y)^2}\, \vpq(z) + \reg \, , \label{Eq:OpeGamma}\\
&\Gq_a(z,x)\Gq_b(w,y) = - \frac{\ri\,\f abc}{x-y} \frac{\Gq_c(z,y)-\Gq_c(w,y)}{z-w} - \frac{\eta_{ab}}{(x-y)^2} \frac{\vpq(z)-\vpq(w)}{z-w} + \reg\,. \notag
\end{align}
The conditions that we impose on the quantum densities $\Sq_{p+1}(z,x)$ are then as follows. Their definition \eqref{Eq:QuantumS} in terms of $\Gq(z,x)$ should be such that, using the OPEs \eqref{Eq:OpeGamma}:
\begin{itemize}
\item[(i)] the singular part of the OPE $\Jq_{\diag}(x) \Sq_{p+1}(z,y)$ is a twisted derivative $D_{z,p}(\dots)$ ;
\item[(ii)] the OPE of any two densities take the form 
\begin{equation}\label{Eq:OpeSS}
\Sq_{p+1}(z,x) \Sq_{q+1}(w,y) = \sum_{k\in\,\mathbb{Z}_{\geq 0}} \frac{\Aq^{(k)}_{p,q}(z,w\,;y)}{(x-y)^{k+1}} + \reg \, ,
\end{equation}
where $\Aq^{(0)}_{p,q}(z,w\,;y)$ is a linear combination $\p_y(\dots) + D_{z,p}(\dots)+D_{w,q}(\dots)$ of a spatial derivative and twisted derivatives\footnote{Technically, one also needs to impose that the quantities of which we take the spatial or twisted derivatives are regular when $z=w$.}.
\end{itemize}

Combined with the fundamental property \eqref{Eq:IntTwisted} of the twisted derivative $D_{z,p}$, the condition (i) implies that the quantity $\Wq_{\gamma,p+1}$, obtained as the integral \eqref{Eq:ConjQ} of $\Pc(z)^{-p/\hv}\Sq_{p+1}(z)$ over $\gamma\in P$, has a regular OPE with the diagonal current $\Jq_{\diag}$. This ensures that $\Wq_{\gamma,p+1}(x)$ is a gauge-invariant local density, or in other words an element of the quantum coset $\Wc$-algebra $\Wh$.

Similarly, the condition (ii) implies that, for any two exponents $p,q\in \Eh$ and any two Pochhammer contours $\gamma,\gamma'\in P$, the OPE $\Wq_{\gamma,p+1}(x)\Wq_{\gamma',q+1}(y)$ has a simple pole $1/(x-y)$ whose coefficient is a total spatial derivative. This ensures that the local charges $\Qq_{\gamma,p}$ and $\Qq_{\gamma',q}$, defined as spatial integrals of $\Wq_{\gamma,p+1}(x)$ and $\Wq_{\gamma',q+1}(y)$, commute as quantum operators, as wanted.\\

The conditions (i) and (ii) discussed here are the quantum equivalent of the ones discussed in Subsection \ref{Sec:RevisitClassIM}, in the sense that they ensure the gauge invariance and commutation of the quantum local charges. This can be made more concrete. Recall that in the classical limit, the quantum twist function $\vpq(z)$ is of order $O(\hbar^{-1})$. The derivative term in the twisted derivative \eqref{Eq:TwistedDer} is thus subdominant compared to the other term. Hence, a twisted derivative $D_{z,p}f(z)$ contributes in the classical limit to a term proportional to $\hbar^{-1}\vp(z)f(z)$. The quantum conditions (i) and (ii) then reduce to the classical ones in the limit $\hbar \to 0$, as expected, and the derivative terms in the twisted derivatives can be interpreted as quantum corrections.

\paragraph{First densities and checks of the conjectures.} Let us finally discuss some first checks of the conjectures discussed in this subsection on the existence of densities $\Sq_{p+1}(z,x)$ defining commuting quantum local charges, by exhibiting the simplest examples of such densities.

The first exponent of an affine algebra is always $p=1$, corresponding to a spin 2 density $\Sq_2(z,x)$. The latter simply reads
\begin{equation}\label{Eq:S2}
\Sq_2(z,x) = \frac{\eta^{ab}}{2} \nor{ \Gq_a(z,x)\Gq_b(z,x)}.
\end{equation}
One easily checks by a direct computation that it satisfies the conditions (i) and (ii). We note that the above expression for $\Sq_2(z,x)$ coincides with the classical density $\Sc_2(z,x)$, up to the normal ordering of the fields: this density thus does not receive quantum corrections. We note however that the local charges defined from it in fact receive such corrections, since they are obtained by integrating $\Sq_2(z,x)$ over Pochhammer contours, rather than by a simple evaluation at a zero $\ze_i$ of $\vp(z)$
as  in the classical case (we will see an explicit example of this phenomenon in the following subsection).\\

For higher-rank affine algebras of type $A$, \textit{i.e.} $\widehat{\mathfrak{sl}}(N)$ for $N \geq 3$, the next exponent is $p=2$. The corresponding spin 3 density $\Sq_3(z,x)$ has been constructed explicitly in~\cite{Lacroix:2018itd} and indeed obeys the conditions (i) and (ii) stated above. We will not need its  expression here.

In this paper, most of the examples that we explore are based on the algebra $\widehat{\mathfrak{sl}}(2)$. For this case, there are no cubic charges and the first higher-degree charge is quartic, built from a spin 4 density $\Sq_4(z,x)$. This charge was not treated explicitly yet in the literature and thus required some additional analysis. The end result of this computation being a bit involved, we give the explicit expression of $\Sq_4(z,x)$ in terms of the Gaudin Lax matrix and the twist function in Appendix \ref{App:S4}, eq.\;\eqref{Eq:QuantS4}.

\subsubsection{Quantum local IMs in the AGM underlying the chiral \texorpdfstring{$\bm{{\rm SU}(2)}$}{SU(2)} Klim\v{c}\'{i}k model}

\paragraph{The quantised AGM.} Let us finally apply the results of this subsection to the AGM underlying the left-moving half of the UV fixed-point of the Klim\v{c}\'{i}k model. At the classical level, this AGM was described in detail in Section \ref{sec555389i}: in particular it possesses 3 punctures. Recall that in this section, we always started the analysis of the quantum model by performing a change of spectral parameter that sends the last puncture to infinity, to make the treatment of the constraint easier. In the present case, where there are two finite punctures left, one can always use translation and dilation of the spectral parameter to send these points to $0$ and $1$. The quantum twist function of the AGM under consideration then takes the simple form
\begin{equation}
\vpq(z) = \frac{k_1}{z} + \frac{k_2}{z-1},
\end{equation}
where $k_1$ and $k_2$ are the quantum levels associated with the punctures $0$ and $1$ respectively, while the level associated with the third puncture at infinity is equal to $k_3=-k_1-k_2-2\hv$, according to eq.\;\eqref{Eq:SumK}. These levels are related,  by the asymptotic property \eqref{Eq:ClassicalLim}, to the classical ones, which we recall are given by eq.\;\eqref{Eq:LevelsL} for the case at hand. In particular, the quantum levels are then such that
\begin{equation}\label{Eq:ClassNu}
\lim_{\hbar\to\, 0}\; \frac{k_1}{k_2} = \nu^2,
\end{equation}
where $\nu$ is the parameter entering the definition of the classical chiral Klim\v{c}\'{i}k model, as in Section \ref{sec31} (this is the main parameter that is needed to relate the classical and quantum physics of the model: indeed, the remaining parameter entering the classical limit of the quantum levels $k_r$ is the coefficient $\Kuv$, which appears as an irrelevant overall prefactor in the classical model).

Similarly, the quantum Gaudin Lax matrix of the AGM is defined in the present case in terms of two Kac-Moody currents $\Jq_1$ and $\Jq_2$ and simply reads
\begin{equation}\label{Eq:GammaKlim}
\Gq(z,x) = \frac{\Jq_1}{z} + \frac{\Jq_2}{z-1}.
\end{equation}

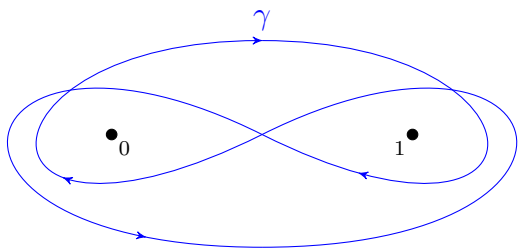
\begin{figure}[h]
\begin{center}
\begin{tikzpicture}[scale=1]
\filldraw (0,0) node [below right=-.5mm]{\scriptsize $0$} circle (2pt);
\filldraw (4,0) node [below left=-.5mm]{\scriptsize $1$} circle (2pt);
\draw[blue, -stealth', postaction={decorate,decoration={markings,mark=at position .4 with {\arrow{stealth'}}}}] (2,0) .. controls (-2,-2) and (-2,1.25) .. (2,1.25) node[above]{$\gamma$};
\draw[blue, postaction={decorate,decoration={markings,mark=at position .8 with {\arrow{stealth'}}}}] (2,1.25) .. controls (6,1.25) and (6,-2) .. (2,0);
\draw[blue, postaction={decorate,decoration={markings,mark=at position .8 with {\arrow{stealth'}}}}] (2,0) .. controls (-2,2) and (-3,-1.5) .. (2,-1.5);
\draw[blue] (2,-1.5) .. controls (7,-1.5) and (6,2) .. (2,0);
\end{tikzpicture}
\caption{Pochhammer contour for two punctures $0$, $1$ in the $z$-plane and a third one at $\infty$.}\vspace{-20pt}
\label{Fig:Poch}
\end{center}
\end{figure}

\paragraph{Function $\bm{\Pc(z)}$ and Pochhammer contour.} Applying eq.\;\eqref{Eq:P}, we find that in the AGM under consideration here, the function $\Pc(z)$ is simply given by
\begin{equation}\label{Eq:PKlim}
\Pc(z) = z^{k_1} (z-1)^{k_2}.
\end{equation}
Of interest for the construction of local charges are the Pochhammer contours, which are defined as the closed paths in $\mathbb{CP}^1$ along which $\Pc(z)$ admits a single-valued branch. In the present case, there exists a unique such contour, represented in Figure \ref{Fig:Poch}, which goes around each puncture twice, in opposite directions (ensuring that any non-trivial phase acquired by $\Pc(z)$ after a revolution around a puncture is cancelled by the reverse one).

\paragraph{Quadratic charge.} We can now describe the quantum local IMs for the AGM under consideration. The first charge is the quadratic one, whose density is obtained by the integral
\begin{equation}\label{Eq:W2Klim}
\Wq_{\gamma,2}(x) = \oint_\gamma\,\Pc(z)^{-1/\hv}\,\Sq_2(z,x) \, \dd z\,,
\end{equation}
where $\gamma$ is the Pochhammer contour described above and $\Sq_2(z,x)$ is defined in terms of $\Gq(z,x)$ by eq.\;\eqref{Eq:S2}. This integral was computed explicitly in~\cite{Lacroix:2018fhf}. We recall the details of this calculation in Appendix \ref{App:3punct} and simply summarise the main steps here. Reinserting the expressions \eqref{Eq:GammaKlim} and \eqref{Eq:PKlim} of $\Gq(z,x)$ and $\Pc(z)$ in the above formula, one reduces the computation to simple integrals of the form
\begin{equation}
B(a,b) = \oint_\gamma z^a (z-1)^b\,\dd z,
\end{equation}
for some real numbers $a$ and $b$. These integrals can be expressed in terms of the Euler Beta function and satisfy a number of useful recursive identities (with respect to $a$ and $b$). Using those, we simply find that
\begin{equation}
\Wq_{\gamma,2}(x) = N_2 \, \Wq_2(x),
\end{equation}
where $\Wq_2$ is the spin 2 W-current defined in eq.\;\eqref{Eq:QuantumW2} and $N_2$ is a constant given by
\begin{equation}
N_2 = \frac{(k_1+k_2)(k_1+k_2+\hv)(k_1+k_2-\hv)}{k_1k_2} \oint_\gamma \Pc(z)^{-1/\hv}\, \dd z.
\end{equation}
In particular, the terms of the type $1/(k_r+\hv)$ in the expression \eqref{Eq:QuantumW2} of $\Wq_2$ are obtained in the present case by integrals over Pochhamer contours. In the classical limit, these terms simply become $1/\ell_r$ and arise from an evaluation at the zero of the twist function. This illustrates the type of quantum corrections that one can get when promoting evaluations at zeros to integrations over Pochhammer contours in the quantum model.
 
As expected from the general discussion in Subsection \ref{Sec:ConjLocal}, the density $\Wq_{\gamma,2}$ belongs to the $\gh_{k_1}\oplus\gh_{k_2}/\gh_{k_1+k_2}$ coset $\Wc$-algebra underlying the AGM. In fact, since there is a unique spin 2 current in this algebra, namely the energy-momentum tensor $\Wq_2$, we could have expected the density of the quadratic charge to be simply proportional to $\Wq_2$.

\paragraph{Quartic charge for $\bm{\mathfrak{sl}(2)}$.} To discuss a less trivial case, let us consider the first higher-order charge for $\g^\C=\mathfrak{sl}(2)$. As explained in Subsection \ref{Sec:ConjLocal}, this charge is quartic and is built from the spin 4 field $\Sq_4(z,x)$ constructed in Appendix \ref{App:S4}, eq.\;\eqref{Eq:QuantS4}. More precisely, the density of this quartic charge is given by
\begin{equation}\label{Eq:W4Int}
\Wq_{\gamma,4}(x) = \oint_\gamma\,\Pc(z)^{-3/2}\,\Sq_4(z,x) \, \dd z\,,
\end{equation}
where we have used that $\hv=2$ for $\mathfrak{sl}(2)$. By construction, this density should then belong to the $\widehat{\mathfrak{sl}}(2)_{k_1}\oplus\widehat{\mathfrak{sl}}(2)_{k_2}/\widehat{\mathfrak{sl}}(2)_{k_1+k_2}$ coset $\Wc$-algebra, \textit{i.e.} the corner-brane algebra. The latter has been discussed extensively in Subsections \ref{Sec:CB} and \ref{Sec:QuantW}. In particular, it possesses three spin 4 fields, the descendants $\nor{\Wq_2^2}$ and $\p^2\Wq_2$ and an additional primary field $\Wq_{4,P}$. For the case of the quartic charge, contrarily to the quadratic one, the density $\Wq_{\gamma,4}$ is thus not fixed solely by its belonging to the $\Wc$-algebra and therefore corresponds to a non-trivial linear combination of $\nor{\Wq_2^2}$ and $\Wq_{4,P}$ (the total derivative $\p^2\Wq_2$ is irrelevant for the present discussion since it does not contribute to the local charge, defined as an integral over the space coordinate).

The integral \eqref{Eq:W4Int} can be performed explicitly using the same type of techniques as for the quadratic case, in particular using the Beta function identities. We then obtain a rather lengthy expression in terms of normal ordered products of components of the currents $\Jq_1$ and $\Jq_2$. As expected form the above discussion, we can rewrite this expresion in terms of the W-currents $\Wq_{4,P}$, $\nor{\Wq_2^2}$ and $\p^2\Wq_2$, finally yielding
\begin{equation}\label{Eq:W4Gaudin}
\Wq_{\gamma,4}(x) = N_4\left( \Wq_{4,P}(x) + \delta_1 \nor{\Wq_2(x)^2} + \delta'_2 \, \p_x^2\Wq_2(x) \right).
\end{equation}
Here the coefficients $N_4$ and $\delta_1$ are defined as
\begin{equation*}
N_4 = -\frac{(k_1 + k_2) (k_1 + k_2 + 2) (3 k_1 + 3 k_2 -2) (3 k_1 + 3 k_2 + 2) (3 k_1 + 3 k_2 + 4)}{4 k_1 (3 k_1 + 2) (3 k_1 + 4) k_2 (3 k_2 + 2) (3 k_2 + 4)} \oint_\gamma\,\Pc(z)^{-3/2} \, \dd z \,,
\end{equation*}
\begin{equation}\label{Eq:Delta1}
\delta_1 = \frac{15 (k_1 + 2) (3 k_1 + 4) (k_2 + 2) (3 k_2 + 4) (k_1 + k_2 + 2)  (3 k_1 + 3 k_2 + 8)}{\beta}\,,
\end{equation}
with $\beta$ given in eq.\;\eqref{Eq:Beta}. The coefficient $\delta'_2$ also possesses an explicit expression in terms of $k_1$ and $k_2$, which however we shall not need here since it multiplies a total derivative, irrelevant for the construction of local charges.\\

This result is to be compared with the density of the local charge computed in Subsection \ref{Sec:CB} using screening charges, in the free-field realisation of the corner-brane $\Wc$-algebra. 
According to eq.\;\eqref{Eq:W4CB},  this density is given by $\mathsf{W}_4^{({\rm PB})}$, which itself
  can be  expressed
in terms of the fields $\Wq_{4,P}$ and $\Wq_2$ and the parameters $n$ and $\varpi$ entering the screening charges
by means of formula \eqref{xcmnhjasnbc}. Recall that in the identification of the corner-brane $\Wc$-algebra with the $\widehat{\mathfrak{sl}}(2)_{k_1}\oplus\widehat{\mathfrak{sl}}(2)_{k_2}/\widehat{\mathfrak{sl}}(2)_{k_1+k_2}$ coset $\Wc$-algebra, these parameters are related to the levels $k_1$ and $k_2$ by eq.\;\eqref{Eq:ParamW}. Using this relation, one easily checks that the two expressions \eqref{Eq:Delta1CB} and \eqref{Eq:Delta1} of $\delta_1$ agree and thus that the density $\Wq_{\gamma,4}$ in eq.\;\eqref{Eq:W4Gaudin} coincides with the one in eq.\;\eqref{Eq:W4CB}, up to a global factor and a total derivative. This thus provides a non-trivial check of the identification of the quantum integrable structures defined from screening charges and from the AGM construction.\footnote{Let us mention that this integrable structure also appeared recently in~\cite{Feigin:2018tor}, where it is studied using the framework of toroidal algebras. Remarkably, it was pointed out in \cite{Feigin:2017tor} that the spectrum of some integrable systems arising from toroidal algebras can be described by Bethe ansatz equations of affine Gaudin type. It would be interesting to explore the potential relations between our results and this framework.}\\

Let us finally comment briefly on the classical limit of the quartic charge. Using the asymptotic properties \eqref{Eq:ClassicalLim}, \eqref{Eq:ClassLimW2} and \eqref{Eq:ClassLimW4} of $k_r$, $\Wq_2$ and $\Wq_4$, one checks that the dominant term in the classical limit of the density \eqref{Eq:W4Gaudin} is proportional to $W_2^2$. We thus recover the fact that, in the classical model, the quartic charge is simply the integral of the energy-momentum tensor squared -- see Sections \ref{sec3899891} and \ref{aaaaaaaaaaaa}. We moreover see that at the quantum level, this is no longer true as the spin 4 density becomes independent of $\Wq_2$, making the quantum integrable structure quite more complicated than the classical one.

\section{Summary and perspectives}\label{Sec:Discussion}

This section starts by offering a brief summary of 
the resulting picture in Subsection \ref{summary}, followed 
by a discussion of some of the emerging perspectives.

\bigskip

In Subsection \ref{genAGM} we point out how the conformal limits
considered here can be
 generalized to the cases with an arbitrary number of punctures.
This indicates that the analysis performed in this work can be generalised
to more general multiparametric NLSMs. 

A key motivation for our work was the perspective to apply the 
ODE/IQFT correspondence to  quantised NLSMs. 
The existence of the 
generalised conformal limits 
suggests that the ODE/IQFT correspondence can be applied to 
such larger classes of models. These perspectives
are discussed in  Subsection \ref{Sec:ODE/IQFT}. 

Another natural direction to explore concerns the study of quantum non-local charges in these models by means of the representation theory of Yang-Baxter algebras. This is discussed in Subsection \ref{YB-persp}.

Some of the most important open questions concern possible existence of 
deformations away from the conformal limits preserving key aspects of the integrable
structures like the ODE/IQFT correspondence and the Yang-Baxter algebra. 
We will briefly discuss some perspectives 
in this direction in Subsection \ref{mass-Def}.

Finally we end in Subsection \ref{Sec:Lambda}
by pointing out the relations between the AGMs studied here to  the 
gauged WZW and $\lambda$-models,
which are related to the  Klim\v{c}\'{i}k model via Poisson-Lie T-duality, 
indicating that the techniques described in this paper can be 
useful for the study of these models as well. 

\subsection{Summary}\label{summary}

The goal of our work has been to make first steps towards the quantisation 
of relativistic AGMs. To this aim we have considered a locus in the parameter
space  expected to admit a conformally 
invariant quantisation. The existence of an extended algebra of conformal symmetry
then provides a basis for the investigation of the quantum integrable structures of these theories.  
These integrable structures 
can be expected to admit a deformation 
away from the conformal locus.

\bigskip

In a first step we considered the
conformal limit of the ${\rm SU}(2)$ Klim\v{c}\'{i}k  model.
This corresponds to sending the parameters of the model to their values at the UV fixed point, for which
the one-loop beta function vanishes.
For 
 the AGM describing the  relativistic field theory, two 
different limiting procedures were introduced. These yielded
two decoupled affine Gaudin models encoding the left-moving and right-moving degrees of freedom of the field theory
at the RG fixed point. 
As a result, we were led to introduce the notion of a ``chiral'' AGM, which should be
clearly distinguished from the relativistic AGM we started from, 
see Section \ref{sec555389i} for a discussion. 

\bigskip

A characteristic feature of 
the chiral AGM is the existence of an infinite number of fields, built
as local combinations of the matrix elements of the Lax connection, which are left/right
moving as a consequence of the classical equations of motion. These form 
a closed (Poisson) algebra which is the  algebra of extended conformal 
 symmetry of the model. For the Klim\v{c}\'{i}k  
NLSM in the conformal limit, we obtained the
 $\widehat\g_{\ell_1} \oplus \widehat\g_{\ell_2}\big/\widehat\g_{\ell_1+\ell_2}$
coset ${\cal W}$-algebra in this way. 
As an illustration, 
we expressed the first few generating currents in terms of the fundamental fields
of the theory, both at the classical and quantum level, in the special case with $G={\rm SU}(2)$.

The formulation as AGM offers a powerful framework for the 
construction of quantum local IMs as integrals over densities formed out
of the generators of the coset ${\cal W}$-algebra \cite{Lacroix:2018fhf,Lacroix:2018itd}.
By applying this framework we  computed 
the first non-trivial local IM. Its expression turned out to be 
consistent with the known results for the Fateev model in the literature, that
were obtained by the method of screening charges.

\bigskip

A gauge-fixed description of the chiral AGM has been
found to be useful for the study of non-local integrals of motion.  
Indeed, it has been shown in Section \ref{sec555389i} that a gauge-fixing 
exists relating the Lax matrix of the AGM to the Lax matrix  studied in 
\cite{Bazhanov:2018xzh} describing the integrable
structure of the conformal limit of the Klim\v{c}\'{i}k model. This should allow us to  apply the
main results and conjectures of \cite{Bazhanov:2018xzh} 
to the quantisation of the non-local IMs in the AGMs  studied here. 
Of particular importance is the conjecture that there exists a quantisation 
of the monodromy matrix constructed from the gauge-fixed Lax matrix 
which satisfies commutation relations of Yang-Baxter type. 
A possible generalization of this was pointed out in Subsection \ref{cxnbbsdbv3},
which would be relevant for the study of chiral AGMs  associated to Lie 
algebras of higher ranks.

\subsection{More general chiral limits of affine Gaudin models} \label{genAGM}

It seems likely to us that much more general affine Gaudin models 
will admit chiral limits similar to the one discussed in this paper. In order to 
substantiate this claim, let us collect some relevant points, referring to Appendix \ref{App:ChiralLim} for more details. To begin with, let us observe that the definition of relativistic AGMs described in Subsection \ref{Sec:Relat} was based on a split of the zeroes of the twist function into 
two subsets of equal cardinality, corresponding to coefficients $s_i$ equal to respectively $+1$ and $-1$ in the definition \eqref{Eq:HamReal} of the Hamiltonian. Here, we will consider a natural class of relativistic AGMs generalising the Klim\v{c}\'{i}k model, which are characterised by a set of zeroes $\{\zeta_1^+,\dots,\zeta_{M}^+\}\sqcup\{\zeta_1^-,\dots,\zeta_{M}^-\}$, where the labels $\pm$ refer to the aforementioned split, and a set of poles $\{z_1^+,\dots,z_{M+1}^+\}\sqcup\{z_1^-,\dots,z_{M+1}^-\}$ also divided into two subsets (in particular, these AGMs possess $2M+2$ punctures). We further use the freedom of performing M\"obius transformations to fix $\ze_M^+=\infty$ and $\ze_M^-=0$. In this setup, the Klim\v{c}\'{i}k model corresponds to the simplest case $M=1$.

Introducing a scaling parameter $\xi$ and representing 
the poles $z_r^\pm$ and zeroes $\ze_i^\pm$ as
$z_r^\pm =\bigl( z^{\cLR}_r / \xi \bigr)^{\pm 1}$
and $\ze_i^\pm =\bigl( \ze^{\cLR}_i / \xi \bigr)^{\pm 1}$, with fixed $z_r^\cLR$ and $\ze_i^\cLR$, define a one-parameter flow in the parameter space of this class of AGMs. Letting $\xi$ tend to $0$ allows us to ensure that the poles $z_r^+$ and zeroes $\zeta_i^+$ will move to infinity, whereas 
the poles $z_r^-$ and zeroes $\zeta_i^-$ will approach zero. The detailed analysis of this limit is treated in Appendix \ref{App:ChiralLim}.
In particular, it may then be shown that the limit of the Lax connection of the theory, keeping fixed a rescaled spectral parameter $z^\cL=\xi\, z$, satisfies
\begin{equation*}
\lim_{\xi\rightarrow\, 0}\,\Lc_+\bigl(z^\cL/\xi\bigr)=\Bc^{\cL}_+ + \Kc_{M}^\cL\,z^\cL + \sum_{i=1}^{M-1} \frac{\Kc_{i}^\cL}{z^\cL-\ze_i^\cL} \qquad 
\text{ and } \qquad \lim_{\xi\rightarrow\, 0} \, \Lc_-\bigl(z^\cL/\xi \bigr) = \Bc^{\cL}_-\,,
\end{equation*}
for some fields $\Bc^\cL_\pm$ and $\Kc_i^\cL$ (independent of $z^\cL$). Working in a gauge where $\Bc^{\cL}_-=0$, one then finds 
that $\mathcal{L}_+(z^\cL/\xi)$ is a left-moving matrix in the limit $\xi\rightarrow 0$, as a consequence of the zero curvature equation. In a similar way, one shows that $\Lc_-(\xi/z^\cR)$ is gauge-equivalent to a right-moving matrix when $\xi\rightarrow 0$. We conclude that chiral limits similar to the one
studied in this paper for the Klim\v{c}\'{i}k model will exist for all AGMs that admit a scaling action of the type 
considered here. Moreover, one can argue (see Appendix \ref{App:ChiralLim} for details) that the left-moving and right-moving Lax matrices obtained through the above procedure arise from chiral realisations of two AGMs, with $M+2$ punctures each. 

\bigskip

In the case of the Klim\v{c}\'{i}k model, corresponding to $M=1$, it has been observed in Subsection \ref{Sec:UVLimAGM} that the 
RG flow of the parameters $z_r^\pm$ is asymptotically equivalent in the UV to the limit $\xi\to 0$ considered above. Based on conjectures and results on the RG flow of general relativistic AGMs~\cite{Delduc:2020vxy,Hassler:2020xyj}, there exists numerical evidence for the next simplest case $M=2$ indicating that a similar
behaviour can be found in this case as well (see Appendix \ref{App:ChiralLim}). We see this as encouraging indications that many aspects of the chiral limits discussed in this paper in the case of the 
Klim\v{c}\'{i}k model have fairly natural generalisations to 
AGMs associated with more complicated twist functions and arbitrary Lie algebras $\mathfrak{g}$. We will 
briefly discuss some of the perspectives for the study
of these theories in the next subsections.

\bigskip

Before that, in view of generalisations to more general classes of models, let us note that we have only focused here on theories which possess gauge symmetries given by an action of the full Lie group $G$ underlying the AGM. It is well known that there exist other types of integrable sigma models which can be described in a gauged formulation where only a specific subgroup $H$ of $G$ is gauged. Typical examples in this class are the sigma models on symmetric spaces $G/H$ and their Yang-Baxter deformations~\cite{Delduc:2013fga}. These can be interpreted as relativistic realisations of a more general type of AGMs, which possess the so-called cyclotomy or dihedrality property~\cite{Vicedo:2017cge}. It would be interesting to study the chiral limits of these coset theories using techniques similar to the ones developed in this paper. For instance, relevant examples in this context might be the conformal limits of Yang-Baxter deformations of the ${\rm O}(N)$ sigma models
and their supersymmetric version, 
which have been the subject of recent works in the literature~\cite{Fateev:2018yos,Litvinov:2018bou,Alfimov:2020jpy}, and 
of the integrable coset sigma models on $G \times G / H$~\cite{Bardakci:1996gs,Georgiou:2016urf,Arutyunov:2020sdo,Levine:2021fof} whose conformal limit is the Guadagnini-Martellini-Mintchev CFT~\cite{Guadagnini:1987ty}.

\subsection{ODE/IQFT correspondence}
\label{Sec:ODE/IQFT}

One may next note that the quantisation of the chiral limits of AGMs considered
in the previous Subsection  \ref{genAGM} are contained in the
class of quantum integrable models 
discussed in \cite{Feigin:2007mr}. The main conjectures proposed in \cite{Feigin:2007mr} imply that the spectra of the models in this class can be described by generalisations
of the ODE/IQFT correspondence. This is motivated by hypothetical 
extensions of the geometric Langlands 
correspondence from the known cases associated to finite-dimensional simple 
Lie algebras $\mathfrak{g}$ (see for instance~\cite{Frenkel:2005pa} for a review) to affine Lie algebras $\widehat{\mathfrak{g}}$.
We thus expect that the general ODE/IQFT correspondence conjectured in \cite{Feigin:2007mr}
describes the spectra of the relativistic 
models considered in Subsection  \ref{genAGM} in the chiral
limits.

\bigskip

In the following we shall focus on a sub-class of 
the integrable models discussed in  \cite{Feigin:2007mr} which are called 
affine Gaudin models with regular singularities, abbreviated as regular AGMs.
These models are 
classified 
by two pieces of data: (i) an affine Lie algebra $\widehat{\mathfrak{g}}$, and (ii) 
a twist function $\varphi^{(\mathrm{qt})}$
of the form $\varphi^{(\mathrm{qt})}(z)=\sum_{r=1}^N\frac{k_r}{z-z_r}$, 
with $k_r$, $r=1,\dots,N$ satisfying $\sum_{r=1}^N k_r=-2\hv$.
The generalisation of the ODE/IQFT correspondence
proposed in 
\cite{Feigin:2007mr} would imply that the eigenvalues of the conserved quantities 
of the regular AGMs
associated to the data $(\widehat{\mathfrak{g}},\varphi^{(\mathrm{qt})})$ 
are encoded in a certain 
class of holomorphic connections on the $N$-punctured Riemann sphere which have a Lie-algebraic definition 
based on the Langlands dual ${}^L\widehat{\mathfrak{g}}$ of the affine 
Lie algebra $\widehat{\mathfrak{g}}$. The connections in this class are called 
affine opers in \cite{Feigin:2007mr}.  
For the sake of illustration, we focus here on the case
$\mathfrak{g}=\mathfrak{sl}(2)$, for which one can characterise the affine opers 
as connections which are gauge equivalent to the form 
\begin{equation}\label{aff-op}
\nabla_z=
\partial_z+
\bigg(\begin{matrix} 0 & v(z)+\chi\, \Pc(z) \\ 1 & 0\end{matrix}\bigg),
\end{equation}
where $\Pc(z)$ is a primitive of the twist function $\varphi^{(\mathrm{qt})}$
satisfying
$\partial_z\log \Pc(z)=\varphi^{(\mathrm{qt})}(z)$, which can be chosen as 
$\Pc(z)=\prod_{r=1}^N(z-z_r)^{k_r}$.
Instead of the connection $\nabla_z$ it is often convenient to consider the 
corresponding second order differential operator 
\begin{equation}\label{PGHO}
\mathcal{D}=-\partial_z^2+v(z)+\chi\, \Pc(z).
\end{equation}
The function 
$v(z)$ is allowed to have second order poles at $z_r$, 
$r=1,\dots,N$,\footnote{Note 
that one puncture is located at infinity in the conventions of 
\cite{Feigin:2007mr}, whereas we will here assume $z_r\neq \infty$ for all $r$ in the following discussion.} and 
an arbitrary number $K$ of apparent singularities at points $x_1,\dots, x_K$. 
The conditions
that the singularities at $x_1,\dots, x_K$ are apparent singularities, equivalent 
to triviality of the monodromy of $\nabla_z$ around these points, imply 
a system of algebraic equations for $x_1,\dots, x_K$ which is expected to 
have a discrete set of solutions. The generalised ODE/IQFT correspondence 
conjectured 
in \cite{Feigin:2007mr} predicts that the elements of this set, 
represented by the corresponding   functions $v(z)$
in \eqref{aff-op}, are in 
one-to-one correspondence with the eigenstates of the 
quantum affine Gaudin model associated to 
$(\widehat{\mathfrak{g}},\varphi^{(\mathrm{qt})})$.

\bigskip

A refinement of these conjectures was proposed in \cite{Lacroix:2018fhf}. It concerns the
relation between the affine opers associated  with the eigenstates of the 
relevant affine Gaudin model on the one hand and the eigenvalues $I_{\gamma,p}$ of the local 
conserved charges $\Qq_{\gamma,p}$ defined in \eqref{Eq:ConjQ} on the other hand. The relation proposed in \cite{Lacroix:2018fhf} can 
be represented in the form 
\begin{equation}\label{extr-eigenval}
I_{\gamma,p} = \oint_{\gamma}\Pc(z)^{-p/h^\vee}v_p(z) \dd z\,,
\end{equation}
which displays a direct correspondence with the definition 
\eqref{Eq:ConjQ} of the local conserved charge $\Qq_{\gamma,p}$.
The densities $v_p(z)$ in \eqref{extr-eigenval} can be 
constructed from the relevant affine opers by an algorithm described
in \cite{Lacroix:2018fhf}. In the $\mathfrak{sl}(2)$ case considered here one obtains a 
family of differential polynomials of the functions $v(z)$ 
characterising the affine opers according to \eqref{aff-op}\footnote{The coefficients of these differential polynomials also involve the function $\Pc(z)$, which depends only on the AGM considered and not on the choice of eigenstate.}. This conjecture
then takes a form equivalent to a previous proposal made in~\cite{Lukyanov:2013wra,Bazhanov:2013cua}
for the UV fixed-point of the ${\rm SU}(2)$ Klim\v{c}\'{i}k model (the explicit relation between the approach of \cite{Lacroix:2018fhf} and the one of~\cite{Lukyanov:2013wra,Bazhanov:2013cua} based on WKB expansions follows from the results of~\cite{Gaiotto:2020dhf}).

\bigskip

The results and conjectures  from \cite{Feigin:2007mr,Lacroix:2018fhf} are directly applicable to the chiral AGMs
discussed in our paper.  The case of the conformal limit of the Klim\v{c}\'{i}k model corresponds to $N=3$.
The  function  $\Pc(z)$ 
 introduced in Subsection \ref{Sec:QuantLocalIM}, eq.\,\eqref{Eq:PKlim}
is easily seen to be the special case of the function $\Pc(z)$ appearing in \eqref{aff-op}
with $z_N=\infty$.

\bigskip 

Strong support for the validity of the ODE/IQFT correspondence in the case of interest
here follows from the detailed investigations described in
 \cite{Lukyanov:2013wra,Bazhanov:2013cua,Bazhanov:2014joa}.
It was proposed in ref.\cite{Fateev:1996ea} that the
Klim\v{c}\'{i}k model for $G={\rm SU}(2)$ possesses a dual Lagrangian representation with Toda like 
interactions.
The ODE/IQFT correspondence for the dual model has been thoroughly investigated in~\cite{Lukyanov:2013wra,Bazhanov:2013cua}.
The differential operator defining the ODE studied in these references is easily 
seen to coincide with the operator $\mathcal{D}$ defined in \eqref{PGHO}.
For the explicit comparison one may note that the term
$\chi\,\Pc(z)$ in \eqref{PGHO} 
corresponds to the function $\lambda^2\Pc_{\mathrm{BL}}^{}(z)$ 
considered in~\cite{Lukyanov:2013wra,Bazhanov:2013cua}, defined as\footnote{$\Pc_{\mathrm{BL}}^{}(z)$ is denoted as $\Pc(z)$ in~\cite{Lukyanov:2013wra,Bazhanov:2013cua}.}
\begin{equation}\label{P-function}
\Pc_{\mathrm{BL}}^{}(z) = \frac{(z_3-z_2)^{a_1}(z_1-z_3)^{a_2}(z_2-z_1)^{a_3}}{(z-z_1)^{2-a_1}(z-z_2)^{2-a_2}(z-z_3)^{2-a_3}}\,,
\end{equation}
with $a_i$, $i=1,2,3$ being the parameters of the dual theory, satisfying $a_1+a_2+a_3=2$. 
Relating the parameters $a_r$ in  the expression \eqref{P-function} 
for $\Pc(z)$  to the levels $k_r$ by the relation
$a_r = k_r + 2$, it is easy to see that 
$\chi \Pc(z)=\lambda^2\Pc_{\mathrm{BL}}(z)$ 
will  hold if $\chi$ is assumed to be a certain function of 
$\lambda$ and $z_1,z_2,z_3$.

\bigskip

The results of our paper clarify the relation between the integrable structures of the 
chiral limit of the Klim\v{c}\'{i}k model and that of a specific relativistic AGM. The existence of chiral limits in a more general
class of relativistic AGMs discussed in Section \ref{genAGM} opens 
the perspective to describe the spectra of these chiral
models with the help of the generalised versions of the 
ODE/IQFT correspondence proposed in \cite{Feigin:2007mr}.

\subsection{The Yang-Baxter algebra in affine Gaudin models}\label{YB-persp}

The problem of quantization of the  non-local conserved charges in the
affine Gaudin model was briefly discussed in Section \ref{sec666}. It was mentioned that,
apart from some early results in ref.  \cite{Feigin:2007mr}, there is no systematic
construction of the higher  quantum non-local IMs, and a proof of
their mutual commutativity is lacking. From the point of view of a
semi-classical quantization of the AGM, one meets two types of problems.
First of all, the matrix elements of the Lax connection
satisfy non-ultralocal Poisson brackets. This makes it not obvious that
there can exist a reasonable definition for the Poisson brackets
of the monodromy matrix that obey natural requirements like the Jacobi identity.  
The other issue is related to the freedom that comes from  
the gauge symmetry of the AGM. It  acts on the Lax connection by conjugation,
inducing a corresponding action on the
monodromy matrices, which could be field dependent.
 The Poisson algebra
generated by the monodromy matrix elements
 will therefore generically
depend on the choice of  gauge.  
For any result of the Poisson bracket  computation,
there is always the possibility that  a different gauge fixing condition would have led to a
 Poisson algebra that is of a simpler form,
e.g., the Sklyanin exchange relations.
\bigskip

As was pointed out in Section \ref{sec666},
a solution to these problems seems to be within reach for
the  AGM representing the conformal limit of
the Klim\v{c}\'{i}k model.
We demonstrated   that
the gauge equivalence class of the classical Lax connection
for $G={\rm SU}(2)$
contains the Lax matrix studied in
\cite{Bazhanov:2018xzh}. This was achieved by means of the
Dirac bracket computation from Section \ref{sec555389i}, where the Poisson algebra of the AGM Lax matrix elements
in a particular gauge  turn out to
coincide with  that for the Lax matrix considered in
ref. \cite{Bazhanov:2018xzh}.
Having established  such a relation, one can apply
to the AGM of our interest
the main conjecture proposed in
that work.
The latter  
 identifies the monodromy
of this Lax connection with the classical limit of an operator-valued monodromy
matrix satisfying the Yang-Baxter algebra. The quantum operator
from  \cite{Bazhanov:2018xzh} may therefore be
regarded as a quantisation of the monodromy of the gauge-fixed AGM Lax connection.
This conjecture would further
imply that there exists a natural definition of the Poisson brackets for the
classical monodromy matrix elements such that the Sklyanin exchange relations
hold true.

\bigskip

The above observations suggest that the
non-local conserved quantities of the  chiral AGM from this paper could be systematically
studied  within the usual paradigm of the QISM based on
Yang-Baxter integrability.  That framework  would also provide  a means of proving the mutual 
commutativity of the non-local IMs. 
A first step in this direction is the rigorous proof of the conjecture from   ref.~\cite{Bazhanov:2018xzh}.
While it only  concerns the $\mathfrak{sl}(2)$ case, it should also be possible
to extend the conjecture  to arbitrary simple Lie algebras $\mathfrak{g}$, 
as was briefly mentioned in  Subsection \ref{cxnbbsdbv3}.
Another question concerns the definition of the Poisson brackets for the matrix elements
of the classical monodromy. It would be interesting to define an explicit prescription for handling the ambiguities
that come from the non-ultralocal bracket of the Lax matrix such that the Jacobi identity holds and
the Sklyanin exchange relations are obeyed. 
The existence of such a regularization procedure would, of course, be a 
corollary once  the proof of the conjecture from ref.\cite{Bazhanov:2018xzh},
and its generalizations, is achieved.

\bigskip

 In view of the underlying Lie-theoretic structures, it seems natural
 to expect that the integrals of motion of more general AGMs can be understood
 in a similar way.
 Quantum monodromy matrices for a so-called generalized affine $\mathfrak{sl}(2)$
Gaudin model  have recently been defined  in ref.~\cite{Kotousov:2021vih}. One may hope that similar
 constructions can be used to study the chiral limits of the AGMs discussed in Subsection \ref{genAGM} above.

\subsection{Deforming away from the chiral limits}\label{mass-Def}

Some of the most interesting questions in this context concern the possibility
that some key features observed in the chiral limits may survive the deformation 
away from the chiral limit. 
We have seen two main manifestations  of the
 integrable structures of the AGMs, related to local and non-local charges,
 respectively.  In this subsection we will briefly discuss prospects
 for deformations of these manifestations away from the chiral limit. 
 
 \bigskip
 
 The ODE/IQFT correspondence appears to describe the 
 eigenvalues of the conserved charges in the chiral limit of the 
 Klim\v{c}\'{i}k model efficiently. It seems very encouraging
 to observe that the correspondence appearing in this context
 admits a natural deformation that would encode the spectrum of
the IMs for the massive Klim\v{c}\'{i}k model.
A key r\^{o}le is played by the
integrable partial differential equation
 of the form \cite{Lukyanov:2013wra,Bazhanov:2013cua}
 \begin{equation}
 \partial _z\partial_{\bar{z}}\eta-\re^{2\eta}+\lambda^2\,\bar\lambda^2\,\Pc(z)\,\bar{\Pc}(\bar{z})\,\re^{-2\eta}=0\,,
 \end{equation}
whose auxiliary linear problem
serves a very similar task  as the ODE does in the chiral limit. 
 It is intriguing to note the re-appearance of the function $\Pc(z)$, which 
 suggests that the deformed ODE/IQFT-correspondence proposed 
 in \cite{Lukyanov:2013wra,Bazhanov:2013cua} admits a natural generalisation to the
 class of models discussed in
 Subsection \ref{genAGM}.
 
 \bigskip
 
 There seem to be a lot of opportunities for generalisations 
 in view of the observation that the ODEs and PDEs which appear to 
 be relevant in this context  are simple examples of a 
 large family of classically integrable equations admitting a Lie-algebraic 
 definition 
 related to the Drinfeld-Sokolov classification.

\bigskip 

It would furthermore be extremely useful if the Yang-Baxter algebraic structures
governing non-local conserved quantities  discussed in Section  \ref{sec666}
would admit a  deformation away from the chiral limits. This possibility
is known to be realised for affine Toda field theories, where integrable lattice
discretisations have been constructed having the property that the structure constants
of the exchange relations satisfied by the quantised monodromy matrices are 
simply mass-independent. This is most transparent in
the lattice light-cone approaches going back to \cite{Faddeev:1992xa,Bazhanov:1995zg}.
Such an approach has been used in 
\cite{Bytsko:2009mg} to investigate the chiral limits of the integrable structures of the 
sinh-Gordon model. These chiral limits 
are given by variants of the 
quantum KdV theory associated to the  Virasoro algebra
with central charge $c>25$ representing the integrable 
structures of the Liouville conformal field theory.
Integrable lattice regularisations based on  Yang-Baxter algebraic structures
have been constructed for 
more general integrable models of affine Toda type in \cite{Ridout:2011wx,Meneghelli:2015sra}. 
Such a framework allows us to understand the integrable structures of  
the massive models as deformations of their chiral limits. Having seen 
evidence for the existence of Yang-Baxter algebraic structures in the chiral
limits of AGMs makes us hope that such structures can also exist 
in the case of relativistic AGMs. 

\subsection[Gauged WZW, $\lambda$-models and Poisson-Lie T-duality]{Gauged WZW, \texorpdfstring{$\bm\lambda$}{lambda}-models and Poisson-Lie T-duality}\label{Sec:Lambda}

The study of integrable deformations of $\sigma$-models has attracted a lot of attention in the past decades. In parallel to the Yang-Baxter deformations, which include the Klim\v{c}\'{i}k model considered in this article, another important class that has been explored is formed by the so-called $\lambda$-deformations. The prototypical member of this class has been introduced by Sfetsos in~\cite{Sfetsos:2013wia} as an integrable deformation of the non-abelian T-dual of the Principal Chiral Model. It was further realised~\cite{Vicedo:2015pna,Hoare:2015gda,Sfetsos:2015nya} that there exists a deep relation between these two types of deformed models, taking the form of the so-called Poisson-Lie T-duality~\cite{Klimcik:1995ux,Klimcik:1995dy}.

In particular, under this correspondence, the Klim\v{c}\'{i}k model is expected to be dual to the $G\! \times\! G/G$ coset $\lambda$-model introduced in~\cite{Sfetsos:2017sep}\footnote{The Klim\v{c}\'{i}k model can in fact be Poisson-Lie T-dualised in two ways, corresponding to its left and right deformed Poisson-Lie symmetries. Dualising with respect to one of these symmetries yields the so-called generalised $\lambda$-model~\cite{Sfetsos:2015nya,Klimcik:2016rov}. A second dualisation with respect to the other Poisson-Lie symmetry is then expected to give the $G\! \times\! G/G$ coset $\lambda$-model.}. In the language of AGMs, these two theories correspond to two different relativistic realisations of the same underlying AGM: in particular, there exists a map between the parameters of the Klim\v{c}\'{i}k model and the $\lambda$-model which allows one to identify their twist function~\cite{Georgiou:2019plp} (more generally, Poisson-Lie T-dual models are canonically equivalent~\cite{Sfetsos:1997pi} and thus share the same underlying Poisson structure). The analysis of the Klim\v{c}\'{i}k model carried out in the present paper should thus possess a rather automatic analogue in the $G\! \times\! G/G$ $\lambda$-model. Of particular interest in this analysis is the study of the UV limit of the theory: for the dual $\lambda$-model, this limit is the well-known $G\! \times\! G/G$ gauged Wess-Zumino-Witten model (more generally, $\lambda$-deformations can be seen as relevant perturbations of gauged WZW models 
as has been found in a variety of cases
\cite{Itsios:2014lca,Appadu:2015nfa,Georgiou:2016zyo,Sfetsos:2017sep}). The techniques developed in this paper should thus allow the study of classical and quantum integrable structures in this conformal model. It would be interesting to explore these aspects further.

A key role in the study of gauged WZW models is played by the parafermions~\cite{Bardakci:1990lbc,Bardakci:1990ad,Karabali:1989dk}, which are non-local chiral fields. Recall that we have encountered such parafermionic fields in Sections \ref{sec31} and \ref{sec555389i}, for instance through an appropriate gauge-fixing of the AGM underlying the chiral Klim\v{c}\'{i}k model. More precisely, we found that this gauge-fixing is described by $\g/\mathfrak{h}$ parafermions and $\dim \mathfrak{h}$ decoupled free fields, where $\mathfrak{h}$ is the Cartan subalgebra of $\g$. At least in the ${\rm SU}(2)$ case, this setup was well-suited for the description of the UV fixed-point of the Klim\v{c}\'{i}k model. In contrast, on the dual side, one expects the $G\! \times\! G/G$ gauged WZW model to be naturally described by $\g\oplus\g/\g$ parafermions~\cite{Bardakci:1990ad}. The latter are obtainable from a different choice of gauge-fixing of the same underlying AGM. We note moreover that they play an important role in the non-conformal $\lambda$-model~\cite{Sfetsos:2017sep}: indeed, the relevant operator driving the perturbation from the conformal point to this model is built as a bilinear in these currents.

\section*{Acknowledgements} 

The authors would like to thank B. Vicedo for useful and interesting discussions. G.K. also acknowledges V. V. Bazhanov and S. L. Lukyanov  for fruitful and stimulating interactions and S.L. similarly thanks F. Delduc, T. Franzini, M. Magro, K. Siampos and C. A. S. Young. The work of G.K. and J.T. is funded by the Deutsche Forschungsgemeinschaft (DFG, German Research Foundation) under Germany's Excellence Strategy -- EXC 2121 ``Quantum Universe" -- 390833306. S.L. is supported by Dr.
Max R\"ossler, the Walter Haefner Foundation and the ETH Z\"urich Foundation.

\appendix

\section{Conventions on Lie algebras and root decompositions}
\label{App:Root}

\paragraph{Root decomposition.} Let us consider the complexified simple Lie algebra $\g^\C$ and its decomposition $\g^\C=\mathfrak{h}\oplus\mathfrak{n}_+\oplus\mathfrak{n}_-$, where $\mathfrak{h}$ is a Cartan subalgebra and $\mathfrak{n}_\pm$ are positive/negative nilpotent subalgebras. We denote by $\Delta = \Delta_+ \sqcup \Delta_- \subset \mathfrak{h}^\ast$ the root system of $\g^\C$, decomposed in positive roots $\Delta_+$ and negative ones $\Delta_-$. For each $\alpha\in\Delta_\pm$, 
let  ${\tt e}_\alpha$ be the corresponding root vector lying in $\mathfrak{n}_\pm$. We then have
\begin{equation}
\mathfrak{n}_\pm = \bigoplus_{\alpha\in\Delta_\pm} \C{\tt e}_\alpha.
\end{equation}
We normalise these vectors such that
\begin{equation}
\langle {\tt e}_\alpha, {\tt e}_\beta \rangle = -\delta_{\alpha+\beta,0}\,, \qquad \forall \, \alpha,\beta\in\Delta,
\end{equation}
where $\langle\cdot,\cdot\rangle$ is our choice of non-degenerate bilinear form on $\g^\C$. If two roots $\alpha,\beta\in\Delta$ are such that $\alpha+\beta$ is also a root, then there exists a non-zero number $N^{\alpha,\beta}$ such that $\bigl[ {\tt e}_\alpha, {\tt e}_\beta \bigr] = N^{\alpha,\beta} \, {\tt e}_{\alpha+\beta}$. If $\alpha+\beta$ is not a root, we set $N^{\alpha,\beta}=0$.

Let us also fix an orthogonal basis $\lbrace {\tt h}_i \rbrace_{i=1,\ldots,\dim\mathfrak{h}}$ of the Cartan subalgebra $\mathfrak{h}$ of $\g^\C$. It will be convenient to normalise it so that
\begin{equation}\label{Eq:NormHi}
\langle {\tt h}_i, {\tt h}_j \rangle = -2\,\delta_{ij}, \qquad \forall\,i,j\in\lbrace 1,\ldots,\dim\mathfrak{h} \rbrace.
\end{equation}
The family $\lbrace {\tt h}_i \rbrace_{i=1,\ldots,\dim\mathfrak{h}} \sqcup \lbrace {\tt e}_\alpha \rbrace_{\alpha\in\Delta}$ then forms a basis of $\g^\C$, in which the Lie bracket reads
\begin{subequations}\label{Eq:ComHE}
\begin{align}
\bigl[ {\tt h}_i, {\tt h}_j \bigr] & = 0 \,, \\
\bigl[ {\tt h}_i, {\tt e}_\alpha \bigr] & = \alpha({\tt h}_i) \,{\tt e}_\alpha \,, \\
\bigl[ {\tt e}_\alpha, {\tt e}_\beta \bigr] &=  N^{\alpha,\beta} \, {\tt e}_{\alpha+\beta} \, , \\
\bigl[ {\tt e}_\alpha, {\tt e}_{-\alpha}\bigr] &= \rho_\alpha^i \,{\tt h}_i \, ,
\end{align}
\end{subequations}
where $i,j\in\lbrace 1,\ldots,\dim\mathfrak{h} \rbrace$ and $\alpha,\beta \in \Delta$ with $\alpha+\beta\neq 0$. In the last equation, a sum over the repeated index $i \in\lbrace 1,\ldots,\dim\mathfrak{h} \rbrace$ is implied and the numbers $\rho_\alpha^i$ are given in our normalisation by $\rho_\alpha^i=\frac{1}{2}\,\alpha({\tt h}_i)$. Finally, the split quadratic Casimir of $\g^\C$ reads
\begin{equation}
\mathsf{C}_2 = -\frac{1}{2} \sum_{i=1}^{\dim\mathfrak{h}} {\tt h}_i \otimes {\tt h}_i - \sum_{\alpha\in\Delta} {\tt e}_\alpha \otimes {\tt e}_{-\alpha}.
\end{equation}

\paragraph{Root levels and Coxeter number.} Let $\lbrace \alpha_i \rbrace_{i=1}^{\dim\mathfrak{h}}$ be the simple roots of $\g^\C$. They form a basis of $\mathfrak{h}^\ast$ such that any positive root $\alpha\in\Delta_+$ can be written as
\begin{equation}
\alpha = \sum_{i=1}^{\dim\mathfrak{h}} m_i\,\alpha_i
\end{equation}
for some non-negative integers $m_i\in\mathbb{Z}_{\geq 0}$. We then define the level of $\alpha$ as
\begin{equation}
\ell(\alpha) = \sum_{i=1}^{\dim\mathfrak{h}} m_i\,.
\end{equation}
One similarly defines the level of a negative root, which is then a negative number.

There exists a unique positive root $\theta\in\Delta_+$ with maximal level, which we call the highest root of $\g^\C$. The so-called Coxeter number $h$ of the algebra is then defined as
\begin{equation}
h = \ell(\theta) + 1\,.
\end{equation}
For any positive root $\alpha\in\Delta_+$, we thus have $0 < \ell(\alpha) < h$.

Let us finally define the Weyl coweight ${\tt h}_\delta$ of the Lie algebra $\g^\C$. It is the unique element ${\tt h}_\delta \in \mathfrak{h}$ in the Cartan subalgebra such that
\begin{equation}
\alpha_i ({\tt h}_\delta) = 1\,, \qquad \forall\, i\in\lbrace 1,\ldots,\dim\mathfrak{h} \rbrace\,.
\end{equation}
By construction, this object then satisfies $\alpha({\tt h}_\delta) = \ell(\alpha)$ for any root $\alpha\in\Delta$. It is thus characterised by the property
\begin{equation}
\bigl[ {\tt h}_\delta, {\tt e}_\alpha \bigr] = \ell(\alpha) {\tt e}_\alpha\,, \qquad \forall\,\alpha\in\Delta\,.
\end{equation}

\paragraph{Bilinear form on $\mathfrak{h}^\ast$.} The restriction of $-\langle\cdot,\cdot\rangle$ to $\mathfrak{h}$ defines a non-degenerate bilinear form. By duality, it then induces a non-degenerate bilinear form on $\mathfrak{h}^\ast$ as well, which we will denote by $(\cdot,\cdot)$. Let us describe it more explicitly. For any linear forms $\alpha,\beta\in\mathfrak{h}^\ast$, we have
\begin{equation}\label{Eq:RootForm}
(\alpha,\beta) = \frac{1}{2} \sum_{i=1}^{\dim\mathfrak{h}} \alpha({\tt h}_i) \beta({\tt h}_i).
\end{equation}
Recall that we have normalised the bilinear form $\langle\cdot,\cdot\rangle$ with respect to the Killing form as (see footnote \ref{ft1})
\begin{equation}
\langle {\tt t}_a,{\tt t}_b\rangle=-\frac{1}{2\hv}\,{f_{ac}}^d\,{f_{bd}}^c\,,
\end{equation}
where $\hv$ is the dual Coxeter number of $\g^\C$. This choice corresponds to asking that $(\theta,\theta)=2$ for $\theta\in\Delta_+$ the highest root of $\g^\C$.

\section{Gauge-fixings and Dirac brackets}
\label{App:Dirac}

The goal of this appendix is to compute Dirac brackets coming from appropriate gauge-fixings of Kac-Moody currents. Before considering this, it will be useful to revisit the case of gauge-fixings in mechanical systems with a finite number of degrees of freedom, to gain some intuition on how to treat the field theory case. This will be the subject of the first subsection. The Dirac brackets of Kac-Moody currents are then covered in the remaining parts of this appendix.

\subsection{Gauge-fixings and Dirac brackets in mechanics}

\paragraph{Constraints and reduction.} Let us consider a finite-dimensional phase space $P$. The algebra of functions $F[P]$ on $P$ is then naturally equipped with a Poisson bracket $\lbrace\cdot,\cdot\rbrace$. We suppose here that $P$ admits a Hamiltonian action of a simple Lie group $G$, generated by a moment-map $\Cc : P \to \g$ (where $\g$ is the Lie algebra of $G$). This moment-map can be seen as an element of $F[P]\otimes\g$ that generates the infinitesimal action of $\g$ on $F[P]$. Namely, the action of $\epsilon\in\g$ on an observable $f\in F[P]$ is given by
\begin{equation}\label{Eq:GAction}
\delta_\epsilon f = \bigl\lbrace \langle \Cc,\epsilon \rangle, f \bigr\rbrace,
\end{equation}
where $\langle \cdot,\cdot \rangle$ is our choice of invariant bilinear form on $\g$. The condition that this infinitesimal action satisfies the commutation relations of $\g$ translates to the fact that $\Cc$ can be made to satisfy the Kirillov-Kostant bracket
\begin{equation}\label{Eq:KK}
\lbrace \Cc \tv \Cc \rbrace = \bigl[ \Id \otimes \Cc, {\mathsf C}_{2} \bigr],
\end{equation}
where we used tensorial notations and where ${\mathsf C}_{2} \in\g\otimes\g$ is the split quadratic Casimir of $\g$.

In this setup, one can consider the Hamiltonian reduction with respect to the $G$-symmetry generated by $\Cc$. For that, we treat this moment-map as a constraint and thus impose $\Cc \approx 0$ (here, we use the sign $\approx$ to denote weak equalities). The Kirillov-Kostant bracket \eqref{Eq:KK} ensures that $\Cc \approx 0$ is a first-class constraint, since $\lbrace \Cc \tv \Cc\rbrace \approx 0$. The action of $G$ is then treated as a gauge symmetry and the reduced phase space $P_{\red}$ is obtained by considering the quotient of the constrained surface by this action. This reduced phase space is equipped with the Marsden-Weinstein symplectic structure.

\paragraph{Gauge-fixing and Dirac bracket.} One concrete way to describe the reduced phase space is to consider a gauge-fixing condition. We treat the latter as an additional constraint $\Fc \equiv 0$ (as in the main text, the symbol $\equiv$ then denotes equalities that hold under both the initial constraint $\Cc \equiv 0$ and the gauge-fixing condition $\Fc \equiv 0$). For this additional constraint to be a good gauge-fixing condition, one needs it to be always attainable by a gauge transformation, \textit{i.e.} each $G$-orbit should contain a representative satisfying this condition. Moreover, we will suppose here that this is a complete gauge-fixing condition, in the sense that there is a unique such representative in each orbit and thus that there are no residual gauge symmetries. Let us note that $\Fc \equiv 0$ should then in fact encode $\dim\g$ gauge-fixing conditions: for simplicity, we will suppose here that these can be organised in such a way that $\Fc$ is a $\g$-valued observable, \textit{i.e.} an element of $F[P]\otimes\g$. It is clear that the reduced phase space $P_{\red}$ is isomorphic to the subset $P_{\text{GF}}$ of points in $P$ which satisfy $\Cc\equiv 0$ and $\Fc \equiv 0$.

In order to describe the reduced symplectic structure in terms of the gauged-fixed space $P_{\text{GF}}$, it is useful to define the notion of a Dirac bracket. For that, let us introduce the decompositions $\Cc = \Cc^a \,\mathsf{t}_a$ and $\Fc = \Fc^a\, \mathsf{t}_a$ of the $\g$-valued observables $\Cc$ and $\Fc$ in a basis $\lbrace \mathsf{t}_a \rbrace$ of $\g$. We then define the matrices of Poisson brackets
\begin{equation}\label{Eq:MDef}
M^{ab} \equiv \lbrace \Cc^a, \Fc^b \rbrace \qquad \text{ and } \qquad \Mt^{\,ab} \equiv \lbrace \Fc^a, \Fc^b \rbrace,
\end{equation}
as well as\footnote{The components $N_{ab}$ and $\Nt_{ab}$ are entries of the inverse of the $(2\dim\g)\times(2\dim\g)$ matrix built from all Poisson brackets of $\Cc^a$ and $\Fc^a$. The top-left block of this matrix is zero since $\lbrace \Cc^a,\Cc^b\rbrace \equiv 0$, so its inverse has a vanishing bottom-right block.}
\begin{equation}\label{Eq:NDef}
N_{ab} \equiv M^{-1}_{ab} \qquad \text{ and } \qquad \Nt_{ab} \equiv \Mt^{\,cd}M^{-1}_{ca}M^{-1}_{db}.
\end{equation}
In particular, the components of $N$ satisfy $M^{ac}N_{cb}\equiv\delta^a_{\;\,b}$ and $N_{ac}M^{cb}\equiv\delta_a^{\;\,b}$. The fact that $M$ is invertible is due to the property that $\Fc \equiv 0$ is a complete gauge-fixing, which implies that the set of constraint $(\Cc^a,\Fc^a)$ is second-class. We note that $\Mt$ and $\Nt$ are skew-symmetric matrices, in the sense that $\Mt^{ab}=-\Mt^{\,ba}$ and $\Nt_{ab}=-\Nt_{ba}$.\\

Using these notations, the Dirac bracket of two observables $f,g\in F[P]$ is defined as
\begin{equation}\label{Eq:Dirac0}
\lbrace f,g \rbrace_D \equiv \lbrace f,g \rbrace + \left( \lbrace f, \Fc^a \rbrace \lbrace g,\Cc^b \rbrace -  \lbrace g, \Fc^a \rbrace \lbrace f,\Cc^b \rbrace \right)N_{ab} +  \lbrace f, \Cc^a \rbrace \lbrace g,\Cc^b \rbrace\,\Nt_{ab}.
\end{equation}
It is clear that it is skew-symmetric and satisfies the Leibniz rule. Moreover one can show that it also obeys the Jacobi identity. The main additional property of this bracket is that the constraints $\Cc^a$ and $\Fc^a$ have a vanishing Dirac bracket with all observables in $F[P]$. Indeed, one easily checks from the above definition and the properties of $N$ and $\Nt$ that
\begin{equation}
\lbrace \Cc^a, f \rbrace_D \equiv \lbrace \Fc^a, f \rbrace_D \equiv 0, \qquad \forall\,a\in\lbrace 1,\dots,\dim\g\rbrace, \quad \forall\, f\in F[P].
\end{equation}
This means that the Dirac bracket is compatible with the constraints $\Cc \equiv \Fc \equiv 0$. It thus descends to the gauge-fixed phase space $P_{\text{GF}}$ and defines a Poisson bracket on the algebra $F[P_{\text{GF}}]$, making $P_{\text{GF}}$ isomorphic to the Marsden-Weinstein quotient $P_{\red}$ as a symplectic manifold.

\paragraph{Reformulation in terms of linear operators.} To gain some useful intuition for the computation of Dirac brackets in affine Gaudin models in the next subsections, it will be useful to reformulate slightly the classical mechanics case described above. For that, let us introduce linear operators $\Nc : \g \to \g$ and $\Nct : \g \to \g$, defined by the following action on the basis $\lbrace \mathsf{t}_a \rbrace$:
\begin{equation}
\Nc(\mathsf{t}_a) = N_{ab}\, \eta^{bc}\, \mathsf{t}_c \qquad \text{ and } \qquad \Nct(\mathsf{t}_a) = \Nt_{ab}\, \eta^{bc}\, \mathsf{t}_c,
\end{equation}
where $\eta^{ab}$ denotes the inverse of the matrix $\eta_{ab}=\langle \mathsf{t}_a, \mathsf{t}_b \rangle$. In other words, the linear operators $\Nc$ and $\Nct$ are characterised by
\begin{equation}
\bigl\langle \Nc(\mathsf{t}_a), \mathsf{t}_b \big\rangle = N_{ab} \qquad \text{ and } \qquad \bigl\langle \Nct(\mathsf{t}_a), \mathsf{t}_b \big\rangle = \Nt_{ab}.
\end{equation}
We note that the skew-symmetry of the coefficients $\Nt_{ab}$ implies $\tp\Nct=-\Nct$, where $t$ denotes the transpose with respect to the bilinear form $\langle\cdot,\cdot\rangle$. In terms of these operators, the Dirac bracket \eqref{Eq:Dirac0} can be rephrased in a basis-independent way:
\begin{equation}\label{Eq:Dirac}
\lbrace f,g \rbrace_D \equiv \lbrace f,g \rbrace - \bigl\langle \lbrace f, \Cc \rbrace, \Nc \bigl( \lbrace g,\Fc \rbrace \bigr) \bigr\rangle +  \bigl\langle \lbrace g, \Cc \rbrace, \Nc \bigl( \lbrace f,\Fc \rbrace \bigr) \bigr\rangle - \bigl\langle \lbrace f, \Cc \rbrace, \Nct \bigl( \lbrace g,\Cc \rbrace \bigr) \bigr\rangle.
\end{equation}

Recall that the coefficients $N_{ab}$ and $\Nt_{ab}$ defining the operators $\Nc$ and $\Nct$ were originally introduced in eq.\;\eqref{Eq:NDef} in terms of the quantities $M^{ab}$ and $\Mt^{\,ab}$ built from Poisson brackets of the constraints $\Cc^a$ and $\Fc^a$. There is also a basis-independent reformulation of this definition. For that, let us introduce the operators $\Mc: \g \to \g$ and $\Mct : \g \to \g$ defined through
\begin{equation}
\Mc(\mathsf{t}_a) = \eta_{ab}\,M^{bc} \, \mathsf{t}_c \qquad \text{ and } \qquad \Mct(\mathsf{t}_a) = \eta_{ab}\, \Mt^{\,bc} \, \mathsf{t}_c.
\end{equation}
Then, we can reformulate eq.\;\eqref{Eq:NDef} as
\begin{equation}\label{Eq:NM}
\Nc = \Mc^{-1} \qquad \text{ and } \qquad \Nct = \Mc^{-1} \circ \Mct\circ \tp\Mc^{-1}.
\end{equation}
Moreover, the definition \eqref{Eq:MDef} of the coefficients $M^{ab}$ and $\Mt^{\,ab}$ can be rephrased in terms of the operators $\Mc$ and $\Mct$ using the Poisson brackets of the constraints $\Cc$ and $\Fc$ in tensorial notations. Namely,
\begin{equation}\label{Eq:PbM}
\lbrace \Cc \tv \Fc \rbrace \equiv \bigl( \Id \otimes \Mc \bigr) \mathsf{C}_2 \qquad \text{ and } \qquad \lbrace \Fc \tv \Fc \rbrace \equiv \bigl( \Id \otimes \Mct \bigr) \mathsf{C}_2,
\end{equation}
where we recall that $\mathsf{C}_2$ is the split quadratic Casimir $\mathsf{C}_2 = \eta^{ab} \, \mathsf{t}_a \otimes \mathsf{t}_b$. We note that the skew-symmetry $\tp\Mct = -\Mct$ of the operator $\Mct$ encodes the skew-symmetry of the bracket $\lbrace \Fc \tv \Fc \rbrace$, using the identity $\bigl( \widehat{O} \otimes \Id \bigr) \mathsf{C}_2 = \bigl( \Id \otimes \tp\widehat{O} \bigr) \mathsf{C}_2$ true for any linear operator $\widehat{O} : \g\to\g$.

To summarise, the formalism introduced in this paragraph allows one to reformulate every step in the definition of the Dirac bracket in a basis-independent way. Concretely, one starts by writing the Poisson bracket of the constraints $\Cc$ and $\Fc$ in the form \eqref{Eq:PbM}, thereby defining the two operators $\Mc$ and $\Mct$. We then introduce the operators $\Nc$ and $\Nct$ through the equation \eqref{Eq:NM}. Finally, we define the Dirac bracket by the formula \eqref{Eq:Dirac}. A similar formulation will be useful to compute Dirac brackets in affine Gaudin models in the next subsections. Let us finally note for completeness that the Kirillov-Kostant bracket \eqref{Eq:KK} can also be naturally written in an ``operatorial'' form, similar to the one of Equation \eqref{Eq:PbM}:
\begin{equation}\label{Eq:PbCFin}
\lbrace \Cc \tv \Cc \rbrace = \bigl( \Id \otimes \ad_{\Cc} \bigr) \mathsf{C}_2 = -\bigl( \ad_{\Cc} \otimes \Id \bigr) \mathsf{C}_2 \approx 0.
\end{equation}
Here, $\ad_{\Cc} : \g \to \g$ is the standard adjoint action of $\Cc$ on $\g$.

\subsection{Preliminaries: Kac-Moody currents and loop algebra}

\paragraph{Loop algebra.} Let us now come back to the case of an affine Gaudin model. Such a theory is described by a Poisson algebra $\Ac$, generated by $N$ Kac-Moody currents $\Jc_r(x)$ with levels $\ell_r$, satisfying the Poisson bracket \eqref{Eq:PbKM}. Moreover, the model is subject to the first class constraint $\Cc(x) = \sum_{r=1}^N \Jc_r(x) \approx 0$. The main difference with the mechanical setup considered in the previous subsection is that here, $\Cc$ is a field on the circle $\mathbb{S}^1$ and the constraint $\Cc(x) \approx 0$ holds for all $x\in \mathbb{S}^1$. To develop an approach similar to the one of the previous subsection in this field theory case, it will be useful to see this constraint in a ``functional'' sense, interpreting the field $\Cc$ as belonging to an infinite-dimensional Lie algebra composed of functions on $\mathbb{S}^1$, called the loop algebra $\Lc\g$: one then formally imposes $\Cc \approx 0$, encoding the infinitely-many constraints $\Cc(x)\approx 0$. We introduce the necessary formalism for that in this preliminary subsection.\\

Formally, we can think of the Kac-Moody current $\Jc_r$ as an element of the space $\Ac \otimes \Lc\g$, \textit{i.e.} as a dynamical observable valued in the loop algebra $\Lc\g$. We define the latter as the space of functions\footnote{More rigorously, one should think of the fields $\Jc_r$ as distributions on $\mathbb{S}^1$ rather than functions, since they possess an infinite Fourier series which does not necessarily converge in the standard functional sense. In particular, the field $\Jc_r$ is more technically valued in a completion of $\Ac \otimes \Lc\g$, which allows for such infinite series (converging with respect to a certain valuation associated with loop modes in $\Ac$ and $\Lc\g$). We will not enter in these subtleties here and refer to~\cite{Vicedo:2017cge} for details on similar considerations.} on the circle $\mathbb{S}^1$ valued in $\g^\C$. An element $A\in\Lc\g$ can be evaluated at points $x\in \mathbb{S}^1$, yielding objects $A(x)$ valued in the finite algebra $\g^\C$. One defines a Lie algebra structure on $\Lc\g$ by extending the Lie bracket of $\g^\C$ pointwise in $\mathbb{S}^1$, \textit{i.e.} such that $[A,B](x)=[A(x),B(x)]$. We will denote by $\p$ the formal derivation operator acting on $\Lc\g$ (such that $(\p A)(x) = \p_x A(x)$ under evaluation of $A\in\Lc\g$ at the point $x\in \mathbb{S}^1$).

\paragraph{Bilinear form on the loop algebra.} The Lie algebra $\Lc\g$ possesses a non-degenerate invariant bilinear form $\llangle\cdot,\cdot\rrangle$, defined in terms of the one $\langle\cdot,\cdot\rangle$ on the finite algebra $\g^\C$ by
\begin{equation}\label{Eq:LoopForm}
\llangle A,B \rrangle = \int_{0}^{2\pi} \bigl\langle A(x), B(x) \bigr\rangle  \; \dd x, \qquad \forall\,A,B \in \Lc\g.
\end{equation}
More algebraically, this bilinear form can be seen as pairing the Fourier coefficient of $e^{inx}$ in $A$ with the Fourier coefficient of $e^{-inx}$ in $B$ (up to a factor $2\pi$).

Let $\widehat{O}:\Lc\g \to \Lc\g$ be a linear operator on the loop algebra. We will denote by $\Tp\widehat{O}$ its transpose with respect to the bilinear form $\llangle\cdot,\cdot\rrangle$, which satisfies $\llangle\widehat{O}A,B\rrangle=\llangle A,\Tp\widehat{O}B\rrangle$ for all $A,B\in\Lc\g$. Here, we used a symbol different from the transpose $t$ on the finite Lie algebra $\g^\C$, to avoid confusion. Note however that if $\widehat{O}$ is a linear operator on $\g^\C$ that we extend to $\Lc\g$ pointwise in $\mathbb{S}^1$, then $\Tp\widehat{O}=\tp\widehat{O}$. Let us finally point out that the derivation operator $\p:\Lc\g\to\Lc\g$ is skew-symmetric with respect to the transpose $\mathsf{T}$. Indeed, one easily checks that $\Tp\p=-\p$ using the functional definition \eqref{Eq:LoopForm} of $\llangle\cdot,\cdot\rrangle$ and an integration by part.

\paragraph{Casimir of the loop algebra.} One can canonically construct a split quadratic Casimir $\Ch$ associated with the non-degenerate bilinear form $\llangle\cdot,\cdot\rrangle$. It is an element of (a completion of) $\Lc\g\otimes\Lc\g$ which satisfies
\begin{equation}\label{Eq:LoopCas}
\llangle[\bigl] \Ch,A \rrangle[\bigr]\ti{2} = A, \qquad \forall\,A\in\Lc\g,
\end{equation}
where the index $\underline{2}$ means that we are applying $\llangle\cdot,A\rrangle$ on the second tensor factor in $\Lc\g\otimes\Lc\g$. Functionally, this loop Casimir can be seen as the distribution
\begin{equation}\label{Eq:CasDelta}
\Ch (x,y) = \eta_{ab}\,\mathsf{t}^a \otimes \mathsf{t}^b \; \left( \frac{1}{2\pi} \sum_{n\in \mathbb{Z}} e^{in(x-y)} \right) = \mathsf{C}_2\, \delta(x-y).
\end{equation}
Here, $(x,y)$ means that we are evaluating the first and second tensor factors in $\Lc\g\otimes\Lc\g$ at the points $x\in \mathbb{S}^1$ and $y\in \mathbb{S}^1$ respectively, $\mathsf{C}_2 \in \g\otimes\g$ is the split Casimir of the finite algebra $\g$ and $\delta(x-y)$ is the Dirac $\delta$-distribution on $\mathbb{S}^1$. This identity can be verified by a direct computation, using the functional expression \eqref{Eq:LoopForm} of the bilinear form $\llangle\cdot,\cdot\rrangle$. Indeed, for any $A\in\Lc\g$, we have
\begin{equation}
\llangle[\bigl] \Ch,A  \rrangle[\bigr]\ti{2}(x) = \int_{0}^{2\pi} \bigl\langle \Ch(x,y), A(y) \bigr\rangle\ti{2} \; \dd y = \int_{0}^{2\pi} \bigl\langle \mathsf{C}_2, A(y) \bigr\rangle_{\underline{2}} \; \delta(x-y)\, \dd y = A(x),
\end{equation}
thus checking the characterisation \eqref{Eq:LoopCas} of $\Ch$ (here, we have used the property $\langle \mathsf{C}_2, \mathsf{a} \rangle\ti{2}  = \mathsf{a}$ of the finite Casimir $\mathsf{C}_2$, true for all $\mathsf{a}\in\g^\C$).

Let us finally note that for any linear operator $\widehat{O}:\Lc\g \to \Lc\g$ on the loop algebra, the loop Casimir satisfies the following identity:
\begin{equation}\label{Eq:CasTp}
\bigl( \widehat{O} \otimes \Id \bigr) \Ch = \bigl( \Id \otimes \Tp\widehat{O} \bigr) \Ch.
\end{equation}

\paragraph{Kac-Moody brackets.} As we will now see, the rather abstract formalism we have just introduced will be useful to describe the algebra of the Kac-Moody currents $\Jc_r$ in a compact way. For that, we introduce the following linear operator on $\Lc\g$:
\begin{equation}
\nabla_r = \ell_r \p + \ad_{\Jc_r},
\end{equation}
where $\ad_{\Jc_r} = [ \Jc_r,\cdot ]$ denotes the adjoint action of $\Jc_r$ on $\Lc\g$. One then checks that the Kac-Moody bracket \eqref{Eq:PbKM} can be rephrased as the following tensorial bracket in $\Ac\otimes\Lc\g\otimes\Lc\g$:\footnote{Let us note for completeness that this reformulation can be naturally interpreted by seeing the Kac-Moody bracket as the Kirillov-Kostant bracket of the affine algebra $\gh$, whose dual is identified with the space $\C\p\rtimes\Lc\g$ of $\g^\C$-connections on $\mathbb{S}^1$.}
\begin{equation}\label{Eq:PbKMLoop}
\lbrace \Jc_r \tv \Jc_s \rbrace = \delta_{rs}(\Id \otimes \nabla_r) \Ch = - \delta_{rs}(\nabla_r \otimes \Id) \Ch.
\end{equation}
The last equality in this equation encodes the skew-symmetry of the Kac-Moody bracket and follows from $\Tp\nabla_r = -\nabla_r$ and the identity \eqref{Eq:CasTp}.

\subsection{Gauge-fixing and Dirac brackets in affine Gaudin models}

\paragraph{Constraint and gauge-fixing in the loop algebra.} The constraint in an affine Gaudin model is given by $\Cc \approx 0$, where we see $\Cc = \sum_r \Jc_r$ as an element of $\Ac\otimes\Lc\g$, \textit{i.e.} an observable valued in the loop algebra $\Lc\g$. Recall that the levels $\ell_r$ of the currents $\Jc_r$ are taken to satisfy the condition $\sum_r \ell_r = 0$ -- see eq.\;\eqref{Eq:SumL}. In the notations introduced in the previous subsection, we then get from the Kac-Moody algebra \eqref{Eq:PbKMLoop} that $\Cc$ satisfies the bracket
\begin{equation}
\lbrace \Cc \tv \Cc \rbrace =(\Id \otimes \ad_\Cc) \Ch = - (\ad_\Cc \otimes \Id) \Ch \approx 0.
\end{equation}
This is the equivalent of eq.\;\eqref{Eq:PbCFin} in the mechanical case, but where the constraint $\Cc$ is now valued in the infinite-dimensional loop algebra $\Lc\g$. We note that in analogy with the equation \eqref{Eq:GAction} in the mechanical case, the infinitesimal gauge symmetry generated by the constraint $\Cc$ acts on an observable $f\in\Ac$ by
\begin{equation}
\delta_\epsilon f = \bigl\lbrace\, \llangle\Cc,\epsilon\rrangle, f \bigr\rbrace = \int_{0}^{2\pi} \bigl\lbrace \langle\Cc(x),\epsilon(x)\rangle, f \bigr\rbrace \;\dd x,
\end{equation}
where $\epsilon\in\Lc\g$ is the infinitesimal local parameter of the transformation.

Let us now consider a gauge-fixing condition for this model. In analogy with the mechanical case, we will suppose here that this condition can be phrased as $\Fc \equiv 0$, for some observable $\Fc \in \Ac\otimes\Lc\g$, built from the Kac-Moody currents $\Jc_r$ and also valued in the loop algebra. We then work in the gauge-fixed algebra $\Ac_{\text{GF}}$, in which we impose $\Cc \equiv \Fc \equiv 0$.

\paragraph{Dirac bracket.}  To describe the Poisson structure of $\Ac_{\text{GF}}$, one needs to define the Dirac bracket. We will do this using a formalism very close to the one developed in terms of linear operators around eq.\;\eqref{Eq:Dirac} for the mechanical case. We first introduce two linear operators $\Mc : \Lc\g \to \Lc\g$ and $\Mct : \Lc\g \to \Lc\g$, defined through the following brackets of the constraint and the gauge-fixing condition, in analogy with eq.\;\eqref{Eq:PbM}:
\begin{equation}\label{Eq:PbMLoop}
\lbrace \Cc \tv \Fc \rbrace \equiv \bigl( \Id \otimes \Mc \bigr) \Ch \qquad \text{ and } \qquad \lbrace \Fc \tv \Fc \rbrace \equiv \bigl( \Id \otimes \Mct \bigr) \Ch.
\end{equation}
One can argue that such operators always exist. We then let
\begin{equation}\label{Eq:NLoop}
\Nc = \Mc^{-1} \qquad \text{ and } \qquad \Nct = \Mc^{-1} \circ \Mct\circ \Tp\Mc^{-1},
\end{equation}
similarly to eq.\;\eqref{Eq:NM} in the mechanical case. Finally, we define the Dirac bracket as the following ``loop'' generalisation of eq.\;\eqref{Eq:Dirac}:
\begin{equation}\label{Eq:DiracLoop}
\lbrace f,g \rbrace_D \equiv \lbrace f,g \rbrace - \llangle[\bigl] \lbrace f, \Cc \rbrace, \Nc \bigl( \lbrace g,\Fc \rbrace \bigr) \rrangle[\bigr] +  \llangle[\bigl] \lbrace g, \Cc \rbrace, \Nc \bigl( \lbrace f,\Fc \rbrace \bigr) \rrangle[\bigr] - \llangle[\bigl] \lbrace f, \Cc \rbrace, \Nct \bigl( \lbrace g,\Cc \rbrace \bigr) \rrangle[\bigr],
\end{equation}
where $f$ and $g$ are observables in $\Ac$. One checks from this definition that this bracket is such that $\lbrace \Cc, f \rbrace_D \equiv \lbrace \Fc, f \rbrace_D \equiv 0$ for all $f\in\Ac$, as expected from a Dirac bracket. This ensures that $\lbrace \cdot,\cdot\rbrace_D$ is compatible with the constraints $\Cc\equiv\Fc\equiv 0$ and thus descends to the algebra $\Ac_{\text{GF}}$.

\paragraph{Linear gauge-fixing.} In all the examples studied in this article, we consider a gauge-fixing condition which is linear in the Kac-Moody currents $\Jc_r$. More precisely, this condition takes the form $\Fc \equiv 0$, with
\begin{equation}\label{Eq:LinearGF}
\Fc = \sum_{r=1}^N \Oh_r (\Jc_r),
\end{equation}
for some non-dynamical linear operators $\Oh_r: \g^\C\to\g^\C$ on the finite Lie algebra (in the examples we consider, these operators $\Oh_r$ will typically be expressed in terms of the Yang-Baxter operator $\hat R$). Starting from the Kac-Moody bracket \eqref{Eq:PbKMLoop}, we find
\begin{equation}
\lbrace \Cc \tv \Fc \rbrace = \sum_{r,s=1}^N (\Id \otimes \Oh_s) \lbrace \Jc_r \tv \Jc_s \rbrace = \sum_{r=1}^N \bigl( \Id \otimes (\Oh_r \circ \nabla_r)  \bigr) \Ch.
\end{equation}
Similarly,
\begin{equation}
\lbrace \Fc \tv \Fc \rbrace = \sum_{r,s=1}^N (\Oh_r \otimes \Oh_s) \lbrace \Jc_r \tv \Jc_s \rbrace = \sum_{r=1}^N \bigl( \Id \otimes (\Oh_r \circ  \nabla_r \circ \Tp\Oh_r) \bigr) \Ch,
\end{equation}
where we used the identity \eqref{Eq:CasTp}. Comparing to eq.\;\eqref{Eq:PbMLoop}, we find
\begin{equation}\label{Eq:MO}
\Mc = \sum_{r=1}^N  \Oh_r \circ \nabla_r \qquad \text{ and } \qquad \Mct = \sum_{r=1}^N \Oh_r \circ  \nabla_r \circ \Tp\Oh_r.
\end{equation}
In principle, we then have all the necessary information to determine the operators $\Nc$ and $\Nct$ in eq.\;\eqref{Eq:NLoop} and thus write down the Dirac bracket \eqref{Eq:DiracLoop}.

\paragraph{Dirac bracket of Kac-Moody currents.} Let us consider two Kac-Moody currents $\Jc_r$ and $\Jc_s$. We would like to compute their Dirac bracket $\lbrace \Jc_r \tv \Jc_s \rbrace_D$ under the linear gauge-fixing introduced above. Since these currents are $\Lc\g$-valued observables themselves, the intermediate computations in the Dirac bracket will involve objects in a 3-fold tensor product $\Lc\g\otimes\Lc\g\otimes\Lc\g$. To simplify the presentation, we will denote, for any $A\in\Lc\g$,
\begin{equation}
A\ti{1} = A \otimes \Id \otimes \Id, \qquad A\ti{2} = \Id \otimes A \otimes \Id, \qquad A\ti{3} = \Id \otimes \Id \otimes A.
\end{equation}
We will use similar notations to express on which tensor factor in $\Lc\g^{\otimes 3}$ linear operators act. Also, we will denote the three possible embeddings of $\Ch\in\Lc\g\otimes\Lc\g$ in $\Lc\g^{\otimes 3}$ as $\Cht{12}$, $\Cht{13}$ and $\Cht{23}$, dropping the index $2$ to lighten the notations. Finally, we indicate by indices $\underline{i}$ along which tensor factors bilinear forms $\llangle\cdot,\cdot\rrangle$ are applied. With these conventions, eq.\;\eqref{Eq:DiracLoop} implies
\begin{align}
\lbrace \Jc_r\null\ti{1}, \Jc_s\null\ti{2} \rbrace_D &\equiv \lbrace \Jc_r\null\ti{1}, \Jc_s\null\ti{2} \rbrace - \llangle[\bigl] \lbrace \Jc_r\null\ti{1}, \Cc\null\ti{3} \rbrace, \Nc\ti{3} \bigl( \lbrace \Jc_s\null\ti{2}, \Fc\null\ti{3} \rbrace \bigr) \rrangle[\bigr]\ti{3} \\
& \hspace{20pt} + \llangle[\bigl] \lbrace \Jc_s\null\ti{2}, \Cc\null\ti{3} \rbrace, \Nc\ti{3} \bigl( \lbrace \Jc_r\null\ti{1}, \Fc\null\ti{3} \rbrace \bigr) \rrangle[\bigr]\ti{3} - \llangle[\bigl] \lbrace \Jc_r\null\ti{1}, \Cc\null\ti{3} \rbrace, \Nct\ti{3} \bigl( \lbrace \Jc_s\null\ti{2}, \Cc\null\ti{3} \rbrace \bigr) \rrangle[\bigl]\ti{3}. \notag
\end{align}
We then use
\begin{equation}
\lbrace \Jc_r\null\ti{i}, \Cc\null\ti{3} \rbrace = - \nabla_r\null\ti{i}\;\Cht{i3} \qquad \text{ and } \qquad \lbrace \Jc_r\null\ti{i}, \Fc\null\ti{3} \rbrace = - (\nabla_r\circ\Tp\Oh_r)\ti{i}\;\Cht{i3},
\end{equation}
where we used the identity \eqref{Eq:CasTp}. By repeated use of this identity again, we can move all operators in the above Dirac bracket to the first and second tensor spaces, yielding
\begin{align}
\lbrace \Jc_r\null\ti{1}, \Jc_s\null\ti{2} \rbrace_D &\equiv \lbrace \Jc_r\null\ti{1}, \Jc_s\null\ti{2} \rbrace - \nabla_r\null\ti{1}(\nabla_s\circ\Tp\Oh_s\circ\Tp\Nc)\ti{2} \, \llangle[\bigl] \Cht{13}, \Cht{23}  \rrangle[\bigr]\ti{3} \\
& \hspace{20pt} + (\nabla_r\circ\Tp\Oh_r\circ\Tp\Nc)\ti{1}\nabla_s\null\ti{2} \, \llangle[\bigl] \Cht{23}, \Cht{13}  \rrangle[\bigr]\ti{3} + \nabla_r\null\ti{1}(\nabla_s\circ\Nct)\ti{2} \, \llangle[\bigl] \Cht{13}, \Cht{23}  \rrangle[\bigr]\ti{3}, \notag
\end{align}
where we also used $\Tp\Nct=-\Nct$. Using the property \eqref{Eq:LoopCas} of the loop Casimir, we then get
\begin{align}
\lbrace \Jc_r\null\ti{1}, \Jc_s\null\ti{2} \rbrace_D &\equiv \lbrace \Jc_r\null\ti{1}, \Jc_s\null\ti{2} \rbrace -  \nabla_r\null\ti{1}(\nabla_s\circ\Tp\Oh_s\circ\Tp\Nc)\ti{2}\,\Cht{12}  \\
& \hspace{20pt} + (\nabla_r\circ\Tp\Oh_r\circ\Tp\Nc)\ti{1}\nabla_s\null\ti{2}\;\Cht{12} + \nabla_r\null\ti{1}(\nabla_s\circ\Nct)\ti{2}\; \Cht{12}. \notag
\end{align}
We can now drop the notation with tensor indices $\underline{i}$ and treat this bracket as we dealt with $\Lc\g\otimes\Lc\g$-valued bracket before. We then rewrite the above equation as
\begin{equation}\label{Eq:DiracKM}
\lbrace \Jc_r \tv \Jc_s \rbrace_D \equiv \lbrace \Jc_r \tv \Jc_s \rbrace +  ( \nabla_r \otimes \nabla_s) \Th^{(rs)},
\end{equation}
where
\begin{equation}
\Th^{(rs)} = \bigl( \Id \otimes (\Nc\circ\Oh_r) - (\Nc\circ\Oh_s) \otimes \Id + \Id\otimes\Nct \bigr)\Ch.
\end{equation}
The computation of the Dirac bracket then mostly amounts to determining this quantity $\Th^{(rs)}$. Recall that $\Nc = \Mc^{-1}$ and $\Nct=\Mc^{-1}\circ\Mct\circ\Tp\Mc^{-1}$. Acting with $\Mc\otimes \Mc$ on the above definition, we see that $\Th^{(rs)}$ is characterised by the equation
\begin{equation}\label{Eq:Trs}
(\Mc\otimes\Mc) \Th^{(rs)} = \bigl( \Mc \otimes \Oh_r - \Oh_s \otimes \Mc + \Id \otimes \Mct \bigr)\Ch.
\end{equation}

\subsection{Parafermionic gauge-fixing}

\paragraph{Parafermionic gauge-fixing condition.} Let us now apply the formalism developed above to the explicit examples of gauge-fixings treated in the main text, for which we consider an affine Gaudin model with $N=3$ Kac-Moody currents. We start with the parafermionic gauge-fixing condition \eqref{Eq:ParaGF}, which we recall here for the reader's convenience\footnote{In the main text, the Kac-Moody currents were denoted as $\Jc_r^\cL$ to emphasize that they correspond to a left-moving affine Gaudin model. To lighten the notations, we have dropped the labels $\cL$ in this appendix.}:
\begin{equation}\label{Eq:ParaGFApp}
\Fc_P = \bigl( \hat{R} - \ri) \Jc_1 - \bigl( \hat{R} + \ri) \Jc_2 \gf 0,
\end{equation}
where $\hat R$ is the standard $R$-matrix \eqref{Rmat1a}. This takes the form of a linear gauge-fixing \eqref{Eq:LinearGF} as considered in the previous subsection, with the operators $\Oh_r$ chosen here to be
\begin{equation}\label{Eq:ParaO}
\Oh_1 = \hat R^-, \qquad \Oh_2 = -\hat R^+ \qquad \text{ and } \qquad \Oh_3 = 0,
\end{equation}
where
\begin{equation}
\hat R^\pm = \hat R \pm \ri\,.
\end{equation}

Let us introduce $\Jc=(\Jc_1-\Jc_2)/2\ri$. The gauge-fixing condition \eqref{Eq:ParaGFApp} and the constraint $\Jc_1+\Jc_2+\Jc_3\equiv 0$ then imply that
\begin{equation}\label{Eq:JK}
\Jc_1 = \hat R^+ \Jc, \qquad \Jc_2 = \hat R^-\Jc \qquad \text{ and } \qquad \Jc_3 = -2\hat R\Jc.
\end{equation}
In particular, $\Jc_1$ and $\Jc_2$ are valued respectively in the Borel subalgebras $\mathfrak{h}\oplus\mathfrak{n}_-$ and $\mathfrak{h}\oplus\mathfrak{n}_+$, while $\Jc_3$ is valued in $\mathfrak{n}_+\oplus\mathfrak{n}_-$.

It will also be useful for us to define the projector $\hat\Pi$ on the Cartan subalgebra $\mathfrak{h}$ in the decomposition $\g^\C = \mathfrak{h} \oplus \mathfrak{n}_+ \oplus \mathfrak{n}_-$. One easily checks that this operator can be expressed in terms of the $R$-matrix as
\begin{equation}
\hat\Pi=\hat R^2+\Id.
\end{equation}
Moreover, it satisfies $\hat\Pi^2=\hat\Pi$ and $\tp\hat\Pi=\hat\Pi$. In particular, this operator can be used to project the split quadratic Casimir $\mathsf{C}_2 \in \g\otimes\g$ of the finite algebra on $\mathfrak{h}\otimes\mathfrak{h}$, defining
\begin{equation}\label{Eq:CasH}
\Hd = (\hat\Pi\otimes\Id) \mathsf{C}_2 = (\Id\otimes\hat\Pi) \mathsf{C}_2.
\end{equation}

\paragraph{Operators $\bm{\Mc}$ and $\bm{\Mct}$.} For the choice of $\Oh_r$ in eq.\;\eqref{Eq:ParaO}, corresponding to the parafermionic gauge-fixing, the operators $\Mc$ and $\Mct$ (given in general by eq.\;\eqref{Eq:MO}) take the form
\begin{subequations}\label{Eq:ParaM}
\begin{align}
\Mc &= \hat R^- \circ \nabla_1 - \hat R^+ \circ \nabla_2, \\
\Mct &= -\hat R^- \circ \nabla_1 \circ \hat R^+ - \hat R^+ \circ \nabla_2 \circ \hat R^-,
\end{align}
\end{subequations}
where we have used $\Tp\hat R^\pm = -\hat R^\mp$. Recall that $\nabla_r = \ell_r \p + \ad_{\Jc_r}$. We then rewrite $\Mct$ as
\begin{equation}
\Mct = - (\ell_1+\ell_2)\hat R^+ \circ \hat R^-\circ \p - \hat R^- \circ \ad_{\hat R^+\Jc} \circ \hat R^+ - \hat R^+ \circ \ad_{\hat R^-\Jc} \circ \hat R^-,
\end{equation}
where we used the gauge-fixed expression \eqref{Eq:JK} of $\Jc_1$ and $\Jc_2$. Using the action \eqref{Rmat1a} of $\hat R$ on the decomposition $\g^{\C} = \mathfrak{h} \oplus \mathfrak{n}_+ \oplus \mathfrak{n}_-$ and the commutation relations $[\mathfrak{h},\mathfrak{h}]=0$ and $[\mathfrak{h}, \mathfrak{n}_\pm] \subset \mathfrak{n}_\pm$, one finds that $\hat R^- \circ \ad_{\hat R^+\Jc} \circ \hat R^+ = \hat R^+ \circ \ad_{\hat R^-\Jc} \circ \hat R^- =  0$. Moreover, one has $\hat R^+\circ\hat R^- = \hat \Pi$. We then simply get
\begin{equation}\label{Eq:Mprime}
\Mct = -(\ell_1+\ell_2)\hat\Pi \circ\p.
\end{equation}
Similarly, we have
\begin{equation}
\Mc = (\ell_1 \hat R^- - \ell_2 \hat R^+)\circ\p + \hat R^- \circ \ad_{\hat R^+\Jc} - \hat R^+ \circ \ad_{\hat R^-\Jc}.
\end{equation}
Reasoning as above, we find that the operators $\hat R^\pm \circ \ad_{\hat R^\mp\Jc}$ act trivially on the Cartan subalgebra $\mathfrak{h}$. Moreover, $\hat R^\pm$ acts as $\pm\ri\,\Id$ on $\mathfrak{h}$. We thus have
\begin{equation}\label{Eq:MCartan}
\Mc A = -\ri(\ell_1+\ell_2)\p A, \qquad \forall \, A\in\Lc\mathfrak{h},
\end{equation}
where $\Lc\mathfrak{h}$ is formed by the elements in the loop algebra $\Lc\g$ valued in the Cartan subalgebra $\mathfrak{h}$. These identities will be useful to compute the Dirac brackets below.

\paragraph{Dirac bracket of $\bm{\Jc_3}$.} Let us now compute the Dirac bracket of $\Jc_3$ with itself under this gauge-fixing. From the general result \eqref{Eq:DiracKM} found in the previous subsection, we have
\begin{equation}
\lbrace \Jc_3 \tv \Jc_3 \rbrace_P \equiv \lbrace \Jc_3 \tv \Jc_3 \rbrace +  ( \nabla_3 \otimes \nabla_3) \Th^{(33)},
\end{equation}
where the tensor $\Th^{(33)}$ is characterised by eq.\;\eqref{Eq:Trs}. In the case at hand, since $\Oh_3=0$ and $\Mct$ is given by \eqref{Eq:Mprime}, the latter simply takes the form
\begin{equation}\label{Eq:T33}
(\Mc\otimes\Mc) \Th^{(33)} = -(\ell_1+\ell_2) \bigl( \Id \otimes \hat\Pi\,\p \bigr)\Ch.
\end{equation}
We now want to solve this equation for $\Th^{(33)}$. From the functional expression \eqref{Eq:CasDelta} of $\Ch$, we find that the evaluation of the right-hand side at $(x,y)\in S^1\times S^1$ is $(\ell_1+\ell_2)\,\Hd\,\p_x\delta(x-y)$, with the tensor $\Hd$ defined as in eq.\;\eqref{Eq:CasH}. It is then natural to take an ansatz for $\Th^{(33)}$ which is itself proportional to $\Hd$. More precisely, we will take the following ansatz:
\begin{equation}
\Th^{(33)}(x,y) = \Hd\,f(x,y),
\end{equation}
where $f(x,y)$ is a distribution to be determined. Recall from eq.\;\eqref{Eq:MCartan} that $\Mc$ acts very simply on Cartan-valued functions. The above equation on $\Th^{(33)}$ then becomes
\begin{equation}
-(\ell_1+\ell_2)^2 \, \Hd\,\p_x\p_y f(x,y) = (\ell_1+\ell_2)\,\Hd\, \p_x\delta(x-y).
\end{equation}
To solve for $f$, we introduce the $\epsilon$-distribution, which satisfies
\begin{equation}\label{Eq:DerEpsilon}
\p_x \epsilon(x-y) = -\p_y \epsilon(x-y) = 2\delta(x-y).
\end{equation}
It is then clear that
\begin{equation}
f(x,y) = \frac{1}{2(\ell_1+\ell_2)} \epsilon(x-y) = \frac{1}{2\Kuv} \epsilon(x-y),
\end{equation}
where the last equality follows from the explicit expression \eqref{Eq:LevelsL} of the levels $\ell_r$ in the affine Gaudin model considered here. To summarise, we then get that the Dirac bracket of $\Jc_3$ simply reads
\begin{equation}\label{Eq:DiracJ3}
\lbrace \Jc_3 \tv \Jc_3 \rbrace_P \equiv \lbrace \Jc_3 \tv \Jc_3 \rbrace + \frac{1}{2\Kuv} ( \nabla_3 \otimes \nabla_3)\, \Hd\,\epsilon,
\end{equation}
where by $\epsilon$ we mean the formal distribution whose evaluation at $(x,y)$ is $\epsilon(x-y)$.

\paragraph{Analysis of the Dirac bracket of $\bm{\Jc_3}$.} Let us now analyse the Dirac bracket \eqref{Eq:DiracJ3}. Evaluating it at $(x,y)$ and reinserting $\nabla_3 = \ell_3\,\p + \ad_{\Jc_3} = - \Kuv\p + \ad_{\Jc_3}$, we get
\begin{equation}
\lbrace \Jc_3(x) \tv \Jc_3(y) \rbrace_P \equiv \lbrace \Jc_3(x) \tv \Jc_3(y) \rbrace + \frac{1}{2\Kuv} \bigl( \Kuv\p_x - \ad_{\Jc_3(x)} \bigr) \otimes \bigl( \Kuv\p_y - \ad_{\Jc_3(y)} \bigr)\, \Hd\,\epsilon(x-y).
\end{equation}
Writing down the initial Kac-Moody bracket of $\Jc_3$ in the first term and developing the various derivatives in the second one, using the identity \eqref{Eq:DerEpsilon} obeyed by the $\epsilon$-distribution, we get
\begin{align}\label{Eq:DiracJ3b}
\lbrace \Jc_3(x) \tv \Jc_3(y) \rbrace_P &\gf \Bigl( \bigl( \Id\otimes\ad_{\Jc_3(x)}\bigr)\mathsf{C}_2 - \bigl( \Id\otimes\ad_{\Jc_3(x)}\bigr) \Hd + \bigl( \ad_{\Jc_3(x)}\otimes\Id\bigr) \Hd \Bigr) \delta(x-y)  \notag\\
& \hspace{20pt} + \Kuv \bigl( \mathsf{C}_2 - \Hd \bigr) \p_x\delta(x-y) + \frac{1}{2\Kuv} \bigl( \ad_{\Jc_3(x)} \otimes \ad_{\Jc_3(y)} \bigr) \Hd\,\epsilon(x-y).
\end{align}
Recall that the gauge-fixing condition \eqref{Eq:ParaGFApp} implies that $\Jc_3=-2\hat R\Jc$ -- see eq.\;\eqref{Eq:JK} -- and thus that $\Jc_3$ is valued in $\mathfrak{n}_+\oplus\mathfrak{n}_-$. One checks that the form of the above bracket ensures that the right-hand side is valued in $(\mathfrak{n}_+\oplus\mathfrak{n}_-)\otimes(\mathfrak{n}_+\oplus\mathfrak{n}_-)$. For instance, the shift by $-\Hd$ in the term proportional to $\p_x\delta(x-y)$ removes the part valued in the Cartan subalgebra. One shows a similar statement for the remaining terms using $[\mathfrak{h},\mathfrak{h}]=0$ and $[\mathfrak{h}, \mathfrak{n}_\pm] \subset \mathfrak{n}_\pm$. This shows the consistency of the gauge-fixing procedure and the computation of the Dirac bracket and is an alternative check of the vanishing of the Dirac bracket on constraints.

As explained in eq.\;\eqref{Eq:BPsi}, we parametrise $\Jc_3$ as $-\sum_{\alpha\in\Delta} \Psi_\alpha\,\tt e_{-\alpha}$, where $\tt e_\alpha$ are root vectors of the algebra $\g^\C$ (see Appendix \ref{App:Root} for conventions). Recall also that we introduced an orthogonal basis $\lbrace\tt h_i\rbrace$ of the Cartan subalgebra $\mathfrak{h}$, with normalisation $\langle {\tt h}_i, {\tt h}_j \rangle = -2\,\delta_{ij}$. We then have
\begin{equation}
\mathsf{C}_2 = -\frac{1}{2} \sum_{i=1}^N {\tt h}_i \otimes {\tt h}_i - \sum_{\alpha\in\Delta} {\tt e}_\alpha \otimes {\tt e}_{-\alpha} \qquad { \text{ and} } \qquad
\Hd = -\frac{1}{2}\, \sum_{i=1}^N {\tt h}_i \otimes {\tt h}_i\,.
\end{equation}
One can then extract the Poisson brackets of the components $\Psi_\alpha$ from the Dirac bracket \eqref{Eq:DiracJ3b}, using the commutation relations \eqref{Eq:ComHE} of the basis $\lbrace \tt e_\alpha, \tt h_i \rbrace$ and the relation \eqref{Eq:RootForm}. We then get
\begin{align}
\bigl\lbrace \Psi_\alpha(x), \Psi_\beta(y) \bigr\rbrace_P &\gf  N^{\alpha,\beta} \, \Psi_{\alpha+\beta}(x)\, \delta(x-y) - \Kuv\,\delta_{\alpha+\beta,0}\,\p_x\delta(x-y) \\
& \hspace{23pt} - \frac{(\alpha,\beta)}{2\Kuv}
\Psi_\alpha(x) \, \Psi_\beta(y) \,\epsilon(x-y) \, . \notag
\end{align}

\paragraph{Dirac bracket of Cartan fields.} Recall that $\Jc_3$ has no Cartan component and is expressed only in terms of the fields $\Psi_\alpha$ associated with root vectors $\tt e_\alpha$. On the other hand, the gauge-fixed current $\Jc_1$ and $\Jc_2$ are in the Borel subalgebras $\mathfrak{h}\oplus\mathfrak{n}_\mp$ and in particular contain fields $D_i$ associated with the Cartan basis $\lbrace \tt h_i \rbrace$ -- see eq.\;\eqref{Eq:BPsi}. Let us then introduce
\begin{equation}\label{Eq:D}
\Dc = \hat\Pi\, \Jc_1 = \frac{1}{2} \sum_{i=1}^{\dim\mathfrak{h}}D_i\,\tt h_i.
\end{equation}
We now want to compute the Dirac brackets of $\Dc$ with itself and the current $\Jc_3$. This can be done in a similar, although slightly more complicated, way than the Dirac brackets computed above. For conciseness, we will not enter into the details of these computations here. Let us mention however a few useful identities, which serve as important steps in the determination of these Dirac brackets. First of all, combining the equations \eqref{Eq:Mprime} and \eqref{Eq:MCartan}, we find
\begin{equation}
\Mct = -\ri\,\Mc\circ\hat\Pi = -(\ell_1+\ell_2)\,\hat\Pi\circ\p.
\end{equation}
Together with the definition \eqref{Eq:NLoop} of $\Nc$ and $\Nct$, this implies
\begin{equation}
\Nct = -\ri\, \hat\Pi\circ\Tp\Nc = \ri\,\Nc\circ\hat\Pi \qquad \text{ and } \qquad \Nc\circ\hat\Pi\circ\p = \frac{\ri}{\ell_1+\ell_2} \hat\Pi\,.
\end{equation}
Moreover, using the fact that the gauge-fixed current $\Jc_1$ is valued in $\mathfrak{h}\oplus\mathfrak{n}_-$ with the commutation relations $[\mathfrak{h},\mathfrak{h}]=0$ and $[\mathfrak{h}, \mathfrak{n}_\pm] \subset \mathfrak{n}_\pm$, one finds
\begin{equation}
\hat R^- \circ \ad_{\Jc_1} \circ \hat\Pi \gf \hat\Pi \circ \ad_{\Jc_1} \circ \hat \Pi \gf 0.
\end{equation}
Finally, the following Poisson brackets are useful:
\begin{equation}
\lbrace \Dc \tv \Cc \rbrace \gf - \bigl( (\hat\Pi \circ \nabla_1) \otimes \Id \bigr) \Ch \qquad \text{ and } \qquad \lbrace \Dc \tv \Fc \rbrace \gf -\ri \,\ell_1\bigl( \Id\otimes(\hat\Pi\circ\p) \bigr) \Ch.
\end{equation}
Combining these results, one computes the Dirac brackets involving $\Dc$. In particular, we find
\begin{equation}
\bigl\lbrace \Dc(x) \tv \Jc_3(y) \bigr\rbrace_P \gf 0 \qquad \text{ and } \qquad \bigl\lbrace \Dc(x) \tv \Dc(y) \bigr\rbrace_P \gf - \frac{\nu^2 \Kuv}{(1+\nu^2)^2} \,\Hd\,\p_x \delta(x-y)\,,
\end{equation}
where the coefficient in the second equation originates from
\begin{equation}
\frac{\ell_1\ell_2}{\ell_1+\ell_2} = \frac{\nu^2 \Kuv}{(1+\nu^2)^2}.
\end{equation}
Using the decomposition \eqref{Eq:D} of $\Dc$ in the basis $\lbrace\tt h_i\rbrace$ and $\Hd = - \frac{1}{2} \sum_i\tt h_i\otimes \tt h_i$, we finally get
\begin{equation}
\bigl\lbrace D_i(x) , \Psi_\alpha(y) \bigr\rbrace_P \gf 0 \qquad \text{ and } \qquad \bigl\lbrace D_i(x) , D_j(y) \bigr\rbrace_P \gf \frac{2\nu^2 \Kuv}{(1+\nu^2)^2}\,\delta_{ij} \,\p_x \delta(x-y) \,.
\end{equation}

\subsection[$\omega$-gauge fixing]{\texorpdfstring{$\bm\omega$}{Omega}-gauge-fixing}

\paragraph{Gauge-fixing condition.} Let us now consider the $\omega$-gauge-fixing condition \eqref{Eq:OmegaGF}. Dropping the $\cL$ labels as in the previous subsection, we will write it in this appendix as
\begin{equation}\label{Eq:OmegaGFApp}
\Fc_\omega = \bigl( \hat R^-\! + \ri\, \mu\, \hat\Pi \bigr) \Jc_2 - \bigl( \hat R^+ + \ri\,\omega\,\hat\Pi \bigr) \Jc_3 \gf 0,
\end{equation} 
where we recall that $\hat R^\pm = \hat R \pm \ri$ and we introduced
\begin{equation}\label{Eq:Mu}
\mu = 2+\nu^2-(1+\nu^2)\omega.
\end{equation}
Here, $\omega$ is a constant parameter and $\hat\Pi$ is the projection on the Cartan subalgebra $\mathfrak{h}$, as above. This gauge-fixing condition implies that the currents $\Jc_2$ and $\Jc_3$ are valued in the Borel subalgebras $\mathfrak{h}\oplus\mathfrak{n}_-$ and $\mathfrak{h}\oplus\mathfrak{n}_+$ respectively. In contrast, the current $\Jc_1$ has non-trivial components in all the factors of the decomposition $\g^{\C} = \mathfrak{h}\oplus\mathfrak{n}_+\oplus\mathfrak{n}_-$, except when $\omega=1+2\nu^{-2}$ in which case $\Jc_1$ has no Cartan component. Outside of this case, all the degrees of freedom in the gauge-fixed currents $\Jc_r$ are then contained in $\Jc_1$ and we can compute the corresponding gauge-fixed Poisson structure by computing the Dirac bracket of $\Jc_1$ with itself.

The condition \eqref{Eq:OmegaGFApp} takes the form of a linear gauge fixing \eqref{Eq:LinearGF}, as considered in the previous subsections, with the operators $\Oh_r$ given by
\begin{equation}
\Oh_1 = 0, \qquad \Oh_2 = \hat R^-\! + \ri\,\mu\, \hat\Pi, \qquad \Oh_3 = -\hat R^+-\ri\,\omega\,\hat\Pi.
\end{equation}
From this data and eq.\;\eqref{Eq:MO}, we can compute the operators $\Mc$ and $\Mct$. By a reasoning similar to the one detailed in the previous subsection for the parafermionic gauge (see eq.\;\eqref{Eq:ParaM} and the discussion below), one finds
\begin{equation}\label{Eq:MPrimeOmega}
\Mct = -\bigl( \ell_2(1-\mu)^2 +\ell_3(1+\omega)^2 \bigr)\hat\Pi \circ\p
\end{equation}
and
\begin{equation}\label{Eq:MCartanOmega}
\Mc A = -\ri\bigl ( \ell_2(1-\mu)+\ell_3(1+\omega) \bigr)\p A, \qquad \forall \, A\in\Lc\mathfrak{h}.
\end{equation}

\paragraph{Dirac bracket of $\bm{\Jc_1}$.} Let us then compute the Dirac bracket of $\Jc_1$ in the $\omega$-gauge. Applying the general result \eqref{Eq:DiracKM}, this bracket is given by
\begin{equation}\label{Eq:DiracJ1}
\lbrace \Jc_1 \tv \Jc_1 \rbrace_\omega \equiv \lbrace \Jc_1 \tv \Jc_1 \rbrace +  ( \nabla_1 \otimes \nabla_1) \Th^{(11)},
\end{equation}
where the tensor $\Th^{(11)}$ is characterised by eq.\;\eqref{Eq:Trs}. In the present case, since $\Oh_1=0$ and $\Mct$ is given by \eqref{Eq:MPrimeOmega}, this equation simply takes the form
\begin{equation}
(\Mc\otimes\Mc) \Th^{(11)} = -\bigl( \ell_2(1-\mu)^2 +\ell_3(1+\omega)^2 \bigr) \bigl( \Id \otimes \hat\Pi\,\p \bigr)\Ch.
\end{equation}
The resolution of this equation is very similar to the one of eq.\;\eqref{Eq:T33} in the parafermionic gauge, using an ansatz for $\Th^{(11)}$ proportional to $\Hd$ together with the equation \eqref{Eq:MCartanOmega}. In the end we find
\begin{equation}
\Th^{(11)} = -\frac{\vartheta}{2}\, \Hd \, \epsilon,
\end{equation}
where $\epsilon(x,y)=\epsilon(x-y)$ and
\begin{equation}
\vartheta = -\frac{\ell_2(1-\mu)^2 +\ell_3(1+\omega)^2}{\bigl(\ell_2(1-\mu)+\ell_3(1+\omega))^2} = \frac{1}{\Kuv} \left( \omega - \frac{\nu^2}{4}(1-\omega)^2 \right) .
\end{equation}
The second equality in the above equation comes from the definition \eqref{Eq:Mu} of $\mu$ and the expression \eqref{Eq:LevelsL} of the levels $\ell_r$ for the affine Gaudin model under consideration here.

\paragraph{Analysis of the Dirac bracket of $\bm{\Jc_1}$.} Reinserting the expression found above for $\Th^{(11)}$ in eq.\;\eqref{Eq:DiracJ1}, one writes down the Dirac bracket of $\Jc_1$ with itself in the $\omega$-gauge. More precisely, we get
\begin{align}\label{Eq:DiracJ1b}
\lbrace \Jc_1(x) \tv \Jc_1(y) \rbrace_\omega &\gf \Bigl( \bigl( \Id\otimes\ad_{\Jc_1(x)}\bigr)\mathsf{C}_2 - \vartheta\ell_1 \bigl( \Id\otimes\ad_{\Jc_1(x)}\bigr) \Hd + \vartheta\ell_1 \bigl( \ad_{\Jc_1(x)}\otimes\Id\bigr) \Hd  \Bigr) \delta(x-y)  \notag\\
& \hspace{20pt} - \ell_1 \bigl( \mathsf{C}_2 - \vartheta \ell_1 \,\Hd \bigr) \p_x\delta(x-y) - \frac{\vartheta}{2} \bigl( \ad_{\Jc_1(x)} \otimes \ad_{\Jc_1(y)} \bigr) \Hd\,\epsilon(x-y).
\end{align}
Let us recall that in the main text, we decompose $\Jc_1$ in the basis $\lbrace\tt h_i, \tt e_\alpha \rbrace$ as in eq.\;\eqref{Eq:J1Xi}, which we recall here for the reader's convenience:
\begin{equation}
\Jc_1 \gf -\frac{1}{2} \sum_{i=1}^{\dim\mathfrak{h}} \Xi_i(x)\,{\tt h_i} - \sum_{\alpha\in\Delta} \Xi_\alpha(x)\,{\tt e}_{-\alpha}\,.
\end{equation}
Reinserting in the above Dirac bracket and using the commutation relations \eqref{Eq:ComHE} and the expression of $\mathsf{C}_2$ and $\Hd$ in the basis $\lbrace\tt h_i, \tt e_\alpha \rbrace$, one extracts the Poisson bracket of the components $\Xi_i$ and $\Xi_\alpha$. More precisely, we find
\begin{subequations}
\begin{align}
\bigl\lbrace \Xi_i(x), \Xi_\alpha(y) \bigr\rbrace_\omega &\gf  \bigl(1-\vartheta\ell_1\bigr)\alpha({\tt h}_i) \,\Xi_\alpha(x) \delta(x-y) \,, \\
\bigl\lbrace \Xi_i(x), \Xi_j(y) \bigr\rbrace_\omega &\gf  2\ell_1\bigl(1-\vartheta\ell_1\bigr)\,\delta_{ij}\,\p_x\delta(x-y) \,, \\
\bigl\lbrace \Xi_\alpha(x), \Xi_\beta(y) \bigr\rbrace_\omega &\gf  N^{\alpha,\beta} \, \Xi_{\alpha+\beta}(x)\, \delta(x-y)   + \frac{1}{2} \, \vartheta \, (\alpha,\beta)\,\Xi_\alpha(x) \, \Xi_\beta(y) \,\epsilon(x-y) \, , \\
\bigl\lbrace \Xi_\alpha(x), \Xi_{-\alpha}(y) \bigr\rbrace_\omega &\gf  \rho_\alpha^i \,\Xi_i(x) + \ell_1\,\p_x\delta(x-y) - \frac{1}{2}\,\vartheta\,(\alpha,\alpha)\, \Xi_\alpha(x) \, \Xi_{-\alpha}(y) \,\epsilon(x-y) \, .
\end{align}
\end{subequations}

\section{Quantum \texorpdfstring{${\bm{\widehat{\mathfrak{sl}}(2)\oplus\widehat{\mathfrak{sl}}(2)/\widehat{\mathfrak{sl}}(2)}}$}{W} coset \texorpdfstring{$\bm\Wc$}{W}-algebra}
\label{App:QuantumW}

Our goal in this appendix is to describe the first fields in the quantum $\widehat{\mathfrak{sl}}(2)\oplus\widehat{\mathfrak{sl}}(2)/\widehat{\mathfrak{sl}}(2)$ coset $\Wc$-algebra in terms of the $\mathfrak{sl}(2)$-Kac-Moody currents $\Jq_1$ and $\Jq_2$.

\paragraph{Segal-Sugawara fields and energy-momentum tensor.} Let us define
\begin{equation}
\mathsf{G}_r(x) = \frac{1}{2} \frac{\eta^{ab}}{k_r+2} \nor{\Jq_{r,a}(x)\,\Jq_{r,b}(x)} \hspace{15pt} \text{ and } \hspace{15pt} \mathsf{G}_{\diag}(x) = \frac{1}{2} \frac{\eta^{ab}}{k_1+k_2+2} \nor{\Jq_{\diag,a}(x)\,\Jq_{\diag,b}(x)} \,,
\end{equation}
for $r\in\lbrace 1,2 \rbrace$. We recognise in this expression the Segal-Sugawara fields associated with the Kac-Moody currents $\Jq_r$ and $\Jq_{\diag}$ (note that the dual Coxeter number of $\mathfrak{sl}(2)$ is given by $\hv=2$). In particular, in terms of these fields, the energy-momentum tensor $\Wq_2$ of the $\widehat{\mathfrak{sl}}(2)\oplus\widehat{\mathfrak{sl}}(2)/\widehat{\mathfrak{sl}}(2)$ coset $\Wc$-algebra, defined in eq.\;\eqref{Eq:QuantumW2}, can simply be rewritten as
\begin{equation}
\Wq_2(x) = \mathsf{G}_1(x) + \mathsf{G}_2(x) - \mathsf{G}_{\diag}(x).
\end{equation}
As explained in the main text, this field $\Wq_2$ is the quantisation of the classical spin 2 current defined in eq.\;\eqref{Eq:W2}.

\paragraph{Quantum spin 4 current.} We now want to describe the quantisation of the classical spin 4 current \eqref{mncxnbsad}. It will be useful for that to first rewrite this classical current in a more concrete form. Let us then introduce the classical commutator field $\mathcal{U} = \bigl[ \Jc_1, \Jc_2 ] = \fu bca \,\Jc_{1,b}\,\Jc_{2,c}\,{\tt t}^a$, where $\fu bca$ are the structure constants of the dual basis $\lbrace {\tt t}^a \rbrace$ of the Lie algebra. In terms of this field, the covariant derivative of the current $\Kc$ considered in Subsection \ref{aaaaaaaaaaaa} reads
\begin{equation}
\nabla_{\!x}\Kc(x) = \ell_2 \p_x \Jc_1(x) - \ell_1 \p_x \Jc_2(x) - \mathcal{U}(x).
\end{equation}
In particular, the classical spin 4 W-current defined in eq.\;\eqref{mncxnbsad} can then be expressed as
\begin{align}\label{Eq:W4Class}
W_4 = & \; \frac{\eta^{ab}}{2\ell_1\ell_2(\ell_1+\ell_2)} \left( \ell_2^2 \, \p\Jc_{1,a}\,\p\Jc_{1,b} + \ell_1^2\, \p\Jc_{2,a}\,\p\Jc_{2,b} - 2\ell_1\ell_2\, \p\Jc_{1,a}\,\p\Jc_{2,b} \right.\\
& \hspace{160pt} \left. - 2\ell_2\, \p\Jc_{1,a}\,\mathcal{U}_b + 2\ell_1\, \p\Jc_{2,a}\,\mathcal{U}_b + \mathcal{U}_a\,\mathcal{U}_b \right)\,. \notag
\end{align}
We define the quantum equivalent of $\mathcal{U}_a$ as
\begin{equation}
\Uq_a(x) = \ri\,\fu bca \nor{\Jq_{1,b}(x)\Jq_{2,c}(x)} \, .
\end{equation}
We then search for the quantum spin 4 W-current in the following form
\begin{align}\label{Eq:W4Quant}
\Wq_4 &= \eta^{ab} \left( \lambda^{(2)}_1 \nor{\p \Jq_{1,a}\,\p \Jq_{1,b}} + \,\lambda^{(2)}_2 \nor{\p \Jq_{2,a}\,\p \Jq_{2,b}} +\, \lambda^{(2)}_3 \nor{\p \Jq_{1,a}\,\p \Jq_{2,b}} + \,\lambda^{(4)} \nor{\Uq_a\,\Uq_b} \right.\\
& \hspace{52pt} \left.  + \,\lambda^{(3)}_1 \nor{\p \Jq_{1,a}\,\Uq_b}  + \,\lambda^{(3)}_{\overline{1}} \nor{\Uq_a\,\p \Jq_{1,b}} + \,\lambda^{(3)}_2 \nor{\p \Jq_{2,a}\,\Uq_b}  + \,\lambda^{(3)}_{\overline{2}} \nor{\Uq_a\,\p \Jq_{2,b}}  \right) + 2\,\Wq^{{(\rm corr)}}_4 \,, \notag 
\end{align}
where the $\lambda^{(n)}_{\bullet}$'s are constant coefficients to be determined (and whose notation will be explained in a future discussion of the classical limit -- see below). The first term in this equation is a natural quantisation of the classical expression \eqref{Eq:W4Class}, taking into account the non-commutativity of the quantum fields. It is however not enough to capture the full quantum spin 4 field, which necessitates the introduction of a quantum correction $\Wq^{{(\rm corr)}}_4$. Here we will take the following ansatz for this correction, in terms of the Segal-Sugawara fields $\mathsf{G}_r$ and $\mathsf{G}_{\diag}$:
\begin{align}\label{Eq:W4corr}
\Wq^{{(\rm corr)}}_4 & =\,\rho^{(2)}_{11} \nor{\mathsf{G}_1^2} + \;\rho^{(2)}_{22} \nor{\mathsf{G}_2^2} + \;\rho^{(2)}_{12} \nor{\mathsf{G}_1 \mathsf{G}_2}  + \;\rho^{(2)}_{13} \nor{\mathsf{G}_1 \mathsf{G}_{\diag}} + \;\rho^{(2)}_{31} \nor{\mathsf{G}_{\diag} \mathsf{G}_1} \\
& \hspace{20pt} + \;\rho^{(2)}_{23} \nor{\mathsf{G}_2 \mathsf{G}_{\diag}} + \;\rho^{(2)}_{32} \nor{\mathsf{G}_{\diag} \mathsf{G}_2} + \,\rho^{(1)}_1 \,\p^2 \mathsf{G}_1 + \,\rho^{(1)}_2 \,\p^2 \mathsf{G}_2 \notag
\end{align}
with some coefficients $\rho^{(n)}_{\bullet}$ to be determined. Note that we did not include the fields $\nor{\mathsf{G}_{\diag}^2}$ and $\p^2\mathsf{G}_{\diag}$ in this ansatz, since these terms can always be eliminated by well-chosen shifts by $\nor{\Wq_2^2}$ and $\p^2\Wq_2$, which are the obvious spin 4 currents in the $\Wc$-algebra.\\

To proceed, we impose that the above ansatz for $\Wq_4$ belongs to the quantum $\Wc$-algebra, \textit{i.e.} is such that the OPE $\Jq_{\diag,a}(x)\Wq_4(y)$ is regular. This translates into a system of linear equations on the coefficients $\lambda^{(n)}_\bullet$ and $\rho^{(n)}_\bullet$, which we can then solve\footnote{For completeness, we should mention that there is in fact a redundancy in the ansatz that we took for $\Wq_4$, since certain terms are not independent. For instance, we have
\begin{equation}
\eta^{ab}\,\nor{\p \Jq_{1,a}\,\Uq_b} \; = \eta^{ab}\,\nor{\Uq_a\,\p \Jq_{1,b}} + 2(k_1+2)\p^2\mathsf{G}_1 + 2(k_2+2)\p^2\mathsf{G}_2 - 2(k_1+k_2+2)\p^2\mathsf{G}_{\diag}.
\end{equation}
Here, we use this redundancy to obtain an expression for the coefficients $\lambda^{(n)}_\bullet$ and $\rho^{(n)}_\bullet$ which is as simple and ``symmetric'' as possible.} up to an overall multiplicative factor that we fix here for convenience. Explicitly, we find
\begin{subequations}\label{Eq:Coeff}
\begin{equation}
\lambda^{(2)}_1 = k_2(3 k_2 - 5), \qquad \lambda^{(2)}_2 = k_1(3 k_1 - 5), \qquad \lambda^{(2)}_3 = -2(3 k_1 k_2 - k_1 - k_2 - 2),
\end{equation}
\begin{equation}
\lambda^{(3)}_1 = \lambda^{(3)}_{\overline{1}} = -3 k_2 + 2, \qquad \lambda^{(3)}_2 = \lambda^{(3)}_{\overline{2}} = 3 k_1 - 2, \qquad \lambda^{(4)} = 3,
\end{equation}
while the coefficients $\rho^{(n)}_{\bullet}$ read
\begin{equation}
\rho^{(2)}_{11} = -4(2 k_1 + k_2 + 4), \qquad \rho^{(2)}_{22} = -4(k_1 + 2k_2 + 4), \qquad \rho^{(2)}_{12} = -8(k_1 + k_2 + 2),
\end{equation}
\begin{equation}
\rho^{(2)}_{13} = -2(k_2-2)(k_1+k_2+2), \qquad \rho^{(2)}_{31} = 2(k_2+2)(k_1+k_2+2)
\end{equation}
\begin{equation}
\rho^{(2)}_{23} = -2(k_1-2)(k_1+k_2+2), \qquad \rho^{(2)}_{32} = 2(k_1+2)(k_1+k_2+2),
\end{equation}
\begin{equation}
\rho^{(1)}_1 = k_2 (5k_1+k_2-1), \qquad \rho^{(1)}_2 = k_1 (k_1+5k_2-1).
\end{equation}
\end{subequations}

\paragraph{Classical limit of $\bm{\Wq_4}$.} Now that we have constructed the quantum spin 4 field $\Wq_4$, let us quickly discuss its classical limit. Recall that when $\hbar\to0$, the quantum Kac-Moody currents $\Jq_r$ and levels $k_r$ are of order $O(\hbar^{-1})$ and more precisely obey the asymptotic \eqref{Eq:ClassicalLim}. Reinserting this in equations \eqref{Eq:W4Quant} and \eqref{Eq:W4corr}, we find that the terms in $\Wq_4$ multiplying the coefficients $\lambda^{(n)}_{\bullet}$ and $\rho^{(n)}_\bullet$ are of order $O(\hbar^{-n})$ (for instance, the field $\nor{\p \Jq_{1,a}\,\p \Jq_{1,b}}$, with coefficient $\lambda^{(2)}_1$ in eq.\;\eqref{Eq:W4Quant}, is of order $O(\hbar^{-2})$). The asymptotic behaviour of the coefficients $\lambda^{(n)}_{\bullet}$ and $\rho^{(n)}_\bullet$ themselves can be computed in a similar way from their expressions \eqref{Eq:Coeff} in terms of $k_r$ (for instance, $\lambda^{(2)}_1=k_2(3k_2-5)$ is of order $O(\hbar^{-2})$). Combining all this, we find that the dominant term in $\Wq_4$ is of order $O(\hbar^{-4})$. More precisely, we get
\begin{equation}\label{Eq:ClassLimW4}
\Wq_4(x) = -\frac{96\pi^4\,\ell_1\ell_2(\ell_1+\ell_2) W_4(x) + O(\hbar)}{\hbar^4},
\end{equation}
with $W_4$ the classical spin 4 field \eqref{Eq:W4Class}, ensuring that the latter is indeed recovered in the classical limit of $\Wq_4$. In particular, during this computation, one finds that the term $\Wq^{{(\rm corr)}}_4$, defined in eq.\;\eqref{Eq:W4corr}, is of order $O(\hbar^{-3})$ and is thus subdominant in the classical limit: this confirms the interpretation of this term as a quantum correction.

\paragraph{Primary spin 4 field.}  Using the explicit expression of the quantum spin 4 field $\Wq_4$, we can determine its OPE with the energy-momentum tensor $\Wq_2$. A direct computation gives
\begin{equation}
\Wq_2(x)\Wq_4(y) = \frac{\alpha_1\,c}{(x-y)^6} + \frac{\alpha_2\,\Wq_2(y)}{(x-y)^4} + \frac{\alpha_1\,\p\Wq_2(y)}{(x-y)^3} +  \frac{4\Wq_4(y)}{(x-y)^2} + \frac{\p \Wq_4(y)}{x-y} + \reg \, ,
\end{equation}
where $c$ is the central charge \eqref{Eq:cGKO} and the coefficients $\alpha_1$ and $\alpha_2$ are defined in terms of $k_1$ and $k_2$ as
\begin{subequations}\label{Eq:Alpha12}
\begin{align}
\alpha_1 &= 4 (k_1 + k_2 + 2) (3 k_1 k_2 + 11 k_1 + 11 k_2 - 39) \, , \\
\alpha_2 &= 8\frac{k_1+k_2+2}{(k_1+2)(k_2+2)} \bigl( (3 k_1 k_2 - 58) (k_1 + k_2) + 6 (k_1^2 + k_2^2) - 26 k_1 k_2 - 128 \bigr)\, .
\end{align}
\end{subequations}
In particular, this means that the W-current $\Wq_4$ is not primary. To construct such a primary field, we consider a linear combination
\begin{equation}
\Wq_{4,P}(x) = \Wq_4(x) + \alpha_3\,\nor{\Wq_2(x)^2} + \alpha_4\,\p^2\Wq_2(x).
\end{equation}
This field is automatically in the $\Wc$-algebra, since both $\Wq_4$ and $\Wq_2$ belong to it. We then ask that it is a primary field by imposing the OPE
\begin{equation}
\Wq_2(x)\Wq_{4,P}(y) = \frac{4\Wq_{4,P}(y)}{(x-y)^2} + \frac{\p \Wq_{4,P}(y)}{x-y} + \reg \, .
\end{equation}
This translates into a system of linear equations on $\alpha_3$ and $\alpha_4$, whose solution reads
\begin{equation}\label{Eq:Alpha34}
\alpha_3 = \frac{6\alpha_1-5\alpha_2}{22+5c} \qquad \text{ and } \qquad \alpha_4 = \frac{3\alpha_2-(8+c)\alpha_1}{22+5c}\,.
\end{equation}

\section{Quantum local IMs in affine Gaudin models}
\label{App:LocalIM}

\subsection[Quartic density for $\mathfrak{sl}(2)$]{Quartic density for \texorpdfstring{$\bm{\mathfrak{sl}(2)}$}{sl(2)}}
\label{App:S4}

In this appendix, we describe the spin 4 density $\Sq_4(z,x)$ from which we construct the quantum quartic charges in AGMs based on the Lie algebra $\g^\C=\mathfrak{sl}(2)$. The classical version of this density is given by
\begin{equation}\label{Eq:ClassicalS4}
\Sc_4(z,x) = \tau_3^{abcd} \, \Gamma_a(z,x)\Gamma_b(z,x)\Gamma_c(z,x)\Gamma_d(z,x),
\end{equation}
where $\tau_3^{abcd}$ is a symmetric tensor of degree 4 on $\mathfrak{sl}(2)$. This tensor is unique up to an overall multiplicative constant and can be written in terms of the degree 2 tensor $\eta^{ab}$:
\begin{equation}
\tau_3^{abcd} = \frac{1}{16} \left( \eta^{ab}\eta^{cd} + \eta^{ac}\eta^{bd} + \eta^{ad}\eta^{bd} \right)\,.
\end{equation}
In particular, this implies that the classical spin 4 density \eqref{Eq:ClassicalS4} is simply propotional to $\langle \Gamma(z,x),\Gamma(z,x) \rangle^2$, \textit{i.e.} to the square of the quadratic one.\\

To search for the quantum spin 4 density $\Sq_4(z,x)$, we start with a general ansatz in terms of the quantum Gaudin Lax matrix $\Gq(z,x)$ and the twist function $\vpq(z)$, formed by the normal-ordered quantum version of eq.\;\eqref{Eq:ClassicalS4} plus corrective terms of the form \eqref{Eq:Corr}. To constrain the coefficients of these terms, we then impose that the density $\Sq_4(z,x)$ satisfies the conditions (i) and (ii) stated after eq.\;\eqref{Eq:OpeGamma}. As it turns out, the condition (i) is sufficient to fix $\Sq_4(z,x)$, up to twisted derivatives with respect to $z$ and a total spatial derivative proportional to $\p^2_x\Sq_2(z,x)$, which do not contribute to the local charges $\Qq_{\gamma,3}$ built from $\Sq_4(z,x)$.\footnote{The fact that the resulting density $\Sq_4(z,x)$ also satisfies the condition (ii) was checked explicitly in the recent work \cite{Franzini:2022duf}, which appeared shortly after the first version of this article. We thank T. Franzini and C.~A.~S. Young for interesting discussions on this subject.}  One can use this freedom to simplify the final expression of $\Sq_4$, which we now present. To declutter this formula, we omit the labels $\qt$ on the quantum Gaudin Lax matrix $\Gq(z,x)$ and the quantum twist function $\vpq(z)$. We then get
{\small\begin{align}\label{Eq:QuantS4}
&\Sq_4(z,x) = \tau_3^{abcd} \, \nor{\Gamma_a(z,x)\Gamma_b(z,x)\Gamma_c(z,x)\Gamma_d(z,x)} + \frac{5\ri}{4}\,f^{abc} \nor{ \p_x\Gamma_a(z,x)\p_z\Gamma_b(z,x)\Gamma_c(z,x) } \\
&+ \frac{\eta^{ab}}{48} \Bigl( 45 \,\vp(z)^2 \nor{ \p_x\Gamma_a(z,x)\p_x\Gamma_b(z,x)} -\, 140 \nor{ \p_z^2\p_x\Gamma_a(z,x)\p_x\Gamma_b(z,x)} -\, 30 \nor{ \p_z\p_x\Gamma_a(z,x)\p_z\p_x\Gamma_b(z,x)} \Bigr)  \notag \\
&+ \frac{5\eta^{ab}}{12} \Bigl( 3 \nor{ \p_z\p_x^2\Gamma_a(z,x) \p_z\Gamma_b(z,x)} - \nor{ \p_z^2\p_x^2\Gamma_a(z,x)\Gamma_b(z,x)} \Bigr) \,. \notag
\end{align}}

\subsection{Quadratic charge for 3 punctures}
\label{App:3punct}

In this appendix, following~\cite{Lacroix:2018fhf}, we perform explicitly the integral \eqref{Eq:W2Klim} over the Pocchammer contour $\gamma$, giving the density $\Wq_{\gamma,2}(x)$ of the quadratic charge in the case of an AGM with 3 punctures at $0$, $1$ and $\infty$. Using the expression \eqref{Eq:S2} of $\Sq_2(z,x)$, we rewrite this integral as
\begin{equation}
\Wq_{\gamma,2}(x) = \frac{\eta^{ab}}{2} \oint_\gamma\,\Pc(z)^{-1/\hv} \nor{ \Gq_a(z,x)\Gq_b(z,x) }  \dd z \, .
\end{equation}
Reinserting eq.\;\eqref{Eq:GammaKlim} and \eqref{Eq:PKlim}, \textit{i.e.} $\Gq(z) = \dfrac{\Jq_1}{z} + \dfrac{\Jq_2}{z-1}$ and $\Pc(z) = z^{k_1} (z-1)^{k_2}$, and developing the products, we get
\begin{align}\label{Eq:W2B}
\Wq_{\gamma,2} &= \frac{\eta^{ab}}{2} \left( B\left(-\frac{k_1}{\hv}-2\,,\,-\frac{k_2}{\hv} \right) \nor{ \Jq_{1,a}\Jq_{1,b} } + \, B\left(-\frac{k_1}{\hv}\,,\,-\frac{k_2}{\hv}-2 \right) \nor{ \Jq_{2,a}\Jq_{2,b} } \right. \\
& \hspace{70pt} \left. + \, 2\,B\left(-\frac{k_1}{\hv}-1\,,\,-\frac{k_2}{\hv}-1 \right) \nor{ \Jq_{1,a}\Jq_{2,b} } \right) \, , \notag
\end{align}
where we defined
\begin{equation}
B(a,b) = \oint_\gamma z^a(z-1)^b\,\dd z \,.
\end{equation}
This integral can be expressed in terms of the Euler Beta function. Here we will not need this explicit expression. Instead, we will use some recursive identities obeyed by $B(a,b)$, which follow either from the above integral expression (see e.g.~\cite[Lemma 6.1]{Lacroix:2018itd}) or from the relation of the Beta function to the Gamma function. For any $a$ and $b$, we have
\begin{equation}
B(a-1,b) = \frac{a+b+1}{a} B(a,b) \qquad \text{ and } \qquad B(a,b-1) = - \frac{a+b+1}{b} B(a,b)\, ,
\end{equation}
from which we get
\begin{equation}\label{Eq:IdentityBeta}
\frac{B(a-2,b)}{B(a,b)} = \frac{(a+b+1)(a+b)}{a(a-1)} \qquad \text{ and } \qquad \frac{B(a-1,b-1)}{B(a,b)} = -\frac{(a+b+1)(a+b)}{ab}\,,
\end{equation}
as well as a similar expression for $B(a,b-2)/B(a,b)$. Let us then define
\begin{equation}
N_2 = \frac{(k_1+k_2)(k_1+k_2+\hv)(k_1+k_2-\hv)}{k_1k_2} \oint_\gamma \Pc(z)^{-1/\hv} \dd z \, .
\end{equation}
Noting that the integral in $N_2$ is equal to $B(-k_1/\hv,-k_2/\hv)$ and using the identities \eqref{Eq:IdentityBeta}, we can rewrite eq.\;\eqref{Eq:W2B} as
\begin{equation}
\Wq_{\gamma,2} = \frac{N_2\,\eta^{ab}}{2(k_1+k_2+\hv)} \left( \frac{k_2}{k_1+\hv} \nor{ \Jq_{1,a}\Jq_{1,b} } \, + \, \frac{k_1}{k_2+\hv} \nor{ \Jq_{2,a}\Jq_{2,b} } - \, 2 \nor{ \Jq_{1,a}\Jq_{2,b} } \right)\,.
\end{equation}
Using $\Jq_{\diag}=\Jq_1+\Jq_2$ to rewrite the crossed term, we find
\begin{equation}
\Wq_{\gamma,2} = \frac{N_2\,\eta^{ab}}{2} \left( \frac{1}{k_1+\hv} \nor{ \Jq_{1,a}\Jq_{1,b} } \,+\, \frac{1}{k_2+\hv} \nor{ \Jq_{2,a}\Jq_{2,b} } \,-\, \frac{1}{k_1+k_2+\hv} \nor{ \Jq_{\diag,a}\Jq_{\diag,b} } \right)\,.
\end{equation}
We recognise here the spin 2 W-current defined in eq.\;\eqref{Eq:QuantumW2}, multiplied by $N_2$.

\section{Chiral limits of general relativistic AGMs}
\label{App:ChiralLim}

We have seen in Subsection \ref{Sec:KlimAGM} that the Klim\v{c}\'{i}k model can be identified as a relativistic realisation of an AGM with 4 punctures and in Section \ref{sec555389i} that its conformal limit splits it into two decoupled AGMs with 3 punctures each, corresponding to the left-moving and right-moving fields of the chiral model. The goal of this appendix is to argue that a similar phenomenom should occur for more general relativistic realisations of AGMs, in particular with a higher number of punctures.

\paragraph{Zeroes of the twist function in a relativistic realisation.} Let us then consider an AGM with $2M+2$ punctures (we restrict here to the case of regular punctures, \textit{i.e.} simple poles, for simplicity; however, most of the discussion can be generalised to the irregular case). Its twist function $\vp(z)\dd z$ then possesses $2M$ zeroes $\ze_i$, which we will suppose here are simple. As explained in Subsection \ref{Sec:Relat}, there exists a general scheme to obtain a relativistic 2-dimensional integrable field theory from this AGM by considering a realisation $\rho$ of its Kac-Moody currents in terms of the canonical fields on a cotangent bundle $T^\ast Q$, with $Q$ of dimension $(M+1)\dim\g$. In particular, recall from eq.\;\eqref{Eq:EpsilonRelat} that the relativistic invariance of this realisation requires that the coefficients $s_i$ entering the Hamiltonian \eqref{Eq:HamReal} are equal to either $+1$ or $-1$. Since these coefficients are naturally associated with the zeroes $\ze_i$ of the twist function, in practice, this has for effect to separate these zeroes into two sets, depending on whether the corresponding coefficient $s_i$ is $+1$ or $-1$. For simplicity, we then adapt slightly our notation and now write these two sets of zeroes as\footnote{Recall also from Subsection \ref{Sec:Relat} that these two sets have the same size, which is thus equal to $M$.} $\lbrace \ze_i^+ \rbrace_{i=1}^M$ and $\lbrace \ze_i^- \rbrace_{i=1}^M$. Acting with a M\"obius transformation, we can further assume that $\ze_M^+=\infty$ and $\ze_M^-=0$ without loss of generality. In particular, the setup considered here is then a natural generalisation of the Klim\v{c}\'{i}k model, which corresponds in the present notation to the case $M=1$ (see Subsection \ref{Sec:KlimAGM}).

The separation of the zeroes into the subsets $\lbrace \ze_i^+ \rbrace_{i=1}^M$ and $\lbrace \ze_i^- \rbrace_{i=1}^M$ reflects the light-cone structure of the relativistic model. For instance, the ``light-cone Hamiltonians'' $\Pc_\pm$, which generate the derivatives $\p_\pm = \frac{1}{2}(\p_t\pm\p_x) = \lbrace \Pc_\pm,\cdot \rbrace$, take the form\vspace{-4pt}
\begin{equation}
\Pc_\pm \approx \pm \sum_{i=1}^M \; \res_{z=\ze_i^\pm} \rho\bigl( \Q(z) \bigr)\dd z\,,\vspace{-4pt}
\end{equation}
with $\Q(z)$ defined by eq.\;\eqref{Eq:Qz}. Thus, only the zeroes $\lbrace \ze_i^+ \rbrace_{i=1}^M$, resp. $\lbrace \ze_i^- \rbrace_{i=1}^M$, contribute to $\Pc_+$, resp. $\Pc_-$. Similarly, from the expressions \eqref{Eq:LxG} and \eqref{Eq:LtG} of the spatial and temporal components of the Lax connection in the relativistic realisation, one finds that the light-cone components take the form
\begin{equation}\label{Eq:Lpm}
\Lc_\pm(z) \approx \Bc_\pm + \Kc_{M,\pm}\,z^{\pm 1} + \sum_{i=1}^{M-1} \frac{\Kc_{i,\pm}}{z-\ze_i^\pm}\,,
\end{equation}
for some $\g^\C$-valued fields $\Bc_\pm$ and $\Kc_{i,\pm}$. In particular, we see that the separation of the zeroes into $\lbrace \ze_i^+ \rbrace_{i=1}^M$ and $\lbrace \ze_i^- \rbrace_{i=1}^M$ directly appears in the analytical structure of the light-cone Lax connection as a meromorphic function of the spectral parameter $z$ and more precisely in the position of its poles (note that the zero $\ze_M^+$, being situated at infinity, contributes to a term linear in $z$ in $\Lc_+(z)$ rather than a simple fraction). For future convenience, it will be useful to write down explicitly the expression of the currents $\Kc_{i,\pm}$ appearing in eq.\;\eqref{Eq:Lpm}:
\begin{equation}\label{Eq:Kpm}
 \Kc_{i,\pm} = \pm \frac{\rho\bigl(\Gamma(\ze_i^\pm)\bigr)}{\vp\,'(\ze_i^\pm)}\qquad\qquad ( 1 \leq i \leq M-1 )\;  ,  
\qquad \qquad\quad \Kc_{M,\pm} \approx \pm \lim_{z\to\infty,0}\; \frac{\rho\bigl(\Gamma(z)\bigr)}{z^{\pm 1}\,\vp(z)}\;.
\end{equation}

\paragraph{Chiral limit: colliding zeroes and poles.} The form \eqref{Eq:Lpm} of the light-cone Lax connection is suggestive of a general scheme to obtain chiral limits of the relativistic realisation. Schematically, since only the zeroes $\lbrace \ze_i^\pm \rbrace$ appear in the analytical structure of the component $\Lc_\pm(z)$, one can intuitively expect that in order to decouple the $(\pm)$-chirality in the AGM, one needs to ``eliminate'' these zeroes. In order to implement this idea, we will consider a limit of the parameters of the AGM in which the punctures collide with the zeroes. More precisely, we ask that the limit is such that half of the poles collide with the zeroes associated with the $(-)$-chirality in a certain choice of spectral parameter $z^{\cL}$, which is built from the initial one $z$ by a M\"obius transformation that is degenerate in the limit, while the remaining poles collide with the zeroes associated with the $(+)$-chirality in another choice of spectral parameter $z^{\cR}$.

This is reminiscent of the conformal limit of the Klim\v{c}\'{i}k model discussed in Subsection \ref{Sec:UVLimAGM}. In the general case considered here, a natural scheme to construct such a limit would thus be to follow the RG flow of the relativistic realisation towards an RG fixed-point (if it exists). In practice, this can be difficult to implement precisely as the exact form of this RG flow can in general be quite complicated. Therefore, we will not follow this approach here and instead will build explicitly a simple flow of the parameters that fits the characteristics described above and which thus sends the classical theory to a chiral one. Although it will not coincide with the RG flow, we expect that it has a similar asymptotic behaviour (at least for well-chosen initial conditions of the parameters), so that the end point of this flow will coincide with a RG fixed-point. We will come back to the RG flow of the theory and its comparison with the chiral limit considered here at the end of this appendix.\\

To define the limit we first split the punctures of the AGM into two subsets, which we will denote $\lbrace z_r^+ \rbrace_{r=1}^{M+1}$ and $\lbrace z_r^- \rbrace_{r=1}^{M+1}$. We then consider the following parametrisation of the punctures $z_r^\pm$ and zeroes $\ze_i^\pm$:
\begin{equation}
z_r^+ = \frac{z_r^{\cL}}{\xi}\,, \;\qquad z_r^- = \frac{\xi}{z_r^{\cR}}\,, \; \qquad \zeta_i^+ = \frac{\zeta_i^{\cL}}{\xi}\,, \; \qquad \ze_i^- = \frac{\xi}{\ze_i^{\cR}}\,,
\end{equation}
in terms of a real positive parameter $\xi$ and $4M$ points $z_1^{({\rm L}/{\rm R})},\ldots,z_{M+1}^{({\rm L}/{\rm R})}$ and $\ze_1^{({\rm L}/{\rm R})},\ldots,\ze_{M-1}^{({\rm L}/{\rm R})}$ which we take to be all distinct, finite and different from 0. We then make the parameter $\xi$ flow to $0$, while keeping these points fixed (more generally, one could consider a limit where the latter are not fixed but simply stay finite and pairwise distinct). It is clear that in the limit $\xi\to 0$, the punctures and zeroes all collide either at $z=\infty$ or at $z=0$. Moreover, we can refine this collision phenomenon by ``zooming-in'' on $0$ or $\infty$ while taking the limit. For instance, working with the rescaled spectral parameter $z^{\cL} = \xi\, z$, the points $\lbrace \xi\, z_r^- \rbrace_{r=1}^{M+1}$ and $\lbrace \xi\, \ze_i^- \rbrace_{i=1}^{M}$ still all collide together at $0$ while the remaining points $\lbrace \xi\, z_r^+ \rbrace_{r=1}^{M+1}$ and $\lbrace \xi\, \ze_i^+ \rbrace_{i=1}^{M}$ stay pairwise distinct. Similarly, for another choice of rescaled spectral parameter $z^{\cR} = \xi/z$, the points $\lbrace \xi / z_r^- \rbrace_{r=1}^{M+1}$ and $\lbrace \xi/ \ze_i^- \rbrace_{i=1}^{M}$ stay pairwise distinct while $\lbrace \xi/ z_r^+ \rbrace_{r=1}^{M+1}$ and $\lbrace \xi/ \ze_i^+ \rbrace_{i=1}^{M}$ all collide. This is very similar to the Klim\v{c}\'{i}k model setup discussed in Subsection \ref{Sec:UVLimAGM}, which we recall corresponds here to $M=1$.

\paragraph{Limits of the twist function.} Let us now consider the limit of the twist function $\vp(z)$ in the rescaled spectral parameters $z^{\cL}$ and $z^{\cR}$. For that, it will be useful to first write $\vp(z)$ in the following way:
\begin{equation}\label{Eq:TwistBeforeLim}
\vp(z) = \frac{K}{\xi^2}\; \frac{z \prod_{i=1}^{M-1} (z-\ze_i^+)(z-\ze_i^-)}{\prod_{r=1}^{M+1} (z-z_r^+)(z-z_r^-)}\,.
\end{equation}
Here, we chose to parametrise the overall coefficient as $K/\xi^2$ to facilitate the limit: indeed, in order to obtain the expected behaviour when $\xi\to 0$, we will need to keep the parameter $K$ fixed, or at least finite and non-zero (this can also be expected from the example of the Klim\v{c}\'{i}k model). Keeping this in mind, let us then take the limit explicitly, focusing first on the spectral parameter $z^{\cL}=\xi\,z$. For this choice of coordinate on $\mathbb{CP}^1$, the twist function before the limit is given by $\xi^{-1} \vp\bigl( \xi^{-1}z^{\cL})$. A direct computation then yields
\begin{equation}
\vp^{\cL}\bigl( z^\cL \bigr) = \lim_{\xi\to\, 0}\, \frac{1}{\xi} \;\vp\left( \frac{z^\cL}{\xi} \right) = K \frac{\prod_{i=1}^{M-1} \bigl(z^\cL - \ze_i^\cL\bigr)}{z^\cL\,\prod_{r=1}^{M+1} \bigl(z^\cL - z_r^\cL\bigr)}\;.
\end{equation}
This corresponds to the twist function of an AGM with $M+2$ punctures $\lbrace z_r^\cL \rbrace_{r=1}^{M+2}$, where we defined $z_{M+2}^\cL=0$. A similar analysis can be performed for the spectral parameter $z^\cR=\xi/z$. Explicitly, we find
\begin{equation*}
\vp^{\cR}\bigl( z^\cR \bigr) = \lim_{\xi\to\, 0}\, -\frac{\xi}{z^{\cR\,2}}\; \vp\left( \frac{\xi}{z^\cR} \right) = -\widetilde{K} \frac{\prod_{i=1}^{M-1} \bigl(z^\cR - \ze_i^\cR\bigr)}{z^\cR\,\prod_{r=1}^{M+1} \bigl(z^\cR - z_r^\cR\bigr)}\;, \quad \widetilde{K} = K \frac{\prod_{i=1}^{M-1} \ze_i^\cL/\ze_i^\cR}{\prod_{r=1}^{M+1} z_r^\cL/z_r^\cR}\,.
\end{equation*}
This also corresponds to the twist function of an AGM with $M+2$ punctures.

Our main goal in the rest of this appendix will be to show that the theory obtained when $\xi\to0$ is chiral, in the sense that it  splits into decoupled left-moving and right-moving sectors, and that the latter are chiral realisations of two AGMs with respective twist functions $\vp^{\cL}\bigl( z^\cL \bigr)$ and $\vp^{\cR}\bigl( z^\cR \bigr)$.

\paragraph{Levels.} It will be useful to study the levels of the model and their behaviour in the limit. In the original relativistic AGM, we have $2M+2$ such levels $\ell_r^\pm$, defined as the residues of $\vp(z)\dd z$ at $z=z_r^\pm$, for $r\in\lbrace 1,\ldots,M+1\rbrace$. From this definition and the expression \eqref{Eq:TwistBeforeLim} of $\vp(z)$, one easily expresses these levels in terms of the parameters $K$, $\xi$, $z_r^{\cLR}$ and $\ze_i^{\cLR}$. Making the parameter $\xi$ flow to $0$ as in the previous paragraph, we find that the levels $\ell_r^\pm$ do not stay fixed in this process, but importantly converge to a finite and non-zero value. In fact, their limits coincide with the residues of the functions $\vp^\cL\bigl(z^\cL\bigr)$ and $\vp^\cR\bigl(z^\cR\bigr)$ obtained above. More precisely, for $r\in\lbrace 1,\ldots,M+1 \rbrace$, we have
\begin{equation}
\res_{z^\cL=z^\cL_r}\; \vp^\cL \bigl( z^\cL \bigr)\, \dd z^\cL = \lim_{\xi\to\,0}\, \ell_r^+ \qquad \text{ and } \qquad \res_{z^\cR=z^\cR_r}\; \vp^\cR \bigl( z^\cR \bigr)\, \dd z^\cR = \lim_{\xi\to\,0}\, \ell_r^-\,.
\end{equation}
We thus expect the limiting values of $\ell_r^+$ and $\ell_r^-$ to define the levels of the left-moving and right-moving chiral AGMs describing the theory at $\xi = 0$.

\paragraph{Limit of the Lax connection.} Let us now discuss the limit of the Lax connection of the theory. For the initial relativistic model, it is given by eq.\;\eqref{Eq:Lpm}. In order to take the limit of this quantity, we will need to study the behaviour of the currents $\Kc_{i,\pm}$ when $\xi\to 0$, using their expression \eqref{Eq:Kpm}. The latter uses the Gaudin Lax matrix $\rho\bigl(\Gamma(z)\bigr)$ in the realisation, which is defined as
\begin{equation}
\rho\bigl(\Gamma(z)\bigr) = \sum_{r=1}^{M+2} \left( \frac{\Jc_{r,+}^\rho}{z-z_r^+} + \frac{\Jc_{r,-}^\rho}{z-z_r^-} \right)\;,
\end{equation}
in terms of the realised Kac-Moody currents $\Jc^\rho_{r,\pm}$. In order to proceed further, we will have to suppose that these currents have a finite and non-zero limit when $\xi\to 0$. This assumption would require a more in-depth analysis but appears sensible thanks to the observation made in the previous paragraph that the levels $\ell_r^\pm$ of these currents stay finite and non-zero in the limit\footnote{Based on the experience of the Klim\v{c}\'{i}k model, we know however that this question can in general be a subtle one. Indeed, in this example, we could consider different ways of parametrising the coordinates and conjugate momenta of the realisation in terms of the parameters of the AGM, giving rise to different limits of the Kac-Moody currents (sometimes requiring fine-tuned conjugations), corresponding to different charts of the target space in the limit. We will not enter into these considerations here.}. Using this hypothesis to study the behaviour of the Gaudin Lax matrix in the limit $\xi\to0$ and combining these results with an analysis of the term $\vp'(\ze_i^\pm)$ in eq.\;\eqref{Eq:Kpm}, a slightly lengthy but straightforward computation shows that the currents $\Kc_{i,+}$ and $\Kc_{i,-}$, $i\in\lbrace 1,\ldots,M-1\rbrace$, respectively diverge and tend to zero in the limit. More precisely, we find
\begin{equation}
\Kc_{i,+} = \frac{1}{\xi} \Bigl( \Kc^\cL_{i} + O( \xi^2 ) \Bigr) \qquad \text{ and } \qquad \Kc_{i,-} = - \frac{\xi}{\ze_i^{\cL\,2}} \Bigl( \Kc^\cR_{i} + O( \xi^2 ) \Bigr)\;,
\end{equation}
for some non-vanishing fields $\Kc^{({\rm L}/{\rm R})}_{i}$, expressed in terms of the limit of the Kac-Moody currents and the parameters $\bigl(K,z_r^\cLR,\ze_i^\cLR\bigr)$ (the sign and prefactor in the second equation have been introduced for further convenience). Similarly, one can study the limit of the currents $\Kc_{M,\pm}$ from the second part of eq.\;\eqref{Eq:Kpm} (in particular the behaviour of the current $\Kc_{M,+}$ is different from the one of the other currents $\Kc_{i,+}$ since it is associated with a zero at infinity). In the end, we find two fields $\Kc^{({\rm L}/{\rm R})}_{M}$ such that
\begin{equation}
\Kc_{M,+} = \xi \Bigl( \Kc^\cL_{M} + O( \xi^2 ) \Bigr) \qquad \text{ and } \qquad \Kc_{M,-} = \xi \Bigl( \Kc^\cR_{M} + O( \xi^2 ) \Bigr)\;.
\end{equation}

Using the above results on the asymptotic behaviour of the currents $\Kc_{\pm,i}$ when $\xi\to 0$, it is then easy to compute the limit of the Lax connection \eqref{Eq:Lpm} in the rescaled spectral parameters $z^\cL = \xi\,z$ and $z^\cR = \xi/z$. Namely,
defining
\begin{equation}
\Lc_\pm^\cL \bigl( z^\cL \bigr) = \lim_{\xi\to\,0}\; \Lc_\pm \left( \frac{z^\cL}{\xi} \right) \qquad \text{ and } \qquad \Lc_\pm^\cR \bigl( z^\cR \bigr) = \lim_{\xi\to\,0}\; \Lc_\pm \left( \frac{\xi}{z^\cR} \right)\;, 
\end{equation}
we get, for some fields $\Bc_\pm^{\cLR}$,
\begin{subequations}
\begin{equation}\label{Eq:LzL}
\Lc_+^\cL \bigl( z^\cL \bigr) \approx \Bc^{\cL}_+ + \Kc_M^\cL\,z^\cL + \sum_{i=1}^{M-1} \frac{\Kc_i^\cL}{z^\cL-\ze_i^\cL}\;, \qquad\quad \Lc_-^\cL \bigl( z^\cL \bigr) \approx \Bc^{\cL}_-\,,
\end{equation}
\begin{equation}\label{Eq:LzR}
\Lc_+^\cR \bigl( z^\cR \bigr) \approx \Bc^{\cR}_+\,, \qquad\quad \Lc_-^\cR \bigl( z^\cR \bigr) \approx \Bc^{\cR}_- + \Kc_M^\cR\,z^\cR + \sum_{i=1}^{M-1} \frac{\Kc_i^\cR}{z^\cR-\ze_i^\cR}\;.
\end{equation}
\end{subequations}

\paragraph{Chirality.} Let us now analyse the properties of the Lax connection $\Lc_\pm^\cL(z^\cL)$ obtained in the limit $\xi\to 0$, using the spectral parameter $z^\cL$. As can be seen in eq.\;\eqref{Eq:LzL}, its main characteristic is the fact that the ``$-$'' component $\Lc_-^\cL(z^\cL)$ becomes independent of the spectral parameter in the limit. This is reminiscent of the situation described in Sections \ref{sec31} and \ref{sec555389i} for the conformal limit of the Klim\v{c}\'{i}k model. As discussed there, the main consequence of this observation is that there exists a gauge transformation $\p_\pm+\Lc^\cL_\pm \mapsto h(\p_\pm+\Lc^\cL_\pm)h^{-1}$ (independent of the spectral parameter) which eliminates the constant term $\Bc^{\cL}_-$ in the component $\Lc_-^\cL(z^\cL)$. In this gauge, the zero curvature equation then simply states that the ``$+$'' component of the Lax connection is a left-moving field. Thus, there exists $h$ such that the gauge-transformed currents $\Bc^{\cL\,h}_+$ and $\Kc_{i}^{\cL\,h}$ are left-moving.

Of course, a similar analysis shows that one can eliminate the ``$+$'' component of the connection $\Lc_\pm^\cR(z^\cR)$ by a gauge transformation $\p_\pm+\Lc^\cR_\pm \mapsto h'\,(\p_\pm+\Lc^\cR_\pm)h'\null^{-1}$. As a result, the gauge-transformed currents $\Bc^{\cR\,h'}_+$ and $\Kc_{i}^{\cR\,h'}$ are right-moving fields. We thus showed that the theory obtained when $\xi\to0$ splits into a left-moving sector and a right-moving one, as expected. Based on this, we will now refer to $\xi\to 0$ as a chiral limit.

\paragraph{Chiral realisations.} In the initial relativistic model, the Lax matrix satisfies a $r\big/s$ bracket with twist function $\vp(z)$, characteristic of the underlying AGM structure. Applying the limit $\xi\to 0$ to this bracket, we find that the limit of the Lax matrix in the spectral parameter $z^\cL$ satisfies a  $r\big/s$  bracket with twist function $\vp^\cL(z^\cL)$. This means that the Lax matrix $\Lc^\cL(z^\cL)$ arises from a realisation of the AGM with this twist function. Moreover, the discussion of the previous paragraph shows that the fields entering this realisation are left-moving, up to a well-chosen gauge transformation. This is the main property entering the condition (LM2) in the definition of a left-moving chiral realisation of an AGM in Subsection \ref{Sec:UVLimAGM}. We thus proved that the limit $\xi\to0$ produces such a left-moving chiral realisation.

Similarly, one can argue that the Lax matrix $\Lc^\cR(z^\cR)$ obtained in the spectral parameter $z^\cR$ arises from a right-moving chiral realisation of the AGM with twist function $\vp^\cR(z^\cR)$. Together, they form a chiral pair of realisations, as defined in Subsection \ref{Sec:UVLimAGM}. In particular, through a careful analysis using similar computations to the one performed earlier in this appendix, one can show that these two realisations indeed satisfy the technical conditions (CP1) and (CP2) entering the definition of a chiral pair\footnote{For the proof of the property (CP2), one has to use the fact that the initial relativistic realisation we are starting with contains all the canonical fields of the theory up to initial values, as assumed in the condition (R2) defining a relativistic realisation -- see Subsection \ref{Sec:Relat}.}.

\paragraph{Sigma models realisations.} So far, we have only used the general properties of the relativistic AGM to discuss its chiral limit, independently of the concrete choice of realisation. Let us then comment on this aspect and the form of the integrable 2-dimensional field theories that we obtain by this construction. We will focus on realisations in terms of canonical fields in $T^\ast Q$ that give rise to a sigma model, corresponding to Kac-Moody currents linear in the momentum fields and the spatial derivative of the coordinate fields (see Subsection \ref{Sec:Relat} for more details). The property (R2) in the definition of a relativistic realisation requires that there are as many degrees of freedom in the realisation that there are in the Kac-Moody currents of the AGM, so that $\dim\,T^\ast Q = 2(M+1)\dim\g$, since the AGM possesses $2(M+1)$ punctures. This describes the unreduced degrees of freedom of the theory: one further needs to impose the constraint and quotient by the corresponding gauge symmetry, effectively eliminating $2\dim\g$ Hamiltonian fields. The physical degrees of freedom of the theory can then be seen as canonical fields on $T^\ast Q_{\red}$, where $Q_{\red}=Q/G$ is of dimension $\dim\,Q_{\red}=M\dim\g$. The space $Q_{\red}$ then plays the role of the target space of the integrable sigma model obtained through the realisation. In the chiral limit discussed above, the AGM underlying this theory splits into a pair of chiral realisations, each possessing $M+2$ punctures. After reduction, these chiral AGMs are then described by $M\dim \g$ Hamiltonian fields each, \textit{i.e.} half of the dimension of $T^\ast Q_{\red}$: this is exactly what we expect, since these two chiral AGMs describe the left-moving and right-moving halves of the theory in the chiral limit.

In principle, there can exist many different realisations of the Kac-Moody currents, which would give rise to various types of target spaces $Q_{\red}$. For concreteness, let us discuss one possible class of interesting examples that naturally generalise the Klim\v{c}\'{i}k model. In the notations of this appendix, these are built by realising separately the $M+1$ pairs of Kac-Moody currents $\bigl(\Jc_{r,+},\Jc_{r,-}\bigr)$ associated with the punctures $(z_r^+,z_r^-)$. More precisely, we will express each of these pairs through a so-called Yang-Baxter realisation, in terms of canonical fields in a cotangent bundle $T^\ast G$, where $G$ is a connected Lie group with Lie algebra $\g$. These realisations are expressed using the Drinfel'd-Jimbo $R$-matrix on $\g$ and coincide with the ones considered in eq.\;\eqref{Eq:CurrentsKlim} for the Klim\v{c}\'{i}k model when the levels $(\ell_r^+,\ell_r^-)$ attached to the punctures are opposite. When these levels are generic, one needs to consider a deformation of this realisation~\cite{Delduc:2014uaa} which involves a field related to the Wess-Zumino term on $G$. Through this construction, we then naturally obtain a realisation in $\Fc[T^\ast Q]$ with $Q=G^{M+1}$. Moreover, the form of the Yang-Baxter realisations is such that the gauge symmetry of the AGM simply takes the form of the diagonal right multiplication on $Q$, so that the physical target space of the sigma model is $Q_{\red} = G^{M+1} / G_{\diag} \simeq G^M$. The theory then describes $M$ coupled $G$-valued fields each possessing a Wess-Zumino term, with non-trivial interactions between the copies and subject to $M+1$ deformations of Yang-Baxter type controlled by the $R$-matrix. This class then consists of coupled multi-copy versions of the Klim\v{c}\'{i}k model. We refer to~\cite{Lacroix:2019xeh,Bassi:2019aaf} for an explicit description of some of these coupled deformed sigma models.

Let us finally mention an interesting subclass of the AGMs considered here, which is defined through a certain symmetry property under the inversion of the spectral parameter $z\mapsto z^{-1}$. More precisely, we will require that the 1-form $\vp(z)\dd z$ is odd under this transformation. In the notations of the appendix, this is achieved by requiring that the poles $z_r^\pm$ and the zeroes $\ze_i^\pm$ of $\vp(z)\dd z$ satisfy $z_r^- = 1/z_r^+$ and $\ze_i^- = 1/\ze_i^+$, or equivalently that $z_r^\cL=z_r^\cR$ and $\ze_i^\cL=\ze_i^\cR$. In particular, this means that the transformation $z\mapsto z^{-1}$ sends the zeroes $\lbrace \ze_i^+ \rbrace_{i=1}^M$ associated with the $(+)$--chirality of the model to the zeroes $\lbrace \ze_i^- \rbrace_{i=1}^M$ associated with the $(-)$--chirality. This transformation then encodes a certain parity symmetry $x\mapsto -x$ of the theory, which exchanges these two chiralities. Another effect of the oddness of $\vp(z)\dd z$ under $z\mapsto z^{-1}$ is that the levels $(\ell_r^+,\ell_r^-)$ are ensured to be opposite one to another. In the class of Yang-Baxter deformed coupled sigma models considered in the previous paragraph, this translates to the absence of Wess-Zumino terms for each of the $G$-valued fields in the theory.

\paragraph{RG flow and conformality.} So far, we have discussed only classical aspects of the chiral limit. However, we expect it to be intimately related to the RG flow of the model and its properties of quantum conformal invariance. Let us finally comment on these aspects. Recently, a conjecture has been proposed in~\cite{Delduc:2020vxy} for a universal formula describing the 1-loop RG flow of relativistic realisations of AGMs in terms of their twist function $\vp(z)$. This conjecture was initially spelled out for AGMs without gauge symmetries, corresponding to the presence of a double pole at infinity in $\vp(z)\dd z$. It has been checked on various examples~\cite{Delduc:2020vxy} and proven in the case of a twist function with simple poles in the complex plane~\cite{Hassler:2020xyj}. There exists a natural generalisation of this conjecture for the case of AGMs with gauge symmetries, which we present here. To phrase it, it will be useful to first describe the inverse of the twist function. By construction, it has poles at the points $\ze_i^\pm$. A simple analysis of its analytical structure then shows that it satisfies the following partial fraction decomposition:
\begin{equation}\label{Eq:PhiInv}
\frac{1}{\vp(z)} = \frac{\xi^2\,z^3}{K} + \frac{\xi^2}{\widetilde{K}\,z} + \sum_{i=1}^{M-1} \left( \frac{1}{\vp\,'(\ze_i^+)} \frac{1}{z-\ze_i^+} + \frac{1}{\vp\,'(\ze_i^-)} \frac{1}{z-\ze_i^-} \right) + a_0 + a_1 z + a_2 z^2,
\end{equation}
for some coefficients $(a_0,a_1,a_2)$ whose explicit expression we will not need. Note that the second term corresponds to the contribution of the zero $\ze_M^-=0$ (in particular, $\vp'(0)=\widetilde{K}/\xi^2$) and the first term to the contribution of the zero $\ze_M^+=\infty$.

We now define a closely related, although slightly different, function $f(z)$:
\begin{equation}\label{Eq:f}
f(z) = \frac{\xi^2\,z^3}{K} - \frac{\xi^2}{\widetilde{K}\,z} + \sum_{i=1}^{M-1} \left( \frac{1}{\vp\,'(\ze_i^+)} \frac{1}{z-\ze_i^+} - \frac{1}{\vp\,'(\ze_i^-)} \frac{1}{z-\ze_i^-} \right) + b_0 + b_1 z + b_2 z^2,
\end{equation}
where, compared to eq.\;\eqref{Eq:PhiInv}, we have changed the sign of the contribution of $\ze_i^-$ and the coefficients of the polynomial terms of degree up to 2. The conjectured 1-loop RG flow of the relativistic realisations of the AGM with twist function $\vp(z)$ then simply reads
\begin{equation}\label{Eq:ConjRG}
\frac{\dd\;}{\dd\tau}\vp(z) = \hbar\hv \frac{\dd\;}{\dd z}\Bigl( f(z)\vp(z) \Bigr) + O(\hbar^2),
\end{equation}
where $\tau=-\frac{1}{2\pi}\,\log(\mu)$ is the RG time, as in Subsection \ref{Sec:RG}. We note that the right-hand side of the RG flow \eqref{Eq:ConjRG} formally takes the form of the variation of $\vp(z)$ under an infinitesimal transformation $z \mapsto z +\hbar\hv\,f(z)\dd\tau$ of the spectral parameter. This can seem surprising at first: indeed, the AGM is invariant under biholomorphic changes of the spectral parameter (see Subsection \ref{Sec:AGM}), and one could thus naively conclude that the above  RG flow leaves the AGM invariant. This is in fact not the case, as one should expect, since $z \mapsto z +\hbar\hv\,f(z)\dd\tau$ is not the infinitesimal version of a biholomorphic transformation on $\mathbb{CP}^1$, \textit{i.e.} of a M\"{o}bius transformation, due to the poles of $f(z)$ at $z=\ze^\pm_i$. The RG flow \eqref{Eq:ConjRG} thus does not integrate to a global invertible transformation $z \mapsto \zt(\tau)$ and hence does not leave the model invariant. We note however that the infinitesimal variations corresponding to M\"{o}bius transformations take the form $\delta z=\epsilon_0 +\epsilon_1 z + \epsilon_2 z^2$: the polynomial term $b_0 + b_1 z + b_2 z^2$ in the definition \eqref{Eq:f} of $f(z)$ then contributes only to a genuine biholomorphism of $z$ in the RG flow \eqref{Eq:ConjRG} and is therefore inconsequential\footnote{The corresponding coefficients $(b_0,b_1,b_2)$ are thus in general arbitrary, except if one has fixed the freedom of performing M\"{o}bius transformations, for instance by sending some of the points $(z_r^\pm,\ze_i^\pm)$ to specific positions. We are in the latter situation here, as we have imposed $(\ze_M^+,\ze_M^-)=(\infty,0)$. In this case, the parameters $(b_0,b_1,b_2)$ are then non-arbitrary and take specific values which ensure that the RG flow \eqref{Eq:ConjRG} preserves this choice.}.\\

Let us now discuss the behaviour of the RG flow \eqref{Eq:ConjRG} in the chiral limit considered in this appendix. For that we will first consider the model in the rescaled spectral parameter $z^\cL=\xi\,z$: we can then apply the RG flow formula \eqref{Eq:ConjRG} for the corresponding twist function $\xi^{-1}\vp\bigl(\xi^{-1}z^\cL\bigr)$, using an equivalent of the function $f(z)$ for this choice of spectral parameter. We will denote by $\vp^\cL\bigl( z^\cL \bigr)$ and $f^\cL\bigl(z^\cL\bigr)$ the limit of these functions when $\xi\to 0$. As explained earlier, the main property of the chiral limit in the coordinate $z^\cL$ is that it eliminates the zeroes $\lbrace \xi\,\ze_i^- \rbrace_{i=1}^{M+1}$ by making them collide with poles. As a result, the contribution of these zeroes to $f^\cL\bigl(z^\cL\bigr)$ disappear in the limit and this function then coincides with the inverse of $\vp^{\cL}\bigl(z^\cL\bigr)$, up to polynomial terms of degrees up to 2, as can be seen from the formulas \eqref{Eq:PhiInv} and \eqref{Eq:f} translated to the spectral parameter $z^\cL$. In other words, we have
\begin{equation}
f^\cL\bigl(z^\cL\bigr) = \vp^\cL\bigl(z^\cL\bigr)^{-1} + a_{L,0}\,z^\cL + a_{L,1}\,z^\cL + a_{L,2}\,\big(z^{\cL}\big)^2
\end{equation}
for some coefficients $(a_{L,0},a_{L,1},a_{L,2})$. Reinserting this is in the RG equation \eqref{Eq:ConjRG} for $\xi\to0$, we find that the term $\vp^\cL\bigl(z^\cL\bigr)^{-1}$ in $f^\cL\bigl(z^\cL\bigr)$ does not contribute to the flow. Since the remaining polynomial term corresponds simply to a M\"{o}bius transformation of $z^\cL$, we then get that the twist function $\vp^\cL\bigl(z^\cL\bigr)$ in the limit has a trivial RG flow and thus that the physical parameters of the left-moving chiral AGM do not run under the flow (at least at 1-loop).

One can apply a similar reasoning in the spectral parameter $z^\cR = \xi/z$, hence proving that the physical parameters of the right-moving chiral AGM also have a vanishing 1-loop $\beta$-function. Together, the pair of chiral AGMs encode all the parameters of the full 2-dimensional theory obtained in the limit and we thus conclude that the latter defines a conformal theory.\\

Although we have argued that the theory in the limit $\xi\to0$ is conformal, \textit{i.e.} defines a fixed-point of the RG, let us recall that the flow in the parameter space that we considered above will in general not coincide exactly with the RG flow itself. Indeed, we have taken here the limit $\xi\to 0$ while keeping the other parameters $\bigl(K,z_r^\cLR,\ze_i^\cLR\bigr)$ fixed. In general, we expect that the RG flow of the theory, at least for well-chosen initial conditions of the parameters, will also send the parameter $\xi$ to zero in the UV, but keeping $\bigl(K,z_r^\cLR,\ze_i^\cLR\bigr)$ finite and non-zero rather than fixed. Let us note moreover that a striking property of the RG equation \eqref{Eq:ConjRG} is that the levels $\ell_r^\pm$ of the model, which are complicated combinations of $\bigl(K,\xi,z_r^\cLR,\ze_i^\cLR\bigr)$, stay fixed under the RG flow. As mentioned earlier, this is not the case for the limit $\xi\to0$: however, the latter was such that $\ell_r^+$ and $\ell_r^-$ converge to finite non-zero values which coincide with the levels of the left-moving and right-moving chiral AGMs respectively. The asymptotic behaviour of the levels is thus similar to the one expected in the UV limit.

More generally, we have mentioned at a few points in this appendix that a limit in which the parameters $\bigl(K,z_r^\cLR,\ze_i^\cLR\bigr)$ stay finite and non-zero rather than fixed would have a qualitative behaviour very similar to the one we considered and that most of the computations would readily apply to this case as well. In other words, we expect this chiral limit to have a similar asymptotic behaviour to the UV limit of the theory (for well-chosen initial conditions). This is motivated by the example of the Klim\v{c}\'{i}k model for $M=1$, which indeed behaves in such a manner. Moreover, we have performed some numerical checks using the conjectured RG flow \eqref{Eq:ConjRG} for a higher number of punctures, namely for $M=2$, showing that this expectation is indeed true when one starts with initial parameters in a well-chosen regime. We thus think that the chiral realisations in these conformal limits can be studied and quantised using techniques similar to the ones discussed in this article for the case of the Klim\v{c}\'{i}k model.

\providecommand{\href}[2]{#2}\begingroup\raggedright\endgroup

\end{document}